\documentclass[a4paper,10pt,oneside,onecolumn,1p,times]{elsarticle}

\newcommand{\ham}{\hat{H}}                          % Hamiltonian 
\newcommand{\Tr}{{\rm Tr}}
\newcommand{\tr}{{\rm Tr }}
\newcommand{\diff}{{\rm d}}
\newcommand{\eul}{{\rm e }}
\newcommand{\imag}{{\rm i }}
\newcommand{\obs}{{\rm obs}}
\newcommand{\fit}{{\rm fit}}
\newcommand{\bra}[1]{\mbox{$\langle #1  |$}} 
\newcommand{\ket}[1]{\mbox{$| #1  \rangle$}} 

\newcommand{\braket}[2]{\mbox{$\langle #1  | #2 \rangle$}} 	% <x|y>
\newcommand{\fat}[1]{\mbox{\boldmath$#1$\unboldmath}}	% fat also for Greek
\newcommand{\syslength}{L}
\newcommand{\Eq}[1]{Eq.~(\ref{#1})}

\begin{document}
\title{The density-matrix renormalization group in the age of matrix product states}
\author{Ulrich Schollw\"{o}ck}
\ead{schollwoeck@lmu.de}
\address{Department of Physics, Arnold Sommerfeld Center for Theoretical Physics and Center for NanoScience, University of Munich, Theresienstrasse 37, 80333 Munich, Germany \\
Institute for Advanced Study Berlin, Wallotstrasse 19, 14159 Berlin, Germany}

\begin{abstract}
The density-matrix renormalization group method (DMRG) has established itself over the last decade as the leading method for the simulation of the statics and dynamics of one-dimensional strongly correlated quantum lattice systems. In the further development of the method, the realization that DMRG operates on a highly interesting class of quantum states, so-called matrix product states (MPS), has allowed a much deeper understanding of the inner structure of the DMRG method, its further potential and its limitations. In this paper, I want to give a detailed exposition of current DMRG thinking in the MPS language in order to make the advisable implementation of the family of DMRG algorithms in exclusively MPS terms transparent. I then move on to discuss some directions of potentially fruitful further algorithmic development: while DMRG is a very mature method by now, I still see potential for further improvements, as exemplified by a number of recently introduced algorithms.   
\end{abstract}

\maketitle 

\tableofcontents

\section{Introduction}
Strongly correlated quantum systems on low-dimensional lattices continue to pose some of the most interesting challenges of modern quantum many-body physics. In condensed matter physics,  correlation effects dominate in quantum spin chains and ladders, in frustrated magnets in one and two spatial dimensions, and in high-temperature superconductors, to name but a few physical systems of interest. More recently, the advent of highly controlled and tunable strongly interacting ultracold atom gases in optical lattices has added an entirely new direction to this field\cite{Bloch08}. 

Both analytically and numerically, these systems are hard to study: only in very few cases exact analytical solutions, for example by the Bethe ansatz in one dimension, are available\cite{Bethe31,Lieb68}. Perturbation theory fails in the presence of strong interactions. Other approaches, such as field theoretical approaches, have given deep insights, for example regarding the Haldane gap physics of integer-spin antiferromagnetic chains\cite{Haldane83}, but make potentially severe approximations that must ultimately be controlled by numerical methods. Such algorithms include exact diagonalization, quantum Monte Carlo, series expansions or coupled cluster methods.

Since its invention in 1992 by Steve White\cite{White92,White93}, the density-matrix renormalization group (DMRG) has firmly established itself as the currently most powerful numerical method
in the study of one-dimensional quantum lattices\cite{Schollwoeck05,Hallberg06}. After initial studies of the static properties (energy, order parameters, $n$-point correlation functions) of low-lying eigenstates, in particular ground states, of strongly correlated Hamiltonians such as the Heisenberg, $t$-$J$ and Hubbard models, the method was extended to the study of 
dynamic properties of eigenstates, such as dynamical structure functions or frequency-dependent conductivities\cite{Hallberg95,Ramasesha97,Kuhner99,Jeckelmann02}. At the same time, its extension to the analysis of two-dimensional classical \cite{Nishino95} and one-dimensional quantum \cite{Wang97,Shibata97} transfer matrices has given access to highly precise finite-temperature information on classical two-dimensional and quantum one-dimensional systems;  more recently the transfer matrix variant of DMRG has also been extended to dynamics at finite temperature\cite{Sirker05}. It has even been extended to the numerically much more demanding study of non-Hermitian (pseudo-) Hamiltonians emerging in the analysis of the relaxation towards classical steady states in one-dimensional systems far from equilibrium\cite{Hieida98,Carlon99,Carlon01,Henkel01}. 

In many applications of DMRG, the accuracy of results is essentially limited only by machine precision, even for modest numerical resources used, quite independent of the detailed nature of the Hamiltonian. It is therefore not surprising that, beyond the extension of the algorithm to more and more problem classes, people wondered about the physical origins of the excellent performance of DMRG and also whether the success story could be extended to the study of real-time dynamics or of two-dimensional systems. 

In fact, both questions are intimately related: as was realized quite soon, DMRG  is only moderately successful when applied to two-dimensional lattices: while relatively small systems can be studied with high accuracy\cite{White96,White98a,White98b,White00,White03,Chernyshev03,Chernyshev05,White07}, the amount of numerical resources needed essentially increases exponentially with system size, making large lattices inaccessible. The totally different behaviour of DMRG in one and two dimensions is, as it turned out, closely related\cite{Vidal03,Latorre04} to the different scaling of quantum entanglement in many-body states in one and two dimensions, dictated by the so-called area laws (for a recent review, see \cite{Eisert10}). 

In this paper, I will stay within the framework of one-dimensional physics; while the generalizations of DMRG to higher dimensions reduce naturally to DMRG in one dimension, the emerging structures are so much richer than in one dimension that they are beyond the scope of this work.  
  
In an originally unrelated development, so-called {\em matrix product states} (MPS) were discovered as an interesting class of quantum states for analytical studies. In fact, the structure is so simple but powerful that it is no surprise that they have been introduced and used under a variety of names over the last fifty or more years (most notably perhaps by Baxter \cite{Baxter68}). In the present context, the most relevant prehistory is arguably given by the exact expression of the seminal one-dimensional AKLT state in this form\cite{Affleck87,Affleck88,Fannes89}, which gave rise to extensive studies of the translationally invariant subclass of MPS known as finitely correlated states\cite{Fannes92}. This form was then subsequently used in a variety of contexts for analytical variational calculations, e.g.\ for spin-1 Heisenberg antiferromagnets\cite{Klumper93,Kolezhuk96a,Kolezhuk96b,Kolezhuk98} and ferrimagnets\cite{Kolezhuk97,Kolezhuk99}.

The connection between MPS and DMRG was made in two steps. In a first step, Ostlund and Rommer \cite{Ostlund95} realized that the block-growth step of the so-called infinite-system DMRG could be expressed as a matrix in the form it takes in an MPS. As in homogeneous systems this block-growth step leads to a fixed point in the thermodynamic limit, they took the fixed point matrix as building block for a translationally invariant MPS. In a further step, it was recognized that the more important finite-system DMRG leads to quantum states in MPS form, over which it variationally optimizes\cite{Dukelsky98}. It was also recognized that in traditional DMRG the state class over which is variationally optimized changes as the algorithm progresses, such that if one demands in some sense ``perfect'' variational character, a small change to the algorithm is needed, which however was found to increase (solvable) metastability problems\cite{Takasaki99,White05}. 

It remains a curious historical fact that only a few of the DMRG practicioners took this development very seriously up to about 2004 when Cirac, Verstraete, Vidal and coworkers started to explore the power of MPS very systematically. While it was considered useful for conceptual purposes, surprisingly little thought was given to rethinking and reexpressing real-life DMRG implementations purely in the  MPS language; arguably, because the overwhelming majority of conventional DMRG applications (i.e.\ ground states for quantum chains with open boundary conditions) hardly profits. What was overlooked is that it easily opens up the way to powerful extensions to DMRG hard to see and express in conventional DMRG language. 

A non-exhaustive list of extensions would list real-time evolutions\cite{Vidal03a,Vidal04b,Daley04,WhiteFeiguin04,VerstraeteRipoll04,Prosen09,Hartmann09,Banuls09}, also at finite temperature\cite{VerstraeteRipoll04,Feiguin05a}, the efficient use of periodic boundary conditions\cite{VerstraetePorras04,Pippan10,Pirvu10}, reliable single-site DMRG\cite{White05}, numerical renormalization group (NRG) applications\cite{Weichselbaum09},  infinite-system algorithms\cite{Vidal07,Orus08,McCulloch08}, continuous systems\cite{Verstraete10}, not talking at all about progress made in higher dimensions starting with \cite{VerstraeteCirac04} using a generalization of the MPS state class\cite{Nishino01}. 

The goal of this paper cannot be to provide a full review of DMRG since 1992 as seen from the perspective of 2010, in particular given the review\cite{Schollwoeck05}, which tries to provide a fairly extensive account of DMRG as of summer 2004. I rather want to limit myself to more recent developments and phrase them entirely in the language of matrix product states, focussing rather on the nuts and bolts of the methods than showing a lot of applications. My hope would be that this review would allow newcomers to the field to be able to produce their own codes quickly and get a firm grasp of the key building blocks of MPS algorithms. It has overlaps with the introductions \cite{VerstraeteMurg08,McCulloch07} in the key methods presented, but focuses on different extensions, some of which arose after these papers, and in many places tries to be more explicit. It takes a different point of view than \cite{Peschel98}, the first comprehensive exposition of DMRG in 1998, because at that time the connection to MPS (though known) and in particular to quantum information was still in many ways unexploited, which is the focus here. Nevertheless, in a first ``historical'' step, I want to remind readers of the original way of introducing DMRG, which does not make use of the idea of matrix product states. This should make older literature easily accessible, but one can jump to Section \ref{sec:matrixproductstates} right away, if one is not interested in that. 
 
In a second step, I will show that any quantum state can be written exactly in a very specific form which is given by the matrix product states already alluded to. In fact, the restriction to one dimension will come from the fact that only in this case MPS are numerically manageable. I will highlight special canonical forms of MPS and establish their connection to the singular value decomposition (SVD) as a mathematical tool and the Schmidt decomposition as a compact representation of quantum states. After this I will explain how MPS are a natural framework for decimation schemes in one dimension as they occur in schemes such as DMRG and Wilson's NRG. As a simple, but non-trivial example, I will discuss the AKLT state in its MPS form explicitly.
We then move on to discuss explicitly operations with MPS: overlaps, normalization, operator matrix elements, expectation values and MPS addition. These are operations one would do with any quantum state; more MPS-specific are methods for bringing them into the computationally convenient canonical forms and for approximating an MPS by another one of smaller dimension. I conclude this exposition of MPS with discussing the relationship and the conversions between the MPS notation I favour here, an alternative notation due to Vidal, and the DMRG way of writing states; this relatively technical section should serve to make the literature more accessible to the reader.

The MPS ideas generalize from states to the representation of operators, so I move on to discuss the use of matrix product operators (MPO)\cite{VerstraeteRipoll04,McCulloch07,VerstraetePirvu08,Crosswhite08,Frowis10}. As far as I can see, all operators of interest to us (ranging from local operators through bond evolution operators to full Hamiltonians) find a very compact and transparent formulation in terms of MPO. This leads to a much cleaner and sometimes even numerically more accurate formulation of DMRG-related algorithms, but their usage is not yet very widely spread.

Admittedly, at this point the patience of the reader may have been stretched quite a bit, as no real-world algorithm e.g.\ for ground state searches or time evolutions has been formulated in MPS language yet; but it will become obvious that a lot of cumbersome numerical details of DMRG algorithms have been hidden away neatly in the MPS and MPO structures. 

I will discuss ground state algorithms, discussing the equivalences and differences between DMRG with one or two center sites and fully MPS-based algorithms, including improvements to avoid trapping. I will focus on finite systems with open boundary conditions, where these methods excel.

After this, I move on to time-dependent methods for dynamics, for pure and mixed states. After a discussion of the basic algorithms and their subtle differences, I will focus on the key problem of extending the time-range of such simulations: The possibility to calculate highly accurate real-time and imaginary-time evolutions of complex quantum many-body states has been particularly exciting for many people, also because it arrived just in time for studying highly tunable ultracold atom systems. While this development has already led to numerous interesting insights and applications, it was quickly realized that the time-range of  time-dependent DMRG and related methods is limited by entanglement growth in quantum states out of equilibrium, such that long-time physics is out of reach. In this context, interesting progress in trying to go beyond has been achieved recently.

The review concludes with two further axes of development. I will start out by discussing the connection between DMRG and Wilson's NRG, showing how NRG can be expressed in a very concise fashion as well as be improved in various directions. This closes an interesting historical loop, as the utter failure of NRG for homogeneous one-dimensional quantum lattices as opposed to quantum impurity models mapped to special non-homogeneous one-dimensional quantum lattices was at the starting point of White's invention of DMRG\cite{WhiteNoack92}.

I continue by looking at infinite-size algorithms using MPS that work directly in the thermodynamic limit, one based on time evolution (iTEBD)\cite{Vidal07}. The other (iDMRG)\cite{McCulloch08} is an extension of infinite-system DMRG algorithm, which has had an interesting history: in many early discussions of DMRG it was presented as the key aspect of DMRG, with finite-system DMRG as a practitioners' add-on to further improve numerical accuracy. Later, it was recognized that applying finite-system DMRG is essential even for qualitative correctness in many cases, and infinite-system DMRG was seen as just a warm-up procedure. Only recently, McCulloch\cite{McCulloch08} pointed out a way how to turn infinite-system DMRG into a highly efficient tool for producing thermodynamic limit states for homogeneous systems. 

Last but not least, I will give an outlook on further applications of MPS that I could not cover here.

\section{Density-matrix renormalization group (DMRG)}
\subsection{Infinite-system and finite-system algorithms}
\begin{figure}
\centering\includegraphics[width=0.8\textwidth]{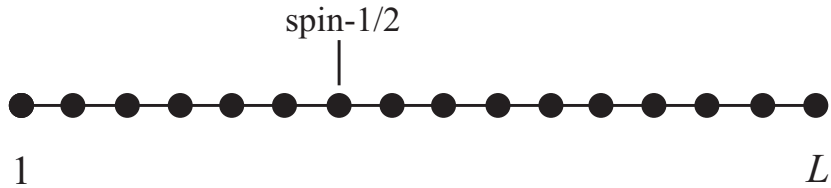}
\caption{Our toy model: a chain of length $L$ with open ends, where a spin-$\frac{1}{2}$ sits on each site and interacts with its nearest neighbours.}
\label{fig:toymodel}
\end{figure}

As a toy model, let us consider an (anisotropic) $S=\frac{1}{2}$ Heisenberg antiferromagnetic ($J=1$) spin chain of length $L$ in one spatial dimension with external magnetic field $h$, 
\begin{equation}
\hat{H} = \sum_{i=1}^{L-1} \frac{J}{2} (\hat{S}^+_{i}  \hat{S}^-_{i+1} + \hat{S}^-_{i} \hat{S}^+_{i+1}) + J^z \hat{S}^z_{i} \hat{S}^z_{i+1} - \sum_{i=1}^L h \hat{S}^z_i .
\label{eq:basicHam}
\end{equation}
We consider {\em open boundary conditions} (Fig.~\ref{fig:toymodel}), which is well worth emphasizing: analytically, periodic boundary conditions are usually more convenient; many numerical methods do not really depend strongly on boundary conditions, and some, like exact diagonalization, even become more efficient for periodic boundary conditions. DMRG, on the other hand, prefers open boundary conditions.

It should also be emphasized that, DMRG being a variational method in a certain state class, it does not suffer from anything like the fermionic sign problem, and can be applied to bosonic and fermionic systems alike. 

The starting point of DMRG is to ask for the ground state and ground state energy of $\hat{H}$. We can ask this question for the thermodynamic limit $L\rightarrow\infty$ or more modestly for finite $L$. In the first case, the answer is provided by {\em infinite-system} DMRG albeit with quite limited precision; in the second case, an answer can be read off from infinite-system DMRG, but it is more adequate to run a two-step procedure, starting with infinite-system DMRG and continuing with {\em finite-system} DMRG.

In any case, the numerical stumbling block is provided by the exponential growth of the Hilbert space dimension, in our example as $d^L$, where $d=2$ is the local state space dimension of a spin-$\frac{1}{2}$. 

\subsection{Infinite-system DMRG}

Infinite-system DMRG deals with this problem by considering a chain of increasing length, usually $L=2$, $4$, $6$, $\ldots$, and discarding a sufficient number of states to keep Hilbert space size manageable. This {\em decimation procedure} is key to the success of the algorithm: we assume that there exists a reduced state space which can describe the relevant physics and that we can develop a procedure to identify it. The first assumption is typical for all variational methods, and we will see that indeed we are lucky in one dimension: for all short-ranged Hamiltonians in 1D there is such a reduced state space that contains the relevant physics!

How is it found? In infinite-system DMRG (Fig.~\ref{fig:DMRGCombined}), the buildup is carried out as follows: we introduce left and right {\em blocks} A and B, which in a first step may consist of one spin (or site) each, such that total chain length is 2. Longer chains are now built iteratively from the left and right end, by inserting pairs of spins between the blocks, such that the chain grows to length 4, 6, and so on; at each step, previous spins are absorbed into the left and right blocks, such that block sizes grow as 1, 2, 3, and so on, leading to exponential growth of the dimension of the full block state space as $2^\ell$, where $\ell$ is the current block size. Our chains then always have a block-site-site-block structure, A$\bullet\bullet$B.

Let us assume that our numerical resources are sufficient to deal with a reduced block state space of dimension $D$, where in practice $D$ will be of $O(100)$ to $O(1000)$, and that for a block A of length $\ell$ we have an effective description of block A in a $D$-dimensional reduced Hilbert space with orthonormal basis $\{ \ket{a_\ell}_A \}$. For very small blocks, the basis dimension may be less than $D$, for larger blocks some truncation must have occurred, which we take for granted at the moment. Let me mention right now that in the literature, $D$ -- which will turn out to be a key number -- comes under a variety of names: in traditional DMRG literature, it is usually referred to as $m$ (or $M$); more recent matrix product state literature knows both $D$ and $\chi$.

Within this framework, we may now ask, (i) what is the ground state of the current chain of length $2\ell +2$, also referred to as {\em superblock}, and (ii) how can we find the reduced Hilbert space of dimension $D$ for the new blocks A$\bullet$ and $\bullet$B. 

Any state of the superblock A$\bullet\bullet$B can be described by 
\begin{equation}
\ket{\psi}= \sum_{a_A \sigma_A \sigma_B a_B} \psi_{a_A \sigma_A \sigma_B a_B} \ket{a}_A \ket{\sigma}_A \ket{\sigma}_B \ket{a}_B \equiv \sum_{i_A j_B} \psi_{i_A j_B} \ket{i}_A \ket{j}_B,
\label{US:eq:DMRGstate}
\end{equation} 
where the states of the site next to A are in the set $\{ \ket{\sigma}_A \}$ of local state space dimension $d$, and analogously those of the site next to B. By numerical diagonalization we find the $\ket{\psi}$ that minimizes the energy
\begin{equation}
E = \frac{\bra{\psi} \hat{H}_{{\rm A}\bullet\bullet{\rm B}} \ket{\psi}}{\braket{\psi}{\psi}}
\end{equation}
with respect to the Hamiltonian of the superblock, answering (i). To this purpose, we need some iterative sparse matrix eigensolver such as provided by the Lanczos or Jacobi-Davidson methods. Given that our Hamiltonian [Eq.~(\ref{eq:basicHam})] is available in the full tensor product space, this minimization of course assumes that it can be expressed readily in the superblock basis. Let us postpone this question, as it is intimately related to growing blocks and decimating their state space. As the matrix dimension is $d^2 D^2$, for typical $D$ the eigenproblem is too large for direct solution, even assuming the use of quantum symmetries. As most matrix elements of short-ranged Hamiltonians are zero, the matrix is sparse, such that the basic matrix-vector multiplication of iterative sparse matrix eigensolvers can be implemented very efficiently. We will discuss this question also further below. 

If we now take states $\{ \ket{i}_A \}$ as the basis of the next larger left block A$\bullet$, the basis dimension grows to $dD$. To avoid exponential growth, we truncate this basis back to $D$ states, using the following procedure for block A$\bullet$ and similarly for $\bullet$B: we consider the {\em reduced density operator} for A$\bullet$, namely
\begin{equation}
\hat{\rho}_{A\bullet} = \Tr_{\bullet B} \ket{\psi}\bra{\psi} 	\quad\quad (\rho_{A\bullet})_{ii'} = \sum_j \psi_{ij}\psi^*_{i'j} .  
\end{equation}
The eigensystem of $\hat{\rho}_{A\bullet}$ is determined by exact diagonalization; the choice is to retain as reduced basis those $D$ orthonormal eigenstates that have the largest associated eigenvalues $w$. If we call them $\ket{b}_A$, the vector entries are simply the expansion coefficients in the previous block-site basis, $_A \braket{a \sigma}{b}_A$. 

After an (approximate) transformation of all desired operators on A$\bullet$ into the new basis, the system size can be increased again, until the final desired size is reached. B is grown at the same time, for reflection-symmetric systems by simple mirroring.

The motivation of the truncation procedure is twofold. The first one, which is on a weaker basis, is that we are interested in the states of A$\bullet$ contributing most to the ground state for A$\bullet$ embedded in the final, much larger, even infinitely large system. We approximate this final system ground state, which we don't know yet, to the best of our knowledge by that of A$\bullet\bullet$B, the largest superblock we can efficiently form. In that sense, we are bootstrapping, and the finite-system DMRG will, as we will see, take care of the large approximations this possibly incurs. 
Let me remark right now that in the MPS formulation of infinite-system DMRG a much clearer picture will emerge. The second motivation for the choice of states is on much firmer foundation: the above prescription follows both from statistical physics arguments or from demanding that the 2-norm distance $\| \ket{\psi} - \ket{\psi}_{{\rm trunc}} \|_2$ between the current ground state $\ket{\psi}$ and its projection onto truncated block bases of dimension $D$, $\ket{\psi}_{{\rm trunc}}$, is minimal. The most transparent proof follows from a singular value decomposition of the matrix $\Psi$ with matrix elements $\Psi_{(a_A \sigma_A), (\sigma_B a_B)}$ formed from the wave function coefficients $\psi_{a_A \sigma_A\sigma_B a_B}$. The overall success of DMRG rests on the observation that even for moderate $D$ (often only a few 100) the total weight of the truncated eigenvalues, given by the {\em truncation error} $\epsilon=1-\sum_{a>D} w_a$ if we assume descending order of the $w_a$, is extremely close to 0, say $10^{-10}$ or less. 

Algorithmically, we may of course also fix a (small) $\epsilon$ we are willing to accept and at each truncation choose $D$ sufficiently large as to meet this requirement. Computational resources are used more efficiently (typically, we will have a somewhat smaller $D$ towards the ends of chains because of reduced quantum fluctuations), the programming effort to obtain this flexibility is of course somewhat higher.

An important aspect of improving the performance of any quantum algorithm is the exploitation of quantum symmetries, ranging from discrete symmetries like a $Z_2$ mirror symmetry if the Hamiltonian is invariant under mirroring from left to right through Abelian symmetries like the $U(1)$ symmetry of particle number (``charge'') or magnetization conservation to non-Abelian symmetries like the {\em SU}(2) symmetry of rotational invariance\cite{McCulloch07,SierraNishino97,McCulloch00,McCulloch01,McCulloch02}. A huge range of these symmetries have been used successfully in numerous applications, with the most common ones being the two $U(1)$ symmetries of charge and magnetization conservation, but also {\em SU}(2) symmetries\cite{McCulloch08a,Smerat09,Langer09,Holzner09}. Let us focus on magnetization and assume that the total magnetization, $\hat{M} = \sum_i \hat{S}^z_i$, commutes with the Hamiltonian, $[\hat{H},\hat{M}]=0$, such that eigenstates of $\hat{H}$ can be chosen to be eigenstates of $\hat{M}$. Let us assume in particular that the good quantum number of the ground state is $M=0$. If the block and site states are eigenstates of magnetization, then $\psi_{a_A \sigma_A \sigma_B a_B} \neq 0$ only if $M (\ket{a}_A) +  M (\ket{\sigma}_A) + M (\ket{\sigma}_B) + M (\ket{a}_B) = 0$;
$M$ is a short-hand for the magnetization of the respective blocks and sites, assuming that the states have magnetization as a good quantum number. This constraint allows to exclude a large number of coefficients from the calculation, leading to a huge speedup of the calculation and (less important) savings in memory.

The decisive point is that if the states $\ket{a}_A$ and $\ket{\sigma}_A$ are eigenstates of magnetization, so will be the eigenstates of the reduced density operator $\hat{\rho}_{A\bullet}$, which in turn will be the states $\ket{a}_A$ of the enlarged block. As the local site states can be chosen to be such eigenstates and as the first block in the growth process consists of one local site, the eigenstate property then propagates through the entire algorithm. To prove the claim, we consider $(\rho_{A\bullet})_{ii'} = \sum_j \psi_{ij} \psi^*_{i'j}$. The states $\ket{i}_A$ and $\ket{j}_B$ are eigenstates of magnetization by construction, hence 
$M (\ket{i}_A) + M(\ket{j}_B) = 0 = M (\ket{i'}_A) + M (\ket{j}_B)$ or $M (\ket{i}_A) = M (\ket{i'}_A)$. The density matrix therefore decomposes into blocks that are formed from states of equal magnetization and can be diagonalized block by block within these sets, such that its eigenstates are also eigenstates of magnetization.

In practical implementations, the use of such good quantum numbers will be done by arranging matrix representations of operators into block structures labelled by good quantum numbers, which are then combined to satisfy local and global constraints. While this is conceptually easy, the coding becomes more complex and will not be laid out here explicitly; hints will be given in the MPS sections.
 
So far, we have postponed the question of expressing operators acting on blocks in the current block bases, in order to construct the Hamiltonian and observables of interest. Let us consider an operator $\hat{O}$ acting on site $\ell$, with matrix elements $O^{\sigma_\ell,\sigma'_\ell}=\bra{\sigma_\ell} \hat{O}_{\ell} \ket{\sigma'_\ell}$. Without loss of generality, assume that site $\ell$ is on the left-hand side and added into block A. Its construction is then initialized when block A grows from $\ell-1 \rightarrow \ell$, as 
\begin{equation}
\bra{a_\ell} \hat{O} \ket{a'_\ell} = \sum_{a_{\ell-1},\sigma_\ell,\sigma'_\ell}
\braket{a_\ell}{a_{\ell-1}\sigma_\ell} \bra{\sigma_\ell} \hat{O} \ket{\sigma'_\ell} \braket{a_{\ell-1}\sigma'_\ell}{a'_\ell} .
\end{equation}
Here, $ \ket{a_\ell}$ and $\ket{a_{\ell-1}}$ are the effective basis states of blocks of length $\ell$ and $\ell-1$ respectively. Of course, updates are necessary during the further growth steps of block A, e.g.\
\begin{equation}
\bra{a_{\ell+1}} \hat{O} \ket{a'_{\ell+1}} = \sum_{a_{\ell},a'_{\ell}\sigma_{\ell+1}}
\braket{a_{\ell+1}}{a_{\ell}\sigma_{\ell+1}} \bra{a_\ell} \hat{O} \ket{a'_\ell} \braket{a'_{\ell}\sigma_{\ell+1}}{a'_{\ell+1}} .
\label{eq:update}
\end{equation}
It is important to realize that the sum in Eq.~(\ref{eq:update}) must be split as 
\begin{equation}
\sum_{a_{\ell}\sigma_{\ell+1}} \braket{a_{\ell+1}}{a_{\ell}\sigma_{\ell+1}} \left( \sum_{a'_{\ell}}
 \bra{a_\ell} \hat{O} \ket{a'_\ell} \braket{a'_{\ell}\sigma_{\ell+1}}{a'_{\ell+1}} \right),
\end{equation}
reducing the calculational load from $O(D^4 d)$ to $2O(D^3 d)$. 

In Hamiltonians, operator products $\hat{O}\hat{P}$ occur. It is tempting, but due to the many truncation steps highly imprecise, to insert the quantum mechanical one in the current block basis and to write
\begin{equation}
\bra{a_\ell} \hat{O}\hat{P} \ket{a'_\ell} = \sum_{\tilde{a}_\ell} \bra{a_\ell} \hat{O} \ket{\tilde{a}_\ell} \bra{\tilde{a}_\ell} \hat{P} \ket{a'_\ell} . \quad\quad {\rm NO!}
\end{equation}
The correct way is to update the first operator (counting from the left for a left block A) until the site of the second operator is reached, and to incorporate it as
\begin{equation}
\bra{a_\ell} \hat{O} \hat{P} \ket{a'_\ell} = \sum_{a_{\ell-1},a'_{\ell-1},\sigma_\ell,\sigma'_\ell}
\braket{a_\ell}{a_{\ell-1}\sigma_\ell}  \bra{a_{\ell-1}} \hat{O} \ket{a'_{\ell-1}} \bra{\sigma_\ell} \hat{P} \ket{\sigma'_\ell} \braket{a'_{\ell-1}\sigma'_\ell}{a'_\ell} .
\label{eq:incorporate}
\end{equation}
The further updates are then as for single operators. Obviously, we are looking at a straightforward sequence of (reduced) basis transformations, with a somewhat cumbersome notation, but it will be interesting to see how much simpler these formulae will look in the MPS language, albeit identical in content.

The ultimate evaluation of expectation values is given at the end of the growth procedure as 
\begin{equation}
\bra{\psi} \hat{O} \ket{\psi} = \sum_{a_A a'_A \sigma_A \sigma_B a_B} \braket{\psi}{a_A \sigma_A \sigma_B a_B} \bra{a_A} \hat{O} \ket{a'_A} \braket{a'_A \sigma_A \sigma_B a_B}{\psi},
\end{equation}
where suitable bracketing turns this into an operation of order $O(D^3 d^2)$. An important special case is given if we are looking for the expectation value of a local operator that acts on one of the free sites $\bullet$ in the final block-site configuration. Then 
\begin{equation}
\bra{\psi} \hat{O} \ket{\psi} = \sum_{a_A \sigma_A \sigma'_A \sigma_B a_B} \braket{\psi}{a_A \sigma_A \sigma_B a_B} \bra{\sigma_A} \hat{O} \ket{\sigma'_A} \braket{a_A \sigma'_A \sigma_B a_B}{\psi},
\end{equation}
which is an expression of order $O(D^2 d^3)$ (for many operators, even $O(D^2 d^2)$). Given that $D \gg d$ in all practical applications, such evaluations are computationally highly advantageous.

Let me conclude these more technical remarks by observing that for an efficient construction of  $\hat{H}$ from such operators, it is essential {\em never} to build it as a full matrix, but to make use of the specific block-site-site-block structure. Assume, for example, a term which contains one operator acting on A and one on B (this is in fact the most complicated case), $\hat{h}=\hat{O}_A \hat{O}_B$. Then
\begin{equation}
\bra{a_A \sigma_A \sigma_B a_B} \hat{h}\ket{\psi} = \sum_{a'_A} 
\bra{a_A}\hat{O}_A\ket{a'_A} \left( \sum_{a'_B} \bra{a_B}\hat{O}_B\ket{a'_B}  \braket{a'_A \sigma_A \sigma_B a'_B}{\psi} \right),
\label{eq:hamfragment}
\end{equation}
which is a sequence of two $O(D^3 d^2)$ multiplications (instead of one naive $O( D^4 d^2)$ calculation) for all coefficients $\bra{a_A \sigma_A \sigma_B a_B} \hat{h}\ket{\psi}$ and similarly for all other terms.

If we stop infinite-system DMRG at some superblock size $L$, we can interpret the final wavefunction in two ways. We can take it as an approximation to the exact state for the superblock of size $L$ and evaluate expectation values. The accuracy is limited not only by the truncations, but also by the fact that the first truncations were carried out for extremely small superblocks: the choice of relevant short block states is likely to be a not too good approximation to those one would have chosen for these short blocks embedded in the final system of length $L$.

Alternatively, we may ignore these boundary effects and focus on the central sites as an approximation to the behaviour of an infinitely large system, provided $L$ is large enough and careful extrapolations of results to $L\rightarrow\infty$ are done. This is not so simple to do in a very controlled fashion, but as we will see, infinite-system DMRG can be used to build highly controlled translationally invariant (modulo, say, a unit cell of length 2) thermodynamic limit states.

\subsection{Finite-system DMRG}

\begin{figure}
\centering\includegraphics[width=250pt]{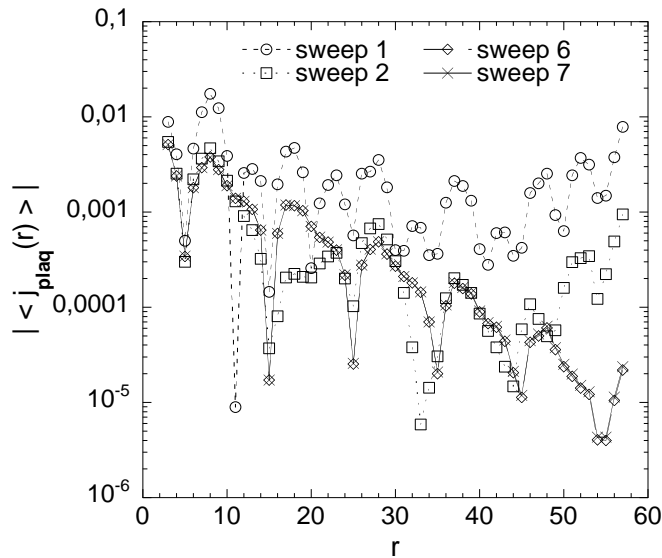}
\caption{Plaquette current on a $t$-$J$-$V$-$V'$-ladder, $J=0.4t$, $V=3t$, $V'=t$ ($V$ nearest, $V'$ next-nearest neighbour interaction) at hole doping $\delta=0.1$, system size $2\times 60$, as induced by a finite boundary current on rung $1$. The absolute current strength is shown;  whereas infinite-system DMRG and the first sweep indicate the generation of a long-ranged pattern, a fully converged calculation (here after 6 to 7 sweeps) reveals an exponential decay into the bulk. Taken from Ref.~\cite{Schollwoeck03}.}
\label{fig:PlaquetteCurrent}
\end{figure}

Once the desired final system size is reached by infinite-system DMRG, it is important in all but the most trivial applications to follow up on it by the so-called finite-system DMRG procedure. This will not merely lead to some slight quantitative improvements of our results, but may change them completely: consider \cite{Schollwoeck03} for an example where even the central physical statement changes: for a $t$-$J$-$V$-$V'$ model on a ladder with moderate hole-doping $\delta=0.1$, an infinite-system DMRG calculation indicates the existence of alternately circulating currents on plaquettes that are triggered by an infinitesimal current at the left end of the ladder, a signal of a so-called $d$-density wave state. Only after applying the finite-system algorithm it becomes obvious that this current is in fact exponentially decaying into the bulk, excluding this type of order (Fig.~\ref{fig:PlaquetteCurrent}). 

\begin{figure}
\centering\includegraphics[width=\textwidth]{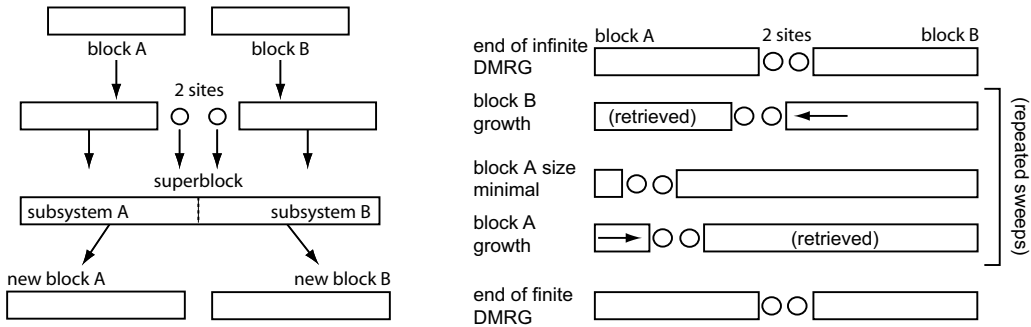}
\caption{The left and right half of the figure present the iterations taken in the infinite-system and finite-system DMRG procedures respectively. In both cases, new blocks are formed from integrating a site into a block, with a state space truncation according to the density-matrix prescription of DMRG. Whereas in the infinite-system version this growth happens on both sides of the chain, leading to chain growth, in the finite-system algorithm it happens only for one side at the expense of the other, leading to constant chain length.}
\label{fig:DMRGCombined}
\end{figure}
 
The finite-system algorithm corrects the choices made for reduced bases in the context of a superblock that was not the system of interest (of final length $L$), but some sort of smaller proxy for it.   
 
What the finite-system algorithm does is the following (Figure \ref{fig:DMRGCombined}): it continues the growth process of (say) block B following the same prescription as before: finding the ground state of the superblock system, determining the reduced density operator, finding the eigensystem, retaining the $D$ highest weight eigenstates for the next larger block. But it does so at the expense of block A, which shrinks (i.e.\ old shorter blocks A are reused). This is continued until A is so small as to have a complete Hilbert space, i.e.\ of dimension not exceeding $D$ (one may also continue until A is merely one site long; results are not affected). Then the growth direction is reversed: A grows at the expense of B, including new ground state determinations and basis choices for A, until B is small enough to have a complete Hilbert space, which leads to yet another reversal of growth direction.

This {\em sweeping} through the system is continued until energy (or, more precisely, the wave function) converges. The intuitive motivation for this (in practice highly successful) procedure is that after each sweep, blocks A or B are determined in the presence of an ever improved embedding.

In practice, this algorithm involves a lot of book-keeping, as all the operators we need have to be maintained in the current effective bases which will change from step to step. This means that the truncated basis transformations determined have to be carried out after each step; operator representations in all bases have to be stored, as they will be needed again for the shrinking block. 

Another important feature is that for finding the ground state $\ket{\psi}$ for each A$\bullet\bullet$B configuration one employs some iterative large sparse matrix eigensolver based on sequential applications of $\hat{H}$ on some initial starting vector. To speed up this most time-consuming part of the algorithm, it is highly desirable to have a good prediction for a starting vector, i.e.\ as close as possible to the ultimate solution. This can be achieved by (approximately) transforming the result of the last step into the shifted  A$\bullet\bullet$B configuration \cite{White96} by applying two basis transformations: e.g.\ A$\bullet\rightarrow$A and B$\rightarrow\bullet$B for a sweep to the right. The explicit formulae (see \cite{Schollwoeck05,White96}) can be derived by writing
\begin{equation}
\ket{\psi} = \sum_{a_\ell\sigma_{\ell+1}\sigma_{\ell+2}b_{\ell+2}} \psi_{a_\ell\sigma_{\ell+1}\sigma_{\ell+2}b_{\ell+2}} \ket{a_\ell}_A \ket{\sigma_{\ell+1}} \ket{\sigma_{\ell+2}} \ket{b_{\ell+2}}_B,
\end{equation}
where $\ket{a_\ell}_A$ and $\ket{b_{\ell+2}}_B$ are the block states for block A comprising sites 1 through $\ell$ and block B comprising sites $\ell+3$ through $L$ (the label of the block states is taken from the label of the bond their ends cut; see Fig.~\ref{fig:labelling}), and inserting twice an approximate identity, namely $\hat{I} = \sum_{a_{\ell+1}} \ket{a_{\ell+1}}_A {\,}_A\bra{a_{\ell+1}}$ and
$\hat{I} = \sum_{\sigma_{\ell+3} b_{\ell+3}} \ket{\sigma_{\ell+3}}\ket{b_{\ell+3}}_B {\, }_B \bra{b_{\ell+3}} \bra{\sigma_{\ell+3}}$.
One then obtains
\begin{equation}
\ket{\psi} = \sum_{a_{\ell+1}\sigma_{\ell+2}\sigma_{\ell+3}b_{\ell+3}} \psi_{a_{\ell+1}\sigma_{\ell+2}\sigma_{\ell+3}b_{\ell+3}} \ket{a_{\ell+1}}_A \ket{\sigma_{\ell+2}} \ket{\sigma_{\ell+3}} \ket{b_{\ell+3}}_B,
\end{equation}
with
\begin{equation}
\psi_{a_{\ell+1}\sigma_{\ell+2}\sigma_{\ell+3}b_{\ell+3}} =
\sum_{a_\ell \sigma_{\ell+1} b_{\ell+2}} 
\psi_{a_\ell\sigma_{\ell+1}\sigma_{\ell+2}b_{\ell+2}}  
\braket{a_{\ell+1}}{a_\ell \sigma_{\ell+1}} \braket{b_{\ell+3}\sigma_{\ell+3}}{b_{\ell+2}} . 
\end{equation}
The basis transformations required in the last equation are all available from previous steps in the DMRG procedure. A similar operation can be carried out for a sweep to the left. As we will see, this astute step, which led to drastic improvements in DMRG performance, is already implicit if one rewrites DMRG in the MPS language, such that we will not discuss it here at length.

An important observation is that both the infinite-system and finite-system algorithm can also be carried out by inserting only a single explicit site $\bullet$, hence one would study superblocks of the form A$\bullet$B, with slightly adapted growth procedures. An advantage of this method would be a speedup by roughly a factor $d$ in the large sparse eigensolver; for example, the application of $\hat{h}$ to a state in Eq.~(\ref{eq:hamfragment}) would then lead to $O(D^3 d)$ operations.
In the infinite-system algorithm an obvious disadvantage would be that superblock lengths oscillate between odd and even; in the finite-system algorithm the question of the relative merits is much more interesting and will be discussed at length in Section \ref{subsec:conventionalDMRGinMPS}.

Obviously, for $D\rightarrow\infty$, no truncations occur and DMRG becomes exact; increasing $D$ reduces truncation and therefore monotonically improves observables, which one extrapolates in $D\rightarrow\infty$ (even better, in the truncation error $\epsilon\rightarrow 0$, for which local observables often show effectively linear error dependence on $\epsilon$) for optimal results.

For more details on DMRG and its applications, I refer to \cite{Schollwoeck05}. 
  
\section{DMRG and entanglement: why DMRG works and why it fails}
The DMRG algorithm quite naturally leads to the consideration of bipartite quantum systems, where the parts are A$\bullet$ and $\bullet$B. For an arbitrary bipartition,  $\ket{\psi} = \sum_{ij} \psi_{ij} \ket{i}_A \ket{j}_B$, where the states $\ket{i}_A$ and $\ket{j}_B$ form orthonormal bases of dimensions $N_A$ and $N_B$ respectively. Thinking of the $\psi_{ij}$ as entries of a rectangular matrix $\Psi$ (dimension $N_A \times N_B$), the reduced density matrices $\rho_A$ and $\rho_B$ take the form
\begin{equation}
\rho_A = \Psi \Psi^\dagger \quad\quad \rho_B = \Psi^\dagger \Psi.
\label{eq:ABdensityoperators}
\end{equation}
If we assume that we know $\ket{\psi}$ exactly, but can approximate it in DMRG only with at most $D$ states per block, the optimal DMRG approximation is provided by retaining as block states the eigenstates belonging to the $D$ largest eigenvalues. If we happen to know the eigenspectra of reduced density operators of $\ket{\psi}$, we can easily assess the quality a DMRG approximation can have; it simply depends on how quickly the eigenvalues $w_a$ decrease. In fact, such analyses have been carried out for some exactly solved systems in one and two dimensions \cite{Peschel99,Okunishi99,Chung00,Chung01,Chan02}. They reveal that in one dimension for gapped systems eigenvalues $w_a$  generically decay exponentially fast (roughly as $\eul^{- c \ln^2 a}$), which explains the success of DMRG, but in two-dimensional stripe geometries of size $L \times W$, $L\gg W$, $c \propto 1/W$, such that with increasing width $W$ (increasing two-dimensionality) the eigenspectrum decay is so slow as to make DMRG inefficient. 

Usually, we have no clear idea about the eigenvalue spectrum; but it turns out that in such cases 
entanglement entropies can serve as ``proxy'' quantities, namely the von Neumann entanglement or entanglement entropy. It is given by the non-vanishing part of the eigenvalue spectrum of $\rho_A$ (identical to that of $\rho_B$, as we will discuss below) as
\begin{equation}
S_{A|B}  = - \Tr\, \rho_A \log_2 \rho_A = - \sum_\alpha w_a \log_2 w_a .
\end{equation}
It would seem as if we have gained nothing, as we don't know the $w_a$, but general laws about entanglement scaling are available. If we consider a bipartitioning A$|$B where AB is in the thermodynamic limit and A of size $L^{\cal D}$, with ${\cal D}$ the spatial dimension, the so-called {\em area laws}
\cite{Eisert10,Bekenstein73,Srednicki93,Callan94,Plenio05} predict that for ground states of short-ranged Hamiltonians with a gap to excitations entanglement entropy is not extensive, but proportional to the surface, i.e. $S(A|B) \sim L^{{\cal D}-1}$, as opposed to thermal entropy. This implies $S \sim $ cst. in one dimension and $S \sim L$ in two dimensions. At criticality, a much richer structure emerges: in one dimension,
$S = \frac{c+\overline{c}}{6} \log_2 L + k$, where $c$ and $\overline{c}$ are the (an)holonomic central charges from conformal field theory\cite{Vidal03,Latorre04}; in two dimensions, bosonic systems seem to be insensitive to criticality (i.e. $S \propto L$)\cite{Srednicki93,Barthel06}, whereas fermionic systems get a logarithmic correction $S \propto L \log_2 L$ for a one-dimensional Fermi surface (with a prefactor proportional to its size), but seem to grow only sublogarithmically if the Fermi surface consists of points \cite{Barthel06,Gioev06}. It should be emphasized that these properties of ground states are highly unusual: in the thermodynamic limit, a random state out of Hilbert space will indeed show extensive entanglement entropy with probability 1.

In a mathematically non-rigorous way one can now make contact between DMRG and the area laws of quantum entanglement: between two $D$-dimensional state spaces for A and B, the maximal entanglement is $\log_2 D$ in the case where all eigenvalues of $\rho_A$ are identical and $D^{-1}$ (such that $\rho_A$ is maximally mixed); meaning that one needs a state of dimension $2^S$ and more to encode entanglement $S$ properly. This implies that for gapped systems in one dimension an increase in system size will not lead to a strong increase in $D$; in two dimensions, $D \sim 2^L$, such that DMRG will fail even for relatively small system sizes, as resources have to grow exponentially (this however does not exclude very precise results for small two-dimensional clusters or quite large stripes). Critical systems in one dimension are borderline cases: $D \sim L^{\frac{c+\overline{c}}{6}}$; this means that the thermodynamic limit is not reachable, but the growth of $D$ is sufficiently slow (usually the power is weak, say $1/3$ or $1/6$, due to typical values for central charges) such that large system sizes ($L \sim O(10^3)$) can be reached; this allows for very precise finite-size extrapolations.

Obviously, this argument implicitly makes the cavalier assumption that the eigenvalue spectrum is close to flat, which leads to maximal entanglement, such that an approximate estimation of $D$ can be made. In practice, the spectrum is dictated by the problem and indeed far from flat: as we have seen, it is in fact usually exponentially decaying. But numerically, it turns out that for standard problems the scaling of the resource $D$ is predicted correctly on the qualitative level. 

It even turns out that in a mathematically strict analysis, von Neumann entanglement does {\em not} allow a general prediction of resource usage: this is because one can construct artificial eigenvalue spectra that allow or forbid efficient simulation, while their von Neumann entanglement would suggest the opposite, following the above argument \cite{Schuch08}. Typical many-body states of interest, however, do not have such ``pathological'' spectra. In fact, Renyi entanglement entropies, a generalization of von Neumann entanglement entropies, do allow mathematically rigourous connections \cite{Schuch08}, but usually are hard to calculate, with criticality in one dimension as an exception due to conformal field theory.  

\section{Matrix product states (MPS)}
\label{sec:matrixproductstates}

If we consider our paradigmatic problem, the one-dimensional Heisenberg antiferromagnet, the key problem is that Hilbert space seems to be exponentially big ($d^L = 2^L$). Looking for the ground state may therefore seem like looking for a needle in the haystack. The claim is that at least for local Hamiltonians with a gap between ground state and first excited state, the haystack is not very big, effectively infinitesimally small compared to the size of the full Hilbert space, as we have already seen from the very peculiar entanglement scaling properties. What is even more important, this relevant corner of Hilbert space can be parametrized efficiently, i.e.\ with modest numerical resources, operated upon efficiently, and efficient algorithms of the DMRG type to solve important questions of quantum physics do exist. This parametrization is provided by the {\em matrix product states (MPS)}.

Maybe the two DMRG algorithms explained above seem to be very cumbersome to implement. But it turns out that if we do quantum mechanics in the restricted state class provided by matrix product states, DMRG and other methods almost force themselves on us. The manipulation of matrix product states seems to be very complicated at first, but in fact can be formalized beautifully, together with a graphical notation that allows to generate permitted operations almost automatically; as any good formalism (such as bra and ket), it essentially enforces correctness. 

\subsection{Introduction of matrix product states}

\subsubsection{Singular value decomposition (SVD) and Schmidt decomposition}
Throughout the rest of this paper, we will make extensive use of one of the most versatile tools of linear algebra, the so-called {\em singular value decomposition} (SVD), which is at the basis of a very compact representation of quantum states living in a bipartite universe AB, the {\em Schmidt decomposition}. Let us briefly recall what they are about.

\begin{figure}
\centering\includegraphics[width=\textwidth]{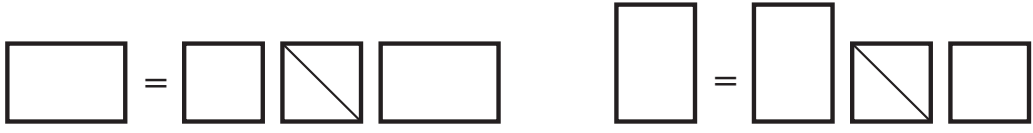}
\caption{Resulting matrix shapes from a singular value decomposition (SVD), corresponding to the two rectangular shapes that can occur. The singular value diagonal serves as a reminder that in $M=USV^{\ \dagger}$ $S$ is purely non-negative diagonal.}
\label{fig:SVD}
\end{figure}

SVD guarantees for an arbitrary (rectangular) matrix $M$ of dimensions $(N_A \times N_B)$ the existence of a decomposition
\begin{equation}
M =   U S V^\dagger ,
\end{equation}
where 
\begin{itemize}
\item $U$ is of dimension $(N_A \times \min (N_A,N_B))$ and has orthonormal columns (the {\em left singular vectors}), i.e. $U^\dagger U = I$; if $N_A \leq N_B$ this implies that it is unitary, and also $UU^\dagger = I$. 
\item $S$ is of dimension $(\min (N_A,N_B) \times \min (N_A,N_B))$, diagonal with non-negative entries $S_{aa} \equiv s_a$. These are the so-called {\em singular values}. The number $r$ of non-zero singular values is the {\em (Schmidt) rank} of $M$. In the following, we assume descending order: $s_1 \geq s_2 \geq \ldots \geq s_r > 0.$
\item $V^\dagger$ is of dimension $(\min (N_A,N_B) \times N_B)$ and has orthonormal rows (the {\em right singular vectors}), i.e.\ $V^\dagger V = I$. If $N_A \geq N_B$ this implies that it is unitary, and also $VV^\dagger  = I$.
\end{itemize}
This is schematically shown in Fig.~\ref{fig:SVD}. Singular values and vectors have many highly interesting properties. One which is of practical importance in the following is the optimal approximation of $M$ (rank $r$) by a matrix $M'$ (with rank $r'<r$) in the Frobenius norm $\| M \|_F^2 = \sum_{ij} |M_{ij}|^2$ induced by the inner product $\braket{M}{N}= \tr\, M^\dagger N$. It is given by 
\begin{equation}
M' = U S' V^\dagger \quad\quad S'={\rm diag} (s_1, s_2, \ldots, s_{r'}, 0,\ldots ),
\end{equation} 
i.e.\ one sets all but the first $r'$ singular values to be zero (and in numerical practice, will shrink the column dimension of $U$ and the row dimension of $V^\dagger$ accordingly to $r'$).

As a first application of the SVD, we use it to derive the {\em Schmidt decomposition} of a general quantum state. Any pure state 
$\ket{\psi}$ on AB
can be written as
\begin{equation}
\ket{\psi} = \sum_{ij} \Psi_{ij} \ket{i}_A \ket{j}_B,
\label{eq:generalquantumstate}
\end{equation}
where $\{ \ket{i}_A \}$ and $\{ \ket{j}_B \}$ are orthonormal bases of A and B with dimension $N_A$ and $N_B$ respectively; we read the coefficients as entries of a matrix $\Psi$. From this representation we can derive the  reduced density operators $\hat{\rho}_A = \tr_B \ket{\psi}\bra{\psi}$ and $\hat{\rho}_B = \tr_A \ket{\psi}\bra{\psi}$, which expressed with respect to the block bases take the matrix form 
\begin{equation} 
\rho_A = \Psi \Psi^\dagger \quad\quad \rho_B = \Psi^\dagger\Psi .
\end{equation}

If we carry out an SVD of matrix $\Psi$ in Eq.~(\ref{eq:generalquantumstate}), we obtain
\begin{eqnarray}
\ket{\psi} &=& \sum_{ij} \sum_{a=1}^{\min(N_A,N_B)} U_{ia} S_{aa} V^*_{ja} \ket{i}_A \ket{j}_B \nonumber \\
&=& \sum_{a=1}^{\min(N_A,N_B)} \left( \sum_i U_{ia} \ket{i}_A \right) s_a \left( \sum_j V_{ja}^* \ket{j} \right) \nonumber \\
&=& \sum_{a=1}^{\min(N_A,N_B)} s_a \ket{a}_A \ket{a}_B .
\end{eqnarray}
Due to the orthonormality properties of $U$ and $V^\dagger$, the sets $\{ \ket{a}_A \}$ and $\{ \ket{a}_B \}$ are orthonormal and can be extended to be orthonormal bases of A and B. If we restrict the sum to run only over the $r \leq \min(N_A,N_B)$ positive nonzero singular values, we obtain the {\em Schmidt decomposition}
\begin{equation}
\ket{\psi} = \sum_{a=1}^r s_a \ket{a}_A \ket{a}_B .
\end{equation}
It is obvious that $r=1$ corresponds to (classical) product states and $r>1$ to entangled (quantum) states.

The Schmidt decomposition allows to read off the reduced density operators for A and B introduced above very conveniently: carrying out the partial traces, one finds
\begin{equation}
\hat{\rho}_A = \sum_{a=1}^r s^2_a \ket{a}_A  \phantom\rangle_A \bra{a} \quad\quad \hat{\rho}_B = \sum_{a=1}^r s^2_a \ket{a}_B \phantom\rangle_B \bra{a} ,
\end{equation}
showing that they share the non-vanishing part of the spectrum, but not the eigenstates. The eigenvalues are the squares of the singular values, $w_a = s^2_a$, the respective eigenvectors are the left and right singular vectors.  The von Neumann entropy of entanglement can therefore be read off directly from the SVD,
\begin{equation}
S_{A|B}(\ket{\psi}) = - \tr\, \hat{\rho}_A \log_2 \hat{\rho}_A = - \sum_{a=1}^r s_a^2 \log_2 s_a^2 .
\end{equation}

In view of the large size of Hilbert spaces, it is also natural to approximate $\ket{\psi}$ by some $\ket{\tilde{\psi}}$ spanned over state spaces of A and B that have dimension $r'$ only. This problem can be related to SVD, because the 2-norm of $\ket{\psi}$ is identical to the Frobenius norm of the matrix $\Psi$,
\begin{equation}
\| \ket{\psi} \|_2^2 = \sum_{ij} | \Psi_{ij} |^2 = \| \Psi \|_F^2 ,
\end{equation}
if and only if the sets $\{ \ket{i} \}$ and $\{ \ket{j} \}$ are orthonormal (which is the case here).
The optimal approximation is therefore given in the 2-norm by the optimal approximation of $\Psi$ by $\tilde{\Psi}$ in the Frobenius norm, where $\tilde{\Psi}$ is a matrix of rank $r'$. As discussed above, $\tilde{\Psi} = US' V^\dagger$, where $S'={\rm diag}(s_1, \ldots, s_{r'}, 0, \ldots)$, constructed from the largest singular values of $\Psi$. Therefore, the Schmidt decomposition of the approximate state reads
\begin{equation}
\ket{\tilde{\psi}} = \sum_{a=1}^{r'} s_a \ket{a}_A \ket{a}_B ,
\label{eq:Schmidtapproximation}
\end{equation}
where the $s_a$ must be rescaled if normalization is desired.

\subsubsection{QR decomposition}
While SVD will be seen to cover all our needs, sometimes it is an overkill: in many cases of the expression $M=USV^\dagger$, we are only interested in the property $U^\dagger U = I$ and the product $SV^\dagger$, for example whenever the actual value of the singular values will not be used explicitly. Then there is a numerically cheaper technique, {\em QR decomposition}, which for an arbitrary matrix $M$ of dimension $(N_A \times N_B)$ gives a decomposition
\begin{equation}
M = QR ,
\label{eq:QRdecomposition}
\end{equation}
hence the name, where $Q$ is of dimension $(N_A \times N_A)$ and unitary, $Q^\dagger Q = I = Q Q^\dagger$, and $R$ is of dimension $(N_A \times N_B)$ and upper triangular, i.e. $R_{ij}=0$ if $i>j$. This {\em full} QR decomposition can be reduced to a {\em thin} QR decomposition: assume $N_A > N_B$: then the bottom $N_A-N_B$ rows of R are zero, and we can write
\begin{equation}
M=Q \cdot \left[ \begin{array}{c} R_1 \\ 0 \end{array} \right] = \left[ \begin{array}{cc} Q_1 & Q_2 \end{array} \right] \left[ \begin{array}{c} R_1 \\ 0 \end{array} \right] = Q_1 R_1 ,
\label{eq:thinQRdecomposition}
\end{equation}
where $Q_1$ is now of dimension $(N_A \times N_B)$, $R_1$ of dimension $(N_B \times N_B)$, and while $Q_1^\dagger Q_1 = I$, in general $Q_1 Q_1^\dagger \neq I$. Whenever I will refer to a QR decomposition in the following, I will imply the thin one. It should also be clear that the matrices $Q$ (or $Q_1$) share properties with $U$ from SVD, but are not the same in general; but as we will see that the MPS representation of states is not unique anyways, this does not matter.

\subsubsection{Decomposition of arbitrary quantum states into MPS}
\label{subsubsec:MPSdecomposition}
Consider a lattice of $L$ sites with $d$-dimensional local state spaces $\{ \sigma_i \}$ on sites $i=1,\ldots,L$. In fact, while we will be naturally thinking of a one-dimensional lattice, the following also holds for a  lattice of arbitrary dimension on which sites have been numbered; however, MPS generated from states on higher-dimensional lattices will not be manageable in numerical practice. The most general pure quantum state on the lattice reads
\begin{equation}
\ket{\psi} = \sum_{\sigma_1,\ldots,\sigma_L} c_{\sigma_1 \ldots \sigma_L} \ket{\sigma_1,\ldots,\sigma_L},
\end{equation}
where we have exponentially many coefficients $c_{\sigma_1 \ldots \sigma_L}$ with quite oblique content in typical quantum many-body problems. Let us assume that it is normalized. Can we find a notation that gives a more local notion of the state (while preserving the quantum non-locality of the state)? Indeed, SVD allows us to do just that. The result may look quite forbidding, but will be shown to relate profoundly to familiar concepts of quantum physics. There are three ways of doing this that are of relevance to us.

{\em (i) Left-canonical matrix product state.} In a first step, we {\em reshape} the state vector with $d^L$ components into a matrix $\Psi$ of dimension $(d \times d^{L-1})$, where the coefficients are related as 
\begin{equation}
\Psi_{\sigma_1, (\sigma_2 \ldots \sigma_L)} = c_{\sigma_1 \ldots \sigma_L}  .
\end{equation}
An SVD of $\Psi$ gives
\begin{equation}
c_{\sigma_1 \ldots \sigma_L}  = \Psi_{\sigma_1, (\sigma_2 \ldots \sigma_L)} = \sum_{a_1}^{r_1} U_{\sigma_1,a_1} S_{a_1,a_1} (V^\dagger)_{a_1, (\sigma_2 \ldots \sigma_L)} \equiv \sum_{a_1}^{r_1}  U_{\sigma_1,a_1} c_{a_1\sigma_2\ldots\sigma_L} ,
\end{equation}
where in the last equality $S$ and $V^\dagger$ have been multiplied and the resulting matrix has been reshaped back into a vector. The rank is $r_1 \leq d$. We now decompose the matrix $U$ into a collection of $d$ row vectors $A^{\sigma_1}$ with entries $A^{\sigma_1}_{a_1}=U_{\sigma_1,a_1}$. At the same time, we reshape $c_{a_1\sigma_2\ldots\sigma_L}$ into a matrix $\Psi_{(a_1\sigma_2),(\sigma_3 \ldots \sigma_L)}$ of dimension $(r_1 d \times d^{L-2})$, to give
\begin{equation}
c_{\sigma_1 \ldots \sigma_L}  = \sum_{a_1}^{r_1} A^{\sigma_1}_{a_1} \Psi_{(a_1\sigma_2), (\sigma_3 \ldots \sigma_L)} .
\end{equation}
$\Psi$ is subjected to an SVD, and we have 
\begin{equation}
c_{\sigma_1 \ldots \sigma_L}  = \sum_{a_1}^{r_1} \sum_{a_2}^{r_2} A^{\sigma_1}_{a_1} U_{(a_1\sigma_2),a_2} S_{a_2,a_2} (V^\dagger)_{a_2, (\sigma_3 \ldots \sigma_L)} =  
\sum_{a_1}^{r_1} \sum_{a_2}^{r_2} A^{\sigma_1}_{a_1} A^{\sigma_2}_{a_1,a_2} \Psi_{(a_2\sigma_3), (\sigma_4 \ldots \sigma_L)} ,
\end{equation}
where we have replaced $U$ by a set of $d$ matrices $A^{\sigma_2}$ of dimension $(r_1 \times r_2)$ with entries $A^{\sigma_2}_{a_1,a_2} = U_{(a_1\sigma_2),a_2}$ and multiplied $S$ and $V^\dagger$, to be reshaped into a matrix $\Psi$ of dimension $(r_2 d \times d ^{L-3})$, where $r_2 \leq r_1 d \leq d^2$. Upon further SVDs, we obtain
\begin{equation}
c_{\sigma_1 \ldots \sigma_L} = \sum_{a_1, \ldots, a_{L-1}} A^{\sigma_1}_{a_1} A^{\sigma_2}_{a_1,a_2} \ldots A^{\sigma_{L_1}}_{a_{L-2},a_{L-1}} A^{\sigma_{L}}_{a_{L-1}}
\end{equation}
or more compactly
\begin{equation}
c_{\sigma_1 \ldots \sigma_L} = A^{\sigma_1} A^{\sigma_2} \ldots A^{\sigma_{L-1}} A^{\sigma_L} ,
\end{equation}
where we have recognized the sums over $a_1$, $a_2$ and so forth as matrix multiplications. The last set of matrices $A^{\sigma_L}$ in fact consists of column vectors. If we wish, dummy indices 1 may be introduced in the first and last $A$ to turn them into matrices, too. In any case, the (arbitrary) quantum state is now represented exactly in the form of a {\em matrix product state:}
\begin{equation}
\ket{\psi} = \sum_{\sigma_1,\ldots,\sigma_L} A^{\sigma_1} A^{\sigma_2} \ldots A^{\sigma_{L-1}} A^{\sigma_L} \ket{\sigma_1,\ldots,\sigma_L} .
\end{equation}

Let us study the properties of the $A$-matrices. The maximal dimensions of the matrices are reached when for each SVD done the number of non-zero singular values is equal to the upper bound (the lesser of the dimensions of the matrix to be decomposed). Counting reveals that the dimensions may maximally be $(1 \times d), (d \times d^2), \ldots, (d^{L/2-1} \times d^{L/2}), (d^{L/2} \times d^{L/2-1}), \ldots, (d^2 \times d), (d \times 1)$, going from the first to the last site (I have assumed $L$ even for simplicity here). This shows that in practical calculations it will usually be impossible to carry out this exact decomposition explicitly, as the matrix dimensions blow up exponentially.

But there is more to it. Because at each SVD $U^\dagger U = I$ holds, the replacement of $U$ by a set of $A^\sigma$ entails the following relationship:
\begin{eqnarray*}
\delta_{a_\ell,a'_\ell} &=& \sum_{a_{\ell-1}\sigma_\ell} (U^\dagger)_{a_\ell, (a_{\ell-1}\sigma_\ell)} U_{(a_{\ell-1}\sigma_\ell), a'_\ell} \\
&=& \sum_{a_{\ell-1}\sigma_\ell} (A^{\sigma_\ell\dagger})_{a_\ell, a_{\ell-1}} A^{\sigma_\ell}_{a_{\ell-1}, a'_\ell} \\
&=& \sum_{\sigma_\ell} ( A^{\sigma_\ell\dagger} A^{\sigma_\ell})_{a_\ell,a'_\ell} 
\end{eqnarray*}
or, more succinctly,
\begin{equation}
\sum_{\sigma_\ell} A^{\sigma_\ell \dagger} A^{\sigma_\ell} = I . 
\end{equation}
Matrices that obey this condition we will refer to as {\em left-normalized}, matrix product states that consist only of left-normalized matrices we will call {\em left-canonical}. In fact, a closer look reveals that on the last site the condition may be technically violated, but as we will see once we calculate norms of MPS this corresponds to the original state not being normalized to 1. Let us ignore this subtlety for the moment.

In view of the DMRG decomposition of the universe into blocks A and B it is instructive to split the lattice into parts A and B, where A comprises sites $1$ through $\ell$ and B sites $\ell+1$ through $L$. We may then introduce states
\begin{eqnarray}
\ket{a_\ell}_A &=& \sum_{\sigma_1,\ldots,\sigma_\ell} (A^{\sigma_1} A^{\sigma_2} \ldots A^{\sigma_\ell})_{1,a_\ell} \ket{\sigma_1,\ldots,\sigma_\ell} \\
\ket{a_\ell}_B &=& \sum_{\sigma_{\ell+1},\ldots,\sigma_L} (A^{\sigma_{\ell+1}} A^{\sigma_{\ell+2}} \ldots A^{\sigma_L})_{a_\ell,1} \ket{\sigma_{\ell+1},\ldots,\sigma_L}
\end{eqnarray}
such that the MPS can be written as
\begin{equation}
\ket{\psi} = \sum_{a_\ell} \ket{a_\ell}_A \ket{a_\ell}_B . 
\end{equation}
This pairing of states looks tantalizingly close to a Schmidt decomposition of $\ket{\psi}$, but this is not the case. The reason for this is that while the $\{ \ket{a_\ell}_A \}$ form an orthonormal set, the $\{ \ket{a_\ell}_B \}$ in general do not. This is an immediate consequence of the left-normality of the $A$-matrices. For part A we find 
\begin{eqnarray*}
\phantom\langle_A \braket{a'_\ell}{a_\ell}_A &=& \sum_{\sigma_1,\ldots,\sigma_\ell} (A^{\sigma_1} \ldots A^{\sigma_\ell})^*_{1,a'_\ell} (A^{\sigma_1} \ldots A^{\sigma_\ell})_{1,a_\ell} \\
&=& \sum_{\sigma_1,\ldots,\sigma_\ell} (A^{\sigma_1} \ldots A^{\sigma_\ell})^\dagger_{a'_\ell,1} (A^{\sigma_1} \ldots A^{\sigma_\ell})_{1,a_\ell} \\
&=& \sum_{\sigma_1,\ldots,\sigma_\ell} (A^{\sigma_\ell\dagger} \ldots A^{\sigma_1\dagger}A^{\sigma_1} \ldots A^{\sigma_\ell})_{a'_\ell,a_\ell} \\
&=& \delta_{a'_\ell,a_\ell} ,
\end{eqnarray*}
where we have iteratively carried out the sums over $\sigma_1$ through $\sigma_\ell$ and used left-normality. On the other hand, the same calculation for part B yields
\begin{eqnarray*}
\phantom\langle_B \braket{a'_\ell}{a_\ell}_B &=& \sum_{\sigma_{\ell+1},\ldots,\sigma_L} (A^{\sigma_{\ell+1}} \ldots A^{\sigma_L})^*_{a'_\ell,1} (A^{\sigma_{\ell+1}} \ldots A^{\sigma_L})_{a_\ell,1} \\
&=& \sum_{\sigma_{\ell+1},\ldots,\sigma_L} (A^{\sigma_L\dagger} \ldots A^{\sigma_{\ell+1}\dagger})_{1,a'_\ell} (A^{\sigma_{\ell+1}} \ldots A^{\sigma_L})_{a_\ell,1} \\
&=& \sum_{\sigma_{\ell+1},\ldots,\sigma_L} (A^{\sigma_{\ell+1}} \ldots A^{\sigma_L}A^{\sigma_L\dagger} \ldots A^{\sigma_{\ell+1}\dagger})_{a'_\ell,a_\ell},
\end{eqnarray*}
which cannot be simplified further because in general $\sum_\sigma A^\sigma A^{\sigma\dagger} \neq I$.

\begin{figure}
\centering\includegraphics[scale=0.6]{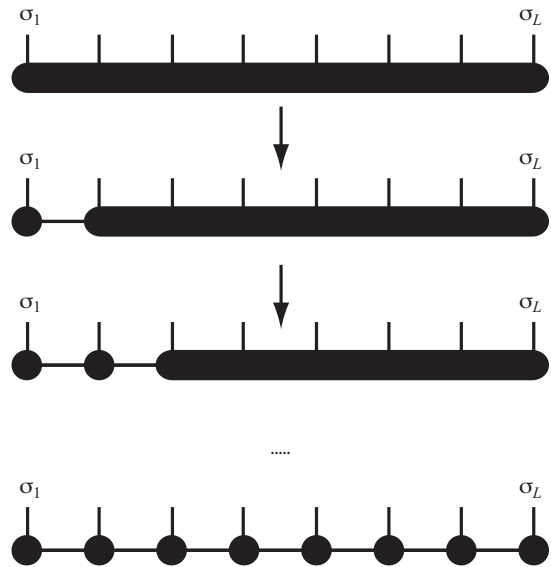}
\caption{Graphical representation of an iterative construction of an exact MPS representation of an arbitrary quantum state by a sequence of singular value decompositions.}
\label{fig:MPSBySVD}
\end{figure}

The change of representation of the state coefficients can also be represented graphically (Fig.~\ref{fig:MPSBySVD}). Let us represent the coefficient $c_{\sigma_1 \ldots \sigma_L}$ as a black box (with rounded edges), where the physical indices $\sigma_1$ through $\sigma_L$ stick out vertically. The result after the first decomposition we represent as in the second line, where we have on the left hand site an object representing $A^{\sigma_1}_{a_1}$, on the right $c_{a_1 \sigma_2 \ldots \sigma_L}$. The auxiliary degrees of freedom ($a_1$) are represented by horizontal lines, and the rule is that {\em connected lines are summed over.} The second step is then obvious, we have 
$A^{\sigma_1}_{a_1}$, then $A^{\sigma_2}_{a_1,a_2}$ and on the right $c_{a_2 \sigma_3 \ldots \sigma_L}$, with all connected lines summed over. In the end, we have arrived at $L$ $A$-matrices multiplied together and labelled by physical indices (last line of the figure).

\begin{figure}
\centering\includegraphics[scale=0.6]{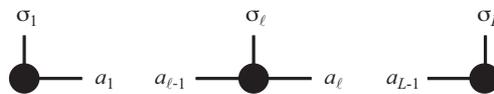}
\caption{Graphical representation of $A$-matrices at the ends and in the bulk of chains: the left diagram represents $A\ ^{\ \sigma_{1}}_{\ 1,a_1}$, the row vector at the left end, the right diagram represents $A\ ^{\sigma_{L}}_{a_L,1}$, the column vector at the right end. In the center there is $A\ ^{\sigma_{\ell}}_{a_{\ell-1},a_\ell}$.}
\label{fig:ABulkEdge}
\end{figure}

The graphical rules for the $A$-matrices, that on the first and last site are row and column vectors respectively, are summarized in Fig.~\ref{fig:ABulkEdge}: a site $\ell$ is represented by a solid circle, the physical index $\sigma_\ell$ by a vertical line and the two matrix indices by horizontal lines.

Let me conclude this exposition by showing the generation of a left-canonical matrix product state by a sequence of QR decompositions. We start as
\begin{equation}
c_{\sigma_1\ldots\sigma_L} = \Psi_{\sigma_1,(\sigma_2\ldots\sigma_L)} =
\sum_{a_1} Q_{\sigma_1,a_1} R_{a_1,(\sigma_2\ldots\sigma_L)} =
\sum_{a_1} A^{\sigma_1}_{1,a_1} \Psi_{(a_1\sigma_2),(\sigma_3\ldots\sigma_L)},
\end{equation}
where we reshape $Q\rightarrow A$ and $R\rightarrow \Psi$ in analogy to the SVD procedure. The next QR decomposition yields
\begin{equation}
c_{\sigma_1\ldots\sigma_L} = \sum_{a_1,a_2} A^{\sigma_1}_{1,a_1} Q_{(a_1\sigma_2),a_2} R_{a_2,(\sigma_3\ldots\sigma_L)} = \sum_{a_1,a_2} A^{\sigma_1}_{1,a_1} A^{\sigma_2}_{a_1,a_2} \Psi_{(a_2\sigma_3),(\sigma_4\ldots\sigma_L)}
\end{equation}
and so on (on the right half of the chain, thin QR is needed, as an analysis of the dimensions shows). $Q^\dagger Q=I$ implies the desired left-normalization of the $A$-matrices. If numerically feasible, this is faster than SVD. What we lose is that we do not see the spectrum of the singular values; unless we use more advanced rank-revealing QR decompositions, we are also not able to determine the ranks $r_1, r_2, \ldots$, unlike in SVD. This means that this decomposition fully exploits the maximal $A$-matrix dimensions.

{\em (ii) Right-canonical matrix product state.} Obviously, there was nothing specific in the decomposition starting from the left, i.e.\ site 1. Similarly, we can start from the right in order to obtain
\begin{eqnarray*}
c_{\sigma_1 \ldots \sigma_L} &=& \Psi_{(\sigma_1 \ldots \sigma_{L-1}), \sigma_L} \\
&=& \sum_{a_{L-1}} U_{(\sigma_1 \ldots \sigma_{L-1}), a_{L-1}} S_{a_{L-1},a_{L-1}} (V^\dagger)_{a_{L-1},\sigma_L} \\
&=& \sum_{a_{L-1}} \Psi_{(\sigma_1 \ldots \sigma_{L-2}),(\sigma_{L-1} a_{L-1})} B^{\sigma_L}_{a_{L-1}} \\
&=& \sum_{a_{L-2}, a_{L-1}} U_{(\sigma_1 \ldots \sigma_{L-2}), a_{L-2}} S_{a_{L-2},a_{L-2}} (V^\dagger)_{a_{L-2},(\sigma_{L-1}a_{L-1})} B^{\sigma_L}_{a_{L-1}} \\
&=& \sum_{a_{L-2}, a_{L-1}} \Psi_{(\sigma_1 \ldots \sigma_{L-3}), (\sigma_{L-2} a_{L-2})} B^{\sigma_{L-1}}_{a_{L-2},a_{L-1}} B^{\sigma_L}_{a_{L-1}} = \ldots \\
&=& \sum_{a_{1},\ldots, a_{L-1}} B^{\sigma_1}_{a_{1}} B^{\sigma_{2}}_{a_{1},a_{2}} \ldots B^{\sigma_{L-1}}_{a_{L-2},a_{L-1}} B^{\sigma_L}_{a_{L-1}}.
\end{eqnarray*}

\begin{figure}
\centering\includegraphics[scale=0.6]{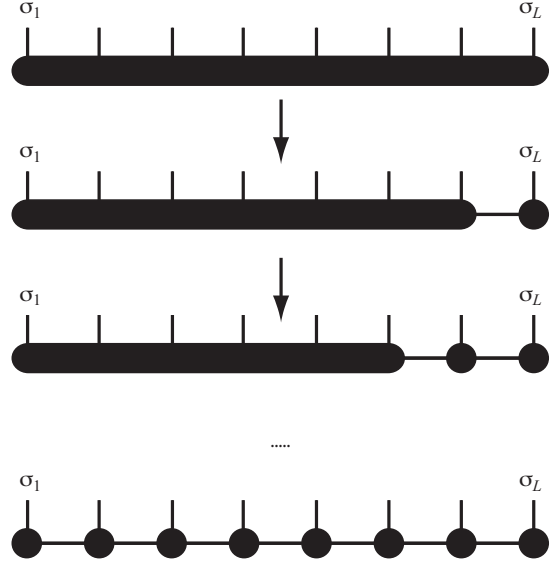}
\caption{Graphical representation of an iterative construction of an exact MPS representation of an arbitrary quantum state by a sequence of singular value decompositions, now starting from the right.}
\label{fig:rightMPSbySVD}
\end{figure}

Here, we have reshaped $(V^\dagger)_{a_{L-1},\sigma_L}$ into $d$ column vectors $B^{\sigma_L}_{a_{L-1}}$, $(V^\dagger)_{(a_{L-2}\sigma_{L-1}), a_{L-1}}$ into $d$ matrices $B^{\sigma_{L-1}}_{a_{L-2},a_{L-1}}$, and so on, as well as multiplied $U$ and $S$ before reshaping into $\Psi$ at each step. The obvious graphical representation is given in Fig.~\ref{fig:rightMPSbySVD}. We do not distinguish in the graphical representation between the $A$- and $B$-matrices to keep notation simple.

We obtain an MPS of the form
\begin{equation}
\ket{\psi} = \sum_{\sigma_1,\ldots,\sigma_L} B^{\sigma_1} B^{\sigma_2} \ldots B^{\sigma_{L-1}} B^{\sigma_L} \ket{\sigma_1,\ldots,\sigma_L}
\end{equation}
where the $B$-matrices can be shown to have the same matrix dimension bounds as the $A$ matrices and also, from $V^\dagger V =I$, to obey
\begin{equation}
\sum_{\sigma_\ell} B^{\sigma_\ell} B^{\sigma_\ell\dagger} = I ,
\end{equation}
such that we refer to them as {\em right-normalized} matrices. An MPS entirely built from such matrices we call {\em right-canonical}. 

Again, we can split the lattice into parts A and B, sites 1 through $\ell$ and $\ell+1$ through $L$, and introduce states 
\begin{eqnarray}
\ket{a_\ell}_A &=& \sum_{\sigma_1,\ldots,\sigma_\ell} (B^{\sigma_1} B^{\sigma_2} \ldots B^{\sigma_\ell})_{1,a_\ell} \ket{\sigma_1,\ldots,\sigma_\ell} \\
\ket{a_\ell}_B &=& \sum_{\sigma_{\ell+1},\ldots,\sigma_L} (B^{\sigma_{\ell+1}} B^{\sigma_{\ell+2}} \ldots B^{\sigma_L})_{a_\ell,1} \ket{\sigma_{\ell+1},\ldots,\sigma_L}
\end{eqnarray}
such that the MPS can be written as
\begin{equation}
\ket{\psi} = \sum_{a_\ell} \ket{a_\ell}_A \ket{a_\ell}_B . 
\end{equation}
This pairing of states looks again tantalizingly close to a Schmidt decomposition of $\ket{\psi}$, but this is again not the case. The reason for this is that while this time the $\{ \ket{a_\ell}_B \}$ form an orthonormal set, the $\{ \ket{a_\ell}_A \}$ in general do not, as can be shown from the right-normality of the $B$-matrices.

Again, the right-normalized form can be obtained by a sequence of QR decompositions. The difference to the left-normalized form is that we do not QR-decompose $\Psi=QR$, but $\Psi^\dagger=QR$, such that $\Psi = R^\dagger Q^\dagger$. This leads directly to the right-normalization properties of the $B$-matrices, if we form them from $Q^\dagger$. Let me make the first two steps explicit; we start from
\begin{equation}
c_{\sigma_1\ldots\sigma_L} = \Psi_{(\sigma_1\ldots\sigma_{L-1}),\sigma_L} =
\sum_{a_{L-1}} R^\dagger_{(\sigma_1\ldots\sigma_{L-1}),a_{L-1}} Q^\dagger_{a_{L-1},\sigma_L} =
\sum_{a_{L-1}} \Psi_{(\sigma_1\ldots\sigma_{L-2}),(\sigma_{L-1}a_{L-1})} B^{\sigma_L}_{a_{L-1},1},
\end{equation}
reshaping $R^\dagger$ into $\Psi$, $Q^\dagger$ into $B$, and continue by a QR decomposition of $\Psi^\dagger$ as 
\begin{equation}
c_{\sigma_1\ldots\sigma_L} = \sum_{a_{L-1},a_{L-2}} R^\dagger_{(\sigma_1\ldots\sigma_{L-2}),a_{L-2}} Q^\dagger_{a_{L-2},(\sigma_{L-1}a_{L-1})}  B^{\sigma_L}_{a_{L-1},1} 
= \sum_{a_{L-1},a_{L-2}} \Psi_{(\sigma_1\ldots\sigma_{L-3}),(\sigma_{L-2}a_{L-2})}
B^{\sigma_{L-1}}_{a_{L-2},a_{L-1}} B^{\sigma_L}_{a_{L-1},1} .
\end{equation}

We have now obtained various different exact representations of $\ket{\psi}$ in the MPS form, which already indicates that the MPS representation of a state is {\em not unique}, a fact that we are going to exploit later on.

{\em (iii) Mixed-canonical matrix product state.} We can also mix the decomposition of the state from the left and from the right. Let us assume we did a decomposition from the left up to site $\ell$, such that
\begin{equation}
c_{\sigma_1\ldots\sigma_L} = \sum_{a_\ell} (A^{\sigma_1} \ldots A^{\sigma_\ell})_{a_\ell} S_{a_\ell,a_\ell} (V^\dagger)_{a_\ell, (\sigma_{\ell+1} \ldots \sigma_L)} .
\end{equation}
We reshape $V^\dagger$ as $\Psi_{(a_\ell \sigma_{\ell+1} \ldots \sigma_{L-1}), \sigma_L}$ and carry out successive SVDs from the right as in the original decomposition from the right, up to and including site $\sigma_{\ell+2}$; in the last SVD $U_{(a_{\ell}\sigma_{\ell+1}),a_{\ell+1}} S_{a_{\ell+1},a_{\ell+1}}$ remains, which we reshape to $B^{\sigma_{\ell+1}}_{a_{\ell}a_{\ell+1}}$. Then we 
obtain
\begin{equation}
(V^\dagger)_{a_\ell, (\sigma_{\ell+1} \ldots \sigma_L)} = \sum_{a_{\ell+1},\ldots,a_{L-1}} B^{\sigma_{\ell+1}}_{a_\ell, a_{\ell+1}} \ldots B^{\sigma_L}_{a_{L-1}} .
\end{equation}
All $B$-matrices are right-normalized. This is simply due to the SVD for sites $\ell+2$ through $L$; on site $\ell+1$, it follows from the property $V^\dagger V = I$:
\begin{eqnarray*}
\delta_{a_\ell,a'_\ell} &=&
\sum_{\sigma_{\ell+1},\ldots} (V^\dagger)_{a_\ell, (\sigma_{\ell+1} \ldots \sigma_L)} V_{( \sigma_{\ell+1} \ldots \sigma_L),a'_\ell} \\
&=& ( \sum_{\sigma_{\ell+1},\ldots} B^{\sigma_{\ell+1}} \ldots B^{\sigma_L} B^{\sigma_L\dagger} \ldots B^{\sigma_{\ell+1}\dagger} )_{a_\ell,a'_\ell} \\
&=& ( \sum_{\sigma_{\ell+1}} B^{\sigma_{\ell+1}} B^{\sigma_{\ell+1}\dagger} )_{a_\ell,a'_\ell} ,
\end{eqnarray*}
where we use in the last line the right-normalization property of all the $B$-matrices on sites $\ell+2,\ldots,L$ to obtain the desired result.

\begin{figure}
\centering\includegraphics[scale=0.6]{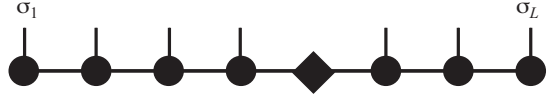}
\caption{Graphical representation of an exact MPS obtained by a sequence of singular value decompositions, starting from the left and right. The diamond represents the diagonal singular value matrix. Matrices to the left are left-normalized, to the right are right-normalized.}
\label{fig:bothMPSbySVD}
\end{figure}

We therefore end up with a decomposition
\begin{equation}
c_{\sigma_1 \ldots \sigma_L} = A^{\sigma_1} \ldots A^{\sigma_\ell} S B^{\sigma_{\ell+1}} \ldots B^{\sigma_L} ,
\end{equation}
which contains the singular values on the bond $(\ell,\ell+1)$ and can be graphically represented as in Fig.~\ref{fig:bothMPSbySVD}.

What is more, the Schmidt decomposition into A and B, where A runs from sites 1 to $\ell$ and B from sites $\ell+1$ to $L$, can now be read off immediately. If we introduce vectors
\begin{eqnarray}
\ket{a_\ell}_A &=& \sum_{\sigma_1,\ldots,\sigma_\ell}
(A^{\sigma_1} \ldots A^{\sigma_\ell})_{1,a_\ell}
 \ket{\sigma_1,\ldots,\sigma_\ell} \\
\ket{a_\ell}_B &=&  \sum_{\sigma_{\ell+1},\ldots,\sigma_L} 
(B^{\sigma_{\ell+1}} \ldots B^{\sigma_L})_{a_\ell,1}
\ket{\sigma_{\ell+1},\ldots,\sigma_L}
\end{eqnarray}
then the state takes the form ($s_a = S_{a,a}$)
\begin{equation}
\ket{\psi} = \sum_{a_\ell} s_a \ket{a_\ell}_A \ket{a_\ell}_B ,
\end{equation}
which is the Schmidt decomposition {\em provided the states on} A {\em and} B {\em are orthonormal respectively.} But this is indeed the case by construction.

(iv) {\em Gauge degrees of freedom.} By now, we have three different ways of writing an arbitrary quantum state as an MPS, all of which present advantages and disadvantages. While these three are arguably the most important ways of writing an MPS, it is important to realise that the degree of non-uniqueness is much higher: MPS are not unique  in the sense that a gauge degree of freedom exists. Consider two adjacent sets of matrices $M^{\sigma_i}$ and $M^{\sigma_{i+1}}$ of shared column/row dimension $D$. Then the MPS is invariant for any invertible matrix $X$ of dimension $(D\times D)$ under
\begin{equation}
M^{\sigma_i} \rightarrow M^{\sigma_i} X, \quad \quad M^{\sigma_{i+1}} \rightarrow X^{-1}M^{\sigma_{i+1}} .
\end{equation}
This gauge degree of freedom can be used to simplify manipulations drastically, our three constructions are just special cases of that.
 
Several questions arise. Is there a connection between this notation and more familiar concepts from many-body physics? Indeed, there is a profound connection to iterative decimation procedures as they occur in renormalization group schemes, which we will discuss in Section \ref{subsubsec:MPSsinglesitedecimation}. 

The matrices can potentially be exponentially large and we will have to bound their size on a computer to some $D$. Is this possible without becoming too inaccurate in the description of the state? Indeed this is possible in one dimension: if we consider the mixed-canonical representation, we see that for the exponentially decaying eigenvalue spectra of reduced density operators (hence exponentially decaying singular values $s_a$) it is possible to cut the spectrum following Eq.~(\ref{eq:Schmidtapproximation}) at the $D$ largest singular values (in the sense of an optimal approximation in the 2-norm) without appreciable loss of precision. This argument can be generalized from the approximation incurred by a single truncation to that incurred by $L-1$ truncations, one at each bond, to reveal that the error is at worst \cite{Verstraete06}
\begin{equation}
\| \ket{\psi} - \ket{\psi_{{\rm trunc}}} \|_2^2 \leq 2 \sum_{i=1}^L \epsilon_i(D) ,
\end{equation}
where $\epsilon_i(D)$ is the truncation error (sum of discarded squared singular values) at bond $i$ incurred by truncating down to the leading $D$ singular values. So the problem of approximability is as in DMRG related to the eigenvalue spectra of reduced density operators, indicating failure in two dimensions, and (a bit more tenuously) to the existence of area laws.

\subsubsection{MPS and single-site decimation in one dimension}
\label{subsubsec:MPSsinglesitedecimation}

\begin{figure}
\centering\includegraphics[scale=0.8]{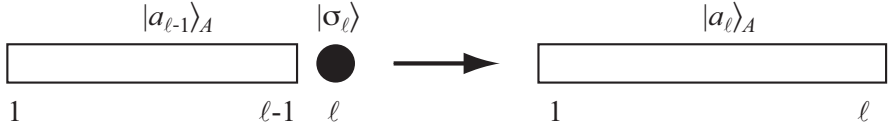}
\caption{A block of length $\ell-1$ is grown towards the right to a block of length $\ell$ by adding a site $\ell$.}
\label{fig:BlockGrowth}
\end{figure}

In order to connect MPS to more conventional concepts, let us imagine that we set up an iterative growth procedure for our spin chain, $\ell \rightarrow \ell + 1$, as illustrated in Fig.~\ref{fig:BlockGrowth}, such that associated state spaces grow by a factor of $d$ at each step. In order to avoid exponential growth, we now demand that state space dimensions have a ceiling of $D$. Once the state space dimension grows above $D$, the state space has to be truncated down by some {\em as of now undefined} procedure.

\begin{figure}
\centering\includegraphics[scale=0.8]{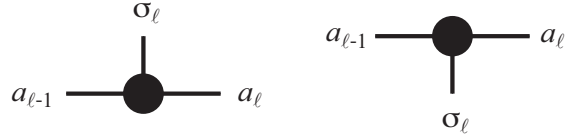}
\caption{Graphical representation of $A$-matrices: the left diagram represents $A\ ^{\sigma_{\ell}}_{a_{\ell-1},a_\ell}$, the right diagram the {\em conjugate} $A\ ^{\sigma_{\ell}*}_{a_{\ell-1},a_\ell}$. The solid circle represents the lattice sites, the vertical line the physical index, the horizontal lines the matrix indices.}
\label{fig:ABulk}
\end{figure}

Assume that somehow we have arrived at such a $D$-dimensional effective basis for our system (or left block A, in DMRG language) of length $\ell-1$, $\{ \ket{a_{\ell-1}}_A \}$. If the $D$ basis states of the (left) block A of length $\ell$ after truncation are $\{ \ket{a_{\ell}}_A \}$ and the local states of the added site  $\{ \ket{\sigma_\ell} \}$, we must have
\begin{equation}
\ket{a_\ell}_A = \sum_{a_{\ell-1}\sigma_\ell} \phantom\langle _A \braket{a_{\ell-1}\sigma_\ell}{a_\ell}_A \,\, \ket{a_{\ell-1}}_A\ket{\sigma_\ell}
\label{eq:basistrafoblocksite}
\end{equation}
for these states, with $\phantom\langle _A \braket{a_{\ell-1}\sigma_\ell}{a_\ell}_A$ as of now unspecified. We now introduce at site $\ell$ $d$ matrices $A^{[\ell]\sigma_\ell}$  of dimension $(D \times D)$ each, one for each possible local state $\ket{\sigma_\ell}$. We can then rewrite Eq.~(\ref{eq:basistrafoblocksite}) as
\begin{equation}
\ket{a_\ell}_A = \sum_{a_{\ell-1}\sigma_\ell} A^{[\ell]\sigma_\ell}_{a_{\ell-1}, a_\ell} \ket{a_{\ell-1}}_A  \ket{\sigma_\ell} 
\label{eq:matrixbyRG}
\end{equation}
where the elements of the matrices $A^{[\ell]\sigma_\ell}$ are given by (see Fig.~\ref{fig:ABulk})
\begin{equation}
A^{[\ell]\sigma_\ell}_{a_{\ell-1}, a_\ell} \equiv \phantom\langle_A \braket{a_{\ell-1}\sigma_\ell}{a_\ell}_A. 
\label{eq:Amatrixelements}
\end{equation}
Let us make a short remark on {\em notations} right here: in $A^{[\ell]\sigma_\ell}$, $[\ell]$ indicates which  set of $A$-matrices is considered, and $\sigma_\ell$ which $A$-matrix in particular. In the present case, this is a notational overkill, because the local state $\ket{\sigma_\ell}$ is taken from the site where the matrices were introduced. In such cases, we drop one of the two $\ell$, usually $[\ell]$:
\begin{equation}
A^{[\ell]\sigma_\ell} \rightarrow A^{\sigma_\ell} .
\end{equation}
We will, however, encounter situations where matrices $A$ are selected by local states {\em not} on the site where they were introduced. In such cases, the full notation obviously has to be restored! 

Similarly, we will shorten $\ket{a_\ell}_A \rightarrow \ket{a_\ell}$, when the fact that the state lives on block A is irrelevant or totally obvious. 

\begin{figure}
\centering\includegraphics[scale=0.8]{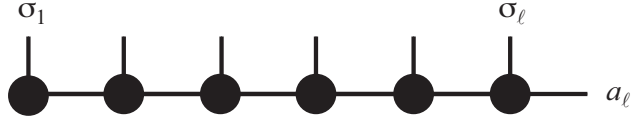}
\caption{Graphical representation of the recursive construction of a state $\ket{a{\,}_\ell}$ by contraction (multiplication) of $A$-matrices. Contractions run over all connected legs.}
\label{fig:BlockByMPS}
\end{figure}

The advantage of the matrix notation, which contains the decimation procedure yet unspecified, is that it allows for a simple {\em recursion} from a block of length $\ell$ to the smallest, i.e.\ vanishing block. Quantum states obtained in this way take a very special form:
\begin{eqnarray}
\ket{a_{\ell}}_A &=& \sum_{a_{\ell-1}} \sum_{\sigma_\ell} A^{\sigma_\ell}_{a_{\ell-1}, a_{\ell}} \ket{a_{\ell-1}}_A \ket{\sigma_{\ell}} \nonumber \\
&=& \sum_{a_{\ell-1},a_{\ell-2}} \sum_{\sigma_{\ell-1},\sigma_\ell} 
A^{\sigma_{\ell-1}}_{a_{\ell-2}, a_{\ell-1}} A^{\sigma_\ell}_{a_{\ell-1}, a_{\ell}} \ket{a_{\ell-2}}_A \ket{\sigma_{\ell-1}}\ket{\sigma_{\ell}} = \ldots \nonumber \\
&=& \sum_{a_1,a_2,\ldots,a_{\ell-1}} \sum_{\sigma_1,\sigma_2,\ldots,\sigma_\ell} A^{\sigma_1}_{1,a_1} A^{\sigma_2}_{a_1,a_2} \ldots A^{\sigma_\ell}_{a_{\ell-1},a_\ell} \ket{\sigma_1}\ket{\sigma_2} \ldots \ket{\sigma_\ell} \nonumber \\
&=& \sum_{\sigma_i \in {\rm A}}  (A^{\sigma_1}A^{\sigma_2}\ldots A^{\sigma_\ell})_{1,a_\ell}  \ket{\sigma_1}\ket{\sigma_2} \ldots \ket{\sigma_\ell} ,
\end{eqnarray}
where $i$ runs through all sites of block A. The index 'A' indicates that we are considering states on the left side of the chain we are building. On the first site, we have, in order to keep in line with the matrix notation, introduced a dummy row index 1: the states of block length 1 are built from the local states on site 1 and the block states of the ``block'' of length 0, for which we introduce a dummy state and index 1. This means that $A^{\sigma_1}$ is in fact a (row) vector (cf.\ Fig.~\ref{fig:ABulkEdge}). We also see that the left or row index of $A$ correspond to states ``left'' of those labelled by the right or column index. Quite generally, we can show this construction as in Fig.~\ref{fig:BlockByMPS}, if -- as before -- we introduce the rule that all connected legs are summed over (contracted). The advantage of the matrix notation is that we can hide summations in matrix multiplications.

\begin{figure}
\centering\includegraphics[width=\textwidth]{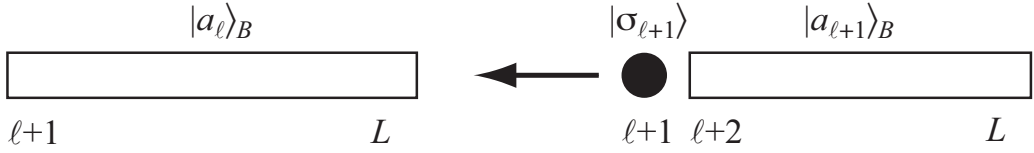}
\caption{A block B of length $L-\ell-1$ is grown towards the left to a block B of length $L-\ell$ by adding site $\ell+1$.}
\label{fig:BlockGrowth_Right}
\end{figure}

Similarly, we can build blocks to grow towards the left instead of to the right (Fig.~\ref{fig:BlockGrowth_Right}): we have
\begin{equation}
\ket{a_\ell}_B = \sum_{a_{\ell+1}\sigma_{\ell+1}} \phantom\langle_B \braket{a_{\ell+1} \sigma_{\ell+1}}{a_\ell}_B \,\, \ket{a_{\ell+1}}_B \ket{\sigma_{\ell+1}} 
\end{equation}
or 
\begin{equation}
\ket{a_\ell}_B = \sum_{a_{\ell+1}\sigma_{\ell+1}}B^{[\ell+1]\sigma_{\ell+1}}_{a_\ell, a_{\ell+1}}  \ket{a_{\ell+1}}_B \ket{\sigma_{\ell+1}} 
\end{equation}
with
\begin{equation}
B^{[\ell+1]\sigma_{\ell+1}}_{a_\ell, a_{\ell+1}} = \phantom\langle_B \braket{a_{\ell+1} \sigma_{\ell+1}}{a_\ell}_B .
\end{equation}
We call matrices $B$ to indicate that they emerge from a growth process towards the left, in DMRG language this would mean block B. Recursion gives
\begin{equation}
\ket{a_{\ell}}_B = \sum_{\sigma_i \in {\rm B}}  (B^{\sigma_{\ell+1}}B^{\sigma_{\ell+2}}\ldots B^{\sigma_L})_{a_{\ell+1},1}  \ket{\sigma_{\ell+1}}\ket{\sigma_{\ell+2}} \ldots \ket{\sigma_L}, 
\end{equation}   
where $i$ runs from $\ell+1$ to $L$, the sites of block B. A similar dummy index as for position 1 is introduced for position $L$, where the $B$-matrix is a (column) vector. 

\begin{figure}
\centering\includegraphics[scale=1.0]{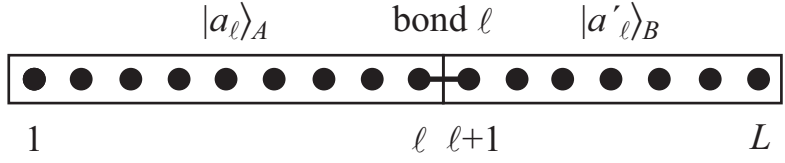}
\caption{Blocks A (sites 1 through $\ell$) and B (sites $\ell+1$ through $L$) are joined at bond $\ell$. States are labelled $\ket{a\ _{\ell}}_A$ and $\ket{a'_\ell}_B$.}
\label{fig:labelling}
\end{figure}

Note a slight asymmetry in the notation compared to the $A$-matrices: in order to be able to match blocks A and B later, we label block states according to the bond at which they terminate: bond $\ell$ connects sites $\ell$ and $\ell+1$, hence a labeling as in Fig.~\ref{fig:labelling}.

\begin{figure}
\centering\includegraphics[scale=1.0]{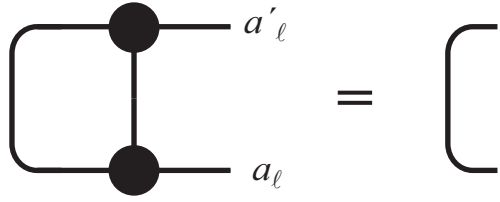}
\caption{If two {\em left}-normalized $A$-matrices are contracted over their {\em left} index and the physical indices, a $\delta\ _{a'_\ell,a_\ell}$ line results.}
\label{fig:LeftNormalization}
\end{figure}

If we introduce $A$-matrices and $B$-matrices in this way, they can be seen to have very special properties. If we consider the growth from the left, i.e.\ $A$-matrices, and demand reasonably that the chosen states should for each block length be orthonormal to each other, we have using Eq.~(\ref{eq:matrixbyRG})
\begin{eqnarray}
\delta_{a'_\ell, a_\ell} = \phantom\langle_A \braket{a'_\ell}{a_\ell}_A &=& \sum_{\sigma'_\ell,\sigma_\ell} \sum_{a'_{\ell-1},a_{\ell-1}} A^{\sigma'_\ell*}_{a'_{\ell-1}, a'_\ell} A^{\sigma_\ell}_{a_{\ell-1},a_\ell} 
 \phantom\langle_A\braket{a'_{\ell-1} \sigma'_\ell}{a_{\ell-1} \sigma_\ell}_A \\
&=& \sum_{\sigma_\ell} \sum_{a_{\ell-1}} A^{\sigma_\ell \dagger}_{a'_\ell,a_{\ell-1}} A^{\sigma_\ell}_{a_{\ell-1},a_\ell} = \sum_{\sigma_\ell}
(A^{\sigma_\ell \dagger}A^{\sigma_\ell})_{a'_{\ell},a_{\ell}} .
\end{eqnarray}
Summarizing we find that the $A$-matrices are {\em left-normalized}:
\begin{equation}
\sum_{\sigma} A^{\sigma\dagger} A^\sigma = I .
\end{equation}
A graphical representation is provided in Fig.~\ref{fig:LeftNormalization}: The multiplication can also be interpreted as the contraction of $A$ and $A^*$ over both $\sigma$ and their left index.

\begin{figure}
\centering\includegraphics[scale=1.0]{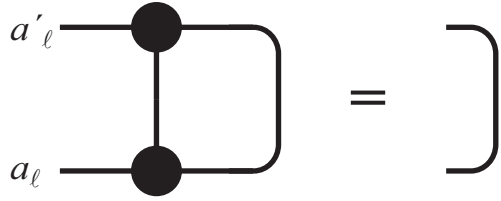}
\caption{If two {\em right}-normalized $B$-matrices are contracted over their {\em right} index and the physical indices, a $\delta\ _{a'_\ell,a_\ell}$ line results.}
\label{fig:RightNormalization}
\end{figure}

Similarly, we can derive for $B$-matrices of blocks B built from the right that the {\em right-normalization} identity
\begin{equation}
\sum_{\sigma} B^\sigma B^{\sigma\dagger} = I
\end{equation}
holds (usually, $A$ and $B$  will be used to distinguish the two cases). See Fig.~\ref{fig:RightNormalization}. This means that orthonormal states can always be decomposed into left- or right-normalized matrices in the MPS sense and that all states constructed from left- or right-normalized matrices form orthonormal sets, provided the type of normalization and the direction of the growth match.

Let us take a closer look at the matrix dimensions. Growing from the left, matrix dimensions go as $(1 \times d)$, $(d \times d^2)$, $(d^2 \times d^3)$, $(d^3 \times D)$, where I have assumed that $d^4 > D$. Then they continue at dimensions $(D \times D)$. At the right end, they will have dimensions  $(D \times D)$, $(D \times d^3)$,  $(d^3 \times d^2)$, $(d^2 \times d)$ and $(d \times 1)$. 

We can now again write down a matrix product state. Putting together a chain of length $L$ from a (left) block A of length $\ell$ (sites $1$ to $\ell$) and a (right) block B of length $L-\ell$ (sites $\ell+1$ to $L$), we can form a general superposition
\begin{equation}
\ket{\psi} = \sum_{a_\ell, a'_\ell} \Psi_{a_\ell, a'_\ell} \ket{a_\ell}_A \ket{a'_\ell}_B .
\end{equation}
Inserting the states explicitly, we find
\begin{equation}
\ket{\psi} = \sum_{\fat{\sigma}} (A^{\sigma_1} \ldots A^{\sigma_\ell})_{1,a_\ell} \Psi_{a_\ell, a'_\ell} 
 (B^{\sigma_{\ell+1}} \ldots B^{\sigma_L})_{a'_\ell,1} \ket{\fat{\sigma}} .
\end{equation}
The bold-faced $\fat{\sigma}$ stands for all local state indices, $\ket{\fat{\sigma}} =\ket{\sigma_1,\sigma_2,\ldots,\sigma_L}$. The notation suggests to interpret $\Psi$ as a matrix; then the notation simplifies to
\begin{equation}
\ket{\psi} = \sum_{\fat{\sigma}} A^{\sigma_1} \ldots A^{\sigma_\ell}\Psi
 B^{\sigma_{\ell+1}} \ldots B^{\sigma_L} \ket{\fat{\sigma}} .
 \label{eq:MPSwithPsi}
\end{equation}

\begin{figure}
\centering\includegraphics[scale=1.0]{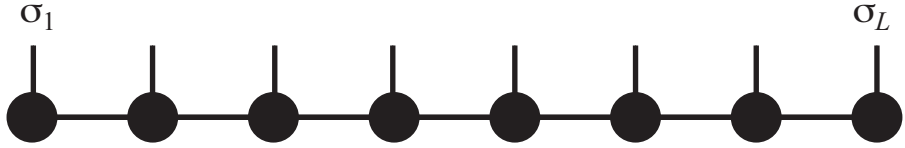}
\caption{Representation of an open boundary condition MPS.}
\label{fig:MPS_OBC}
\end{figure}

If we allow general matrices and don't worry about left, right or no normalization, we can simply multiply the $\Psi$-matrix into one of the adjacent $A$ or $B$ matrices, such that the general MPS for open boundary conditions appears (see Fig.~\ref{fig:MPS_OBC}):
\begin{equation}
\ket{\psi} = \sum_{\fat{\sigma}} M^{\sigma_1} \ldots M^{\sigma_L} \ket{\fat{\sigma}}  \quad{\rm (MPS \ for \ OBC)},
\label{eq:MPSOBC}
\end{equation}
where {\em no} assumption about the normalization is implied (which is why I call matrices $M$). Due to the vectorial nature of the first and last matrices the product results in a scalar. This is exactly the form of an MPS already discussed in the last section. 

\begin{figure}
\centering\includegraphics[scale=0.85]{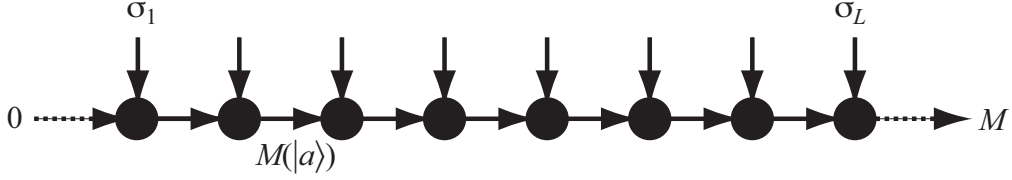}
\caption{Representation of an open boundary condition MPS with good (additive) quantum numbers. Physical states and bonds become directed, such that the quantum numbers on the ingoing lines equal those on the outgoing lines. For the dummy bonds before the first and after the last site we set suitable values to fix the global good quantum number.}
\label{fig:MPS_OBC_quantumnumbers}
\end{figure}

At this point it is easy to see how a matrix product state can exploit good quantum numbers. Let us focus on magnetization and assume that the global state has magnetization $M$. This Abelian quantum number is additive, $M = \sum_i M_i$. We choose local bases $\{ \sigma_i \}$ whose states are eigenstates of local magnetization. Consider now the growth process from the left. If we choose the states $\ket{a_1}$ to be eigenstates of local magnetization (e.g.\ by taking just the $\ket{\sigma_1}$), then Eq.~(\ref{eq:basistrafoblocksite}) allows us to construct by induction states $\ket{a_\ell}$ that are eigenstates of magnetization, provided the matrices $A^{\sigma_\ell}_{a_{\ell-1}, a_\ell}$ obtain a block structure such that for each non-zero matrix element 
\begin{equation}
M(\ket{a_{\ell-1}}) + M(\ket{\sigma_\ell}) = M(\ket{a_\ell})
\end{equation}
holds. This can be represented graphically easily by giving directions to the lines of the graphical representation (Fig.~\ref{fig:MPS_OBC_quantumnumbers}), with ingoing and outgoing arrows. The rule is then simply that the sum of the magnetizations on the ingoing lines equals that on the outgoing lines. In order to enforce some global magnetization $M$, we may simply give magnetization values 0 and $M$ to the ingoing and outgoing dummy bonds before the first and after the last site. We may envisage that the indices of the MPS matrices are multiindices for a given magnetization allowing degeneracy, leading to elegant coding representation. An inversion of the bond arrows would directly tie in with the structure of $B$-matrices from the growth from the right, but proper book-keeping gives us lots of freedom for the arrows: an inversion means that the sign has to be reversed.

In order to use good quantum numbers in practice, they have to survive under the typical operations we carry out on matrix product states. It turns out that all operations that are not obviously unproblematic and maintain good quantum numbers can be expressed by SVDs. An SVD will be applied to matrices like $A_{(a_{i-1}\sigma_i),a_i}$. If we group states $\ket{a_{i-1}\sigma_i}$ and $\ket{a_i}$ according to their good quantum number, $A$ will consist of blocks; if we rearrange labels appropriately, we can write $A = A_1 \oplus A_2 \oplus \ldots = U_1 S_1 V^\dagger_1 \oplus U_2 S_2 V^\dagger_2 \oplus \ldots$ or $A=USV^\dagger$ where $U= U_1 \oplus U_2 \oplus \ldots$ and so forth. But this means that the new states generated from $\ket{a_{i-1}\sigma_i}$ via $U$ will also have good quantum numbers. When the need for truncation arises, this property of course still holds for the retained states. If we replace SVD by QR where possible and carry it out on the individual blocks, $A_i =Q_i R_i$, the unitary matrices $Q_i$ transform within sets of states of the same quantum numbers, hence they remain good quantum numbers.

Let us now assume that our lattice obeys periodic boundary conditions. At the level of the state coefficients $c_{\sigma_1\ldots\sigma_L}$ there is no notion of the boundary conditions, hence our standard form of an MPS is capable to describe a state that reflects periodic boundary conditions. In that sense it is in fact wrong to say that Eq.~(\ref{eq:MPSOBC}) holds only for open boundary conditions. It is true in the sense that the anomalous structure of the matrices on the first and last sites is not convenient for periodic boundary conditions; indeed, the entanglement across the bond $(L,1)$ must be encoded as stretching through the entire chain. This leads to numerically very inefficient MPS representations.

\begin{figure}
\centering\includegraphics[scale=1.0]{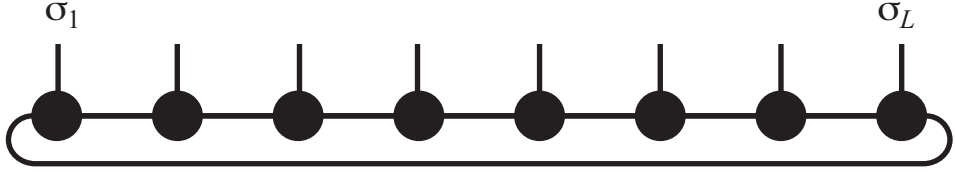}
\caption{Representation of a periodic boundary condition MPS; the long line at the bottom corresponds to the trace operation.}
\label{fig:MPS_PBC}
\end{figure}

For periodic boundary conditions the natural generalization of the MPS form is to make all matrices of equal dimensions $(D \times D)$; as site $L$ connects back to site $1$, we make the MPS consistent with matrix multiplications on all bonds by taking the trace (see Fig.~\ref{fig:MPS_PBC}):
\begin{equation}
\ket{\psi} = \sum_{\fat{\sigma}} \tr (M^{\sigma_1} \ldots M^{\sigma_L}) \ket{\fat{\sigma}}  \quad{\rm (MPS \ for \ PBC)}.
\label{eq:MPSforPBC}
\end{equation}
While {\em a priori} not more accurate than the other, it is much better suited and computationally  far more efficient.

In this section, our emphasis has been on approximate representations of quantum states rather than on usually unachievable exact representations. While we have no prescription yet how to construct these approximate representations, some remarks are in order.

Even an approximate MPS is still a linear combination of all states of the Hilbert space, no product basis state has been discarded. The limiting constraint is rather on the form of the linear combinations: instead of $d^L$ coefficients, $dL$ matrices of dimension $(D\times D)$ with a matrix-valued normalization constraint that gives $LD^2$ scalar constraints have $(d-1)LD^2$ independent parameters only, generating interdependencies of the coefficients of the state.
 
The quality of the optimal approximation of any quantum state for given matrix dimensions will improve monotonically with $D$: take $D<D'$, then the best approximation possible for $D$ can be written as an MPS with $D'$ with $(D \times D)$ submatrices in the $(D' \times D')$ matrices and all additional rows and columns zero. They give further parameters for improvement of the state approximation.

Product states (with Schmidt rank 1 for any Schmidt decomposition) can be written exactly using $D=1$ MPS. Real quantum physics with entangled states starts at $D=2$. Given the exponential number of coefficients in a quantum state, it may be a surprise that even in this simplest non-trivial case interesting quantum physics can be done exactly! But there are important quantum states that find a compact exact expression in this new format.

\subsubsection{The AKLT state as a matrix product state} 
\begin{figure}
\centering\includegraphics[scale=1.0]{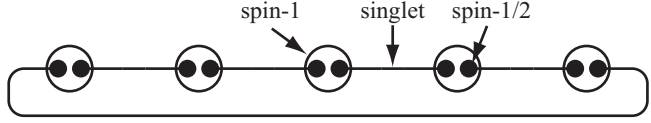}
\caption{The Affleck-Kennedy-Lieb-Tasaki (AKLT) state is built from expressing the local spin-1 as two totally symmetrized spin-$\frac{1}{2}$ particles which are linked across sites by singlet states. The picture shows the case of PBC, for OBC the ``long'' bond is cut, and two single spins-$\frac{1}{2}$ appear at the ends.}
\label{fig:akltstate}
\end{figure}

In order to make the MPS framework less abstract, let us construct the MPS representation of a non-trivial quantum state. One of the most interesting quantum states in correlation physics is the {\em Affleck-Kennedy-Lieb-Tasaki state} introduced in 1987, which is the ground state of  the AKLT Hamiltonian\cite{Affleck87,Affleck88}
\begin{equation}
\ham = \sum_{i} {\bf S}_i \cdot {\bf S}_{i+1} + 
\frac{1}{3} ({\bf S}_i \cdot {\bf S}_{i+1})^2 , 
\label{eq:aklt_hamiltonian}
\end{equation}
where we deal, exceptionally in this paper, with $S=1$ spins. It can be shown that the ground state of this Hamiltonian can be constructed as shown in Fig.~\ref{fig:akltstate}. Each individual spin-1 is replaced by a pair of spin-$\frac{1}{2}$ which are completely symmetrized, i.e. of the four possible states we consider only the three triplet states naturally identified as $S=1$ states:
\begin{eqnarray}
\ket{+}&=&\ket{\uparrow\uparrow} \nonumber \\
\ket{0}&=&\frac{\ket{\uparrow\downarrow}+\ket{\downarrow\uparrow}}{\sqrt{2}} 
\label{eq:tripletmapping} \\
\ket{-}&=&\ket{\downarrow\downarrow} \nonumber
\end{eqnarray}
On neighbouring sites, adjacent pairs of spin-$\frac{1}{2}$ are linked in a singlet state
\begin{equation}
\frac{\ket{\uparrow\downarrow}-\ket{\downarrow\uparrow}}{\sqrt{2}} .
\end{equation}

As it turns out, this state can be encoded by a matrix product state of the lowest non-trivial dimension $D=2$ and contains already lots of exciting physics \cite{Affleck87,Affleck88,Fannes89}. In the language of the auxiliary $2\syslength$ spin-$\frac{1}{2}$ states on a chain of length $\syslength$ any state is given as
\begin{equation}
    \ket{\psi} = \sum_{{\bf a}}\sum_{{\bf b}} c_{{\bf ab}} \ket{{\bf 
    ab}}
\end{equation}
with $\ket{{\bf a}}=\ket{a_1,\ldots,a_\syslength}$ and $\ket{{\bf b}}=\ket{b_1,\ldots,b_\syslength}$ representing the first and second spin-$\frac{1}{2}$ on each site. We now encode the singlet bond $\Sigma_i$ on bond $i$ connecting sites $i$ and $i+1$ as 
\begin{equation}
\ket{\Sigma^{[i]}} = \sum_{b_i a_{i+1}} \Sigma_{ba} \ket{b_i}\ket{a_{i+1}}
\end{equation}
introducing a $2 \times 2$ matrix $\Sigma$
\begin{equation}
\Sigma = \left[ 
\begin{array}{cc} 0 & \frac{1}{\sqrt{2}} \\ -\frac{1}{\sqrt{2}} & 0 \end{array}
\right] .
\end{equation}
Then the state with singlets on all bonds reads
\begin{equation}
\ket{\psi_\Sigma} = \sum_{{\bf a}}\sum_{{\bf b}} \Sigma_{b_1 a_2} \Sigma_{b_2 a_3} \ldots \Sigma_{b_{\syslength-1} a_\syslength}\Sigma_{b_\syslength a_1} \ket{{\bf 
    ab}} 
\end{equation}
for periodic boundary conditions. If we consider open boundary conditions, $\Sigma_{b_\syslength a_1}$ is omitted and the first and last spin-$\frac{1}{2}$ remain single.

Note that this state is a product state factorizing upon splitting any site $i$ into its two constituents. We now encode the identification of the symmetrized states of the auxiliary spins with the physical spin by introducing a {\em mapping} from the 
states of the two auxiliary spins-$\frac{1}{2}$, $\ket{a_i}\ket{b_i} \in \{ 
\ket{\uparrow},\ket{\downarrow}\}^{\otimes 2}$ to the states of the 
physical spin-1, $\ket{\sigma_i} \in \{ \ket{+},\ket{0},\ket{-} \}$. To represent \Eq{eq:tripletmapping}, we introduce $M_{ab}^{\sigma} 
\ket{\sigma} \bra{ab}$, with $\ket{ab}$ and $\ket{\sigma}$ representing the auxiliary spins and the physical spin on site $i$. Writing $M_{ab}^{\sigma}$ as three $2\times 2$  matrices, one for each value of $\ket{\sigma}$, with rows and column indices standing for the values of $\ket{a}$ and $\ket{b}$, we find
\begin{equation}
    M^+ = \left[ \begin{array}{cc} 1 & 0 \\ 0 & 0 \end{array} \right] \quad
    M^0 = \left[ \begin{array}{cc} 0 & \frac{1}{\sqrt{2}} \\ 
    \frac{1}{\sqrt{2}} & 0 \end{array} \right] \quad
    M^- = \left[ \begin{array}{cc} 0 & 0 \\ 0 & 1 \end{array} \right] .
\end{equation}
The mapping on the spin-1 chain Hilbert space $\{ \ket{\fat{\sigma}} \}$ then reads
\begin{equation} 
\sum_{\fat{\sigma}} \sum_{{\bf a}{\bf b}} M^{\sigma_1}_{a_1 b_1} M^{\sigma_2}_{a_2 b_2} \ldots M^{\sigma_\syslength}_{a_\syslength b_\syslength}
\ket{\fat{\sigma}} \bra{{\bf a}{\bf b}} .
\end{equation}
$\ket{\psi_\Sigma}$ therefore is mapped to 
\begin{equation}
\sum_{\fat{\sigma}} \sum_{{\bf a}{\bf b}} M^{\sigma_1}_{a_1 b_1} \Sigma_{b_1 a_2}
M^{\sigma_2}_{a_2 b_2} \Sigma_{b_2 a_3} \ldots \Sigma_{b_{\syslength-1} a_\syslength} 
M^{\sigma_\syslength}_{a_\syslength b_\syslength} \Sigma_{b_\syslength a_1} 
\ket{\fat{\sigma}}
\end{equation}
or
\begin{equation}
    \ket{\psi} = \sum_{\fat{\sigma}} \tr (M^{\sigma_{1}} \Sigma
    M^{\sigma_{2}} \Sigma \ldots M^{\sigma_{\syslength}} \Sigma] ) \ket{\fat{\sigma}} ,
\end{equation}
using the matrix notation. To simplify further, we introduce $\tilde{A}^\sigma= M^\sigma \Sigma$, such that
\begin{equation}
    \tilde{A}^+ = \left[ \begin{array}{cc} 0 & \frac{1}{\sqrt{2}} \\ 0 & 0 \end{array} \right] \quad
    \tilde{A}^0 = \left[ \begin{array}{cc} -\frac{1}{2} & 0 \\ 
    0 & +\frac{1}{2} \end{array} \right] \quad
    \tilde{A}^- = \left[ \begin{array}{cc} 0 & 0 \\ -\frac{1}{\sqrt{2}} & 0 \end{array} \right] .
    \label{eq:unnormalizeda}
\end{equation}
The AKLT state now takes the form 
\begin{equation}
    \ket{\psi} = \sum_{\fat{\sigma}} \tr (\tilde{A}^{\sigma_{1}} 
    \tilde{A}^{\sigma_{2}} \ldots \tilde{A}^{\sigma_{\syslength}}) \ket{\fat{\sigma}} .
    \label{eq:akltstate}
\end{equation}
Let us left-normalize the $\tilde{A}^\sigma$. $\sum_\sigma \tilde{A}^{\sigma\dagger}\tilde{A}^\sigma = \frac{3}{4}I$, which implies that the matrices $\tilde{A}$  should 
be rescaled by $\frac{2}{\sqrt{3}}$, such that we obtain normalized matrices $A$,
\begin{equation}
    A^+ = \left[ \begin{array}{cc} 0 & \sqrt{\frac{2}{3}} \\ 0 & 0 \end{array} \right] \quad
    A^0 = \left[ \begin{array}{cc} -\frac{1}{\sqrt{3}} & 0 \\ 
    0 & \frac{1}{\sqrt{3}} \end{array} \right] \quad
    A^- = \left[ \begin{array}{cc} 0 & 0 \\ -\sqrt{\frac{2}{3}} & 0 \end{array} \right] .
    \label{eq:normalizeda}
\end{equation}
This normalizes the state in the thermodynamic limit: we have
\begin{eqnarray*}
\braket{\psi}{\psi} &=& \sum_{\fat{\sigma}} \tr (A^{\sigma_1} \ldots A^{\sigma_\syslength})^* \tr (A^{\sigma_1} \ldots A^{\sigma_\syslength}) \\
&=&  \tr \left( \sum_{\sigma_1} A^{\sigma_1*} \otimes A^{\sigma_1}\right) \ldots \left( \sum_{\sigma_L} A^{\sigma_\syslength*}\otimes A^{\sigma_\syslength} \right) \\
&=& \tr E^L = \sum_{i=1}^4 \lambda_i^L .
\end{eqnarray*}
In this expression, the $\lambda_i$ are the 4 eigenvalues of 
\begin{equation}
E = \sum_\sigma A^{\sigma*} \otimes A^{\sigma} = \left[ \begin{array}{cccc} 
\frac{1}{4} & 0 & 0 & \frac{1}{2} \\
0 & -\frac{1}{4} & 0 & 0 \\
0 & 0 & -\frac{1}{4} & 0 \\
\frac{1}{2} & 0 & 0 & \frac{1}{4}
\end{array} \right] ,
\end{equation}
namely $1,-\frac{1}{3},-\frac{1}{3},-\frac{1}{3}$. But then $\braket{\psi}{\psi} = 1 + 3(-1/3)^L \rightarrow 1$ for $L\rightarrow\infty$.

The methods of the next section can now be used to work out analytically the correlators of the AKLT state:  antiferromagnetic correlators are decaying exponentially, $\langle S^z_i S^z_j \rangle \propto (-1/3)^{i-j}$, whereas the string correlator $\langle S^z_i \eul^{\imag \pi\sum_{i<k<j} S^z_k}S^z_j \rangle =-4/9$ for $j-i>2$, indicating hidden order.

To summarize, it has been possible to express the AKLT state as a $D=2$ matrix product state, the simplest non-trivial MPS! In fact, the projection from the larger state space of the auxiliary spins which are linked by maximally entangled states (here: singlets) onto the smaller physical state space can also be made the starting point for the introduction of MPS and their higher-dimensional generalizations\cite{VerstraetePorras04,VerstraeteCirac04}. 

\subsection{Overlaps, expectation values and matrix elements}

Let us now turn to operations with MPS, beginning with the calculation of expectation values.
Expectation values are obviously special cases of general matrix elements, where states $\ket{\psi}$ and $\ket{\phi}$ are identical. Staying in the general case, let us consider an overlap between states $\ket{\psi}$ and $\ket{\phi}$, described by matrices $M$ and $\tilde{M}$, and focus on open boundary conditions. 

\begin{figure}
\centering\includegraphics[width=\textwidth]{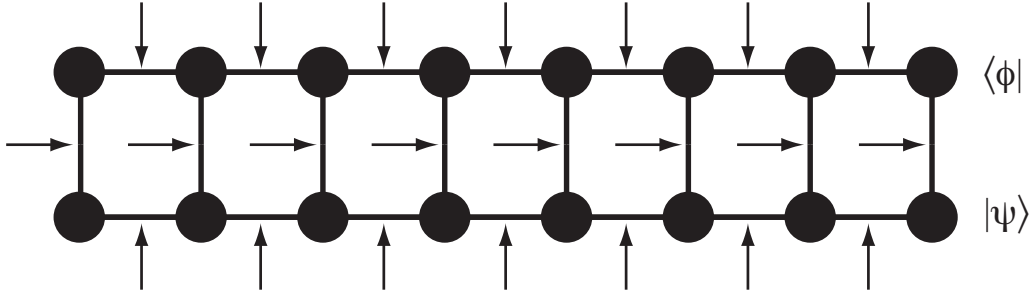}
\caption{Overlap between two states $\ket{\phi}$ and $\ket{\psi}$. All contractions (sums) over same indices are indicated by arrows.}
\label{fig:overlap}
\end{figure}

Taking the adjoint of $\ket{\phi}$, and considering that the wave function coefficients are scalars, the overlap reads
\begin{equation}
\braket{\phi}{\psi} = \sum_{\fat{\sigma}} \tilde{M}^{\sigma_1*} \ldots \tilde{M}^{\sigma_L*} M^{\sigma_1} \ldots M^{\sigma_L} .
\end{equation}
Transposing the scalar formed from the $\tilde{M} \ldots \tilde{M}$ (which is the identity operation), we arrive at adjoints with reversed ordering:
\begin{equation}
\braket{\phi}{\psi} = \sum_{\fat{\sigma}} \tilde{M}^{\sigma_L\dagger} \ldots \tilde{M}^{\sigma_1\dagger} M^{\sigma_1} \ldots M^{\sigma_L} .
\label{eq:finaloverlap}
\end{equation}
In a pictorial representation (Fig.~\ref{fig:overlap}), this calculation becomes much simpler, if we follow the rule that all bond indices are summed over.

\subsubsection{Efficient evaluation of contractions}

\begin{figure}
\centering\includegraphics[width=\textwidth]{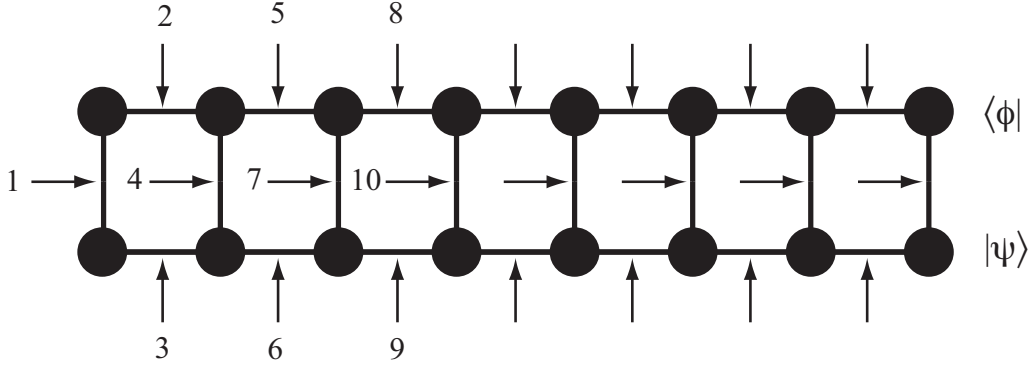}
\caption{Overlap between two states $\ket{\phi}$ and $\ket{\psi}$ with indication of the optimal sequence of contractions, running like a zipper through the chain.}
\label{fig:overlaporder}
\end{figure}

Evaluating expression (\ref{eq:finaloverlap}) in detail shows the importance of finding the {\em right (optimal) order of contractions} in matrix or more generally tensor networks. We have contractions over the matrix indices implicit in the matrix multiplications, and over the physical indices. If we decided to contract first the matrix indices and then the physical indices, we would have to sum over $d^L$ strings of matrix multiplications, which is exponentially expensive. But we may regroup the sums as follows:
\begin{equation}
\braket{\phi}{\psi} = \sum_{\sigma_L} 
\tilde{M}^{\sigma_L\dagger} \left( 
\ldots \left( 
\sum_{\sigma_2}  \tilde{M}^{\sigma_2 \dagger} \left(
\sum_{\sigma_1} \tilde{M}^{\sigma_1\dagger} M^{\sigma_1}\right) 
M^{\sigma_2}\right) 
 \ldots  \right) 
 M^{\sigma_L} .
\end{equation}
This means, in the (special) first step we multiply the column and row vectors $\tilde{M}^{\sigma_1\dagger}$ and $M^{\sigma_1}$ to form a matrix and sum over the (first) physical index. In the next step, we contract a three-matrix multiplication over the second physical index, and so forth (Fig.~\ref{fig:overlaporder}). The important observation is that from the second step onwards the complexity does not grow anymore. Also, it is of course most efficient to decompose matrix multiplications $ABC$ as $(AB)C$ or $A(BC)$. Then we are carrying out $(2L-1)d$ multiplications, each of which is of complexity $O(D^3)$, ignoring for simplicity that matrices are non-square in general at the moment. The decisive point is that we go from exponential to weak polynomial complexity, with total operation count $O(LD^3 d)$.

\begin{figure}
\centering\includegraphics[width=\textwidth]{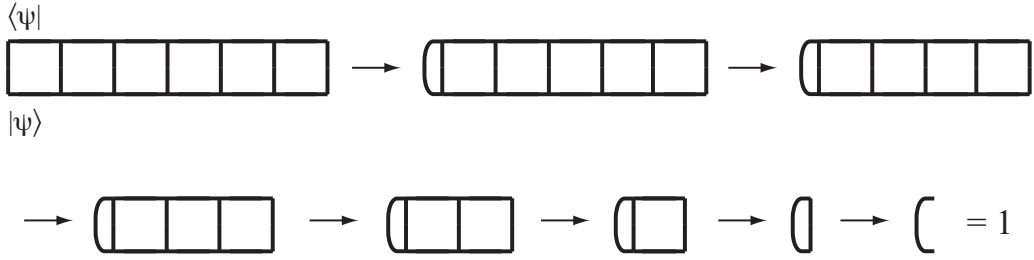}
\caption{Steps of a norm calculation for a {\em left normalized} state $\ket{\psi}$ by subsequent applications of the contraction rule for left-normalized $A$-matrices.}
\label{fig:overlaplnormalized}
\end{figure}

What is also immediately obvious, is that for a norm calculation $\braket{\psi}{\psi}$ and OBC having a state in left- or right-normalized form immediately implies that it has norm 1. In the calculation above it can be seen that for left-normalized matrices $A$, the innermost sum is just the left-normalization condition, yielding $I$, so it drops out, and the next left-normalization condition shows up, until we are through the chain (Fig.~\ref{fig:overlaplnormalized}):
\begin{eqnarray*}
\braket{\psi}{\psi} &=& \sum_{\sigma_L} A^{\sigma_L\dagger} \left( \ldots \left( \sum_{\sigma_2}  A^{\sigma_2 \dagger} \left(\sum_{\sigma_1} A^{\sigma_1\dagger} A^{\sigma_1}\right) A^{\sigma_2}\right)  \ldots  \right) A^{\sigma_L} \\
&=&  \sum_{\sigma_L} A^{\sigma_L\dagger} \left( \ldots \left( \sum_{\sigma_2}  A^{\sigma_2 \dagger}  A^{\sigma_2}\right)  \ldots  \right) A^{\sigma_L} = \ldots \\
&=& \sum_{\sigma_L} A^{\sigma_L\dagger} A^{\sigma_L} = 1.
\end{eqnarray*}

\begin{figure}
\centering\includegraphics[scale=1.0]{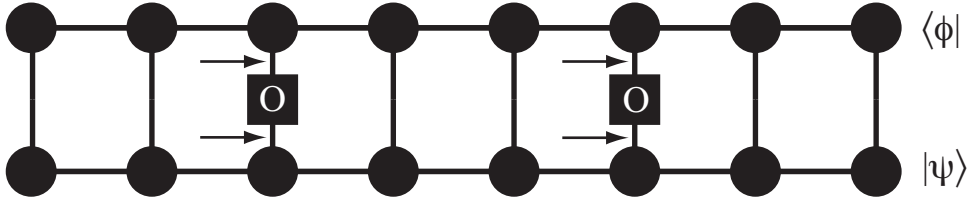}
\caption{Matrix elements between two states $\ket{\phi}$ and $\ket{\psi}$ are calculated like the overlap, with the operators inserted at the right places, generating a double sum of physical indices there, as indicated by the arrows.}
\label{fig:matrixelement}
\end{figure}

To calculate general matrix elements, we consider $\bra{\phi} \hat{O}^{[i]} \hat{O}^{[j]} \ldots \ket{\psi}$, tensored operators acting on sites $i$ and $j$. The matrix elements of such operators are taken from 
\begin{equation}
 \hat{O}^{[\ell]} = \sum_{\sigma_\ell,\sigma'_\ell} O^{\sigma_\ell,\sigma'_\ell} \ket{\sigma_\ell} \bra{\sigma'_\ell} .
\end{equation}
Let us extend this to an operator on every site, which in practice will be the identity on almost all sites, e.g. for local expectation values or two-site correlators. We are therefore considering operator matrix elements $O^{\sigma_1,\sigma'_1} O^{\sigma_2,\sigma'_2} \cdots  O^{\sigma_L,\sigma'_L}$.
In the analytical expression, we again transpose and distribute the (now double) sum over local states (matrix multiplications for the $M$-matrices are as before): 
\begin{eqnarray*}
& & \sum_{\fat{\sigma},\fat{\sigma}'} \tilde{M}^{\sigma_1*} \ldots \tilde{M}^{\sigma_L*} 
O^{\sigma_1,\sigma'_1} O^{\sigma_2,\sigma'_2} \cdots  O^{\sigma_L,\sigma'_L}
M^{\sigma'_1} \ldots M^{\sigma'_L} \\
&=& \sum_{\sigma_L,\sigma'_L} O^{\sigma_L,\sigma'_L} \tilde{M}^{\sigma_L\dagger}  \left( \ldots \left( \sum_{\sigma_2,\sigma'_2}  O^{\sigma_2,\sigma'_2} \tilde{M}^{\sigma_2 \dagger} \left(\sum_{\sigma_1,\sigma'_1} O^{\sigma_1,\sigma'_1} \tilde{M}^{\sigma_1\dagger} M^{\sigma'_1}\right) M^{\sigma'_2}\right)  \ldots  \right) M^{\sigma'_L}
\end{eqnarray*}
This amounts to the same calculation as for the overlap, with the exception that formally the single sum over the physical index turns into a double sum (Fig.~\ref{fig:matrixelement}). For typical correlators the double sum will trivially reduce to a single sum on most sites, as for most sites only the identity acts, $\hat{O}^{[i]}=\hat{I}$; on the few non-trivial sites, of the up to $d^2$ matrix elements, most will be zero for conventional operators, strongly restricting the number of terms, so essentially the operational count is $O(LD^3 d)$ again.

\begin{figure}
\centering\includegraphics[scale=1.0]{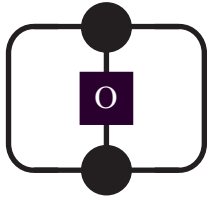}
\caption{$\bra{\psi} \hat{O}^{[\ell]} \ket{\psi}$ for a state $\ket{\psi}$ with left- and right-normalized matrices to the left and right of site $\ell$.}
\label{fig:localoverlap}
\end{figure}

Important simplifications for expectation values $\bra{\psi} \hat{O}^{[\ell]} \ket{\psi}$ are feasible and should be exploited whenever possible: Assume that we look at a local operator $\hat{O}^{[\ell]}$ and that normalizations are such that to the left of site $\ell$ all matrices are left-normalized and to the right of site $\ell$ all matrices are right-normalized; the status of site $\ell$ itself is arbitrary. Then left- and right-normalization can be used to contract the network as in Fig.~\ref{fig:overlaplnormalized} without explicit calculation, such that just two matrices remain (Fig.~\ref{fig:localoverlap}). The remaining calculation is just
\begin{equation}
\bra{\psi} \hat{O}^{[\ell]} \ket{\psi} = \sum_{\sigma_\ell \sigma'_\ell} O^{\sigma_\ell \sigma'_\ell} \tr
(M^{\sigma_\ell \dagger} M^{\sigma'_\ell}),
\end{equation} 
an operation of order $O(D^2 d^2)$, saving one order of $D$ in calculation time. As we will encounter algorithms where the state is in such mixed-canonical representation, it makes sense to calculate observables ``on the sweep''. This is just identical to expectation values on the explicit sites of DMRG.
    
A frequently used notation for the calculation of overlaps, expectation values and matrix elements is provided by reading the hierarchy of brackets as an iterative update of a matrix, which eventually gives the scalar result. We now introduce matrices $C^{[\ell]}$, where $C^{[0]}$ is a dummy matrix, being the scalar 1. Then an overlap $\braket{\phi}{\psi}$ can be carried out iteratively by running $\ell$ from 1 through $L$:
\begin{equation}
C^{[\ell]} = \sum_{\sigma_\ell} \tilde{M}^{\sigma_{\ell}\dagger} C^{[\ell-1]} M^{\sigma_{\ell}} ,
\end{equation}
where $C^{[L]}$ will be a scalar again, containing the result. For operators taken between the two states, the natural extension of this approach is 
\begin{equation}
C^{[\ell]} = \sum_{\sigma_\ell,\sigma'_\ell} O^{\sigma_\ell,\sigma'_\ell} \tilde{M}^{\sigma_{\ell}\dagger} C^{[\ell-1]} M^{\sigma'_{\ell}} .
\end{equation}
Again, the right order of evaluating the matrix products makes a huge difference in efficiency:
\begin{equation}
C^{[\ell]}_{a_\ell,a'_\ell} = \sum_{\sigma_\ell,a_{\ell-1}}  \tilde{M}^{\sigma_{\ell}*}_{a_{\ell-1},a_\ell} \left( \sum_{\sigma'_\ell} O^{\sigma_\ell,\sigma'_\ell} 
\left( \sum_{a'_{\ell-1}} C^{[\ell-1]}_{a_{\ell-1},a'_{\ell-1}} 
M^{\sigma'_{\ell}}_{a'_{\ell-1},a'_\ell} \right) \right) 
\end{equation}
reduces an operation $O(D^4 d^2)$ to $O(D^3 d) + O(D^2d^2) + O(D^3 d)$.

Of course, we can also proceed from the right end, introducing matrices $D^{[\ell]}$, starting with
$D^{[L]}=1$, a scalar. To this purpose, we exchange the order of scalars in $\braket{\phi}{\psi}$,
\begin{equation}
\braket{\phi}{\psi} = \sum_{\fat{\sigma}}M^{\sigma_1} \ldots M^{\sigma_L}  \tilde{M}^{\sigma_1*} \ldots \tilde{M}^{\sigma_L*}  ,
\end{equation}
and transpose again the $\tilde{M}$-matrices, leading to a hierarchy of bracketed sums, with the sum over $\sigma_L$ innermost. The iteration running from $\ell=L$ to $\ell=1$ then reads:
\begin{equation}
D^{[\ell-1]} = \sum_{\sigma_\ell}  M^{\sigma_{\ell}} D^{[\ell]} \tilde{M}^{\sigma_{\ell}\dagger} ,
\end{equation}
which can be extended to the calculation of matrix elements as
\begin{equation}
D^{[\ell-1]} = \sum_{\sigma_\ell,\sigma'_\ell}  O^{\sigma_\ell,\sigma'_\ell} M^{\sigma'_{\ell}} D^{[\ell]} \tilde{M}^{\sigma_{\ell}\dagger} .
\end{equation}
$D^{[0]}$ then contains the result.

This approach is very useful for book-keeping, because in DMRG we need operator matrix elements for left and right blocks, which is just the content of the $C$- and $D$-matrices for blocks A and B. As blocks grow iteratively, the above sequence of matrices will be conveniently generated along with block growth. 

\subsubsection{Transfer operator and correlation structures}
\label{subsubsec:transferoperator}
Let us formalize the iterative construction of $C^{[\ell]}$-matrices of the last section a bit more, because it is useful for the understanding of the nature of correlations in MPS to introduce a transfer (super)operator $\hat{E}^{[\ell]}$,  which is a mapping from operators defined on block A with length $\ell-1$ to operators defined on block A with length $\ell$,
\begin{equation}
\{ \ket{a_{\ell-1}} \bra{a'_{\ell-1}} \} \rightarrow \{ \ket{a_{\ell}} \bra{a'_{\ell}} \}, 
\end{equation}
and defined as
\begin{equation}
\hat{E}^{[\ell]} = \sum_{a_{\ell-1},a'_{\ell-1}} \sum_{a_\ell,a'_\ell} \left( \sum_{\sigma_\ell} M^{[\ell]\sigma_\ell *} \otimes M^{[\ell]\sigma_\ell} \right)_{(a_{\ell-1}a'_{\ell-1}),(a_\ell,a'_\ell)} (\ket{a_{\ell-1}} \bra{a'_{\ell-1}}) (\ket{a_{\ell}} \bra{a'_{\ell}}),
\end{equation}
where we read off the expression in brackets as the matrix elements of $E^{[\ell]}$ of dimension $(D^2_{\ell-1} \times D^2_{\ell})$, the $M$-matrix dimensions at the respective bonds. It generalizes to the contraction with an interposed operator at site $\ell$ as
\begin{equation}
E^{[\ell]}_O = \sum_{\sigma_\ell,\sigma'_\ell} O^{\sigma_\ell,\sigma'_\ell} M^{[\ell]\sigma_\ell *} \otimes M^{[\ell]\sigma'_\ell} .
\end{equation}
How does $\hat{E}^{[\ell]} $ act? From the explicit notation
\begin{equation}
E^{[\ell]}_{(a_{\ell-1}a'_{\ell-1}),(a_{\ell}a'_{\ell})} = \sum_{\sigma_\ell} M^{[\ell]\sigma_\ell *}_{a_{\ell-1},a_\ell} \cdot M^{[\ell]\sigma_\ell}_{a'_{\ell-1},a'_\ell}
\end{equation}
we can read $\hat{E}^{[\ell]}[ \hat{O}^{[\ell-1]}]$ as an operation on a matrix $O^{[\ell-1]}_{a,a'}$ as 
\begin{equation}
E^{[\ell]} [O^{[\ell-1]}]  = \sum_{\sigma_\ell} M^{\sigma_\ell\dagger} O^{[\ell-1]} M^{\sigma_\ell}
\end{equation} 
or on a row vector of length $D_{\ell-1}^2$ with coefficients $v_{aa'} = O^{[\ell-1]}_{a,a'}$ multiplied from the left,
\begin{equation}
\sum_{\sigma_\ell} \sum_{a_{\ell-1},a'_{\ell-1}} v_{a_{\ell-1}a'_{\ell-1}} M^{\sigma_\ell*}_{a_{\ell-1},a_\ell} M^{\sigma_\ell}_{a'_{\ell-1},a'_\ell} .
\end{equation} 
The $C$-matrices of the last section are then related as
\begin{equation}
C^{[\ell]} = E^{[\ell]} [C^{[\ell-1]}]  ,
\end{equation}
but we will now take this result beyond numerical convenience: If $D_{\ell-1}=D_\ell$, we can also ask for eigenvalues, eigenmatrices and (left or right) eigenvectors interchangeably. In this context we obtain the most important property of $E$, namely that if it is constructed from left-normalized matrices $A$ or right-normalized matrices $B$, all eigenvalues $|\lambda_k| \leq 1$. 

In fact, for $\lambda_1=1$ and left-normalized $A$-matrices, the associated left eigenvector $v_{aa'}=\delta_{aa'}$, as can be seen by direct calculation or trivially if we translate it to the identity matrix:
\begin{equation}
E[I] = \sum_\sigma A^{\sigma\dagger} \cdot I \cdot A^{\sigma} = 1 \cdot I .
\end{equation}
The right eigenvector for $E$ constructed from left-normalized $A$ matrices is non-trivial, but we will ignore it for the moment. For right-normalized $B$ matrices, the situation is reversed: explicit calculation shows that $v_{aa'}=\delta_{aa'}$ is now right eigenvector with $\lambda_1=1$, and the left eigenvector is non-trivial. 

To show that 1 is the largest eigenvalue\cite{Weichselbaum05}, we consider $C' = E[C]$. The idea is that then one can show that $s'_1 \leq s_1$ for the largest singular values of $C'$ and $C$, if $E$ is constructed from either left- or right-normalized matrices. This immediately implies that all eigenvalues of $E$, $|\lambda_i| \leq 1$: $C'=\lambda_i C$ implies $s'_1 = |\lambda_i| s_1$, such that $|\lambda_i|>1$ would contradict the previous statement. The existence of several $|\lambda_i|=1$ cannot be excluded.
The proof runs as follows (here for left-normalized matrices): consider the SVD $C = U^\dagger S V$. $C$ is square, hence $U^\dagger U = UU^\dagger = V^\dagger V = VV^\dagger = I$. We can then write 
\begin{equation}
C' = \sum_\sigma A^{\sigma\dagger} U^\dagger S V A^{\sigma} = 
\left[ \begin{array}{ccc} (UA^1)^\dagger & \ldots  & (UA^d)^\dagger \end{array} \right]
\left[ \begin{array}{ccc} S & & \\ & S & \\ & & S \end{array} \right]
\left[ \begin{array}{c} VA^1 \\ \vdots \\ VA^d \end{array} \right] = P^\dagger \left[ \begin{array}{ccc} S & & \\ & S & \\ & & S \end{array} \right] Q .
\end{equation}
We have $P^\dagger P = I$ and $Q^\dagger Q = I$ (however $PP^\dagger \neq I$, $QQ^\dagger \neq I$), if the $A^\sigma$ are left-normalized: $P^\dagger P =\sum_\sigma A^{\sigma\dagger} U^\dagger U A^\sigma = \sum_\sigma A^{\sigma\dagger} A^\sigma = I$ and similarly for $Q$; they therefore are reduced basis transformations to orthonormal subspaces, hence the largest singular value of $C'$ must be less or equal to that of $S$, which is $s_1$.

Independent of normalization, the overlap calculation becomes
\begin{equation}
\braket{\psi}{\psi} = E^{[1]} E^{[2]} E^{[3]} \ldots E^{[L-2]} E^{[L-1]} E^{[L]}  ,
\end{equation}
and expectation value calculations before proper normalization by $\braket{\psi}{\psi}$ would read
\begin{equation}
\bra{\psi} \hat{O}^{[1]} \hat{O}^{[2]} \ldots \hat{O}^{[L-1]} \hat{O}^{[L]}\ket{\psi} = E^{[1]}_{O_1} E^{[2]}_{O_2} E^{[3]}_{O_3} \ldots E^{[L-2]}_{O_{L-2}} E^{[L-1]}_{O_{L-1}} E^{[L]}_{O_L}  .
\end{equation}
Numerically, this notation naively taken is not very useful, as it implies $O(D^6)$ operations; of course, if its internal product structure is accounted for, we return to $O(D^3)$ operations as previously discussed. But analytically, it reveals very interesting insights. Let us assume a translationally invariant state with left-normalized site-independent $A$-matrices (hence also site-independent $E$) with periodic boundary conditions. Then we obtain in the limit $L\rightarrow\infty$
\begin{eqnarray*}
\bra{\psi} \hat{O}^{[i]}\hat{O}^{[j]} \ket{\psi} &=& \tr E^{[1]} \ldots E^{[i-1]} E^{[i]}_O E^{[i+1]} \ldots E^{[j-1]} 
E^{[j]}_O E^{[j+1]} \ldots E^{[L]} \\
&=& \tr E^{[i]}_O E^{j-i-1} E^{[j]}_O E^{L-j+i-1} \\
&=& \sum_{l,k} \bra{l} E^{[i]}_O \ket{k} \lambda_k^{j-i-1} \bra{k} E^{[j]}_O \ket{l} \lambda_l^{L-j+i-1}\\
&=& \sum_k \bra{1} E^{[i]}_O \ket{k} \lambda_k^{j-i-1} \bra{k} E^{[j]}_O \ket{1} \quad (L\rightarrow\infty)
\end{eqnarray*}
where $\lambda_k$ are the eigenvalues of $E$. We have used $|\lambda_k|\leq 1$ for $E$ from normalized matrices and that $\lambda_1=1$ is the only eigenvalue of modulus 1; but relaxing the latter (not necessarily true) assumption would only introduce a minor modification. $\ket{k}$ and $\bra{k}$ are the right and left eigenvectors of (non-hermitian) $E$ for eigenvalues $\lambda_k$. $\bra{1}$ corresponds to the eigenoperator $I$ for $E$ from left-normalized $A$.

The decisive observation is that correlators can be long-ranged (if the matrix elements  $ \bra{1} E^{[i]}_O \ket{1}$ are finite) or are a superposition of exponentials with decay length $\xi_k = -1/ \ln \lambda_k$, such that MPS two-point correlators take the generic form
\begin{equation}
\frac{\bra{\psi} \hat{O}^{[i]} \hat{O}^{[j]} \ket{\psi}}{\braket{\psi}{\psi}} = c_1 + \sum_{k=2}^{D^2} c_k \eul^{-r/\xi_k} ,
\end{equation}
where $r=|j-i-1|$ and $c_k = \bra{1} E^{[i]}_O \ket{k} \bra{k} E^{[j]}_O \ket{1}$ for $i<j$. 

A simple example can be extracted from the AKLT state of the previous section. The eigenvalues of $E$ were already found to be $1,-1/3,-1/3,-1/3$.  For the spin-operators, the matrix elements for long-range order vanish, such that the correlation $\langle \hat{S}_i^z \hat{S}_j^z \rangle = (12/9)(-1)^{j-i} \eul^{-(j-i)\ln 3}$ for $j>i$, a purely exponential decay with correlation length $\xi = 1/\ln 3= 0.9102$. On the other hand, for the string correlator, the long-range matrix elements are finite, and long-range order emerges in the string correlator.

The form of correlators of MPS has important consequences: the usual form correlators take in one dimensional quantum systems in the thermodynamic limit is either the Ornstein-Zernike form
\begin{equation}
\langle O_0 O_x \rangle \sim \frac{\eul^{-x/\xi}}{\sqrt{x}}
\end{equation}
or the critical power-law form (maybe with logarithmic corrections),
\begin{equation}
\langle O_0 O_x \rangle \sim x^{-\alpha} .
\end{equation}
The AKLT state belongs to a very special state class whose correlation functions mimic a quantum system in a lower spatial dimension (so-called dimensional reduction), which removes the $\sqrt{x}$-term; the AKLT state sits on a so-called disorder line, where such phenomena occur \cite{Garel96}. 

Any finite-dimensional MPS therefore will only approximate the true correlator by a superposition of exponentials. It turns out that this works very well on short distances, even for power laws. What one observes numerically is that the true correlation function will be represented accurately on increasingly long length scales as $D$ is increased. Eventually, the slowest exponential decay will survive, turning the correlation into a pure exponential decay with $\xi =  -1/\ln \lambda$, where $\lambda$ is the largest eigenvalue of $E$ that contributes to the correlator. The comparison of curves for various $D$ is therefore an excellent tool to gauge the convergence of correlation functions and the length scale on which it has been achieved.  

\subsubsection{MPS and reduced density operators}
As we have already seen for DMRG, the concept of reduced density operators is of importance in various ways. Let us express them using the MPS notation. We have
\begin{equation}
\ket{\psi}\bra{\psi} = \sum_{\fat{\sigma},\fat{\sigma}'} A^{\sigma_1} \ldots A^{\sigma_L} A^{\sigma'_1*} \ldots A^{\sigma'_L*} \ket{\fat{\sigma}} \bra{\fat{\sigma}'} = 
\sum_{\fat{\sigma},\fat{\sigma}'} A^{\sigma_1} \ldots A^{\sigma_L} A^{\sigma'_L\dagger} \ldots A^{\sigma'_1\dagger} \ket{\fat{\sigma}} \bra{\fat{\sigma}'} .
\label{eq:fullstateprojector}
\end{equation}
We now bipartition into AB, where A contains sites 1 through $\ell$ and B sites $\ell+1$ through $L$. Tracing out the degrees of freedom of B in the last expression we obtain
\begin{equation}
\hat{\rho}_A^{[\ell]} = \tr_B \ket{\psi}\bra{\psi} = \sum_{\fat{\sigma},\fat{\sigma}'\in A} 
A^{\sigma_1} \ldots A^{\sigma_\ell} \rho_A^{[\ell]} A^{\sigma'_\ell\dagger} \ldots A^{\sigma'_1\dagger} \ket{\fat{\sigma}} \bra{\fat{\sigma}'} ,
\end{equation}
where
\begin{equation}
\rho_A^{[\ell]} =  \sum_{\fat{\sigma}\in B} A^{\sigma_{\ell+1}} \ldots A^{\sigma_L} A^{\sigma_L\dagger} \ldots A^{\sigma_{\ell+1}\dagger} .
\end{equation}
This equation immediately implies a recursion relation between different reduced density matrices, namely
\begin{equation}
\rho_A^{[\ell-1]} = \sum_{\sigma_\ell} A^{\sigma_\ell} \rho_A^{[\ell]} A^{\sigma_\ell\dagger} .
\label{eq:rhoarecursion}
\end{equation}
In the thermodynamic limit $L\rightarrow\infty$, $\ell\rightarrow\infty$ of a translationally invariant system, we may therefore ask whether a fixed point relationship
\begin{equation}
\rho_A^f = \sum_{\sigma} A^{\sigma} \rho_A^{f} A^{\sigma\dagger} 
\end{equation}
is fulfilled.

All these equations hold even if the matrices of the MPS are not left-normalized. In the case that they are, we can directly express the density operator in the orthonormal basis generated by the $A$-matrices, namely
\begin{equation}
\hat{\rho}_A^{[\ell]} = \sum_{a_\ell, a'_\ell} (\rho_A^{[\ell]})_{a_\ell, a'_\ell} \ket{a_\ell}_A \phantom{{}}_A\bra{a'_\ell} .
\end{equation}

Similar relationships hold for the reduced density operator of B, where (using $B$-matrices now) we obtain
\begin{equation}
\hat{\rho}_B^{[\ell]} = \tr_A \ket{\psi}\bra{\psi} = \sum_{\fat{\sigma},\fat{\sigma}'\in B} 
B^{\sigma'_L\dagger} \ldots B^{\sigma'_{\ell+1}\dagger} \rho_B^{[\ell]} B^{\sigma_{\ell+1}} \ldots B^{\sigma_L} \ket{\fat{\sigma}} \bra{\fat{\sigma}'} ,
\end{equation}
where
\begin{equation}
\rho_B^{[\ell]} =  \sum_{\fat{\sigma}\in A} B^{\sigma_{\ell}\dagger} \ldots B^{\sigma_1\dagger} B^{\sigma_1} \ldots B^{\sigma_{\ell} }
\end{equation}
and the recursion relationship
\begin{equation}
\rho_B^{[\ell]} = \sum_{\sigma_\ell} B^{\sigma_\ell\dagger} \rho_B^{[\ell-1]} B^{\sigma_\ell} .
\label{eq:rhobrecursion}
\end{equation}
giving rise to a potential fixed point relationship
\begin{equation}
\rho_B^f = \sum_{\sigma} B^{\sigma\dagger} \rho_B^{f} B^{\sigma} .
\end{equation}
Again, all these relationships would hold for arbitrary MPS matrices, but if they are right-normalized, we again get an expression in an orthonormal basis, now generated by the $B$-matrices,
\begin{equation}
\hat{\rho}_B^{[\ell]} = \sum_{a_\ell, a'_\ell} (\rho_B^{[\ell]})_{a_\ell, a'_\ell} \ket{a_\ell}_B \phantom{{}}_B\bra{a'_\ell} .
\end{equation}
In the case of a mixed-canonical state $\ket{\psi}=\sum_{\fat{\sigma}} A^{\sigma_1}\ldots A^{\sigma_\ell}\Psi B^{\sigma_{\ell+1}} \ldots B^{\sigma_L} \ket{\fat{\sigma}}$ a rerun of the calculation shows that 
\begin{equation}
\hat{\rho}^{[\ell]}_A = \Psi \Psi^\dagger
\end{equation}
and
\begin{equation}
\hat{\rho}^{[\ell]}_B = \Psi^\dagger\Psi ,
\end{equation}
expressed in an orthonormal basis.

\subsection{Adding two matrix product states}

An operation that one needs comparatively rarely in practice is the addition of two MPS. Let us first consider the PBC case, which is easier. Taking two MPS, with no normalization assumed,
\begin{equation}
\ket{\psi} = \sum_{\fat{\sigma}} \tr (M^{\sigma_1} \ldots M^{\sigma_L}) \ket{\fat{\sigma}} 
\quad\quad
\ket{\phi} = \sum_{\fat{\sigma}} \tr (\tilde{M}^{\sigma_1} \ldots \tilde{M}^{\sigma_L}) \ket{\fat{\sigma}} 
\end{equation}
we can write down
\begin{equation}
\ket{\psi} + \ket{\phi} = \sum_{\fat{\sigma}} \tr (N^{\sigma_1} \ldots N^{\sigma_L}) \ket{\fat{\sigma}}
\end{equation}
where 
\begin{equation}
N^{\sigma_i} =  M^{\sigma_i} \oplus \tilde{M}^{\sigma_i} .
\end{equation} 
This means that we simply take $M$ and $\tilde{M}$ as diagonal blocks of a matrix $N$. The diagonal block structure implies that upon multiplying the $N$ matrices the result is again diagonal, with $MMMMM\ldots$ in the first and $\tilde{M}\tilde{M}\tilde{M}\tilde{M}\tilde{M}\ldots$ in the second block. Then the trace can be split, and we are back at the original states:
\begin{equation}
\tr (NNNNNN) = \tr \left( \begin{array}{cc} MMMMMM & 0 \\ 0 & \tilde{M}\tilde{M}\tilde{M}\tilde{M}\tilde{M}\tilde{M} \end{array} \right) = \tr (MMMMMM) + \tr (\tilde{M}\tilde{M}\tilde{M}\tilde{M}\tilde{M}\tilde{M}).
\end{equation} 

In the case of OBC, we can proceed in exactly the same fashion. On the first and last sites, something special has to happen: naively, the first and last dimensions would go up to 2, and the scalar nature be lost. Physically, these indices are dummies anyways. So what we have to do (and a simple calculation shows that this works) is to form a row vector $[ M\ \tilde{M}]$ and a column vector $[M\ \tilde{M}]^T$ on the last sites, from the row and column vectors of the original states.

Addition of MPS therefore leads to new matrices with dimension $D_N = D_M + D_{\tilde{M}}$, such that MPS of a certain dimension are not closed under addition. It is also obvious that in many cases this way of describing a new state is uneconomical: the extreme case would be adding $\ket{\psi}+\ket{\psi}$, where the resulting state is the same, just with a prefactor 2, so matrix dimensions should not increase. So after additions it is worthwhile to consider {\em compressing} the MPS again to some lower dimension, which depending on the states added may or may not (like in the example) incur a loss of information.

\subsection{Bringing a matrix product state into canonical form}

For a general matrix product state, no particular demands are placed on the matrices $M^{\sigma_i}$ except that their dimensions must match appropriately. Certain classes of matrices are to be preferred, namely left- and right-normalized matrices, leading to left- and right-canonical MPS: certain contractions become trivial, orthonormal reduced bases are generated automatically.

In order to bring an arbitrary MPS to canonical form we exploit that SVD generates either unitary matrices or matrices with orthonormal rows and columns which can be shown to obey the left- or right normalization condition.

\subsubsection{Generation of a left-canonical MPS}
Setting out from a general MPS, without normalization assumption, making the contractions explicit,
\begin{equation}
\ket{\psi}= \sum_{\fat{\sigma}} \sum_{a_1,\ldots} M^{\sigma_ 1}_{1,a_1} M^{\sigma_2}_{a_1,a_2} M^{\sigma_3}_{a_2,a_3} \ldots \ket{\fat{\sigma}}
\end{equation}
we reshape $M^{\sigma_ 1}_{1,a_1}$ by grouping physical and left (row) index to carry out an SVD on the new $M$, yielding $M=ASV^\dagger$:
\begin{eqnarray}
& & \sum_{\fat{\sigma}} \sum_{a_1,\ldots} M_{(\sigma_ 1,1),a_1} M^{\sigma_2}_{a_1,a_2} M^{\sigma_3}_{a_2,a_3} \ldots \ket{\fat{\sigma}} \nonumber \\
&=&  \sum_{\fat{\sigma}} \sum_{a_1,\ldots} \sum_{s_1} A_{(\sigma_ 1,1),s_1} S_{s_1,s_1} V^\dagger_{s_1,a_1}  M^{\sigma_2}_{a_1,a_2}  \ldots \ket{\fat{\sigma}} \nonumber \\
&=& \sum_{\fat{\sigma}} \sum_{a_2,\ldots} \sum_{s_1} A^{\sigma_1}_{1,s_1} \left( \sum_{a_1} S_{s_1,s_1} V^\dagger_{s_1,a_1}  M^{\sigma_2}_{a_1,a_2} \right) M^{\sigma_3}_{a_2,a_3} \ldots \ket{\fat{\sigma}} \nonumber \\
&=& \sum_{\fat{\sigma}} \sum_{a_2,\ldots} \sum_{s_1} A^{\sigma_1}_{1,s_1} \tilde{M}^{\sigma_2}_{s_1,a_2} M^{\sigma_3}_{a_2,a_3} \ldots \ket{\fat{\sigma}} .
\end{eqnarray}
As $A^\dagger A = I$ due to SVD, after reshaping to $A^{\sigma_1}$, left-normalization holds for  $A^{\sigma_1}$. The remaining two matrices of the SVD are multiplied into $M^{\sigma_2}$, such that a new MPS with
$\tilde{M}^{\sigma_2}_{s_1,a_2} = \sum_{a_1} S_{s_1,s_1} V^\dagger_{s_1,a_1}  M^{\sigma_2}_{a_1,a_2}$ is generated. 

\begin{figure}
\centering\includegraphics[scale=0.6]{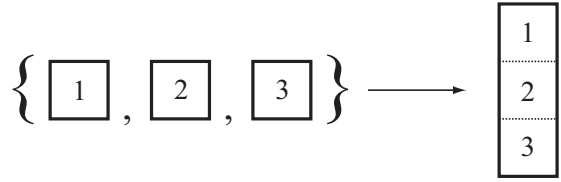}
\caption{For canonization, sets of matrices on a given site are brought together in a single matrix.}
\label{fig:Mreshaping}
\end{figure}

Now the procedure can be iterated: $\tilde{M}^{\sigma_2}_{s_1,a_2}$ is reshaped to $\tilde{M}_{(\sigma_2,s_1),a_2}$ (Fig.~\ref{fig:Mreshaping}), singular value decomposed as $ASV^\dagger$, generating $A_{(\sigma_2,s_1),s_2}$, reshaped to a left-normalized $A^{\sigma_2}_{s_1,s_2}$. The right two matrices of the SVD are again multiplied into the next ansatz matrix, and so forth. After the last step, left-normalized matrices $A^{\sigma_i}_{s_{i-1},s_i}$ live on all sites. $S_{1,1} (V^\dagger)_{1,1}$, a scalar as $A^{\sigma_L}$ is a column vector, survive at the last site, but this scalar is nothing but the norm of $\ket{\psi}$. We may keep it separately if we want to work with non-normalized states.

This procedure extends trivially to normalization: we identify the prefactor of the state, but instead of storing it, we simply set it to 1. As we do not use the singular values explicitly, the above procedure can be easily reformulated using QR decompositions, along the lines of Section \ref{subsubsec:MPSdecomposition}. Standard QR, however, does not show us whether the matrices used are bigger than necessary, i.e. have singular values that are zero, such that matrices can be trimmed to a smaller size without loss of accuracy; this would only be possible using rank revealing QR; for many MPS it is however clear from the underlying physics that the spectrum of singular values has a long tail, such that this issue does not arise. The same argumentation holds also for the generation of a right-canonical MPS, which we turn to in the following. 
    
\subsubsection{Generation of a right-canonical MPS}
The same procedure can be applied to arrive at a state of right-normalized matrices, by carrying out a sequence of SVDs starting from the right on reshaped matrices $\{ M^\sigma \} \rightarrow M=USB$, splitting $B$ into matrices $B^\sigma$ that are right-normalized (due to $BB^\dagger = I$), and multiplying $U$ and $S$ to the left, creating the matrix to be singular value decomposed next:

\begin{eqnarray}
& & \sum_{\fat{\sigma}} \sum_{\ldots,a_{L-1}}  \ldots M^{\sigma_{L-2}}_{a_{L-3},a_{L-2}} M^{\sigma_{L-1}}_{a_{L-2},a_{L-1}} M^{\sigma_{L}}_{a_{L-1},1}  \ket{\fat{\sigma}} \nonumber \\
&=& \sum_{\fat{\sigma}} \sum_{\ldots,a_{L-1}}  \ldots M^{\sigma_{L-2}}_{a_{L-3},a_{L-2}} M^{\sigma_{L-1}}_{a_{L-2},a_{L-1}} M_{a_{L-1},(\sigma_{L},1)}  \ket{\fat{\sigma}} \nonumber \\
&=&  \sum_{\fat{\sigma}} \sum_{\ldots a_{L-1}} \sum_{s_{L-1}}  \ldots M^{\sigma_{L-2}}_{a_{L-3},a_{L-2}} M^{\sigma_{L-1}}_{a_{L-2},a_{L-1}} 
U_{a_{L-1},s_{L-1}} S_{s_{L-1},s_{L-1}} B_{s_{L-1},(\sigma_L,1)}   \ket{\fat{\sigma}} \nonumber \\
&=& \sum_{\fat{\sigma}} \sum_{\ldots a_{L-2}} \sum_{s_{L-1}} 
\ldots M^{\sigma_{L-2}}_{a_{L-3},a_{L-2}} \left( \sum_{a_{L-1}} M^{\sigma_{L-1}}_{a_{L-2},a_{L-1}} U_{a_{L-1},s_{L-1}} S_{s_{L-1},s_{L-1}}\right) 
B_{s_{L-1},(\sigma_L,1)}   \ket{\fat{\sigma}} \nonumber \\
&=& \sum_{\fat{\sigma}} \sum_{\ldots a_{L-2}} \sum_{s_{L-1}}
\ldots M^{\sigma_{L-2}}_{a_{L-3},a_{L-2}} \tilde{M}^{\sigma_{L-1}}_{a_{L-2},s_{L-1}} B^{\sigma_L}_{s_{L-1},1}   \ket{\fat{\sigma}} ,
\end{eqnarray}
proceeding as before, with the sole differences that (i) the direction is reversed and (ii) reshaping now groups the physical index with the column instead of the row index: $\tilde{M}^{\sigma_i}_{a_{i-1},s_{i}} \rightarrow \tilde{M}_{a_{i-1},(\sigma_i s_i)} \rightarrow B_{s_{i-1},(\sigma_i s_i)} \rightarrow B^{\sigma_i}_{s_{i-1},s_i}$.

\subsection{Approximate compression of an MPS}

The (rare) addition of MPS and various algorithms that can be formulated with MPS lead to a substantial increase in matrix dimensions of the result. It is therefore a recurrent issue how to approximate a given MPS with matrix dimensions $(D'_i \times D'_{i+1})$ by another MPS with 
matrix dimensions $(D_i \times D_{i+1})$, where $D_i < D'_i$, as closely as possible. 

Fundamentally, two procedures are available, {\em SVD compression} and {\em variational compression}. Both have advantages and disadvantages: for small degrees of compression, $D \sim D'$, SVD is fast, but it is never optimal; it becomes very slow if $D' \gg D$. Variational compression is optimal, but slow if the starting point is chosen randomly, can however be greatly speeded up by providing a good trial state e.g.\ from the SVD approach. Generally, issues of getting stuck in a non-optimal compression may arise in the variational ansatz.
 
Depending on the specific nature of the state to be compressed, procedures can be optimized, for example if the MPS to be compressed is a sum of MPS or if it is the result of the application of a matrix product operator (MPO; Sec.\ \ref{sec:MPO}) to an MPS.

\subsubsection{Compressing a matrix product state by SVD}
Let us consider an MPS in mixed canonical representation,
\begin{equation}
\ket{\psi} = \sum_{\fat{\sigma}} A^{\sigma_1} A^{\sigma_2} \ldots A^{\sigma_{\ell}} \Lambda^{[\ell]} B^{\sigma_{\ell+1}} \ldots B^{\sigma_{L-1}} B^{\sigma_L}  \ket{\fat{\sigma}} ,
\label{eq:compressionstart}
\end{equation}
from which we read off the Schmidt decomposition $\ket{\psi} = \sum_{a_\ell=1}^{D'} s_{a_\ell} \ket{a_\ell}_A \ket{a_\ell}_B$, where the states on A and B form orthonormal sets respectively; this follows from the canonical construction. Let us suppose there are $D'$ states each for this decomposition. We now look for the state $\ket{\tilde{\psi}}$ that approximates $\ket{\psi}$ best in the 2-norm and can be spanned by $D$ states each in A and B. We have shown that SVD provides the result by retaining the $D$ largest singular values, and the compressed state simply reads $\ket{\psi} = \sum_{a_\ell=1}^{D} s_{a_\ell} \ket{a_\ell}_A \ket{a_\ell}_B$, providing a simple truncation prescription: retain the first $D$ columns of $A^{\sigma_{\ell}}$, the first $D$ rows of  $B^{\sigma_{\ell+1}}$, and the first $D$ rows and columns of  $\Lambda^{[\ell]}$. If normalization is desired, the retained singular values must be rescaled.

This procedure rests on the orthonormality of the states on A and B, therefore can only be carried out at one bond. In order to shrink all matrices, we have to work our way through all mixed canonical representations, say from right to left, truncate, and shift the boundary between left- and right-normalized matrices by one site to the left, using techniques from canonization.

After the first step of right-canonization of a left-canonical state, it reads:
\begin{equation}
\ket{\psi^{(L-1)}} = \sum_{\fat{\sigma}} A^{\sigma_1} \ldots A^{\sigma_{L-1}} U S B^{\sigma_L} \ket{\fat{\sigma}} ,
\end{equation}
where I have already reshaped $B$, which is right-normalized and guarantees that states formed as $\ket{a_{L-1}}_B = \sum_{\sigma_L} (B^{\sigma_L})_{a_{L-1},1} \ket{\sigma_L}$ are orthonormal. But so are the states 
\begin{equation}
\ket{a_{L-1}}_A = \sum_{\sigma_1 \ldots \sigma_{L-1}} (A^{\sigma_1} \ldots A^{\sigma_{L-1}} U)_{1,a_{L-1}} \ket{\sigma_1 \ldots \sigma_{L-1}},
\end{equation}
as SVD guarantees $U^\dagger U = 1$: we are just doing a basis transformation within the orthonormal basis set constructed from the left-normalized $A^{\sigma_i}$. Hence, we have a correct Schmidt decomposition as
\begin{equation}
\ket{\psi^{(L-1)}} = \sum_{a_{L-1}} s_{a_{L-1}} \ket{a_{L-1}}_A \ket{a_{L-1}}_B .
\end{equation}
The difference to a right canonization is now the truncation: matrices $U$, $S$ and $B^{\sigma_L}$ are truncated (and singular values possibly renormalized) to $\tilde{U}$, $\tilde{S}$ and $\tilde{B}^{\sigma_L}$ just as explained before: retain the $D$ largest singular values. $\tilde{B}^{\sigma_L}$ is still right-normalized. The next $A^{\sigma_{L-1}}$ to the left, $\tilde{U}$ and $\tilde{S}$ are multiplied together to form $M^{\sigma_{L-1}}$. By reshaping, SVD and reshaping as 
\begin{equation}
M^{\sigma}_{ij} = M_{i, (\sigma j)} = \sum_k U_{ik} S_{kk} B_{k,(\sigma j)} = \sum_k U_{ik} S_{kk} B^{\sigma}_{kj} 
\end{equation} 
we obtain right-normalized $B^{\sigma_{L-1}}$, truncate $U$, $S$ and $B^{\sigma_{L-1}}$ to $\tilde{U}$, $\tilde{S}$ and $\tilde{B}^{\sigma_{L-1}}$, and the procedure continues:
\begin{eqnarray*}
\ket{\psi} &=& \sum_{\fat{\sigma}}  A^{\sigma_1} \ldots A^{\sigma_{L-3}}A^{\sigma_{L-2}} \left( A^{\sigma_{L-1}} U S \right)B^{\sigma_L}   \ket{\fat{\sigma}} \\
&\rightarrow& \sum_{\fat{\sigma}} A^{\sigma_1} \ldots A^{\sigma_{L-3}}A^{\sigma_{L-2}}\left( A^{\sigma_{L-1}} \tilde{U} \tilde{S} \right) \tilde{B}^{\sigma_L}   \ket{\fat{\sigma}} \\
&=&  \sum_{\fat{\sigma}} A^{\sigma_1} \ldots A^{\sigma_{L-3}}A^{\sigma_{L-2}} M^{\sigma_{L-1}} \tilde{B}^{\sigma_L}   \ket{\fat{\sigma}} \\
&=& \sum_{\fat{\sigma}} A^{\sigma_1} \ldots A^{\sigma_{L-3}} \left( A^{\sigma_{L-2}} U S \right) B^{\sigma_{L-1}} \tilde{B}^{\sigma_L}   \ket{\fat{\sigma}} \\
&\rightarrow& \sum_{\fat{\sigma}} A^{\sigma_1} \ldots A^{\sigma_{L-3}} \left( A^{\sigma_{L-2}} \tilde{U} \tilde{S} \right) \tilde{B}^{\sigma_{L-1}} \tilde{B}^{\sigma_L}   \ket{\fat{\sigma}} \\
&=& \ldots
\end{eqnarray*}
At the end, the compressed MPS $\ket{\tilde{\psi}}$ is right-normalized and given by $\tilde{B}$-matrices. As we will see, this compression procedure is just the truncation that is carried out by (time-dependent) DMRG or TEBD as they sweep through the chain. Both methods at each bond have correctly normalized matrices (i.e.\ orthonormal states) to the left and right, carry out the cut and proceed.

The disadvantage of the procedure is that a one-sided interdependence of truncations occurs: the matrix $M$ always contains a truncated $\tilde{U}$ from the previous step, hence is truncation-dependent. Generally, truncations cannot be independent of each other because for each decomposition (\ref{eq:compressionstart}) truncations of the $A$- and $B$-matrices affect the orthonormal systems, but here the dependence is one-sided and ``unbalanced'': truncations further to the left depend on those to the right, but not vice versa. If the truncation is small -- which it usually is for small time steps in time-dependent DMRG -- the introduced additional inaccuracy is minor; however, for cases where large truncations may occur, the dependence might become too strong and the truncation far from optimal. 

A second concern regards efficiency: for matrix dimensions $(m\times n)$, $m\ge n$, the cost of SVD is $O(m n^2)$. This means that the SVD cost $O((D')^2 dD)$ if $D' \le dD$ and $O(D' d^2 D^2)$ otherwise; the matrix multiplications cost $O(dD(D')^2)$. In many applications, $D'\gg D$; then this method becomes quite slow. The situation is even worse if the original state is not in canonical form and has to be brought to that form first by a sequence of SVDs, that are of order $O((D')^3)$.

Let me conclude this section with the remark that of course we can of course also compress by imposing some $\epsilon$ which at each truncation we accept as maximal 2-norm distance between the original and the compressed state (given by the sum of the squares of the discarded singular values), implicitly defining $D$. 

\subsubsection{Compressing a matrix product state iteratively}
The optimal approach is to start from an ansatz MPS of the desired reduced dimension, and to minimize its distance to the MPS to be approximated iteratively, i.e.\ by changing its $\tilde{M}^\sigma$ matrices iteratively. The matrices play the role of variational parameters.

The mathematically precise form of optimal compression of $\ket{\psi}$ from dimension $D'$ to $\ket{\tilde{\psi}}$ with dimension $D$ is to minimize $\| \ket{\psi} - \ket{\tilde{\psi}} \|_2^2$, which means that we want to minimize $\braket{\psi}{\psi} - \braket{\tilde{\psi}}{\psi} -\braket{\psi}{\tilde{\psi}} + \braket{\tilde{\psi}}{\tilde{\psi}}$ with respect to $\ket{\tilde{\psi}}$. Let us call the matrices $M$ and $\tilde{M}$ respectively, to emphasize that we make no assumption about the normalization. Expressed in the underlying matrices $\tilde{M}$, this is a highly nonlinear optimization problem.

But this can be done iteratively as follows. Start with an initial guess for $\ket{\tilde{\psi}}$, which could be an SVD-compression of $\ket{\psi}$, arguably not optimal, but a good starting point. Then we sweep through the set of $\tilde{M}^{\sigma_i}$ site by site, keeping all other matrices fixed and choosing the new $\tilde{M}^{\sigma_i}$, such that distance is minimized. The (usually justified hope) is that repeating this sweep through the matrices several times will lead to a converged optimal approximation.

\begin{figure}
\centering\includegraphics[scale=0.6]{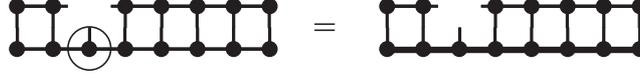}
\caption{Linear equation system to be solved for iterative compression of an MPS. The fatter lines correspond to the state to be compressed, the thinner lines to the compressed state. The unknown matrix is circled.}
\label{fig:iterativecompression}
\end{figure}

The new $\tilde{M}^{\sigma_i}$ is found by extremizing with respect to $\tilde{M}^{\sigma_i*}_{a_{i-1},a_{i}}$, which only shows up in $ - \braket{\tilde{\psi}}{\psi} + \braket{\tilde{\psi}}{\tilde{\psi}}$. We find
\begin{eqnarray*}
& & \frac{\partial}{\partial \tilde{M}^{\sigma_i*}_{a_{i-1},a_i}} ( \braket{\tilde{\psi}}{\tilde{\psi}}
- \braket{\tilde{\psi}}{\psi} ) = \\
& & \sum_{\fat{\sigma}*} (\tilde{M}^{\sigma_1*} \ldots \tilde{M}^{\sigma_{i-1}*})_{1, a_{i-1}}(\tilde{M}^{\sigma_{i+1}*} \ldots \tilde{M}^{\sigma_L*})_{a_i,1} \tilde{M}^{\sigma_1} \ldots \tilde{M}^{\sigma_i} \ldots \tilde{M}^{\sigma_L} -  \\
& & \sum_{\fat{\sigma}*}
(\tilde{M}^{\sigma_1*} \ldots \tilde{M}^{\sigma_{i-1}*})_{1, a_{i-1}}(\tilde{M}^{\sigma_{i+1}*} \ldots \tilde{M}^{\sigma_L*})_{a_i,1} M^{\sigma_1} \ldots M^{\sigma_i} \ldots M^{\sigma_L}= 0.
\end{eqnarray*}
The sum over $\fat{\sigma}*$ runs over all physical sites except $i$. This system looks complicated, but is in fact quite easy. Keeping the matrix to be found, $\tilde{M}^{\sigma_i}$, explicit, we may rewrite this equation as
\begin{equation}
\sum_{a'_{i-1}a'_i} \tilde{O}_{a_{i-1}a_i,a'_{i-1}a'_i} \tilde{M}^{\sigma_i}_{a'_{i-1}a'_i} = O^{\sigma_i}_{a_{i-1}a_i}.
\end{equation}
If, for each $\sigma_i$, we interpret the matrix $\tilde{M}^{\sigma_i}$ as a vector $v$ of length $D^2$, $\tilde{O}$ as a matrix $P$ of dimension $(D^2 \times D^2)$ and $O^{\sigma_i}$ as a vector $b$ of length $D^2$, we have a linear equation system
\begin{equation}
P v = b. 
\end{equation}
The result $v$ can then taken to be the matrix we are looking for. As this system is usually too big for a direct solution, an iterative solver has to be used, such as a conjugate gradient method. The system is Hermitian, as can be seen from the construction of $P$: unconjugated and conjugated $\tilde{M}$ simply reverse their role under transposition. Once again, the graphical representation is simplest (Fig.~\ref{fig:iterativecompression}). 

\begin{figure}
\centering\includegraphics[scale=0.6]{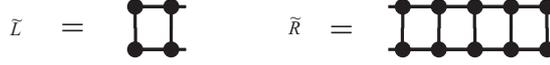}
\caption{Iteratively constructed objects $\tilde{L}$ and $\tilde{R}$ for compression.}
\label{fig:normalizediterativecompressionLR}
\end{figure}

As in the later case of finding ground states variationally, it is important to realize that the cost of the matrix-vector multiplications in the conjugate gradient method is not $O(D^4)$ as dimensions would naively suggest. There is an obvious factorization, which becomes particularly obvious graphically, that then leads to a cost of $O(D^3)$:
\begin{equation}
\tilde{O}_{a_{i-1}a_i,a'_{i-1}a'_i} = \tilde{L}_{a_{i-1},a'_{i-1}} \cdot \tilde{R}_{a_{i},a'_{i}}
\end{equation}
where 
\begin{equation}
\tilde{L}_{a_{i-1},a'_{i-1}} = \left( \sum_{\sigma_{i-1}}  \tilde{M}^{\sigma_{i-1}\dagger} \left( \ldots \left( \sum_{\sigma_1}  
\tilde{M}^{\sigma_{1}\dagger} \tilde{M}^{\sigma_{1}} \right) \ldots \right) \tilde{M}^{\sigma_{i-1}} \right)_{a_{i-1},a'_{i-1}}
\end{equation}
and similarly $\tilde{R}$. In the graphical representation (Fig.~\ref{fig:normalizediterativecompressionLR}), they are simply the contracted objects to the left and right of the circled $\tilde{M}$-matrix we are solving for.

Then 
\begin{equation}
\sum_{a'_{i-1}a'_i} \tilde{O}_{a_{i-1}a_i,a'_{i-1}a'_i} \tilde{M}^{\sigma_i}_{a'_{i-1}a'_i} =
\sum_{a'_{i-1}} \tilde{L}_{a_{i-1},a'_{i-1}} \left( \sum_{a'_i} \tilde{R}_{a_{i},a'_{i}}
 \tilde{M}^{\sigma_i}_{a'_{i-1}a'_i} \right),
\end{equation}
two operations of cost $O(D^3)$.

\begin{figure}
\centering\includegraphics[scale=0.6]{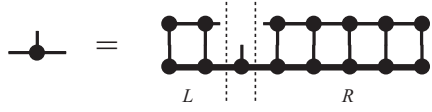}
\caption{Equation for iterative compression of an MPS for a suitably normalized state. The fatter lines correspond to the state to be compressed, the thinner lines to the compressed state.}
\label{fig:normalizediterativecompression}
\end{figure}

A similar decomposition simplifies the calculation of the vector $b$, which is formed from $O^{\sigma_i}_{a_{i-1}a_i}$ as
\begin{equation}
\sum_{a'_{i-1}a'_i} L_{a_{i-1},a'_{i-1}} M^{\sigma_i}_{a'_{i-1}a'_i} R_{a_{i},a'_{i}} , 
\end{equation}
with $L$ and $R$ as indicated in Fig.~\ref{fig:normalizediterativecompression}.
In fact, calculating $L$ and $R$ is nothing but carrying out the first steps of an overlap calculation, starting from left or right. The result would then be the $C$ matrix produced there at intermediate steps. If one sweeps through the system from left to right and back one can build $L$ and $R$ iteratively from previous steps, which is the most efficient way. 

We can however {\em drastically} simplify the compression procedure if we exploit the canonical form! Assume that $\ket{\tilde{\psi}}$ is in mixed canonical form $\tilde{A}^{\sigma_1} \ldots \tilde{A}^{\sigma_{i-1}} \tilde{M}^{\sigma_i} \tilde{B}^{\sigma_{i+1}}  \ldots \tilde{B}^{\sigma_L}$ and we want to update $\tilde{M}^{\sigma_i}$: to the left of the matrix to be updated, everything is left-normalized, to the right everything is right-normalized and the form of $\tilde{M}^{\sigma_i}$ does not matter, as it will be recalculated anyways. 

Then $\tilde{L}_{a_{i-1},a'_{i-1}} = \delta_{a_{i-1},a'_{i-1}}$ because of left normalization, and similarly $\tilde{R}_{a_{i},a'_{i}} = \delta_{a_{i},a'_{i}}$ because of right normalization, hence 
$\tilde{O}_{a_{i-1}a_i,a'_{i-1}a'_i} = \delta_{a_{i-1},a'_{i-1}} \delta_{a_{i},a'_{i}}$. In the linear equation system this means that $P=I$ and we have trivially
\begin{equation}
v = b,
\end{equation}
so there is no need to solve a large equation system (Fig.~\ref{fig:normalizediterativecompression}). 

To make this work for an entire chain, we have to shift the boundary between the left and right normalized matrices as we move through the chain. Assume we begin with all left-normalized matrices. Then we move through the chain from right to left, start by solving on the last site for $\tilde{M}^{\sigma_L}$, right-normalize it via SVD (or, as we do not need the singular values, more cheaply by QR) as before, to obtain $\tilde{A}^{\sigma_1} \ldots \tilde{A}^{\sigma_{L-2}} \tilde{M}^{\sigma_{L-1}} \tilde{B}^{\sigma_{L}}$ where $\tilde{M}^{\sigma_{L-1}}$ is in general without normalization. It is now optimized as
\begin{equation}
\tilde{M}^{\sigma_i}_{a_{i-1},a_i} =  \sum_{a'_{i-1}} L_{a_{i-1},a'_{i-1}} \left( \sum_{a'_i} R_{a_{i},a'_{i}}
M^{\sigma_i}_{a'_{i-1},a'_i} \right), 
\label{eq:compressionrhs}
\end{equation}
where 
\begin{equation}
L_{a_{i-1},a'_{i-1}} = \left( \sum_{\sigma_{i-1}}  \tilde{M}^{\sigma_{i-1}\dagger} \left( \ldots \left( \sum_{\sigma_1}  
\tilde{M}^{\sigma_{1}\dagger} M^{\sigma_{1}} \right) \ldots \right) M^{\sigma_{i-1}} \right)_{a_{i-1},a'_{i-1}}
\end{equation}
and similarly $R$, the result is right-normalized, and so on as we go through the chain. At the end, all matrices are right-normalized, and we restart from the left.

In order to assess convergence, we can monitor at each step $\| \ket{\psi} - \ket{\tilde{\psi}} \|^2$, and observe the convergence of this value; if necessary, $D$ has to be increased. The calculation may seem costly, but isn't. If we keep $\ket{\tilde{\psi}}$ in proper mixed normalization, and use Eq.~(\ref{eq:compressionrhs}) to simplify the overlap $\braket{\psi}{\tilde{\psi}}$, we find
\begin{equation}
\| \ket{\psi} - \ket{\tilde{\psi}} \|^2 = 1 - \sum_{\sigma_i} \tr (\tilde{M}^{\sigma_i\dagger}\tilde{M}^{\sigma_i}) , 
\end{equation}
which is easy to calculate. The subtracted sum is just $\braket{\tilde{\psi}}{\tilde{\psi}}$; at the end, this allows us to normalize the state $\ket{\tilde{\psi}}$ by simple rescaling. 

\begin{figure}
\centering\includegraphics[scale=0.6]{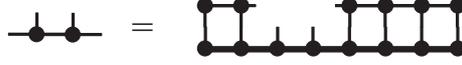}
\caption{Equation for iterative compression of an MPS in a two-site approach.}
\label{fig:twositecompression}
\end{figure}

As already hinted at for single-site DMRG -- and we will discuss this issue at length in Sec. \ref{sec:groundstates} -- there is a danger that this variational ansatz gets stuck in a non-global minimum for the distance between the compressed and the original state. Often (but not always) it is helpful to consider two sites at the same time, by analogy to two-site DMRG, for optimization. An operation count shows that this is somewhat slower. Assume the compressed $\ket{\tilde{\psi}}$ is in a mixed-canonical representation
\begin{equation}
\ket{\tilde{\psi}} = \sum_{{\fat \sigma}} \tilde{A}^{\sigma_1} \ldots \tilde{A}^{\sigma_{\ell-1}} \tilde{M}^{\sigma_{\ell}\sigma_{\ell+1}} \tilde{B}^{\sigma_{\ell+2}} \ldots \tilde{B}^{\sigma_L}  \ket{{\fat \sigma}}.
\end{equation}
Running through the same arguments as before, optimizing with respect to $\tilde{M}^{\sigma_{\ell}\sigma_{\ell+1}*}_{a_{\ell-1},a_{\ell+1}}$, yields an equation as in Fig.~\ref{fig:twositecompression} for $\tilde{M}^{\sigma_{\ell}\sigma_{\ell+1}}_{a_{\ell-1},a_{\ell+1}}$. The major change occurs now: we reshape the new $\tilde{M}^{\sigma_{\ell}\sigma_{\ell+1}}$ as $\tilde{M}_{(a_{\ell-1}\sigma_{\ell}),(\sigma_{\ell+1}a_{\ell+1})}$, carry out an SVD to obtain 
\begin{equation}
\sum_{a_\ell} \tilde{U}_{(a_{\ell-1}\sigma_\ell), a_\ell} S_{a_\ell} (V^\dagger)_{a_\ell, (\sigma_{\ell+1}a_{\ell+1})} = \sum_{a_\ell} \tilde{M}^{\sigma_\ell}_{a_{\ell-1}, a_\ell} B^{\sigma_{\ell+1}}_{a_\ell, a_{\ell+1}} ,
\end{equation}
where the $\tilde{M}$ is formed from reshaping $\tilde{U} S$. In fact, it is discarded because we shift one site towards the left, looking for $\tilde{M}^{\sigma_{\ell-1}\sigma_{\ell}}$. We can also use a cheaper QR decomposition of $\tilde{M}^\dagger$ instead, to obtain $B$ from $Q^\dagger$.

Let me conclude this section on compression by discussing the relative merits of the methods. If the compression is only small, the interdependency of the SVD approach will not matter too much. Still, the variational ansatz is superior; its only weakness is that because of its iterative nature one has to provide an initial guess for the compressed state. Taken randomly, the method will waste a lot of time on just getting into the right vicinity. Therefore, the smart proposal is to take the SVD-compressed state as the first input into the iterative method. How can we avoid the potentially high costs due to $D'$ at least partially? 

In practice, compressions occur mainly in two situations: (i) MPS have been added, hence the matrix dimensions have been added; (ii) a matrix product operator (MPO) has been applied to an MPS; we will see that this leads to the multiplication of the matrix dimensions of MPS and MPO.

In the first case, the variational compression can be speeded up by using the fact that $\ket{\psi} = \ket{\phi_1} + \ket{\phi_2} + \ldots \ket{\phi_n}$. Then we may rewrite the variational equation as
\begin{equation}
 \frac{\partial}{\partial \tilde{M}^{\sigma_i*}_{a_{i-1}a_i}} ( \braket{\tilde{\psi}}{\tilde{\psi}}
- \braket{\tilde{\psi}}{\psi} ) =  \frac{\partial}{\partial \tilde{M}^{\sigma_i*}_{a_{i-1}a_i}} ( \braket{\tilde{\psi}}{\tilde{\psi}} - \braket{\tilde{\psi}}{\phi_1}
- \braket{\tilde{\psi}}{\phi_2} - \ldots - \braket{\tilde{\psi}}{\phi_n}) = 0 .
\end{equation}
If we work out the equation (assuming mixed canonical representation), the right-hand side consists now of a sum of $n$ overlaps involving $D$-dimensional matrices, instead of one overlap involving $D$- and $nD$-dimensional matrices (costing up to $O(n^2D^3)$ and $O(nD^3)$ in the two types of matrix multiplications occurring in the overlap) we now have $n$ overlaps costing $O(D^3)$. For large $n$, the ``decomposed'' approach should be up to $n$ times faster.

The second case we postpone for details until we have discussed MPOs. The idea is to carry out an SVD compression, but without the particularly costly step of previously ensuring correct normalization; if for some reason the block states are almost orthonormal nevertheless, the outcome should be quite reasonable (and can be brought into canonical form, which is cheap after compression) or can at least serve as a reasonable input for the variational method \cite{Stoudenmire10}.

\subsection{Notations and conversions}
So far, we have explored an MPS notation based on one set of matrices per site; special normalization properties for these matrices were exploited to arrive at MPS with attractive additional features (like the generation of orthonormal sets of states or the encoding of a Schmidt decomposition). If we consider our lattice with sites $1$ through $L$, it would be useful in view of the DMRG construction to be able to access easily all $L-1$ possible bipartitionings of the system AB that can be obtained with a single cut. 

\begin{figure}
\centering\includegraphics[width=\textwidth]{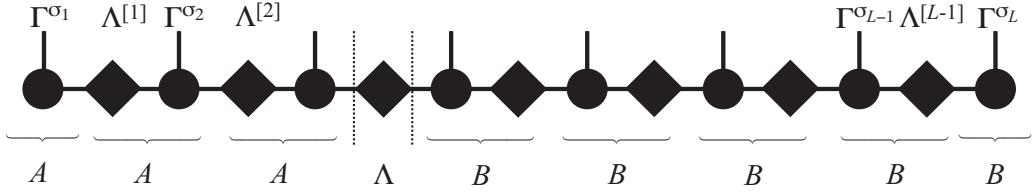}
\caption{Representation of an MPS in Vidal's notation. Singular values remain explicit on bonds (diamonds). $\Lambda$ sit on bonds, $\Gamma$ on sites. By construction, adjacent $\Lambda$ and $\Gamma$ can be contracted to $A$ or $B$ matrices, that are either left- or right-normalized. The state can be trivially grouped into a string of $A$ (giving orthonormal block states), a singular value matrix, and a string of $B$ (giving orthonormal block states). }
\label{fig:vidalmps}
\end{figure}

Such a notation has been introduced by Vidal\cite{Vidal03a} and takes the following form:
\begin{equation}
\ket{\psi} = \sum_{\sigma_1,\ldots,\sigma_L} \Gamma^{\sigma_1} \Lambda^{[1]}  \Gamma^{\sigma_2} \Lambda^{[2]}  \Gamma^{\sigma_3} \Lambda^{[3]} \ldots  \Gamma^{\sigma_{L-1}} \Lambda^{[L-1]}  \Gamma^{\sigma_L} \ket{\sigma_1,\ldots,\sigma_L} ,
\label{eq:canonicalform}
\end{equation} 
where we introduce on each site $\ell$ a set of $d$ matrices $\Gamma^{\sigma_\ell}$ and on each bond $\ell$ one diagonal matrix $\Lambda^{[\ell]}$. The matrices are specified by the demand that for arbitrary $1\leq \ell < L$ we can read off the Schmidt decomposition
\begin{equation}
\ket{\psi} = \sum_{a_\ell} s_{a_\ell} \ket{a_\ell}_A \ket{a_\ell}_B
\end{equation}
where the Schmidt coefficients are the diagonal elements of $\Lambda^{[\ell]}$, $s_{a_\ell} = \Lambda^{[\ell]}_{a_\ell,a_\ell}$ and the states on A and B are given as
\begin{eqnarray}
\ket{a_\ell}_A &=& \sum_{\sigma_1,\ldots,\sigma_\ell}
 (\Gamma^{\sigma_1}\Lambda^{[1]}\Gamma^{\sigma_2} \ldots \Lambda^{[\ell-1]} \Gamma^{\sigma_\ell})_{a_{\ell}}
\ket{\sigma_1,\ldots,\sigma_\ell}, \\
\ket{a_\ell}_B &=& \sum_{\sigma_{\ell+1},\ldots,\sigma_L} 
(\Gamma^{\sigma_{\ell+1}}\Lambda^{[\ell+1]}\Gamma^{\sigma_{\ell+2}} \ldots \Lambda^{[L-1]} \Gamma^{\sigma_L})_{a_\ell}
\ket{\sigma_{\ell+1},\ldots,\sigma_L},
\end{eqnarray}
where the states on A and on B are orthonormal respectively, reminding of similar constructions from $A$- and $B$-matrices. Graphically, the new notation can be represented as in Fig.~\ref{fig:vidalmps}. It is obviously a more explicit version of the $A$-matrix notation with the advantage of keeping explicit reference to the singular values, reduced density matrix eigenvalues and entanglement: Cutting bond $\ell$, the reduced density operators $\hat{\rho}_A$ and $\hat{\rho}_B$ read in eigenbasis representation
\begin{equation}
\rho_A^{[\ell]} = \rho_B^{[\ell]} = (\Lambda^{[\ell]})^2 ,
\end{equation}
more precisely (but irrelevant for real and diagonal $\Lambda^{[\ell]}$) $\rho_A^{[\ell]} = \Lambda^{[\ell]}\Lambda^{[\ell]\dagger}$ and $\rho_B^{[\ell]} = \Lambda^{[\ell]\dagger}\Lambda^{[\ell]}$,
where the eigenstates of $\rho_A^{[\ell]}$ and $\rho_B^{[\ell]}$ are given by $\{ \ket{a_\ell}_A \}$ and $\{ \ket{a_\ell}_B \}$ respectively. The von Neumann entropy of entanglement can be read off directly from $\Lambda^{[\ell]}$ as $S_{A|B} = -\tr (\Lambda^{[\ell]})^2 \log_2 (\Lambda^{[\ell]})^2$. 

Before exploring the connections to other notations, let us first show that any quantum state can indeed be brought into that form by a procedure in close analogy to the one that decomposed $\ket{\psi}$ into a product of $A$-matrices (or $B$-matrices, for that matter). Starting from coefficients $c_{\sigma_1\ldots\sigma_L}$, we reshape to $\Psi_{\sigma_1, (\sigma_2 \ldots \sigma_L)}$, which is SVDed iteratively. We label the singular value matrices $\Lambda^{[i]}$. After the first SVD, we rename $A^{\sigma_1}$ to $\Gamma^{\sigma_1}$. In the subsequent SVDs, as before we form the next matrix to be SVDed by multiplying $\Lambda$ and $V^\dagger$ into $\Psi$, reshaping such that there is always one $a$- and one $\sigma$-index for the rows. Using the reshaping of $U_{(a_{\ell-1}\sigma_\ell), a_{\ell}} \rightarrow A^{\sigma_\ell}_{a_{\ell-1},a_\ell}$ already used, we obtain
\begin{eqnarray*}
c_{\sigma_1\ldots\sigma_L} &=& \Psi_{\sigma_1, (\sigma_2 \ldots \sigma_L)} \\
&=& \sum_{a_1} A^{\sigma_1}_{a_1} \underbrace{\Lambda^{[1]}_{a_1,a_1} (V^\dagger)_{a_1, (\sigma_2 \ldots \sigma_L)}} \\
&=& \sum_{a_1} \Gamma^{\sigma_1}_{a_1} \Psi_{(a_1\sigma_2), (\sigma_3 \ldots \sigma_L)} \\
&=& \sum_{a_1,a_2} \Gamma^{\sigma_1}_{a_1} A^{\sigma_2}_{a_1,a_2} \underbrace{\Lambda^{[2]}_{a_2,a_2} (V^\dagger)_{a_2, (\sigma_3 \ldots \sigma_L)}} \\
&=& \sum_{a_1,a_2} \Gamma^{\sigma_1}_{a_1} \overbrace{\Lambda^{[1]}_{a_1,a_1} \Gamma^{\sigma_2}_{a_1,a_2}} \Psi_{(a_2\sigma_3), (\sigma_4 \ldots \sigma_L)} \\
&=& \sum_{a_1,a_2,a_3} \Gamma^{\sigma_1}_{a_1} \Lambda^{[1]}_{a_1,a_1} \Gamma^{\sigma_2}_{a_1,a_2} A^{\sigma_3}_{a_2,a_3}  \underbrace{\Lambda^{[3]}_{a_3,a_3} (V^\dagger)_{a_3, (\sigma_4 \ldots \sigma_L)}} \\ 
&=& \sum_{a_1,a_2,a_3} \Gamma^{\sigma_1}_{a_1} \Lambda^{[1]}_{a_1,a_1} \Gamma^{\sigma_2}_{a_1,a_2} \overbrace{\Lambda^{[2]}_{a_2,a_2} \Gamma^{\sigma_3}_{a_2,a_3}}  \Psi_{(a_3\sigma_4), (\sigma_5 \ldots \sigma_L)} 
\end{eqnarray*}
and so on. The crucial difference to the decomposition into $A$-matrices is that each $A$ is decomposed, using the knowledge of $\Lambda^{[\ell-1]}$ obtained in the previous step, into 
\begin{equation}
A^{\sigma_\ell}_{a_{\ell-1},a_\ell} = \Lambda^{[\ell-1]}_{a_{\ell-1},a_{\ell-1}} \Gamma^{\sigma_\ell}_{a_{\ell-1},a_\ell} ,
\label{eq:agammalambda}
\end{equation}
which implies a division by the diagonal elements of $\Lambda^{[\ell-1]}$. If in our SVD we keep only the non-zero singular values, this is a mathematically valid operation, albeit potentially fraught with numerical difficulty. Ignoring this issue in this conceptual demonstration, we do arrive at a decomposition of the desired form; in order to prove that it is indeed correct, we have to show that at each iteration we indeed obtain a Schmidt decomposition. But this is easy to see: The matrices to the left of any $\Lambda^{[\ell]}$ can all be grouped into (or rather, have been generated from) left-normalized $A$-matrices, which generate a set of orthonormal states on the part of the lattice ranging from site 1 to $\ell$. On the right hand side of any $\Lambda^{[\ell]}$, there is a matrix $V^\dagger$ with orthonormal rows, which means that the states $\ket{a_\ell}_B = \sum_{\sigma_{\ell+1},\ldots}
(V^\dagger)_{a_\ell, \sigma_{\ell+1} \ldots} \ket{\sigma_{\ell+1} \ldots}$ are also orthonormal. Hence, the SVD giving $\Lambda^{[\ell]}$ indeed leads to a valid Schmidt decomposition.

An alternative way of obtaining this notation would be to carry out a standard left-normalized decomposition, and store all singular value matrices generated (and previously discarded) as $\Lambda^{[i]}$, and to insert afterwards the identities $\Lambda^{[i]}(\Lambda^{[i]})^{-1}$ between all neighbouring $A$-matrices $A^{\sigma_i}$ and $A^{\sigma_{i+1}}$. Then using Eq.~(\ref{eq:agammalambda}) leads to the same result. 

Similarly, starting the decomposition from the right using the right-normalization of $B$-matrices the same state is obtained with a grouping
\begin{equation}
B^{\sigma_\ell}_{a_{\ell-1},a_{\ell}} = \Gamma^{\sigma_\ell}_{a_{\ell-1},a_\ell} \Lambda^{[\ell]}_{a_\ell,a_\ell} ,
\label{eq:bgammalambda}
\end{equation}
where for notational simplification for this and for the corresponding equation for the $A$-matrix, Eq.~(\ref{eq:agammalambda}), it is useful to introduce dummies $\Lambda^{[0]}$ and $\Lambda^{[L]}$ that are both scalar 1.

The groupings for $A$ and $B$-matrices allow to reexpress the left- and right-normalization conditions in the $\Gamma\Lambda$-language: The left-normalization condition reads 
\begin{equation}
I = \sum_{\sigma_i} A^{\sigma_i\dagger} A^{\sigma_i} = \sum_{\sigma_i} \Gamma^{\sigma_i\dagger} \Lambda^{[i-1]\dagger} \Lambda^{[i-1]} \Gamma^{\sigma_i}
\end{equation}
or, more compactly,
\begin{equation}
\sum_{\sigma_i}  \Gamma^{\sigma_i\dagger} \rho_B^{[i-1]} \Gamma^{\sigma_i} = I .
\label{eq:leftnormalizevidal}
\end{equation}
The right-normalization condition reads
\begin{equation}
\sum_{\sigma_i} \Gamma^{\sigma_i} \rho_A^{[i]} \Gamma^{\sigma_i\dagger} = I . 
\label{eq:rightnormalizevidal}
\end{equation}
Interestingly, Eqns.~(\ref{eq:leftnormalizevidal}) and (\ref{eq:rightnormalizevidal}) also arise if we translate the density operator recursions Eqns.~(\ref{eq:rhoarecursion}) and (\ref{eq:rhobrecursion}) using Eqns.~(\ref{eq:agammalambda}) and  (\ref{eq:bgammalambda}). 
A matrix product state in the form of Eq.~(\ref{eq:canonicalform}) which meets the constraints Eq.~(\ref{eq:leftnormalizevidal}) and Eq.~(\ref{eq:rightnormalizevidal}) is called {\em canonical}.

Conversions between the $AB$-notation, the $\Gamma\Lambda$-notation and also the block-site notation of DMRG are possible, albeit fraught with some numerical pitfalls.

{\em Conversion} $\Gamma\Lambda \rightarrow A,B$: The conversion from $\Gamma\Lambda \rightarrow A,B$ is easy. If one introduces an additional dummy scalar $\Lambda^{[0]} = 1$ as a ``matrix'' to the very left of $\ket{\psi}$, we can use the above defining Eq.~(\ref{eq:agammalambda}) to group 
\begin{equation}
(\Lambda^{[0]} \Gamma^{\sigma_1})
(\Lambda^{[1]} \Gamma^{\sigma_2})(\Lambda^{[2]} \Gamma^{\sigma_3})
 \ldots \rightarrow A^{\sigma_1}A^{\sigma_2}A^{\sigma_3}\ldots 
 \end{equation}
or using Eq.~(\ref{eq:bgammalambda}) 
\begin{equation}
(\Gamma^{\sigma_1}\Lambda^{[1]})
(\Gamma^{\sigma_2}\Lambda^{[2]} )(\Gamma^{\sigma_3}\Lambda^{[3]})
 \ldots \rightarrow B^{\sigma_1}B^{\sigma_2}B^{\sigma_3}\ldots 
 \end{equation}

In view of what DMRG and other MPS methods actually do, it is interesting to consider {\em mixed} conversions. Consider bond $\ell$ between sites $\ell$ and $\ell+1$. We could contract $\Lambda \Gamma \rightarrow A$ starting from the left, giving left-normalized matrices, and $\Gamma\Lambda \rightarrow B$ from the right, giving right normalized matrices, leaving out just $\Lambda^{[\ell]}$ in the center (Fig.~\ref{fig:dmrg0mps}):
\begin{figure}
\centering\includegraphics[width=\textwidth]{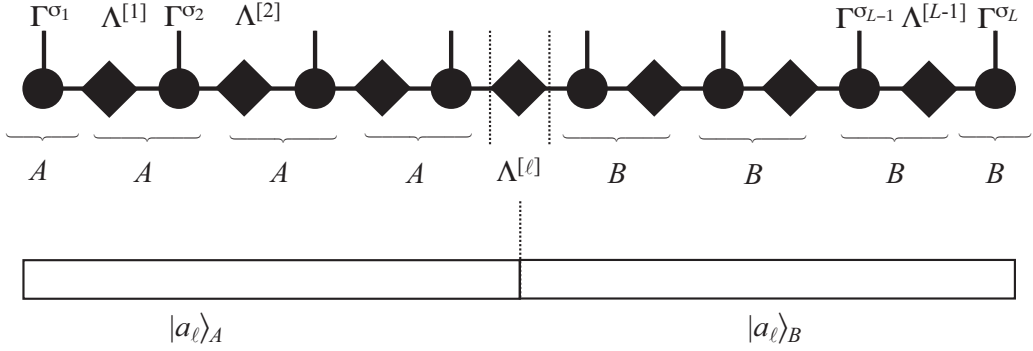}
\caption{Vidal's MPS notation, $A$, $B$-matrix MPS notation, and DMRG block notation. The $A$-matrices generate the left block states, the $B$ matrices generate the right block states. The matrix $\Lambda^{[\ell]}$ connects them via singular values.}
\label{fig:dmrg0mps}
\end{figure}

\begin{figure}
\centering\includegraphics[width=\textwidth]{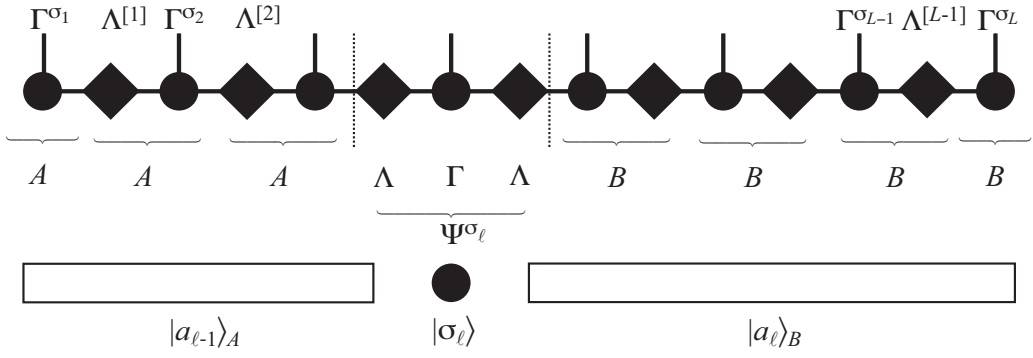}
\caption{Representation of a state in single-site DMRG: translating Vidal's MPS notation and $A,B$-matrix MPS notation into DMRG block notation. The $A$-matrices generate the left block states, the $B$ matrices generate the right block states. The matrix elements of $\Psi^{\sigma_\ell}$ are just the coefficients of the DMRG state.}
\label{fig:dmrg1mps}
\end{figure}

\begin{equation}
(\Lambda^{[0]}\Gamma)(\Lambda^{[1]}\Gamma)\ldots(\Lambda^{[\ell-2]}\Gamma)(\Lambda^{[\ell-1]}\Gamma) \Lambda^{[\ell]} (\Gamma\Lambda^{[\ell+1]})(\Gamma\Lambda^{[\ell+2]}) \ldots (\Gamma\Lambda^{[L]}) .
\end{equation}
As the bracketing to the left of bond $\ell$ generates left-normalized $A$-matrices and right-normalized matrices $B$ on the right, we can multiply them out as done in the recursions of the previous Section to arrive at orthonormal block bases for A and B, hence at a Schmidt decomposition
\begin{equation}
\ket{\psi} = \sum_{a_{\ell}} \ket{a_\ell}_A s_{a_\ell} \ket{a_\ell}_B .
\end{equation}
What is more, we can also take one site ($\ell$) or two sites ($\ell,\ell+1$) and multiply all matrices into one there (Fig.~\ref{fig:dmrg1mps} and Fig.~\ref{fig:dmrg2mps}):
\begin{figure}
\centering\includegraphics[width=\textwidth]{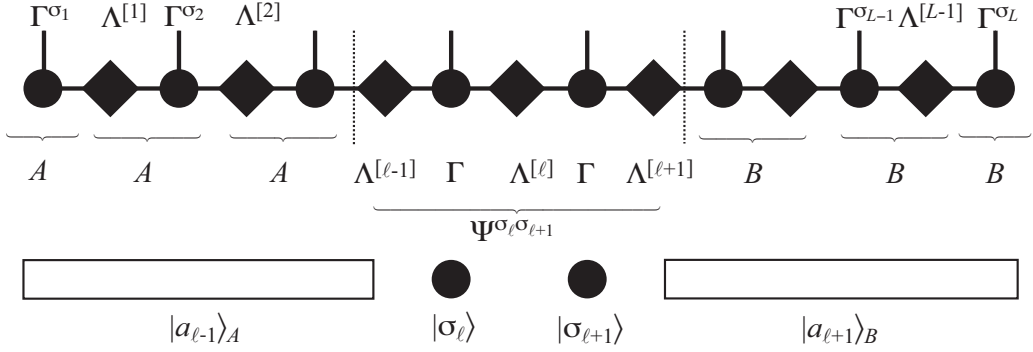}
\caption{Representation of a state in two-site DMRG: translating Vidal's MPS notation and $A,B$-matrix MPS notation into DMRG block notation. The $A$-matrices generate the left block states, the $B$ matrices generate the right block states. The elements of matrix $\Psi^{\sigma_{\ell}\sigma_{\ell+1}}$ are just the coefficients of the DMRG state.}
\label{fig:dmrg2mps}
\end{figure}

\begin{equation}
(\Lambda^{[0]}\Gamma)(\Lambda^{[1]}\Gamma)\ldots(\Lambda^{[\ell-2]}\Gamma)(\Lambda^{[\ell-1]}\Gamma \Lambda^{[\ell]}) (\Gamma\Lambda^{[\ell+1]})(\Gamma\Lambda^{[\ell+2]}) \ldots (\Gamma\Lambda^{[L]}) .
\end{equation}
Calling the central matrix $\Psi^{\sigma_\ell}= \Lambda^{[\ell-1]}\Gamma^{\sigma_\ell} \Lambda^{[\ell]}$, we can write 
\begin{equation}
\ket{\psi} = \sum_{\fat{\sigma}} A^{\sigma_1} \ldots A^{\sigma_{\ell-1}} \Psi^{\sigma_\ell} B^{\sigma_{\ell+1}} \ldots B^{\sigma_L}
\ket{\fat{\sigma}}
\end{equation}
or, again building block bases,
\begin{equation}
\ket{\psi} = \sum_{a_{\ell-1},a_{\ell},\sigma_\ell} \ket{a_{\ell-1}}_A \Psi^{\sigma_\ell}_{a_{\ell-1}a_\ell} \ket{a_\ell}_B .
\end{equation}
If we group even two sites, we have
\begin{equation}
(\Lambda^{[0]}\Gamma)(\Lambda^{[1]}\Gamma)\ldots(\Lambda^{[\ell-2]}\Gamma)(\Lambda^{[\ell-1]}\Gamma \Lambda^{[\ell]} \Gamma\Lambda^{[\ell+1]})(\Gamma\Lambda^{[\ell+2]}) \ldots (\Gamma\Lambda^{[L]}) 
\end{equation}
or, with central matrix $\Psi^{\sigma_\ell \sigma_{\ell+1}}=\Lambda^{[\ell-1]}\Gamma^{\sigma_\ell} \Lambda^{[\ell]} \Gamma^{\sigma_{\ell+1}}\Lambda^{[\ell+1]}$,
\begin{equation}
\ket{\psi} = \sum_{\fat{\sigma}} A^{\sigma_1} \ldots A^{\sigma_{\ell-1}} \Psi^{\sigma_\ell \sigma_{\ell+1}} B^{\sigma_{\ell+2}} \ldots B^{\sigma_L} \ket{\fat{\sigma}}
\end{equation}
or, using block bases,
\begin{equation}
\ket{\psi} = \sum_{a_{\ell-1},a_{\ell+1},\sigma_\ell,\sigma_{\ell+1}} \ket{a_{\ell-1}}_A \Psi^{\sigma_\ell \sigma_{\ell+1}}_{a_{\ell-1}a_{\ell+1}} \ket{a_{\ell+1}}_B .
\end{equation}
These are just the states considered by ``single-site'' and the original ``two-site'' DMRG, which keep one or two sites explicit between two blocks.

{\em Conversion} $A,B \rightarrow \Gamma\Lambda$: Conversion in the other direction $A,B \rightarrow \Gamma\Lambda$ is more involved. The idea is to obtain iteratively the Schmidt decompositions (hence singular values) of a state, hence the diagonal matrices $\Lambda^{[\ell]}$, from which the $\Gamma^{\sigma}$-matrices can be calculated. The procedure is in fact very similar to that used to show the existence of a canonical $\Gamma\Lambda$ representation for an arbitrary quantum state.

Let us assume that $\ket{\psi}$ is right-normalized, then state coefficients take the form $B^{\sigma_1}B^{\sigma_2}B^{\sigma_3}B^{\sigma_4}\ldots$. Then we can proceed by a sequence of SVDs as
\begin{eqnarray*}
& & B^{\sigma_1}B^{\sigma_2}B^{\sigma_3}B^{\sigma_4} \ldots \\
&=& (A^{\sigma_1} \Lambda^{[1]} V^\dagger) B^{\sigma_2}B^{\sigma_3}B^{\sigma_4} \ldots \\
&=& \Gamma^{\sigma_1} M^{\sigma_2} B^{\sigma_3}B^{\sigma_4} \ldots \\
&=& \Gamma^{\sigma_1} (A^{\sigma_2}  \Lambda^{[2]} V^\dagger) B^{\sigma_3}B^{\sigma_4} \ldots \\
&=& \Gamma^{\sigma_1} \Lambda^{[1]} \Gamma^{\sigma_2} M^{\sigma_3} B^{\sigma_4} \ldots \\
\end{eqnarray*} 
and so forth. Here, the to-be-SVDed matrices $M^{\sigma_\ell} = \Lambda^{[\ell-1]} V^\dagger B^{\sigma_\ell}$. The $\Gamma^{\sigma_\ell}$-matrices are obtained from the $A^{\sigma_\ell}$-matrices by remembering $\Lambda^{[\ell-1]}$ and using Eq.~(\ref{eq:agammalambda}), which implies a division by singular values.

The division by singular values is a numerical headache as they can and will often be very small, in particular if a high-precision calculation is attempted and even very small singular values will be carried along. It is numerically advisable to proceed as in the calculation of the (pseudo)inverse of an almost singular matrix and set all $s_a < \epsilon$, with, say, $\epsilon=10^{-8}$, to 0 and exclude them from all sums (e.g.\ in a Schmidt decomposition). As we order $s_a$ by size, this implies shrinking matrices $U$ and $V^\dagger$ accordingly. These small singular values carry little weight in the reduced density operators (their square), hence the loss of accuracy in the state description is very small compared to the numerical pitfalls. In fact, the problem is that at various places in algorithms we implicitly rely on (ortho)normality assumptions that may no longer hold after a ``wild'' division.

\begin{figure}
\centering\includegraphics[scale=0.5]{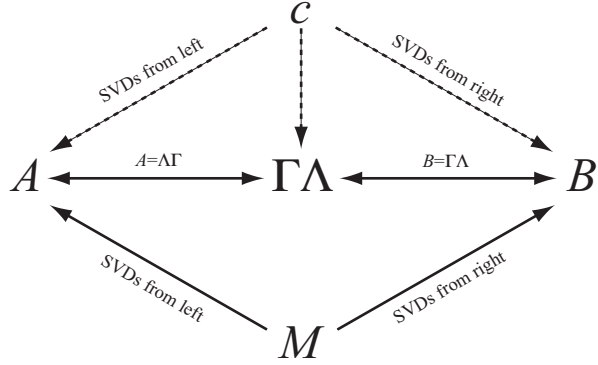}
\caption{Conversions of representations: $c$ is the explicit representation by the exponentially large number of state coefficients; $A$, $B$ and $\Gamma\Lambda$ stand for left-canonical, right-canonical and canonical MPS; $M$ stands for an arbitrary MPS. Solid lines indicate computationally feasible conversions, dashed lines more hypothetical ones.}
\label{fig:conversions}
\end{figure}

Let me conclude this long section by summarizing the various conversion and canonization procedures in a diagram (Fig.~\ref{fig:conversions}), where it should however be kept in mind that some conversions are only possible in theory, not in numerical practice.

\section{Matrix product operators (MPO)}
\label{sec:MPO}
\begin{figure}
\centering\includegraphics[width=250pt]{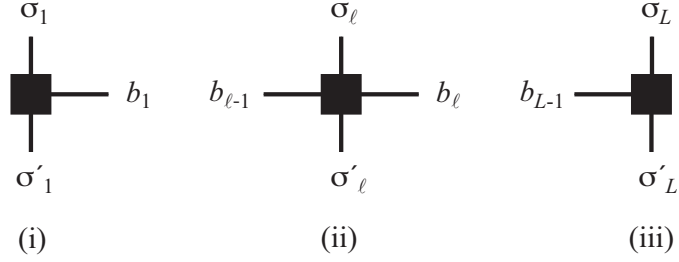}
\caption{Elements of a matrix product operator: (i) a corner matrix operator $W^{[1]\sigma_1\sigma'_1}_{1,b_1}$ at the left end of the chain; (ii)  a bulk matrix operator $W^{[\ell]\sigma_\ell\sigma'_\ell}_{b_{\ell-1},b_{\ell}}$; (iii) a corner operator $W^{[L]\sigma_L\sigma'_L}_{b_{L-1},1}$ at the right end: the physical indices points up and down, the matrix indices are represented by  horizontal lines.}
\label{fig:singleMPO}
\end{figure}

\begin{figure}
\centering\includegraphics[width=250pt]{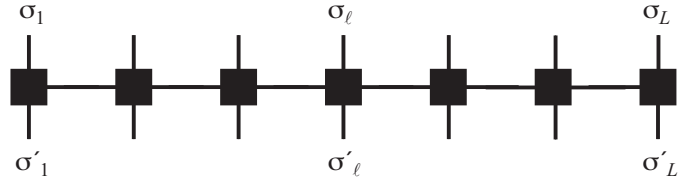}
\caption{A matrix product operator acting on an entire chain: the horizontal matrix indices are contracted, and the MPO is ready to be applied to an MPS by simple contraction of vertical (physical) indices.}
\label{fig:MPO}
\end{figure}

If we consider a single coefficient $\braket{\fat{\sigma}}{\psi}$ of an MPS, 
\[
\braket{\fat{\sigma}}{\psi} = M^{\sigma_1}M^{\sigma_2}\ldots M^{\sigma_{L-1}}M^{\sigma_L},
\]
it is a natural generalization to try to write coefficients $\bra{\fat{\sigma}} \hat{O} \ket{\fat{\sigma}'}$ of operators as \cite{VerstraeteRipoll04,McCulloch07,VerstraetePirvu08,Crosswhite08,Frowis10}
\begin{equation}
\bra{\fat{\sigma}} \hat{O} \ket{\fat{\sigma}'} = W^{\sigma_1\sigma'_1}W^{\sigma_2\sigma'_2} \ldots
W^{\sigma_{L-1}\sigma'_{L-1}}W^{\sigma_L\sigma'_L}
\end{equation}
where the $W^{\sigma\sigma'}$ are matrices just like the $M^{\sigma}$, with the only difference that as representations of operators they need both outgoing and ingoing physical states:
\begin{equation}
\hat{O} = \sum_{\fat{\sigma},\fat{\sigma}'} W^{\sigma_1\sigma'_1}W^{\sigma_2\sigma'_2} \ldots
W^{\sigma_{L-1}\sigma'_{L-1}}W^{\sigma_L\sigma'_L} \ket{\fat{\sigma}}\bra{\fat{\sigma}'} ,
\label{eq:mpoform}
\end{equation}
with the same extension to periodic boundary conditions as for MPS. The pictorial representation introduced for MPS can be extended in a straightforward fashion: instead of one vertical line for the physical state in the representation of $M$, we now have two vertical lines, one down, one up, for the ingoing and outgoing physical state in $W$ (Fig.~\ref{fig:singleMPO}). The complete MPO itself then looks as in Figure \ref{fig:MPO}. If we want to use good quantum numbers, the methods for MPS translate directly: we introduce an ingoing local state quantum number from the top, an outgoing one towards the bottom, and an ingoing quantum number from the left and an outgoing one to the right. The rule is, as for MPS, that the total sum of ingoing and outgoing quantum numbers must be equal, or $M(\ket{\sigma_i}) + M(\ket{b_{i-1}}) = M(\ket{\sigma'_i}) + M(\ket{b_{i}})$, where I have interpreted the bond labels as states for the notation. We may also think about dummy indices before the first and after the last site as in an MPS, which reflect in which (definite!) way the operator changes the total quantum number. For a Hamiltonian, which commutes with the corresponding operator, the change is zero, and we can ignore the dummies. The MPOs we are going to build can all be shown to have good quantum numbers on the bonds, because they originate either from SVDs (e.g. for time evolutions) or from rules that involve operators with well-defined changes of quantum numbers (e.g. for MPOs for Hamiltonians).

In fact, {\em any} operator can be brought into the form of Eq.~(\ref{eq:mpoform}), because it can be written as
\begin{eqnarray}
\hat{O} &=& \sum_{\sigma_1,\ldots,\sigma_L,\sigma'_1,\ldots,\sigma'_L} c_{(\sigma_1\ldots\sigma_L)(\sigma'_1\ldots\sigma'_L)} \ket{\sigma_1,\ldots,\sigma_L} \bra{\sigma'_1,\ldots,\sigma'_L}  \nonumber \\ 
&=& \sum_{\sigma_1,\ldots,\sigma_L,\sigma'_1,\ldots,\sigma'_L} c_{(\sigma_1\sigma'_1) \ldots (\sigma_L \sigma'_L)} \ket{\sigma_1,\ldots,\sigma_L} \bra{\sigma'_1,\ldots,\sigma'_L}
\end{eqnarray}
and we can decompose it like we did for an MPS, with the double index $\sigma_i \sigma'_i$ taking the role of the index $\sigma_i$ in an MPS.

As for MPS, we have to ask how we operate with them and how they can be constructed in practice, because the naive decomposition might be exponentially complex. As it turns out, most operations run in perfect analogy to the MPS case.

\subsection{Applying an MPO to an MPS}
The application of a matrix product operator to a matrix product state runs as 
\begin{eqnarray*}
\hat{O}\ket{\psi} &=& \sum_{\fat{\sigma},\fat{\sigma}'} (W^{\sigma_1,\sigma'_1} W^{\sigma_2,\sigma'_2}\ldots)(M^{\sigma'_1} M^{\sigma'_2}\ldots) \ket{\fat{\sigma}} \\
&=&  \sum_{\fat{\sigma},\fat{\sigma}'}\sum_{\fat{a},\fat{b}} (W^{\sigma_1,\sigma'_1}_{1,b_1} W^{\sigma_2,\sigma'_2}_{b_1,b_2}\ldots)(M^{\sigma'_1}_{1,a_1} M^{\sigma'_2}_{a_1,a_2} \ldots) \ket{\fat{\sigma}} \\
&=& \sum_{\fat{\sigma},\fat{\sigma}'}\sum_{\fat{a},\fat{b}} (W^{\sigma_1,\sigma'_1}_{1,b_1} M^{\sigma'_1}_{1,a_1})(W^{\sigma_2,\sigma'_2}_{b_1,b_2}M^{\sigma'_2}_{a_1,a_2}) \ldots \ket{\fat{\sigma}} \\
&=& \sum_{\fat{\sigma}}\sum_{\fat{a},\fat{b}} N^{\sigma_1}_{(1,1),(b_1,a_1)} N^{\sigma_2}_{(b_1,a_1),(b_2,a_2)} \ldots \ket{\fat{\sigma}} \\
&=& \sum_{\fat{\sigma}} N^{\sigma_1} N^{\sigma_2} \ldots \ket{\fat{\sigma}}
\end{eqnarray*}

\begin{figure}
\centering\includegraphics[width=250pt]{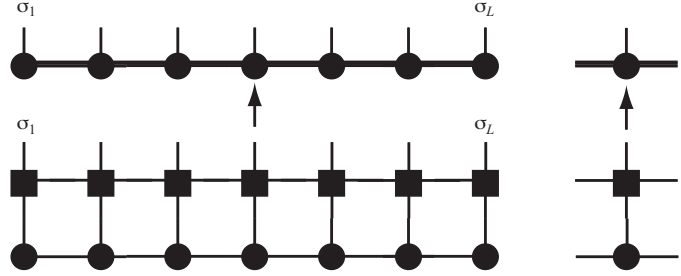}
\caption{A matrix product operator acting on a matrix product state: matching physical (vertical) indices are contracted, a new matrix product state emerges, with multiplied matrix dimensions and product structure in its matrices.}
\label{fig:MPOtimesMPS}
\end{figure}

The beauty of an MPO is that it leaves the form of the MPS invariant, at the prize of an increase in matrix size: the new MPS dimension is the product of that of the original MPS and that of the MPO (Fig.~\ref{fig:MPOtimesMPS}).

The result can be summarized as $\ket{\phi}=\hat{O}\ket{\psi}$ with $\ket{\phi}$ an MPS built from matrices $N^{\sigma_i}$ with 
\begin{equation}
N^{\sigma_i}_{(b_{i-1},a_{i-1}),(b_i,a_i)} = \sum_{\sigma_i'} W^{\sigma_i\sigma_i'}_{b_{i-1}b_i} M^{\sigma_i'}_{a_{i-1}a_i} .
\end{equation}
If we use (additive) good quantum numbers, one can show from the sum rules at each tensor that they are additive on the in- and outgoing horizontal bonds. 

Once again, a seemingly exponentially complex operation (sum over exponentially many $\fat{\sigma}$) is reduced to a low-cost operation: the operational count is of order 
$Ld^2 D_W^2 D^2$, $D_W$ being the dimension of the MPO.

\subsection{Adding and multiplying MPOs}

Operations with MPOs follow very much the lines of MPS. If we consider the addition of two operators, $\hat{O}$ and $\hat{P}$, that have MPO representations $W^{\sigma_i\sigma'_i}$ and 
$\tilde{W}^{\sigma_i\sigma'_i}$, then the resulting MPO is formed exactly as in the case of MPS, by the direct sum $W^{\sigma_i\sigma'_i} \oplus \tilde{W}^{\sigma_i\sigma'_i}$ for all sites $1 < i < L$, with the same special rules for sites $1$ and $L$. In essence, we (again) just have to consider $\sigma_i$ and $\sigma'_i$ as one ``big'' physical index. 

The multiplication of (or rather, subsequent operation with) two operators, $\hat{P} \hat{O}$, can also be understood easily: the application of $\hat{O}$ to some state $\ket{\psi}$ leads to a new MPS with 
matrices $N^{\sigma_i}_{(b_{i-1}a_{i-1}),(b_i a_i)} = \sum_{\sigma'_i} W^{\sigma_i\sigma'_i}_{b_{i-1}b_i} M^{\sigma'_i}_{a_{i-1}a_i}$. Then the subsequent operation of $\hat{P}$ gives a new MPS with $K^{\sigma_i}_{(\tilde{b}_{i-1}b_{i-1}a_{i-1}),(\tilde{b}_i b_i a_i)} =  \sum_{\sigma'_i} \tilde{W}^{\sigma_i\sigma'_i}_{\tilde{b}_{i-1} \tilde{b}_i} N^{\sigma'_i}_{(b_{i-1}a_{i-1}),(b_i a_i)}$. But from this we can read off right away (and this is also obvious from the graphical representation of a sequence of two MPOs applied to a state) that the new MPO (with matrices $V^{\sigma_i\sigma'_i}$) is given by
\begin{equation}
V^{\sigma_i\sigma'_i}_{(\tilde{b}_{i-1}b_{i-1}),(\tilde{b}_{i}b_{i})} = \sum_{\sigma''_i} 
\tilde{W}^{\sigma_i\sigma''_i}_{\tilde{b}_{i-1} \tilde{b}_i} W^{\sigma''_i\sigma'_i}_{b_{i-1}b_i} .
\end{equation}
Hence, MPO dimensions simply multiply as for tensors. If we consider an MPS as an MPO with dummy indices in one physical direction, the rule for applying an MPO to an MPS follow as a special case. 
 
\subsection{Compressing MPOs and MPO-MPS products}

As for MPS, the question of compressing an MPO may arise. This should be obvious from the last section, where MPO dimensions summed up or multiplied. If it is no option to shift the issue of compression to the application of an MPO to an MPS (then of dimension $D_W D$ and a natural candidate for compression), we have to compress the MPO. A typical example would be given by the representation of a longer-ranged Hamiltonian in MPO form, which quickly leads to large dimensions.

But we can apply the same techniques as for compressing MPS, both by SVD and iteratively, in order to approximate the exact MPO by one with smaller $D_W$. The only change is that instead of one physical index $\sigma$, we have now two physical indices $\sigma,\sigma'$, which we may take as one single index $(\sigma,\sigma')$. The approximation is then done in the Frobenius norm, which naturally extends the 2-norm of vectors we used for MPS approximations.

At this point it is worthwhile mentioning that it has been proposed \cite{Stoudenmire10} that in the special case of compressing an MPO-MPS product, an important speedup over the standard methods may be achieved: SVD may be very slow if normalization has to be carried out first at a cost $O(D_W^3 D^3)$, but a good starting point for the variational method would be essential to have. But the proposed solution from SVD compression may not be bad if the block states are almost orthonormal and it seems that in the MPO-MPS product case this is essentially true if both the MPO and the MPS were in canonical form (for the MPO again formed by looking at the double index as one big index), which can be achieved at much lower cost ($O(d D^3)$ and $O(d^2 D_W^3)$, where $D_W \ll D$ usually, versus $O(d D^3 D_W^3)$). Even if the proposed compression is not too good, it will still present a much better starting point for the variational compression. So the procedure would be: (i) bring both MPO and MPS in the same canonical form; (ii) do SVD compression, of course only multiplying out MPO and MPS matrices on the fly; (iii) use this as variational input if you don't trust the result too much.

\section{Ground state calculations with MPS}
\label{sec:groundstates}
Assume we want to find the ground state of some Hamiltonian $\hat{H}$.  In order to find the optimal approximation to it, we have to find the MPS $\ket{\psi}$ of some dimension $D$ that minimizes
\begin{equation}
E = \frac{\bra{\psi} \hat{H} \ket{\psi}}{\braket{\psi}{\psi}} .
\end{equation} 
The most efficient way of doing this (in particular compared to an imaginary time evolution starting from some random state, which is also possible) is a variational search in the MPS space. In order to make this algorithm transparent, let us first express $\hat{H}$ as an MPO.

\subsection{MPO representation of Hamiltonians}
Due to the product structure inherent in the MPO representation, it might seem a hopeless task  -- despite its guaranteed existence -- to explicitly construct a compact MPO representation for a Hamiltonian such as
\begin{equation}
\hat{H} = \sum_{i=1}^{L-1} \frac{J}{2}\hat{S}^+_i \hat{S}^-_{i+1} + \frac{J}{2}\hat{S}^-_i \hat{S}^+_{i+1} + J^z \hat{S}^z_i \hat{S}^z_{i+1} - h\sum_{i} \hat{S}^z_i .
\end{equation}
This common notation is of course an abbreviation for sums of tensor products of operators:
\begin{eqnarray*}
\hat{H} &=& J^z\hat{S}^z_1 \otimes \hat{S}^z_2 \otimes \hat{I} \otimes \hat{I} \otimes \hat{I} \ldots + \\
 & & \hat{I} \otimes J^z\hat{S}^z_2 \otimes \hat{S}^z_3 \otimes \hat{I} \otimes \hat{I} \ldots + \\
 & & \ldots
 \end{eqnarray*}
It is however surprisingly easy to express this sum of tensor products in MPO form \cite{McCulloch07} -- to this purpose it is convenient to reconsider the building block $W^{\sigma\sigma'}_{bb'}$ combined with its associated projector $\ket{\sigma}\bra{\sigma'}$ to become an operator-valued matrix $\hat{W}_{bb'} = \sum_{\sigma\sigma'} W^{\sigma\sigma'}_{bb'} \ket{\sigma}\bra{\sigma'}$ such that the MPO takes the simple form
\begin{equation}
\hat{O} = \hat{W}^{[1]}\hat{W}^{[2]} \ldots \hat{W}^{[L]} .
\end{equation}
Each $\hat{W}^{[i]}$ acts on a different local Hilbert space at site $i$, whose tensor product gives the global Hilbert space. Multiplying such operator-valued matrices yields sums of tensor products of operators such that expressing $\hat{H}$ in a compact form seems feasible. 

To understand the construction, we move through an arbitrary operator string appearing in $\hat{H}$: starting from the right end, the string contains unit operators, until at one point we encounter one of (in our example) 4 non-trivial operators. For the field operator, the string part further to the left may only contain unit operators; for the interaction operators, the complementary operator must follow immediately to complete the interaction term, to be continued by unit operators further to the left. For book-keeping, we introduce 5 corresponding states of the string at some given bond: state 1: only units to the right,
states 2,3,4: one $\hat{S}^+$, $\hat{S}^-$, $\hat{S}^z$ just to the right, state 5: completed interaction or field term $-h\hat{S}^z$ somewhere to the right. Comparing the state of a string left and right of one site, only a few transitions are allowed: $1 \rightarrow 1$ by the unit operator $\hat{I}$, $1 \rightarrow 2$ by $\hat{S}^+$, $1 \rightarrow 3$ by $\hat{S}^-$, $1 \rightarrow 4$ by $\hat{S}^z$, $1 \rightarrow 5$ by $-h\hat{S}^z$. Furthermore $2\rightarrow 5$ by $(J/2)\hat{S}^-$, $3 \rightarrow 5$ by $(J/2)\hat{S}^+$ and $4 \rightarrow 5$ by $J^z\hat{S}^z$, to complete the interaction term, and $5 \rightarrow 5$ for a completed interaction by the unit operator $\hat{I}$. Furthermore all string states must start at 1 to the right of the last site and end at 5 (i.e. the dimension $D_W$ of the MPO to be) to the left of the first site. This can now be encoded by the following operator-valued matrices:

\begin{equation}
\hat{W}^{[i]} = \left[
\begin{array}{ccccc}
\hat{I} & 0 & 0 & 0 & 0 \\
\hat{S}^+ & 0 & 0 & 0 & 0 \\
\hat{S}^- & 0 & 0 & 0 & 0 \\
\hat{S}^z & 0 & 0 & 0 & 0 \\
-h\hat{S}^z & (J/2)\hat{S}^- & (J/2)\hat{S}^+ & J^z\hat{S}^z & \hat{I}
\end{array}
\right]
\end{equation}
and on the first and last sites
\begin{equation}
\hat{W}^{[1]} = \left[
\begin{array}{ccccc}
-h\hat{S}^z & (J/2)\hat{S}^- & (J/2)\hat{S}^+ & J^z\hat{S}^z & \hat{I} 
\end{array}
\right]
\quad\quad 
\hat{W}^{[L]} = \left[
\begin{array}{c}
\hat{I} \\ \hat{S}^+ \\ \hat{S}^- \\ \hat{S}^z \\ -h\hat{S}^z 
\end{array}
\right] .
\end{equation}
Indeed, a simple multiplication shows how the Hamiltonian emerges. Inserting the explicit operator representations gives the desired $W^{\sigma\sigma'}$-matrices for the MPO. It is therefore possible to express Hamiltonians exactly in a very compact MPO form; a similar set of rules leading to the same result has been given by \cite{Crosswhite08a}.

For longer-ranged Hamiltonians, further ``intermediate states'' have to be introduced. Let us consider a model with just $\hat{S}^z \hat{S}^z$-interactions, but between nearest and next-nearest neighbours,
\begin{equation}
\hat{H} = J_1 \sum_{i} \hat{S}^z_i \hat{S}^z_{i+1} + J_2 \sum_{i} \hat{S}^z_i \hat{S}^z_{i+2} .
\end{equation}
Then the bulk operator would read
\begin{equation}
\hat{W}^{[i]} = \left[
\begin{array}{cccc}
\hat{I} & 0 & 0 & 0 \\
\hat{S}^z & 0 & 0 & 0 \\
0 & \hat{I} & 0 & 0 \\
0 & J_1\hat{S}^z & J_2\hat{S}^z & \hat{I}
\end{array}
\right] .
\end{equation}
While the $J_1$-interaction can be encoded as before (moving as $1 \rightarrow 2 \rightarrow 4$), for the next-nearest neighbour interaction, one has to insert an additional step between 2 and 4, an intermediate state 3, where exactly one identity is inserted (moving as $1 \rightarrow 2 \rightarrow 3 \rightarrow 4$). It merely serves as a book-keeping device. Similarly, one can construct longer-ranged interactions. Except the top-left and botton-right corner, the non-vanishing parts of $\hat{W}^{[i]}$ are all below the diagonal by construction.

It might seem that for longer-ranged interactions the dimension $D_W$ will grow rapidly as more and more intermediate states are needed (one additional state per unit of interaction range and per interaction term). While this is true in general, important exceptions are known which can be formulated much more compactly \cite{Frowis10,Crosswhite08a}; consider for example the following exponentially decaying interaction strength $J(r)=J e^{-r/\xi} = J \lambda^r$, where $r>0$ and $\lambda = \exp(-1/\xi)$. An interaction term $\sum_r J(r)\hat{S}^z_i \hat{S}^z_{i+r}$ would be captured by a bulk operator
\begin{equation}
\hat{W}^{[i]} = \left[
\begin{array}{ccc}
\hat{I} & 0 & 0  \\
\hat{S}^z & \lambda\hat{I} & 0  \\
0 & J\lambda \hat{S}^z & \hat{I}
\end{array}
\right] .
\end{equation}

But even if such a simplification does not occur, it turns out that MPOs with quite small dimensions and moderate loss of accuracy can be found, either by approximating an arbitrary interaction function $J(r)$ by a sum of exponentials coded as above\cite{VerstraetePirvu08,Crosswhite08a}, minimizing the $L_2$ distance $\| J(r) - \sum_{i=1}^n \alpha_i \lambda_i^r \|$ in $\alpha_i, \lambda_i$, where $n$ is given by the $D_W$ and loss of accuracy one is willing to consider. Alternatively \cite{Frowis10}, one can of course construct the exact MPO where feasible and compress it by adapting MPS compression techniques to an acceptable $D_W$ (and loss of accuracy). 

\subsection{Applying a Hamiltonian MPO to a mixed canonical state}
\label{subsec:Hamiltonianmixedcanonical}
\begin{figure}
\centering\includegraphics[scale=0.7]{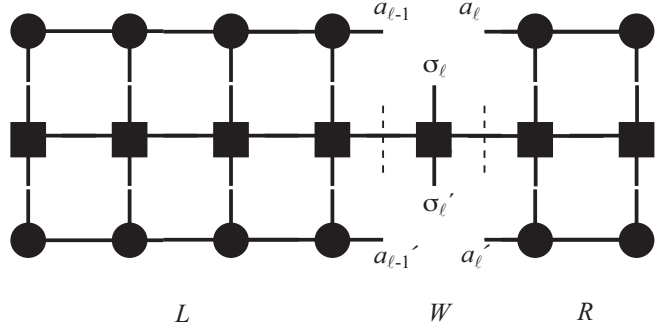}
\caption{Representation of the DMRG expression $\bra{a_{\ell-1} \sigma_{\ell} a_{\ell}} \hat{H}  \ket{a'_{\ell-1} \sigma'_{\ell} a'_{\ell}}$ in MPO/MPS language. The Hamiltonian MPO is contracted with four block state expansions in MPS form (two bras, two kets, two on block A, two on block B). The contracted network decouples into parts $L$, $W$ and $R$, corresponding to blocks A and B and the center site.}
\label{fig:mpodmrghamiltonian}
\end{figure}

Let us consider $\ket{\psi}$ in the following mixed canonical representation, identical to the single-site DMRG representation,
\begin{equation}
\ket{\psi} = \sum_{\fat{\sigma}} A^{\sigma_1} \ldots A^{\sigma_{\ell-1}} \Psi^{\sigma_{\ell}} B^{\sigma_{\ell+1}} \ldots B^{\sigma_L} \ket{\fat{\sigma}} 
\end{equation}
or 
\begin{equation}
\ket{\psi} = \sum_{a_{\ell-1},a_{\ell}} \ket{a_{\ell-1}}_A \Psi^{\sigma_{\ell}}_{a_{\ell-1}, a_{\ell}} \ket{a_{\ell}}_B .
\end{equation}
Let us now look at the matrix elements $\bra{a_{\ell-1} \sigma_{\ell} a_{\ell}} \hat{H}  \ket{a'_{\ell-1} \sigma'_{\ell} a'_{\ell}}$ obtained using the MPO representation for $\hat{H}$. By inserting twice the identity $\hat{I} = \sum_{\fat{\sigma}} \ket{\fat{\sigma}}\bra{\fat{\sigma}}$, we obtain (the sums with a star exclude site $\ell$)
\begin{eqnarray*}
& & \bra{a_{\ell-1} \sigma_{\ell} a_{\ell}} \hat{H}  \ket{a'_{\ell-1} \sigma'_{\ell} a'_{\ell}} \\
&=& \sum_{\fat{\sigma}} \sum_{\fat{\sigma}'} W^{\sigma_1,\sigma'_1} \ldots W^{\sigma_L,\sigma'_L} 
\braket{a_{\ell-1} \sigma_{\ell} a_{\ell}}{\fat{\sigma}} \braket{\fat{\sigma}'}{a'_{\ell-1} \sigma'_{\ell} a'_{\ell}} \\
&=& \sum_{\fat{\sigma}*}\sum_{\fat{\sigma}'*} W^{\sigma_1,\sigma'_1} \ldots W^{\sigma_{\ell},\sigma'_{\ell}} \ldots W^{\sigma_L,\sigma'_L} \\
& & \braket{a_{\ell-1}}{\sigma_1 \ldots \sigma_{\ell-1}} 
\braket{a_{\ell}}{\sigma_{\ell+1} \ldots \sigma_L}  \braket{\sigma'_1 \ldots \sigma'_{\ell-1}}{a'_{\ell-1}}
\braket{\sigma'_{\ell+1} \ldots \sigma'_L}{a'_{\ell}} \\
&=& \sum_{\fat{\sigma}*}\sum_{\fat{\sigma}'*} W^{\sigma_1,\sigma'_1} \ldots W^{\sigma_{\ell},\sigma'_{\ell}} \ldots W^{\sigma_L,\sigma'_L} \\
& & (A^{\sigma_1} \ldots A^{\sigma_{\ell-1}})^*_{1,a_{\ell-1}} (B^{\sigma_{\ell+1}} \ldots B^{\sigma_L})^*_{a_{\ell},1} (A^{\sigma'_1} \ldots A^{\sigma'_{\ell-1}})_{1,a'_{\ell-1}} (B^{\sigma'_{\ell+1}} \ldots B^{\sigma'_L})_{a'_{\ell},1} \\
&=& \sum_{\{ a_i, b_i, a'_i\} } \left( \sum_{\sigma_1\sigma'_1} A^{\sigma_1*}_{1,a_1} W^{\sigma_1,\sigma'_1}_{1,b_1} A^{\sigma'_1}_{1,a'_1} \right) \left( \sum_{\sigma_2\sigma'_2} A^{\sigma_2*}_{a_1,a_2} W^{\sigma_2,\sigma'_2}_{b_1,b_2} A^{\sigma'_2}_{a'_1,a'_2} \right) \ldots 
\times W^{\sigma_{\ell},\sigma'_{\ell}}_{b_{\ell-1},b_{\ell}} \times \\
& &  \left( \sum_{\sigma_{\ell+1}\sigma'_{\ell+1}} B^{\sigma_{\ell+1}*}_{a_{\ell},a_{\ell+1}} W^{\sigma_{\ell+1},\sigma'_{\ell+1}}_{b_{\ell},b_{\ell+1}} B^{\sigma'_{\ell+1}}_{a'_{\ell},a'_{\ell+1}} \right) \ldots \left( \sum_{\sigma_L\sigma'_L} B^{\sigma_L*}_{a_{L-1},1} W^{\sigma_L,\sigma'_L}_{b_{L-1},1} B^{\sigma'_L}_{a'_{L-1},1} \right) .
\end{eqnarray*}   
All the beauty of the MPO formulation seems gone, but a graphical representation restores it (Fig.~\ref{fig:mpodmrghamiltonian}). It can be understood most easily from the second or third line of the explicit expressions above: the Hamilton MPO (expressed in the product basis) is projected on the block states of A and B, which have an expansion in the $\fat{\sigma}$-basis.

In fact, we can also encode the obvious tripartite structure of the expression as
\begin{equation}
\bra{a_{\ell-1} \sigma_{\ell} a_{\ell}} \hat{H}  \ket{a'_{\ell-1} \sigma'_{\ell} a'_{\ell}} 
= \sum_{b_{\ell-1}, b_{\ell}} L^{a_{\ell-1},a'_{\ell-1}}_{b_{\ell-1}}  W^{\sigma_{\ell},\sigma'_{\ell}}_{b_{\ell-1},b_{\ell}} R^{a_{\ell},a'_{\ell}}_{b_{\ell}} ,
\end{equation}
where $L$ and $R$ contain the contracted left and right parts of the graphical network: 
\begin{eqnarray}
L^{a_{\ell-1},a'_{\ell-1}}_{b_{\ell-1}} &=& \sum_{\{ a_i, b_i, a'_i; i<\ell-1\} } \left( \sum_{\sigma_1\sigma'_1} A^{\sigma_1*}_{1,a_1} W^{\sigma_1,\sigma'_1}_{1,b_1} A^{\sigma'_1}_{1,a'_1} \right) \ldots \left( \sum_{\sigma_{\ell-1}\sigma'_{\ell-1}} A^{\sigma_{\ell-1} *}_{a_{\ell-2},a_{\ell-1}} W^{\sigma_{\ell-1},\sigma'_{\ell-1}}_{b_{\ell-2},b_{\ell-1}} A^{\sigma'_{\ell-1}}_{a'_{\ell-2},a'_{\ell-1}} \right) 
\label{eq:LdefineinMPO} \\
R^{a_{\ell},a'_{\ell}}_{b_{\ell}} &=& \sum_{\{ a_i, b_i, a'_i; i>\ell\} } 
\left( \sum_{\sigma_{\ell+1}\sigma'_{\ell+1}} B^{\sigma_{\ell+1}*}_{a_{\ell},a_{\ell+1}} W^{\sigma_{\ell+1},\sigma'_{\ell+1}}_{b_{\ell},b_{\ell+1}} B^{\sigma'_{\ell+1}}_{a'_{\ell},a'_{\ell+1}} \right) \ldots \left( \sum_{\sigma_L\sigma'_L} B^{\sigma_L*}_{a_{L-1},1} W^{\sigma_L,\sigma'_L}_{b_{L-1},1} B^{\sigma'_L}_{a'_{L-1},1} \right) 
\end{eqnarray}
We can now write the action of $\hat{H}$ on a state $\ket{\psi}$ in the mixed canonical or single-site DMRG representation as
\begin{equation}
\hat{H}\ket{\psi} = \sum_{b_{\ell-1}, b_{\ell}} \sum_{a'_{\ell-1}, \sigma'_{\ell}, a'_{\ell}}
 L^{a_{\ell-1},a'_{\ell-1}}_{b_{\ell-1}}  W^{\sigma_{\ell},\sigma'_{\ell}}_{b_{\ell-1},b_{\ell}} R^{a_{\ell},a'_{\ell}}_{b_{\ell}} \Psi^{\sigma'_{\ell}}_{a'_{\ell-1},a'_{\ell}} \ket{a_{\ell-1}}_A \ket{\sigma_{\ell}} \ket{a_{\ell}}_B .
\end{equation}
As we will discuss in an instant, $\hat{H}\ket{\psi}$ is the key operation in an iterative ground state search. Evaluating this expression naively is inacceptably slow; it can be drastically accelerated on two counts: first, $L$ and $R$ can be built iteratively in order to maximally reuse available information; this involves an optimal arrangement of a network contraction. Moreover, the final action of $L$, $R$ and $W$ on $\ket{\psi}$ can also be arranged highly efficiently.

Let us first consider building $L$ and $R$. In actual applications, we will never carry out the full network contraction that stands behind them, because in the spirit of DMRG we are looking at blocks that are growing and shrinking in size site by site. The construction of $L$ and $R$, however, is iterative in a way that directly matches block growth and shrinkage. I will illustrate it for $L$, using $A$-matrices; left-normalization will be exploited explicitly for further simplification at one point only such that the formulae are generic. We start by considering the block of size 1: we contract $A^{[1]}$ and $A^{[1]\dagger}$ with $W^{[1]}$.  The block basis representation is then given by 
\begin{equation}
F^{[1]}_{a_1,b_1;a'_1} = \sum_{\sigma_1,\sigma_1',a_0,b_0,a'_0} W^{[1]\sigma_1\sigma'_1}_{b_0,b_1} (A^{[1]\sigma_1\dagger})_{a_1,a_0} F^{[0]}_{a_0,b_0,a'_0} A^{[1]\sigma'_1}_{a'_0,a'_1}
\end{equation}
where we have introduced a dummy scalar $F^{[0]}_{a_0,b_0,a'_0}=1$, and where $a_0,b_0,a'_0$ can just take the value 1; this is just to make the first step more consistent with all that follow. The resulting object is a tensor $F^{[1]}_{a_1,b_1,a'_1}$, corresponding to the three legs sticking out. 

\begin{figure}
\centering\includegraphics[scale=0.7]{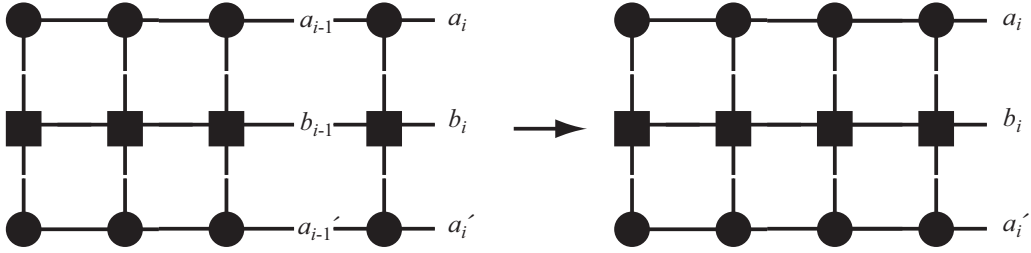}
\caption{Update from $F^{[i-1]}$ to $F^{[i]}$ by contracting with $A^{[i]*}$, $W^{[i]}$ and $A^{[i]}$. While it makes sense mathematically to consider the three added tensors as one object, in numerical practice, they are contracted into the network sequentially for efficiency.}
\label{fig:constructionEmatrices}
\end{figure}

We can now simply continue to contract $A$, $A^\dagger$ and $W$ on the next site, and the contraction update reads
\begin{equation}
F^{[i]}_{a_i,b_{i},a'_i} =  \sum_{\sigma_i,\sigma_i',a_{i-1},b_{i-1},a'_{i-1}} W^{[i]\sigma_i\sigma'_i}_{b_{i-1},b_{i}}  
(A^{[i]\sigma_i\dagger})_{a_i,a_{i-1}} F^{[i-1]}_{a_{i-1},b_{i-1},a'_{i-1}} A^{[i]\sigma'_i}_{a'_{i-1},a'_i} 
\end{equation}
and can be represented pictorially as in Fig.~\ref{fig:constructionEmatrices}. 

This construction can be calculated most efficiently by optimal bracketing as 
\begin{equation}
F^{[i]}_{a_i,b_{i},a'_i} = \sum_{\sigma_i,a_{i-1}}
(A^{[i]\sigma_i\dagger})_{a_i,a_{i-1}} 
\left( \sum_{\sigma_i',b_{i-1}}
W^{[i]\sigma_i\sigma'_i}_{b_{i-1},b_{i}}
\left( \sum_{a'_{i-1}} F^{[i-1]}_{a_{i-1},b_{i-1},a'_{i-1}} A^{[i]\sigma'_i}_{a'_{i-1},a'_i}  \right) \right) .
\end{equation}
Here, we have contracted the three new tensors into the network one by one, at operational counts $O(dD^3 D_W)$ in the innermost bracket, then $O(d^2 D^2 D_W^2)$ and last $O(d D^3 D_W)$. In fact, the second operation is faster in practice, as we know that most operators in $\hat{W}$ are simply zero; the remaining ones also often have a simple structure. Another acceleration is possible in the case of building $L$ from left-normalized matrices for indices $b_i=D_W$, if we build $\hat{H}$ following the rules outlined in the previous section: we know that in this case only identities operate towards the left, implying that $F^{[i]}_{a_i,D_W,a'_i} = \delta_{a_i,a'_i}$, simplifying both the innermost bracket and the outermost operation. The same idea applies for indices $b_i=1$ on the right side for building $R$ from right-normalized matrices.

Note that this construction is a generalization of the representation update explained for DMRG: a typical situation is the representation of $\hat{O}_i \hat{O}_j$, where the two operators act locally on sites $i$ and $j$ respectively. Then the MPOs are of dimension $(1\times 1)$ everywhere and $W^{\sigma\sigma'} = \delta_{\sigma,\sigma'}$ everywhere but on sites $i$ and $j$, where they read $W^{\sigma_i\sigma'_i} = O^{[i]\sigma_i,\sigma'_i}$ and similarly for $j$. Pushing forward the contractions, $F^{[i-1]}$ is still a scalar 1. Then
\begin{equation}
F^{[i]} =  \sum_{\sigma_i,\sigma_i'} O^{[i]\sigma_i,\sigma'_i} A^{[i]\sigma_i\dagger} A^{[i]\sigma'_i}
\end{equation}
where $F^{[i]}$, $A^{[i]\sigma_i\dagger}$ and $A^{[i]\sigma'_i}$ are matrices and a multiplication $A^{[i]\sigma_i\dagger} A^{[i]\sigma'_i}$ is implied. The $F^{[i]}$-matrix is just the operator representation in the block basis, comprising sites 1 through $i$. 

The update up to site $j-1$ then simplifies to
\begin{equation}
F^{[k]} =  \sum_{\sigma_k} A^{[k]\sigma_k\dagger} F^{[k-1]} A^{[k]\sigma_k} ,\quad (i<k<j)
\end{equation}
matrix multiplications implied, and at site $j$ we get again a non-trivial step,
\begin{equation}
F^{[j]} = \sum_{\sigma_j,\sigma'_j} O^{[j]\sigma_j,\sigma'_j} A^{[j]\sigma_j\dagger} F^{[j-1]} A^{[j]\sigma'_j},
\end{equation}
after which updates continue as on the previous sites. Making the matrix multiplications explicit, one sees that this is just the construction discussed for the DMRG algorithm.
  
In the end, $\hat{H}\ket{\psi}$ can be bracketed advantageously as follows:
\begin{equation}
\hat{H}\ket{\psi} = \sum_{b_{\ell-1}, a'_{\ell-1}} 
 L^{a_{\ell-1},a'_{\ell-1}}_{b_{\ell-1}} 
\left( \sum_{b_{\ell} \sigma'_{\ell}} 
W^{\sigma_{\ell},\sigma'_{\ell}}_{b_{\ell-1},b_{\ell}}
 \left( \sum_{a'_{\ell}} R^{a_{\ell},a'_{\ell}}_{b_{\ell}} \Psi^{\sigma'_{\ell}}_{a'_{\ell-1},a'_{\ell}} \right) \right) \ket{a_{\ell-1}}_A \ket{\sigma_{\ell}} \ket{a_{\ell}}_B ,
\end{equation}
which scales at worst as $O(D^3)$. More precisely, the innermost operation is $O(D^3 D_W d)$; the next one is $O(D^2 D_W^2 d^2)$, after this we have a sum of cost $O(D^3 D_W^2 d^2)$. It is advantageous to keep track of the structure of $W$, namely exploiting for which $(b_{\ell-1},b_{\ell})$ configurations it is zero and nothing has to be calculated (usually, for most of them), and to use the simplifications for $L$ and $R$ just discussed if the state is in mixed-canonical form.
 
\subsection{Iterative ground state search}

\begin{figure}
\centering\includegraphics[scale=0.7]{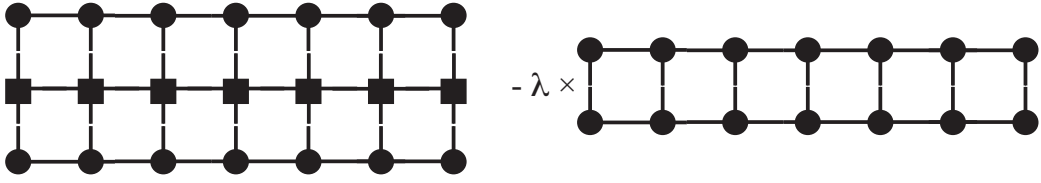}
\caption{Network to be contracted to obtain the functional to be extremized to find the ground state and its energy. The left-hand side represents the term $\bra{\psi} \hat{H} \ket{\psi}$, the right-hand side the squared norm $\braket{\psi}{\psi}$.}
\label{fig:extremizationexpression}
\end{figure}

Let us now turn to the algorithm. Assume $\hat{H}$ given in MPO form and consider a class of MPS with predefined matrix dimensions (simply think about a random MPS with matrices $M^\sigma$ of desired shape and size, but no normalization assumed for the moment). In order to find the optimal approximation to the ground state within this class, we have to find the MPS $\ket{\psi}$ that minimizes
\begin{equation}
E = \frac{\bra{\psi} \hat{H} \ket{\psi}}{\braket{\psi}{\psi}} .
\end{equation} 
It turns out that this can be turned into a ground state algorithm much more efficient than imaginary time evolution from some random state. In order to solve this problem, we introduce a Lagrangian multiplier $\lambda$, and extremize 
\begin{equation}
\bra{\psi} \hat{H} \ket{\psi} - \lambda {\braket{\psi}{\psi}} ;
\label{eq:Lagrange}
\end{equation}
in the end, $\ket{\psi}$ will be the desired ground state and $\lambda$ the ground state energy. The MPS network that represents Eq.~(\ref{eq:Lagrange}) is shown in Fig.~\ref{fig:extremizationexpression}. 

The problem with this approach is that the variables (the matrix elements $M^\sigma_{aa'}$) appear in the form of products, making this a highly non-linear optimization problem. But it can be done iteratively, too, and this is the idea that also drives DMRG: while keeping the matrices on all sites but one ($\ell$) constant, consider only the matrix entries $M^{\sigma_\ell}_{a_{\ell-1}a_\ell}$ on site $\ell$ as variables. Then the variables appear in Eq.~(\ref{eq:Lagrange}) only in quadratic form, for which the determination of the extremum is a benign linear algebra problem. This will lower the energy, and find a variationally better state, but of course not the optimal one. Now one continues to vary the matrix elements on another site for finding a state again lower in energy, moving through all sites multiple times, until the energy does not improve anymore.

Let us first consider the calculation of the overlap, while keeping the chosen $M^{\sigma_\ell}$ explicit. We find 
\begin{equation}
\braket{\psi}{\psi} = \sum_{\sigma_\ell} \sum_{a_{\ell-1}a_{\ell}} \sum_{a'_{\ell-1}a'_{\ell}} \Psi^A_{a_{\ell-1},a'_{\ell-1}} M^{\sigma_\ell*}_{a_{\ell-1}, a_{\ell}} M^{\sigma_\ell}_{a'_{\ell-1}, a'_{\ell}} \Psi^B_{a_\ell, a_\ell'},
\end{equation}
where 
\begin{eqnarray}
\Psi^A_{a_{\ell-1},a'_{\ell-1}} &=& \sum_{\sigma_1,\ldots,\sigma_{\ell-1}} (M^{\sigma_{\ell-1}\dagger} \ldots M^{\sigma_1\dagger} M^{\sigma_1} \ldots M^{\sigma_{\ell-1}})_{a_{\ell-1},a'_{\ell-1}} \\
\Psi^B_{a_\ell, a'_\ell} &=& \sum_{\sigma_{\ell+1},\ldots,\sigma_L} (M^{\sigma_{\ell+1}} \ldots M^{\sigma_L} M^{\sigma_L\dagger} \ldots M^{\sigma_{\ell+1}\dagger})_{a'_\ell,a_\ell} .
\end{eqnarray}
As is particularly clear in the graphical representation, for obtaining the last two expressions the same rules about smart contracting apply as for overlaps; moreover, if we move through sites $\ell$ from neighbour to neighbour, they can be updated iteratively, minimizing computational cost. In the case where sites 1 through $\ell-1$ are left-normalized and sites $\ell+1$ through $L$ right-normalized, normalization conditions lead to a further simplification, namely
\begin{equation} 
\Psi^A_{a_{\ell-1},a'_{\ell-1}} = \delta_{a_{\ell-1},a'_{\ell-1}} \quad\quad  \Psi^B_{a_\ell a'_\ell} = \delta_{a_\ell a'_\ell} .
\end{equation}

Let us now consider $\bra{\psi} \hat{H} \ket{\psi}$, with $\hat{H}$ in MPO language. Taking into account the analysis of $\hat{H}\ket{\psi}$ in the last section, we can immediately write
\begin{equation}
\bra{\psi} \hat{H} \ket{\psi} = \sum_{\sigma_\ell,\sigma'_\ell} \sum_{a'_{\ell-1}a'_{\ell}}  \sum_{a_{\ell-1}a_{\ell}} \sum_{b_{\ell-1},b_\ell} L^{a_{\ell-1},a'_{\ell-1}}_{b_{\ell-1}} W^{\sigma_\ell,\sigma'_\ell}_{b_{\ell-1},b_\ell} R^{a_\ell,a'_\ell}_{b_\ell} M^{\sigma_\ell*}_{a_{\ell-1},a_\ell} M^{\sigma'_\ell}_{a'_{\ell-1},a'_\ell} 
\end{equation}
with $L$ and $R$ as defined before; how such an expression can be evaluated efficiently has been discussed previously.

\begin{figure}
\centering\includegraphics[scale=0.7]{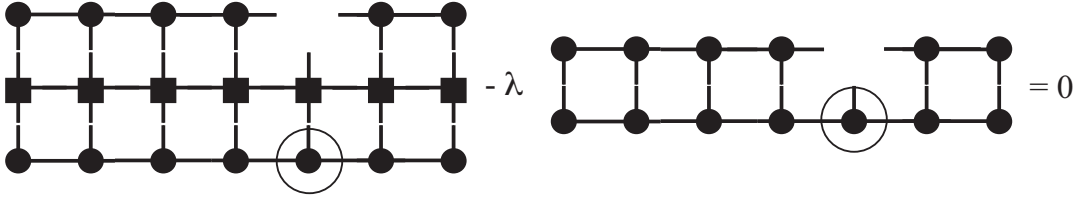}
\caption{Generalized eigenvalue problem for the optimization of $M^{\ \sigma_\ell}_{\ a_{\ell-1},a_\ell}$. The unknown matrix is circled on the left and right networks.}
\label{fig:equationsystem}
\end{figure}

If we now take the extremum of Eq.~(\ref{eq:Lagrange}) with respect to $M^{\sigma_\ell*}_{a_{\ell-1},a_\ell}$ we find
\begin{equation}
\sum_{\sigma'_\ell}  \sum_{a'_{\ell-1}a'_{\ell}} \sum_{b_{\ell-1},b_\ell} L^{a_{\ell-1},a'_{\ell-1}}_{b_{\ell-1}} W^{\sigma_\ell,\sigma'_\ell}_{b_{\ell-1},b_\ell} R^{a_\ell,a'_\ell}_{b_\ell}  M^{\sigma'_\ell}_{a'_{\ell-1},a'_\ell} 
 - \lambda  \sum_{a'_{\ell-1}a'_{\ell}} \Psi^A_{a_{\ell-1},a'_{\ell-1}} \Psi^B_{a_\ell a'_\ell} M^{\sigma_\ell}_{a'_{\ell-1}, a'_{\ell}} = 0.
 \label{eq:vMPSeigenequation}
\end{equation}
This is in fact a very simple eigenvalue equation; if we introduce matrices $H$ and $N$ by reshaping $H_{(\sigma_\ell a_{\ell-1} a_\ell), (\sigma'_\ell a'_{\ell-1} a'_\ell)} = \sum_{b_{\ell-1},b_\ell} L^{a_{\ell-1},a'_{\ell-1}}_{b_{\ell-1}} W^{\sigma_\ell,\sigma'_\ell}_{b_{\ell-1},b_\ell} R^{a_\ell,a'_\ell}_{b_\ell} $ and 
$N_{(\sigma_\ell a_{\ell-1} a_\ell), (\sigma'_\ell a'_{\ell-1} a'_\ell)} = \Psi^A_{a_{\ell-1},a'_{\ell-1}} \Psi^B_{a_\ell, a'_\ell} \delta_{\sigma_\ell,\sigma'_\ell}$ as well as a vector $v$ with $v_{\sigma_\ell a_{\ell-1} a_\ell} = M^{\sigma_\ell}_{a_{\ell-1}, a_{\ell}}$, we arrive at a {\em generalized eigenvalue problem}
of matrix dimension $(dD^2 \times dD^2)$,
\begin{equation}
H v - \lambda N v = 0,
\end{equation}
represented in Fig.~\ref{fig:equationsystem}.
Solving for the lowest eigenvalue $\lambda_0$ gives us a $v^0_{\sigma_\ell a_{\ell-1} a_\ell}$, which is reshaped back to $M^{\sigma_\ell}_{a_{\ell-1}a_{\ell}}$, $\lambda_0$ being the current ground state energy estimate. 

A few remarks are in order.

\begin{figure}
\centering\includegraphics[scale=0.7]{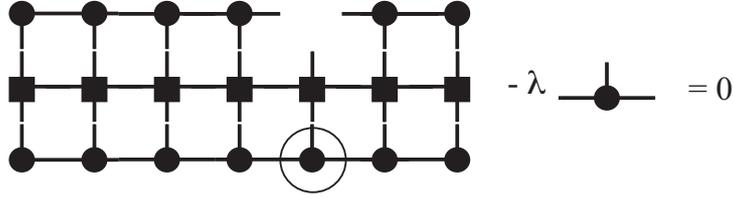}
\caption{Standard eigenvalue problem for the optimization of $M^{\ \sigma_\ell}_{\ a_{\ell-1},a_\ell}$. The unknown matrix is circled on the left network.}
\label{fig:normalizedequationsystem}
\end{figure}

\begin{itemize}
\item The problem is Hermitian; both $H$ and $N$ are Hermitian, as can be seen from the construction and the Hermiticity of the MPO employed.
\item In general, $dD^2$ is too large for an exact diagonalization, but as we are only interested in the lowest eigenvalue and eigenstate, an iterative eigensolver that aims for the ends of the spectrum will do. Typical methods are the Lanczos or Jacobi-Davidson large sparse matrix solvers. The speed of convergence of such methods ultimately rests on the quality of the initial starting or guess vector. As this eigenproblem is part of an iterative approach to the ground state, the current $M^{\sigma_\ell}$ is a valid guess that will dramatically speed up calculations close to convergence.
\item Generalised eigenvalue problems can be numerically very demanding, if the condition number of $N$ becomes bad. But this is no issue for open boundary conditions, if one ensures that the state is left-normalized up to site $\ell-1$ and right-normalized from site $\ell+1$ onwards. Then 
the simplifications for $\Psi^A$ and $\Psi^B$ imply that $N$ is just the identity matrix $I$. The eigenvalue problem then simplifies to a standard one,
\begin{equation}
\sum_{\sigma'_\ell} \sum_{a'_{\ell-1}a'_{\ell}} \sum_{b_{\ell-1},b_\ell} 
L^{a_{\ell-1},a'_{\ell-1}}_{b_{\ell-1}} W^{\sigma_\ell,\sigma'_\ell}_{b_{\ell-1},b_\ell} R^{a_\ell,a'_\ell}_{b_\ell}
M^{\sigma'_\ell}_{a'_{\ell-1}, a'_\ell} - \lambda  M^{\sigma_\ell}_{a_{\ell-1}, a_{\ell}} = 0 .
\end{equation}
or $Hv-\lambda v=0$, as represented in Fig.~\ref{fig:normalizedequationsystem}. The evaluation of the sums will be done using the optimal bracketing for $\hat{H}\ket{\psi}$.
To achieve this simplification, one will sweep the position $\ell$ from right to left and vice versa through the chain, such that the optimal normalization configuration can be maintained by a single step of the left or right canonization procedure after each minimization. 
\end{itemize}

The optimal algorithm then runs as follows. 
\begin{itemize}
\item Start from some initial guess for $\ket{\psi}$, which is right-normalized, i.e. consists of $B$-matrices only. 
\item Calculate the $R$-expressions iteratively for all site positions $L-1$ through 1 iteratively.
\item {\em Right sweep:} Starting from site $\ell=1$ through site $L-1$, sweep through the lattice to the right as follows: solve the standard eigenproblem by an iterative eigensolver for $M^{\sigma_\ell}$, taking its current value as starting point. Once the solution is obtained, left-normalize $M^{\sigma_\ell}$ into $A^{\sigma_\ell}$ by SVD (or QR) to maintain the desired normalization structure. The remaining matrices of the SVD are multiplied to the $M^{\sigma_{\ell+1}}$ to the right, which will be the starting guess for the eigensolver for the next site. Build iteratively the $L$ expression by adding one more site. Move on by one site, $\ell \rightarrow \ell+1$, and repeat.
\item {\em Left sweep:} Starting from site $\ell=L$ through site $2$, sweep through the lattice to the left as follows: solve the standard eigenproblem by an iterative eigensolver for $M^{\sigma_\ell}$, taking its current value as starting point. Once the solution is obtained, right-normalize $M^{\sigma_\ell}$ into $B^{\sigma_\ell}$ by SVD (or QR) to maintain the desired normalization structure. The remaining matrices of the SVD are multiplied to the $M^{\sigma_{\ell-1}}$ to the left, which will be the starting guess for the eigensolver for the next site. Build iteratively the $R$ expression by adding one more site. Move on by one site, $\ell \rightarrow \ell-1$, and repeat.
\item Repeat right and left sweeps, until convergence is achieved. Convergence is achieved if energy converges, but the best test is (using MPO) to consider
$\bra{\psi} \hat{H}^2 \ket{\psi} - (\bra{\psi} \hat{H} \ket{\psi})^2$ to see whether an eigenstate has been reached; this expression should approach 0 as closely as possible.
\end{itemize}

If we call matrices $A$, $B$, $M$ depending on their normalization ($M$ always being the one on the site currently attended to), and giving them an subscript index $i$ to label the number of updates by the eigensolver they have undergone, the algorithm would formalize as
\begin{eqnarray*}
& & M_0 B_0 B_0 B_0 B_0 B_0 \\
&\stackrel{diag}{\rightarrow}& M_1 B_0 B_0 B_0 B_0 B_0 
\stackrel{SVD}{\rightarrow} A_1 M_0 B_0 B_0 B_0 B_0 \\
&\stackrel{diag}{\rightarrow}& A_1 M_1 B_0 B_0 B_0 B_0 
\stackrel{SVD}{\rightarrow} A_1 A_1 M_0 B_0 B_0 B_0 \\
&\stackrel{diag}{\rightarrow}& A_1 A_1 M_1 B_0 B_0 B_0 
\stackrel{SVD}{\rightarrow} A_1 A_1 A_1 M_0 B_0 B_0 \\
& & \ldots \\
&\stackrel{diag}{\rightarrow}& A_1 A_1 A_1 A_1 M_1 B_0 
\stackrel{SVD}{\rightarrow} A_1 A_1 A_1 A_1 A_1 M_0 \\
&\stackrel{diag}{\rightarrow}& A_1 A_1 A_1 A_1 A_1 M_1 
\stackrel{SVD}{\rightarrow} A_1 A_1 A_1 A_1 M_1 B_1 \\
&\stackrel{diag}{\rightarrow}& A_1 A_1 A_1 A_1 M_2 B_1 
\stackrel{SVD}{\rightarrow} A_1 A_1 A_1 M_1 B_2 B_1 \\
& & \ldots \\
&\stackrel{diag}{\rightarrow}& A_1 M_2 B_2 B_2 B_2 B_1 
\stackrel{SVD}{\rightarrow} M_1 B_2 B_2 B_2 B_2 B_1 \\
\end{eqnarray*}
and again moving from left to right, starting with a diagonalization step.

In this iterative process, the energy can only go down, as we continuously improve by varying the parameters. Two problems occur: starting from a random state, the guesses for the $M^{\sigma_\ell}$ in the iterative eigensolvers will be very bad in the initial sweeps, leading to large iteration numbers and bad performance. Moreover, we cannot guarantee that the global minimum is actually reached by this procedure instead of being stuck in a non-global minimum.

One way of addressing the first issue is to start out with infinite-system DMRG to produce an initial guess; an optimal MPS version of infinite-system DMRG is discussed in Section~\ref{sec:infinite}. While this initial guess may be far from the true solution, it will usually fare much better than a random starting state. Moreover, one can try to balance the number of iterations (high in the first sweeps) by starting with small $D$, converge in that ansatz class, enlarge $D$ and add zeros in the new matrix entries, converge again, and so on. When $D$ gets large, the guess states will hopefully be so close to the final state that only very few iterations will be needed. It turns out, however, that starting with too small $D$ may land us in a non-global minimum that we will not get out of upon increasing $D$. Quite generally, as in DMRG, one should never calculate results for just a single $D$, but increase it in various runs until results converge (they are guaranteed to be exact in the $D\rightarrow\infty$ limit).

If we are looking for low-lying excited states instead of a ground state, two typical situations occur: 
(i) The excited state is known to be the ground state of another sector of the Hilbert space decomposed according to some good quantum number. Then the calculation is just a ground state calculation in that different sector. (ii) The excited state is the first, second, or higher excitation in the sector of the ground state. Then we have to calculate these excitations iteratively, and orthonormalize the state with respect to the lower-lying states already identified; this clearly limits the approach to a few low-lying excitations. The place where the algorithm is to be modified is in the iterative eigensolver; e.g.\ in the Lanczos iterations, the next Lanczos state generated is orthonormalized not only with respect to the previous Lanczos states, but also already constructed eigenstates of the Hamiltonian. This is a standard extension of the Lanczos algorithm.
 
The variational MPS algorithm just introduced is quite prone to getting stuck. How this is going to happen, actually depends a bit on how initial states are chosen in the procedure: Assume that, as is the case for the anisotropic Heisenberg chain, there is a quantum symmetry with some commuting operator $[\hat{H},\hat{O}]=0$, in this case the total magnetization operator $\hat{M}=\sum_i \hat{S}^z_i$, giving rise to magnetization $M$ as a good quantum number. Then initial states fall into two categories, whether they are eigenstates of $\hat{M}$ or not. The latter will generally be the case if the state is chosen randomly; the former is the case if it is generated by infinite-system DMRG or its MPS variant. 

Decomposing the Hilbert space into eigenstates of magnetisation, ${\mathcal H} = \oplus_M {\mathcal H}_M$, we can write initial states as 
\begin{equation}
\ket{\psi} = \sum_M \psi_M \ket{M} \quad\quad \ket{M} \in {\mathcal H}_M .
\end{equation}
The ground state we are looking for is $\ket{\psi_0} \in {\mathcal H}_{\tilde{M}}$; the initial state will then have either arbitrary $\psi_M$ or $\psi_M = 0$ if $M \neq \tilde{M}$ (assuming that we don't run into the desaster of offering an initial state in the wrong symmetry sector). Let us assume that in the first case, sweeping will eliminate contributions from the wrong symmetry sectors; if they don't, the variationally optimal state can never be reached anyways because wrong admixtures survive. As an iterative ground state search by e.g.\  Lanczos is an optimized version of the power method $\lim_{n\rightarrow\infty} \hat{H}^n \ket{\psi}$ for finding the largest eigenvalue and associated eigenstate, one can show that in the full Hilbert space wrong symmetry sectors will definitely be projected out.  In our algorithm, this iterative projection proceeds in a highly constrained state space and might not be as efficient, as it looks at various wave function components sequentially. As random starting states are very inefficient, I cannot report on a lot of practical experience here.
In any case, once we arrive in a well-defined symmetry sector, we will have, for any Schmidt decomposition $\ket{\psi} = \sum_{a_\ell} s_{a_\ell} \ket{a_\ell}_A \ket{a_\ell}_B$, that each of the states will have a good quantum number (superpositions of different quantum numbers lead immediately to a contradiction to a global good quantum number), namely $m_a^A$ and $m_a^B$ such that $m_a^A + m_a^B = \tilde{M}$, where I have simplified indices. Taking the $m_a^B$, for example, they will be distributed over some range, say 1 state with magnetization $m$, 3 states with magnetization $m'$, 5 states with magnetization $m''$ and so forth. As I will show next, this distribution stays {\em fixed} in further sweeps. This means that if it does not correspond to the distribution that the variationally optimal state would yield, it can never reach that state. In the random state approach one may hope that the slow elimination of other total magnetizations ``eases'' us into the right distributions but there is no guarantee; in the infinite-system approach one has to hope that this warm-up scheme produces the right distribution right away, which is quite unlikely to happen. 

The reason why the distribution stays fixed can be seen from the SVD of $M^{\sigma_{\ell}}_{a_{\ell-1},a_{\ell}}$ to carry out one (for example) left-normalization step: reshaping matrices $M^{\sigma_{\ell}}$ into some $\Psi$ and applying an SVD gives at most $D$ non-vanishing singular values; the right-singular vectors in $V^\dagger$ are nothing but the eigenvectors of  $\Psi^\dagger \Psi$, which is block-diagonal because the states $\ket{a_{\ell}}_B$ have good quantum numbers. The right singular vectors (eigenvectors) therefore encode a basis transformation within blocks of the same quantum number, hence the number of states with a given quantum number remains the same, and so does the number of states with a given quantum number in the other part of the system because of the matching of quantum numbers required in the Schmidt decomposition. 

Various ways of getting out of this potential trap have been proposed. The first one is to modify the algorithm to consider two sites at the same time, just as in conventional (two-site) DMRG; we will discuss its MPS implementation in the next section. While this approach is slower (roughly by a factor of $d$), it offers a slightly enlarged ansatz space with a subsequent truncation that allows the algorithm to be more robust against the danger of getting stuck in local energy minima in ground state searches.  In particular, the enlarged ansatz space of the two-site algorithm allows a reshuffling of the quantum number distribution due to the truncation. Once this is converged, one may switch to the single-site algorithm, as proposed by Takasaki et al. \cite{Takasaki99}, although it is not at all clear that this leads strictly to the optimal outcome\cite{McCulloch08}.

Much better, there is a procedure by White \cite{White05} that protects reasonably against trapping and ensures reshuffling. It is crucial for a reliable single-site DMRG (or variational MPS, which we will show to be identical) algorithm and turns it into the state of the art form of the method. It starts from the observation that quantum numbers of a subsystem A are changed by quantum fluctuations due to those parts of the Hamiltonian that connect A to the rest of the system.  We therefore have to consider the structure of $\hat{H}\ket{\psi}$ in more detail. 

\begin{figure}
\centering\includegraphics[scale=0.7]{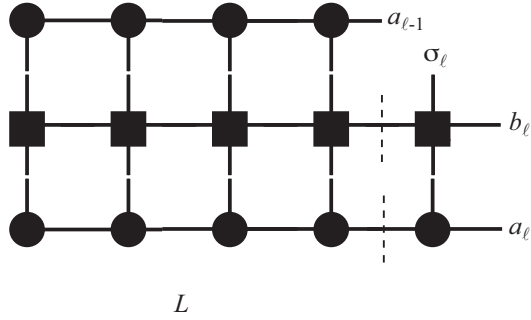}
\caption{$(\hat{H}^{A\bullet}_{b_\ell}\ket{\psi})^{a_{\ell-1},\sigma_{\ell},a_{\ell}}$ represented graphically: with the exception of one $W$-tensor and one $\Psi$-tensor, all the contractions have already been computed to obtain $L^{a_{\ell-1},a'_{\ell-1}}_{b_{\ell-1}}$. }
\label{fig:Whiteimprovement}
\end{figure}

Consider $\psi^{\sigma_{\ell}}_{a_{\ell-1},a_{\ell}}$ {\em after energy optimization} in the single-site algorithm. We can also write
\begin{equation}
\hat{H} = \sum_{b_{\ell}} \hat{H}^{A\bullet}_{b_{\ell}} \hat{H}^B_{b_{\ell}} 
\end{equation}
where
\begin{equation}
\hat{H}^{A\bullet}_{b_{\ell}} = \sum_{\fat{\sigma}, \fat{\sigma}'\in A\bullet} ( W^{\sigma_1,\sigma'_1} \ldots W^{\sigma_{\ell},\sigma'_{\ell}} )_{b_{\ell}} \quad
\hat{H}^B_{b_{\ell}} = \sum_{\fat{\sigma}, \fat{\sigma}'\in B} ( W^{\sigma_{\ell+1},\sigma'_{\ell+1}} \ldots W^{\sigma_L,\sigma'_L} )_{b_{\ell}} ,
\end{equation}
such that there are $D_W$ terms in this sum. If we think in terms of block states, we would like to know which new states can be reached on A$\bullet$ by the action of $\hat{H}^{A\bullet}_{b_{\ell}}$. 
Projecting the result of this action onto the A$\bullet$B basis it will read
\begin{eqnarray}
(\hat{H}^{A\bullet}_{b_{\ell}} \ket{\psi})^{a_{\ell-1}, \sigma_{\ell},a_{\ell}} &=&
\sum_{\sigma'_{\ell}} \sum_{a'_{\ell-1},b_{\ell-1}} \sum_{\{ a_i, b_i, a'_i; i<\ell-1\} } \left( \sum_{\sigma_1\sigma'_1} A^{\sigma_1*}_{1,a_1} W^{\sigma_1,\sigma'_1}_{1,b_1} A^{\sigma'_1}_{1,a'_1} \right) \ldots \times \nonumber \\
& & \left( \sum_{\sigma_{\ell-1}\sigma'_{\ell-1}} A^{\sigma_{\ell-1} *}_{a_{\ell-2},a_{\ell-1}} W^{\sigma_{\ell-1},\sigma'_{\ell-1}}_{b_{\ell-2},b_{\ell-1}} A^{\sigma'_{\ell-1}}_{a'_{\ell-2},a'_{\ell-1}} \right)  W^{\sigma_{\ell},\sigma'_{\ell}}_{b_{\ell-1},b_{\ell}} \Psi^{\sigma'_{\ell}}_{a'_{\ell-1},a_{\ell}} 
\end{eqnarray}
which is just 
\begin{equation}
(\hat{H}^{A\bullet}_{b_{\ell}} \ket{\psi})^{a_{\ell-1}, \sigma_{\ell},a_{\ell}} =
\sum_{\sigma'_{\ell}} \sum_{a'_{\ell-1},b_{\ell-1}}
L^{a_{\ell-1},a'_{\ell-1}}_{b_{\ell-1}} W^{\sigma_{\ell},\sigma'_{\ell}}_{b_{\ell-1},b_{\ell}} \Psi^{\sigma'_{\ell}}_{a'_{\ell-1},a_{\ell}} 
\end{equation}
using $L^{a_{\ell-1},a'_{\ell-1}}_{b_{\ell-1}}$ from Eq.~(\ref{eq:LdefineinMPO}), as can be seen graphically in Fig.~\ref{fig:Whiteimprovement}. This indicates that the actual cost of computation is very low, because we have already done the most complicated part. 

Now we would like to include the states generated by $\hat{H}^{A\bullet}_{b_{\ell}}$ into the search for a good basis for A$\bullet$. Here, DMRG offers the possibility of multiple-state targetting. The conventional algorithm would now proceed by calculating $\hat{\rho}^{A\bullet} = \tr_B \ket{\psi}\bra{\psi}$ or $\rho_{(a_{\ell-1} \sigma_{\ell}),(a'_{\ell-1}  \sigma'_{\ell})} = \sum_{a_{\ell}} \Psi^{\sigma_{\ell}}_{a_{\ell-1},a_{\ell}}\Psi^{\sigma'_{\ell}\dagger}_{a_{\ell},a'_{\ell-1}}$, finding the eigenvalues (squares of the singular values), the eigenvectors (left singular vectors), truncation and so on. But we can look at a modified density matrix, which takes also into account the new terms as
\begin{equation}
\hat{\rho}^{A\bullet} = \tr_B \ket{\psi}\bra{\psi} + \alpha \sum_{b_{\ell}} \tr_B \hat{H}^{A\bullet}_{b_{\ell}} \ket{\psi} \bra{\psi} \hat{H}^{A\bullet}_{b_{\ell}} ,
\end{equation}
where $\alpha$ is a small number, say $10^{-4}$, giving a little weight to contributions that the conventional algorithm may miss. The price paid is that at the end of the spectrum a few very-small weight states from $\ket{\psi}\bra{\psi}$ will drop out. Upon multiple sweeping, $\alpha$ will be taken slowly to zero.

The new density matrix is diagonalized, and truncation according to the $D$ largest eigenvalues is carried out, yielding a $(dD \times D)$ matrix of orthonormal columns, $U_{(a_{\ell-1}\sigma_{\ell}), a_{\ell}} \rightarrow A^{\sigma_{\ell}}_{a_{\ell-1}, a_{\ell}}$, and we continue on the next site; as we are not using the eigenvalues of the modified density matrix beyond their relative ordering, it does not matter that they do not sum up to 1. For predicting the next $\Psi^{\sigma_{\ell+1}}$ for the large sparse matrix solver, we use the DMRG prediction formula derived in the next section,
\begin{equation}
\Psi^{\sigma_{\ell+1}}_{a_{\ell}a_{\ell+1}} = 
\sum_{a_{\ell-1}\sigma_{\ell} a_{\ell}'} 
A^{\sigma_{\ell} \dagger}_{a_{\ell},a_{\ell-1}} \Psi^{\sigma_{\ell}}_{a_{\ell-1}a_{\ell}'}  B^{\sigma_{\ell+1}}_{a_{\ell}', a_{\ell+1}} .
\end{equation}
Otherwise, everything remains the same. The additional numerical cost and programming effort is minimal for an algorithm that often converges much faster and is much less prone to getting stuck at a non-optimal result.

\subsection{Conventional DMRG in MPS language: the subtle differences}   
\label{subsec:conventionalDMRGinMPS}
How can the previous approach be related to conventional DMRG? The essential answer is that the MPS approach is identical to finite-size DMRG for OBC, albeit only if we shift one site instead of two,
i.e.\ consider ``single-site'' instead of ``two-site'' DMRG, where we consider a block-site-block configuration A$\bullet$B instead of a block-site-site-block configuration A$\bullet\bullet$B.

Let us first remind ourselves of the key steps of the algorithms, assuming that we are sweeping to the right: (i) given some configuration A$\bullet$B (or corresponding configuration $AAAMBB$) and a Hamiltonian $\hat{H}$, the ground state is found by a large sparse matrix eigensolver looking for the optimal $\psi_{a_{\ell-1} \sigma_\ell a_\ell}$ (in DMRG) or $M^{\sigma_\ell}_{a_{\ell-1}, a_\ell}$ (in MPS) respectively; analogously for A$\bullet\bullet$B. (ii) Given the ground state, MPS derives a set of left-normalized $A$-matrices, whereas DMRG finds new block states whose structure can be encoded by left-normalized $A$-matrices. (iii) All algorithms switch to a new A$\bullet$B, A$\bullet\bullet$B or $AAAAMB$ configuration, where the active center is shifted by one site to the right and provide an initial guess for the calculation of the next ground state, taking us back to step (i).

Step (i): Results must be identical if we use the same state configuration and the same Hamiltonian. As DMRG grows the blocks A and B from left and right, and as each block growth step A$\bullet \rightarrow$A can be encoded by $A$-matrices and similarly $\bullet$B $\rightarrow$ B, we conclude that all matrices on A are left-normalized and those on B right-normalized, hence the two-site DMRG state takes the form
\begin{equation}
\ket{\psi} = \sum_{a_{\ell-1} \sigma_\ell \sigma_{\ell+1} a_{\ell+1}} \Psi^{\sigma_\ell\sigma_{\ell+1}}_{a_{\ell-1}, a_{\ell+1}} \ket{a_{\ell-1}}_A \ket{\sigma_\ell} \ket{\sigma_{\ell+1}} \ket{a_{\ell+1}}_B = \sum_{\fat{\sigma}} A^{\sigma_1} \ldots A^{\sigma_{\ell-1}} 
\Psi^{\sigma_\ell \sigma_{\ell+1}} B^{\sigma_{\ell+2}} \ldots B^{\sigma_L} \ket{\fat{\sigma}},
\end{equation}
with the obvious change for a single-site DMRG state, $\Psi^{\sigma_\ell\sigma_{\ell+1}} \rightarrow \Psi^{\sigma_\ell}$. This is in perfect agreement with the mixed-canonical states of the variational MPS approach and we are looking at the same state structure. 

It remains to show that the Hamiltonians are identical, too. Strictly speaking, this is not the case: 
The MPO representation of $\hat{H}$ we just used is clearly exact. On the other hand, the representation of $\hat{H}$ in DMRG contains a series of reduced basis transformations, hence is inherently inexact. So, the two representations seem unrelated, with an advantage on the MPO side because it is exact. But a more careful analysis reveals that  on the level of calculating expectation values $\bra{\psi} \hat{H} \ket{\psi}$ as they appear in MPS and DMRG ground state searches both representations give identical results (they are not identical for higher moments, such as  $\bra{\psi} \hat{H}^2 \ket{\psi}$, where the MPO representation is demonstrably more accurate at a numerical cost, see below).

Both the DMRG and the MPO Hamiltonian contain all terms of the exact Hamiltonian. As we have already seen in the application of a Hamiltonian MPO to a mixed canonical state (Sec.~\ref{subsec:Hamiltonianmixedcanonical}), the evaluation of the $L$ and $R$-objects appearing in the large sparse eigenproblem Eq.~(\ref{eq:vMPSeigenequation}) is nothing but the sequence of reduced basis transformations occuring in DMRG up to the current A$\bullet$B configuration. Hence, for $\bra{\psi} \hat{H} \ket{\psi}$ (but in general only for this!), both approaches are identical.

Moreover, the calculation $\hat{H}\ket{\psi}$ appearing in the eigenproblem does not have a worse operational count than in the corresponding DMRG procedure. To see this, let us focus on our example MPO for an anisotropic nearest-neighbour Heisenberg chain. There seems to be a difference in efficiency when we consider the double sum over $b_{\ell-1}, b_{\ell}$. From the structure of the $W$-matrix it is clear that for most of the $D_W^2$ (in the example 25) entries we find zeros, such that we can strongly restrict the sum. But this would still give the following count: setting the field 0, for the Heisenberg Hamiltonian there are 8 contributions in the DMRG setup: one each for $\hat{H}_A$ and $\hat{H}_B$, the parts of the Hamiltonian that act strictly on A and B, and three per block for the three operator combinations linking a block and the site. All of them are diagonal in the other block, so there are altogether 8 operations of cost $O(D^3)$. In the MPO calculation, the double matrix-matrix multiplication would naively suggest 16 operations of cost $O(D^3)$, for the 8 non-vanishing entries of $W^{\sigma_{\ell},\sigma'_{\ell}}$. But then we can exploit the following rules: if $b_{\ell-1}=d_W$, then there are no operations to the left, $L^{a_{\ell-1}, a'_{\ell-1}}_{d_W}= \delta_{a_{\ell-1}, a'_{\ell-1}}$, and one operation drops out. Similarly, if $b_{\ell}=1$, then there are no operations to the right and $R^{a_{\ell},a'_{\ell}}_{b_{\ell}}=\delta_{a_{\ell},a'_{\ell}}$, and again one operation drops out. Looking at the structure of $W$, all non-vanishing entries meet one or the other condition, and the count halves down to 8 operations. Only longer-ranged interactions do not fit this picture, but they would be of cost $2O(D^3)$ in DMRG as well.

Step (ii): After energy minimization, variational MPS and DMRG produce (identical) $M^{\sigma_\ell}$ and $\Psi^{\sigma_\ell}$, or $\Psi^{\sigma_\ell \sigma_{\ell+1}}$. Both methods now seem to proceed differently with the result, but in fact do the same: in variational MPS one just shifts one site to the right after an SVD to ensure left-normalization, to continue minimizing on the next site. In DMRG one previously carries out a density matrix analysis to determine a new (truncated) block basis. But if one carries out the corresponding SVD, the number of non-zero singular values (hence non-zero density matrix eigenvalues) is limited by $D$, the matrix dimension: 
\begin{equation}
\Psi^{\sigma_\ell}_{a_{\ell-1},a_ \ell} \rightarrow \Psi_{(a_{\ell-1} \sigma_ \ell), a_ \ell} = \sum_{k=1}^{\min(dD,d)=D} A^{\sigma_ \ell}_{a_{\ell-1},k} S_{kk} (V^\dagger)_{k, a_ \ell} .
\end{equation}
Hence, no truncation happens, and we are just doing a unitary transformation to obtain orthonormal states for the new larger block A (which is just the left-normalization in the MPS because of the link between SVD and density matrix diagonalization). Both formalisms act identically; as no truncation occurs, thin QR would do, too.

On the other hand, in two-site DMRG the same step reads
\begin{equation}
\Psi^{\sigma_ \ell\sigma_{\ell +1}}_{a_{\ell-1},a_{\ell +1}} \rightarrow \Psi_{(a_{\ell-1} \sigma_ \ell), (\sigma_{\ell +1} a_{\ell +1})} = \sum_{k=1}^{\min(dD,dD)=dD} A^{\sigma_ \ell}_{a_{\ell-1},k} S_{k,k} (V^\dagger)_{k,(\sigma_{\ell +1} a_{\ell +1})} .
\end{equation}
But we can only keep $D$ states in the new block, hence truncation has to occur! Here is the {\em only} difference between variational MPS and single-site DMRG on the one and two-site DMRG on the other hand.

Step (iii): In DMRG, after completion of one iteration, the free site(s) are shifted by one, leading to block growth of A and shrinkage of B. Here, all methods agree again: in variational MPS, the shrinkage of B is simply reflected in the states being formed from a string of $B$-matrices where the leftmost one has dropped off. The growth of A is given by a similar string, where one $A$-matrix has been added. The matrix on the free sites is to be determined in all approaches, so nothing is to be said about its normalization.

Minimization of ground state energy is, as we have seen, a costly large sparse matrix problem. As the methods are iterative, a good initial guess is desirable. DMRG has provided some ``state prediction'' for that \cite{White96}. In fact, it turns out that the result of the prediction is just what one gets naturally in variational MPS language without the intellectual effort involved to find state prediction. 

Let us assume that for single-site DMRG we just optimized $\Psi^{\sigma_\ell}$, deriving a new $A^{\sigma_ \ell}$. Then in MPS language the next $\Psi^{\sigma_{\ell +1}}= SV^{\dagger} B^{\sigma_{\ell +1}}$, where $S$ and $V^\dagger$ are from the SVD. In DMRG language, we take
\begin{equation}
\ket{\psi} = \sum_{a_{\ell-1}\sigma_ \ell a_{\ell }'} \Psi^{\sigma_ \ell}_{a_{\ell-1}, a_{\ell }} \ket{a_{\ell-1}}_A \ket{\sigma_ \ell} \ket{a_{\ell}'}_B
\end{equation}
and insert twice approximate identities $I= \sum \ket{a_ \ell}_A\phantom\langle_A\bra{a_ \ell}$ and 
$I= \sum \ket{\sigma_{\ell +1}}\ket{a_{\ell+1}}_B \phantom\langle_B\bra{a_{\ell +1}} \bra{\sigma_{\ell+1}}$. Expressing the matrix elements by $A$ and $B$ matrices, the state now reads
\begin{equation}
\ket{\psi} = \sum_{a_\ell\sigma_{\ell +1} a_{\ell +1}} \left( \sum_{a_{\ell-1}\sigma_ \ell  a_{\ell}'} 
A^{\sigma_ \ell \dagger}_{a_ \ell,a_{\ell-1}} \Psi^{\sigma_ \ell}_{a_{\ell-1}, a_{\ell}'}  B^{\sigma_{\ell +1}}_{a_{\ell}' a_{\ell +1}} \right) 
\ket{a_{\ell}}_A \ket{\sigma_{\ell +1}} \ket{a_{\ell +1}}_B
\end{equation}
So the  prediction reads
\begin{equation}
\Psi^{\sigma_{\ell +1}}_{a_{\ell}, a_{\ell+1}} = 
\sum_{a_{\ell-1}\sigma_ \ell a_{\ell}'} 
A^{\sigma_ \ell \dagger}_{a_ \ell,a_{\ell-1}} \Psi^{\sigma_ \ell}_{a_{\ell-1}, a_{\ell}'}  B^{\sigma_{\ell +1}}_{a_{\ell}' a_{\ell +1}} .
\end{equation}
But this is exactly the MPS ansatz for the next eigenproblem, as $ A^{\sigma_ \ell \dagger} \Psi^{\sigma_ \ell}  B^{\sigma_{\ell +1}} =
A^{\sigma_ \ell \dagger} A^{\sigma_ \ell } S V^{\dagger} B^{\sigma_{\ell +1}}$. But this is just 
$S V^{\dagger} B^{\sigma_{\ell +1}}$ because in the ansatz, $\sigma_ \ell $ is summed over and left-normalization holds. Two-site DMRG proceeds by analogy and is left as an exercise for the reader.

While this clarifies the relationship between variational MPS, single-site DMRG (the same) and two-site DMRG (different), it is important to note that the different ways of storing information more implicitly or more explicitly implies differences even if the algorithms are strictly speaking identical -- the fact that in one formulation prediction is trivial and in the other is not already gave us an example. But there is more.

(i) In DMRG, the effective bases for representing the states and the Hamiltonian or other operators are tied up. This is why concepts such as targetting multiple states arise, if we consider several different states like the ground state and the first excited state at the same time. One then considers mixed reduced density operators
\begin{equation}
\hat{\rho}_A = \sum_{i} \alpha_i \tr_B \ket{\psi_i}\bra{\psi_i}
\end{equation}
with $\ket{\psi_i}$ the target states and $0 < \alpha_i \leq 1$, $\sum_i \alpha_i =1$, to give a joint set of bases for all states of interest. This can of course only be done at a certain loss of accuracy for given numerical resources and for a few states only. At the price of calculating the contractions anew for each state, in the MPO/MPS formulation, the state bases are only tied up at the level of the exact full basis. MPO/MPS formulations therefore acquire their full potential versus conventional DMRG language once multiple states get involved.

\begin{figure}
\centering\includegraphics[scale=0.7]{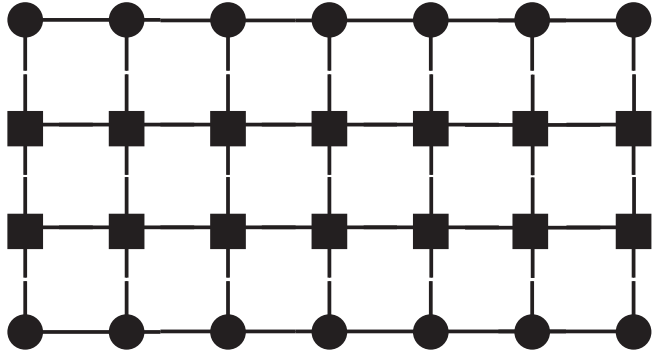}
\caption{``Exact'' calculation of the expectation value of $\hat{H}^2$: the Hamiltonian MPO is repeated twice and sandwiched between $\ket{\psi}$ at the bottom and $\bra{\psi}$ at the top.}
\label{fig:calculationH2}
\end{figure}

(ii) Another instance where the MPO/MPS formulation is superior, albeit at elevated numerical cost, is the calculation of the expression $\bra{\psi} \hat{H}^2 \ket{\psi}$, which is interesting e.g.\ in the context of estimating how accurately a ground state has been obtained. In the MPO formalism, it can be done exactly up to the inherent approximations to $\ket{\psi}$ by contracting the network shown in Fig.~\ref{fig:calculationH2}. It would of course be most economical for the programmer to calculate $\hat{H}\ket{\psi}$ and take the norm, two operations which at this stage he has at hand. The operational cost of this would be $O(LD^2D_W^2 d^2)$ for the action of the MPO and $O(LD^3 D_W^3 d)$ for the norm calculation. The latter is very costly, hence it is more efficient to do an iterative construction as done for $\bra{\psi} \hat{H} \ket{\psi}$. Let me make the important remark that dimension $D_W^2$ is only the worst case for $\hat{H}^2$\cite{Frowis10}: Writing out the square and introducing rules for the expression leads to more efficient MPOs, whose optimality can be checked numerically by doing an SVD compression and looking for singular values that are zero. Our anisotropic Heisenberg Hamiltonian takes $D_W=9$ instead of 25 for $\hat{H}^2 $. For higher powers, the gains are even more impressive, and can be obtained numerically by compressing an explicit MPO for $\hat{H}^n$ with discarding only zeros among the singular values. 

\begin{figure}
\centering\includegraphics[scale=0.7]{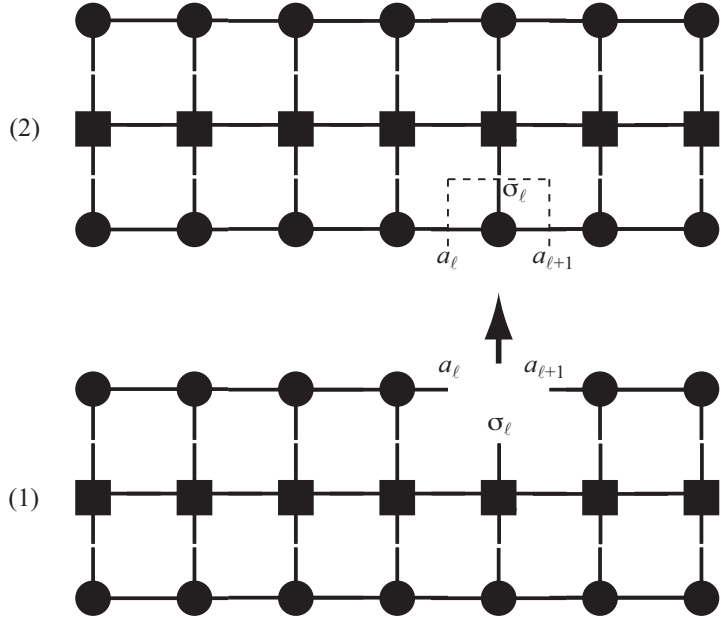}
\caption{DMRG calculation of the expectation value of $\hat{H}^2$: the Hamiltonian MPO is applied once to $\ket{\psi}$ in step (1) in the DMRG basis, i.e. the result is projected onto the reduced bases, yielding some $\Phi^{\, \sigma_{\ell}}$. This in turn replaces $\Psi^{\, \sigma_{\ell}}$ in the second application of $\hat{H}$ in step (2). Ultimately, the result is projected on the block bases.}
\label{fig:DMRGcalculationH2}
\end{figure}

In a DMRG calculation, there would be a sequence $\hat{H}(\hat{H} \ket{\psi})$, in the DMRG block-site basis as shown in Fig.~\ref{fig:DMRGcalculationH2}. The point is that before the second application of $\hat{H}$, a projection onto the reduced block bases happens, which is not the identity and loses information.

What does the comparison MPS and DMRG imply algorithmically? First of all, the truncation error of conventional DMRG, which has emerged as a highly reliable tool for gauging the quality of results, is nothing but an artefact of the somewhat anomalous two-site setup. In variational MPS or single-site DMRG it has to be replaced by some other criterion, like the variance of the energy. Second, while all the approaches are variational in the sense that they are looking for the lowest energy that can be achieved in a given type of ansatz, it varies from site to site in two-site DMRG (because of the $\Psi^{\sigma\sigma}$ anomaly in the ansatz), the ansatz stays the same all the time in single-site DMRG, which is conceptually nicer. That this comes at the expense of potential trapping serves as a reminder that the mathematically most beautiful does not have to be the most practical.
 
\section{Time evolutions (real and imaginary) with MPS}
The calculation of the action of operators like $\eul^{-\imag \hat{H}t}$ or $\eul^{-\beta\hat{H}}$ on quantum states is of central interest in quantum mechanics, for real-time evolutions of quantum states and for quantum statistical mechanics; $\beta$ can be interpreted as an imaginary time. It is one of the most attractive features of MPS that such real or imaginary time evolutions can be encoded very neatly and efficiently. This holds both for pure and mixed states, important at finite temperature. In the following, I will focus on time evolution based on a Trotter decomposition of the evolution operators \cite{Vidal03a,Vidal04b,Daley04,WhiteFeiguin04,VerstraeteRipoll04}, explaining first the Trotter decomposition and the structure of the algorithm for pure states, then the representation of the Trotter decomposition by MPOs. After this, I will discuss the changes necessary for the simulation of the dynamics of mixed states.
\subsection{Conventional time evolution: pure states}

\subsubsection{Trotter decompositions of time evolution}
Let us assume that $\hat{H}$ consists of nearest-neighbour interactions only, i.e.\ $\hat{H} = \sum_i \hat{h}_i$, where $\hat{h}_i$ contains the interaction between sites $i$ and $i+1$. We can then discretize time as $t = N \tau$ with $\tau\rightarrow 0$, $N \rightarrow \infty$ and (in the most naive approach) do a first-order Trotter decomposition as
\begin{equation}
\eul^{-\imag \hat{H}\tau} = \eul^{-\imag  \hat{h}_1 \tau}\eul^{-\imag  \hat{h}_2 \tau}\eul^{-\imag  \hat{h}_3 \tau} \ldots  \eul^{-\imag  \hat{h}_{L-3} \tau}\eul^{-\imag  \hat{h}_{L-2} \tau}\eul^{-\imag  \hat{h}_{L-1} \tau} + O(\tau^2),
\end{equation}
which contains an error due to the noncommutativity of bond Hamiltonians, $[ \hat{h}_i,\hat{h}_{i+1}] \neq 0$ in general; higher order decompositions will be discussed later in this section. All time evolutions on odd ($\eul^{-\imag \hat{H}_{{\rm odd}} \tau}$) and even ($\eul^{-\imag \hat{H}_{{\rm even}} \tau}$) bonds respectively commute among each other, and can be carried out at the same time. So we are looking for an MPO doing an infinitesimal time step on the odd bonds and for another MPO doing the same on the even bonds.

As any operator is guaranteed to be MPO-representable, let us assume for a moment that indeed we can construct these representation of infinitesimal time steps efficiently (see next section for the explicit construction). As we will see, the maximum bond dimension of the infinitesimal time step MPOs is $d^2$ because the dimension of $\eul^{-\imag \hat{h} \tau}$ is $(d^2 \times d^2)$. The application of the infinitesimal time step MPOs thus increases the bond dimensions from $D$ up to $d^2 D$. Repeated applications of the infinitesimal time evolution MPOs leads to an exponential growth of the matrix dimensions, which therefore have to be truncated after time steps.

The resulting time evolution algorithm takes a very simple form: starting from $\ket{\psi(t=0)}$, repeat the following steps:
\begin{itemize}
\item Apply the MPO of the odd bonds to $\ket{\psi(t)}$.
\item Apply the MPO of the even bonds to $\eul^{-\imag \hat{H}_{{\rm odd}}\tau}\ket{\psi(t)}$.
\item Compress the MPS $\ket{\psi(t+\tau)}=\eul^{-\imag \hat{H}_{{\rm even}}\tau}\eul^{-\imag \hat{H}_{{\rm odd}}\tau}\ket{\psi(t)}$ from dimensions $d^2 D$ to $D$, monitoring the error. Obviously, one may also allow for some compression error (state distance) $\epsilon$ and choose a time-dependent $D$: it will typically grow strongly with time, limiting the reachable timescale. By analogy to the ground state calculations, all results should be extrapolated in $D\rightarrow\infty$ or $\epsilon\rightarrow 0$.
\end{itemize}

After each time step, we may evaluate observables in the standard way, $\langle O(t) \rangle = \bra{\psi(t)} \hat{O} \ket{\psi(t)}$. But we can do more: we can calculate time-dependent correlators as
\begin{equation}
\langle \hat{O}(t) \hat{P} \rangle = \bra{\psi} \eul^{+i\hat{H}t} \hat{O} \eul^{-\imag \hat{H}t} \hat{P} \ket{\psi}
= \bra{\psi(t)} \hat{O} \ket{\phi(t)}
\end{equation}
where $\ket{\psi(t)} = \eul^{-\imag \hat{H}t}\ket{\psi}$ and $\ket{\phi(t)} = \eul^{-\imag \hat{H}t} \hat{P} \ket{\psi}$. If we take e.g.\ $\hat{O} = \hat{S}^z_i$ and $\hat{P} = \hat{S}^z_j$, we can calculate 
$\langle \hat{S}^z_i(t) \hat{S}^z_j \rangle$ and by a double Fourier transformation the structure function
\begin{equation}
S^{zz}(k,\omega) \propto \int dt \sum_{n} \langle \hat{S}^z_i(t) \hat{S}^z_{i+n} \rangle \eul^{\imag kn}\eul^{-\imag \omega t},
\end{equation}
where I have assumed translational invariance and infinite extent of the lattice for simplicity of the formula.

A simple improvement on the algorithm given above is to do a second-order Trotter decomposition 
\begin{equation}
\eul^{-\imag  \hat{H}\tau} = \eul^{-\imag  \hat{H}_{{\rm odd}} \tau/2}\eul^{-\imag  \hat{H}_{{\rm even}} \tau}\eul^{-\imag  \hat{H}_{{\rm odd}} \tau/2} + O(\tau^3),
\end{equation}
where the error per timestep is reduced by another order of  $\tau$. If we do not do evaluations after each time step, we can group half steps, and work at no additional expense compared to a first-order Trotter decomposition.

A very popular implementation of a fourth order-Trotter decomposition that originates in quantum Monte Carlo would be given by the following formula due to Suzuki\cite{Suzuki76,Suzuki91}:
\begin{equation}
\eul^{-\imag  \hat{H}\tau} = \hat{U}(\tau_1) \hat{U}(\tau_2) \hat{U}(\tau_3) \hat{U}(\tau_2) \hat{U}(\tau_1),  
\end{equation}
where 
\begin{equation}
\hat{U} (\tau_i) = \eul^{-\imag  \hat{H}_{{\rm odd}} \tau_i/2}\eul^{-\imag  \hat{H}_{{\rm even}} \tau_i}\eul^{-\imag  \hat{H}_{{\rm odd}}
\tau_i/2}
\end{equation}
and
\begin{equation}
\tau_1 = \tau_2 = \frac{1}{4-4^{1/3}} \tau  \quad\quad \tau_3 = \tau -2 \tau_1 - 2\tau_2 .
\end{equation}
Even smaller errors can be achieved at similar cost using less symmetric formulae\cite{McLachlan95}. This completes the exposition of the algorithm (an error analysis will be given after the other methods for time evolution have been explained), and we now have to construct the MPOs.

\subsubsection{MPO for pure state evolution}
\begin{figure}
\centering\includegraphics[width=250pt]{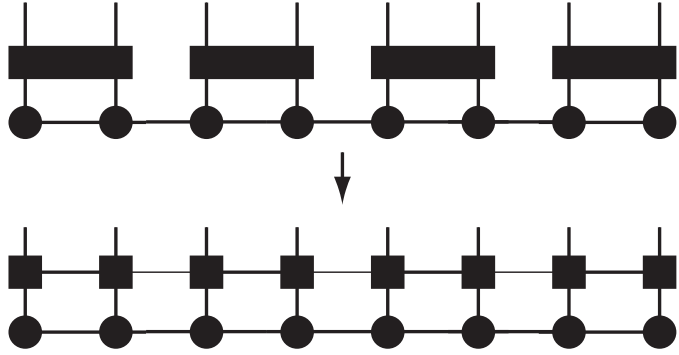}
\caption{A Trotter step: On all odd bonds, an (infinitesimal) bond time evolution is carried out. This merges two sites, such that the simple product form of MPS is lost at first sight, but the time evolution can be translated into MPOs. As the time evolution factorizes, the MPOs have dimension 1 on all even bonds (thin lines).}
\label{fig:trotterstep}
\end{figure}

Let us consider the Trotter step for all odd bonds of a chain:
\begin{equation}
\eul^{-\imag  \hat{h}_1 \tau} \otimes \eul^{-\imag  \hat{h}_3 \tau} \otimes \ldots \otimes \eul^{-\imag  \hat{h}_{L-1} \tau} \ket{\psi} ;
\label{eq:oddevolution}
\end{equation}
each bond-evolution operator like $\eul^{-\imag  \hat{h}_1 \tau}$ takes the form $\sum_{\sigma_1 \sigma_2, \sigma'_1 \sigma'_2} O^{\sigma_1\sigma_2,\sigma'_1\sigma'_2} \ket{\sigma_1\sigma_2}\bra{\sigma'_1\sigma'_2}$. Both in the pictorial and the explicit mathematical representation it is obvious that this operator destroys the MPS form (Fig.~\ref{fig:trotterstep}).

It would therefore be desirable to have $O^{\sigma_1\sigma_2,\sigma'_1\sigma'_2}$ in some form containing tensor products $O^{\sigma_1, \sigma'_1} \otimes O^{\sigma_2, \sigma'_2}$, to maintain the MPS form. To this purpose, we carry out the procedure for decomposing an arbitrary state into an MPS, adapted to an operator (two indices per site). It works because there are so few indices. One reorders $O$ to group local indices and carries out a singular value decomposition:
\begin{eqnarray*}
O^{\sigma_1 \sigma_2, \sigma'_1 \sigma'_2} &=& P_{(\sigma_1 \sigma'_1), (\sigma_2 \sigma'_2)} \\
&=& \sum_k U_{\sigma_1 \sigma'_1,k} S_{k,k} (V^\dagger)_{k, (\sigma_2 \sigma'_2)} \\
&=& \sum_k U^{\sigma_1 \sigma'_1}_{k} \overline{U}^{\sigma_2 \sigma'_2}_{k} \\
&=& \sum_k U^{\sigma_1 \sigma'_1}_{1,k} \overline{U}^{\sigma_2 \sigma'_2}_{k,1}
\end{eqnarray*}
where $U^{\sigma_1 \sigma'_1}_{k} =  U_{(\sigma_1 \sigma'_1),k}\sqrt{S_{k,k}}$ and 
$\overline{U}^{\sigma_2 \sigma'_2}_{k} = \sqrt{S_{k,k}} (V^\dagger)_{k, (\sigma_2 \sigma'_2)}$. In the very last step of the derivation, we have introduced a dummy index taking value 1 to arrive at the form of an MPO matrix. The index $k$ may run up to $d^2$, giving the bond dimension $D_W$ of the MPO.

\begin{figure}
\centering\includegraphics[width=250pt]{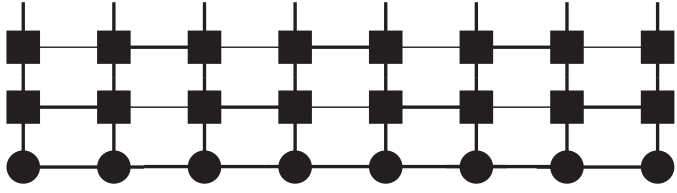}
\caption{A complete first-order Trotter time step (odd and even bonds). Fat and thin lines correspond to dimension1 and $>1$ on MPO bonds. The MPOs in the top line on the first and last site are trivial scalar identities 1.}
\label{fig:oddevenTrotter}
\end{figure}

The MPO representing the operator in Eq.~(\ref{eq:oddevolution}), $U^{\sigma_1\sigma'_1} \overline{U}^{\sigma_2\sigma'_2}U^{\sigma_3\sigma'_3} \overline{U}^{\sigma_4\sigma'_4}\ldots$, factorizes on every second bond, as do the original unitaries. If one site does not participate in any bond evolution, we simply assign it the identity unitary as a $(1\times 1)$-matrix: $I^{\sigma,\sigma'}_{1,1}=\delta_{\sigma,\sigma'}$. Then the global MPO can be formed trivially from local MPOs. The MPO for time evolution on all odd bonds would read
$U\overline{U}U\overline{U}U\overline{U} \ldots$, whereas the even-bond time step reads $IU\overline{U}U\overline{U}U\overline{U}\ldots I$ (Fig.~\ref{fig:oddevenTrotter}).

\subsection{Conventional time evolution: mixed states}

\subsubsection{Purification of mixed states}
\begin{figure}
\centering\includegraphics[scale=0.7]{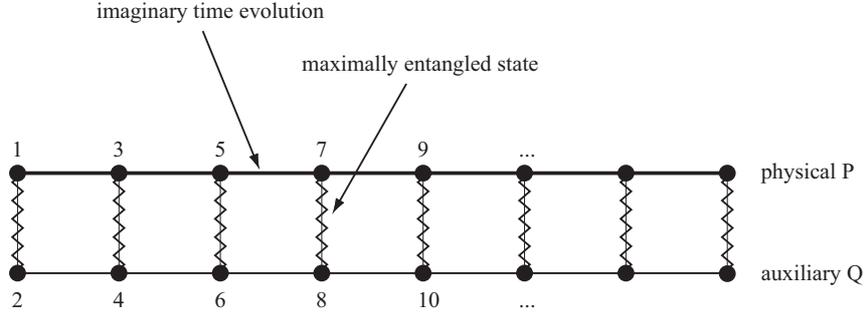}
\caption{Schematic representation of finite-temperature simulations: instead of a chain, one sets up a ladder with an identical copy of the chain. Physical sites have odd, auxiliary sites have even labels. Equivalent sites on the physical and auxiliary leg are linked by maximally entangled states. To reach inverse temperature $\beta$, an imaginary time evolution is carried out up to ``time'' $-\imag\beta/2$.}
\label{fig:finitetemperaturesketch}
\end{figure}

Finite temperature calculations can be carried out based on the purification of an arbitrary mixed quantum state\cite{VerstraeteRipoll04}: if we consider a mixed state in physical space P formed from orthonormal states, we can interpret it as the result of a partial trace over a Schmidt decomposition of a pure state on PQ, where Q is an auxiliary space:
\begin{equation}
\hat{\rho}_P = \sum_{a=1}^r s_a^2 \ket{a}_P \bra{a}_P \rightarrow \ket{\psi} = \sum_{a=1}^r s_a \ket{a}_P \ket{a}_Q \quad \quad \hat{\rho}_P = \tr_Q \ket{\psi}\bra{\psi} .
\end{equation}
The auxiliary state space can simply be taken as a copy of the original one, so finite-temperature density operators on a chain can be expressed as pure states on a ladder (see Fig.~\ref{fig:finitetemperaturesketch}).

To calculate a thermal density operator $\hat{\rho}_\beta= Z(\beta)^{-1} \eul^{-\beta \hat{H}}$, $Z(\beta) = \tr_P \eul^{-\beta \hat{H}}$, we write
\begin{equation}
\hat{\rho}_\beta = Z(\beta)^{-1} \eul^{-\beta \hat{H}} = Z(\beta)^{-1} \eul^{-\beta \hat{H}/2} \cdot \hat{I} \cdot \eul^{-\beta \hat{H}/2} .
\end{equation}
The identity $\hat{I}$ is nothing but $Z(0) \hat{\rho}_0$, the infinite temperature density operator times the infinite temperature partition function. Assume we know the purification of $\hat{\rho}_0$ as an MPS, $\ket{\psi_{\beta=0}}$. Then
\begin{equation}
\hat{\rho}_\beta = (Z(0)/Z(\beta)) \eul^{-\beta \hat{H}/2} \cdot \tr_Q \ket{\psi_0}\bra{\psi_0} \cdot \eul^{-\beta \hat{H}/2} = (Z(0)/Z(\beta)) \tr_Q \eul^{-\beta \hat{H}/2}\ket{\psi_0} \bra{\psi_0}\eul^{-\beta \hat{H}/2} .
\end{equation}
The trace over Q can be pulled out as the Hamiltonian does not act on Q. But the result means that we have to do an imaginary time evolution
\begin{equation}
\ket{\psi_\beta} = \eul^{-\beta \hat{H}/2}\ket{\psi_0} .
\end{equation}
Expectation values are given by 
\begin{equation}
\langle \hat{O} \rangle_\beta = \tr_P \hat{O}\hat{\rho}_\beta =
(Z(0)/Z(\beta)) \tr_P \hat{O} \tr_Q \ket{\psi_\beta} \bra{\psi_\beta}
= (Z(0)/Z(\beta)) \bra{\psi_\beta} \hat{O} \ket{\psi_\beta} .
\end{equation} 
$(Z(0)/Z(\beta)) = (d^L/Z(\beta))$ may seem difficult to obtain, but follows trivially from the expectation value of the identity, 
\begin{equation}
1 = \langle \hat{I} \rangle_\beta = \tr_P \hat{\rho}_\beta =
(Z(0)/Z(\beta)) \tr_P \tr_Q \ket{\psi_\beta} \bra{\psi_\beta}
= (Z(0)/Z(\beta)) \braket{\psi_\beta}{\psi_\beta} ,
\end{equation} 
hence $Z(\beta)/Z(0)= \braket{\psi_\beta}{\psi_\beta}$, or, in complete agreement with standard quantum mechanics,
\begin{equation}
\langle \hat{O} \rangle_\beta = \frac{ \bra{\psi_\beta} \hat{O}\ket{\psi_\beta}}{\braket{\psi_\beta}{\psi_\beta}} .
\end{equation}
But this takes us right back to expressions we know how to calculate. All we have to do is to find $\ket{\psi_0}$, carry out imaginary time evolution up to $-\imag\beta/2$, and calculate expectation values as for a pure state. We can even subject the purified state $\ket{\psi_\beta}$ to subsequent real time evolutions, to treat time dependence at finite $T$.

We can also do thermodynamics quite simply, as $Z(\beta)/Z(0)$ is given by the square of the norm of $\ket{\psi_\beta}$ and $Z(0)=d^L$. This means that we can obtain $Z(\beta)$ by keeping the purified state normalized at all temperatures and by accumulating normalization factors as temperature goes down and $\beta$ increases. From $Z(\beta)$, we have $F(\beta)=-\beta^{-1} \ln Z(\beta)$. At the same time, $U(\beta) = \langle \hat{H} \rangle_\beta = \bra{\psi_\beta} \hat{H} \ket{\psi_\beta}$. But this in turn gives us $S(\beta) = \beta(U(\beta)-F(\beta))$. Further thermodynamic quantities follow similarly.

The purification of the infinite temperature mixed state is a simple MPS of dimension 1, because it factorizes (if we take one ladder rung as a big site):
\begin{equation}
\hat{\rho}_0 = \frac{1}{d^L} \hat{I} = \left(\frac{1}{d} \hat{I}\right)^{\otimes L} .
\end{equation}
As $\hat{\rho}_0$ factorizes, we can now purify the local mixed state on each physical site as a pure state on rung $i$ , to get $\ket{\psi_{i0}}$, then 
$\ket{\psi_0} = \ket{\psi_{1,0}}\ket{\psi_{2,0}}\ket{\psi_{3,0}} \ldots$, a product state or an MPS of dimension 1. If we consider some rung $i$ of the ladder, with states $\ket{\sigma}_P$ and $\ket{\sigma}_Q$ on the physical site $2i-1$ and the auxiliary site $2i$, we can purify as follows:
\begin{equation}
\frac{1}{d} \hat{I} = \sum_{\sigma} \frac{1}{d} \phantom\langle_P\ket{\sigma}\bra{\sigma}_P = \tr_Q \left[ \left( \sum_{\sigma} \frac{1}{\sqrt{d}} \ket{\sigma}_P \ket{\sigma}_Q \right) \left( \sum_{\sigma} \frac{1}{\sqrt{d}} \bra{\sigma}_P \bra{\sigma}_Q  \right) \right] .
\end{equation}
Hence the purification is given by a maximally entangled state (entanglement entropy is $\log_2 d$),
\begin{equation}
\ket{\psi_{i0}} = \sum_{\sigma} \frac{1}{\sqrt{d}} \ket{\sigma}_P \ket{\sigma}_Q .
\end{equation}
It is easy to see that one can carry out local unitary transformations on both P and Q separately that leave that structure invariant. For example, for the purification of a spin-1/2 chain it is  advantageous to use the singlet state as local purification,
\begin{equation}
\ket{\psi_{i,0}} = \frac{1}{\sqrt{2}} [ \ket{\uparrow_P\downarrow_Q} - \ket{\downarrow_P\uparrow_Q} ],
\end{equation}
in case the program knows how to exploit good quantum numbers: this state would allow to conserve total $S=0$ and $S^z=0$ at the same time. In this case, the four $A$-matrices would read
\begin{equation}
A^{\uparrow_P\uparrow_Q} =0 \quad 
A^{\uparrow_P\downarrow_Q} =1/\sqrt{2} \quad 
A^{\downarrow_P\uparrow_Q} =-1/\sqrt{2} \quad 
A^{\downarrow_P\downarrow_Q} =0  ,
\end{equation}
and the purified starting state $\ket{\psi_0}$ for $\beta=0$ is now given by a product of singlet bonds on a ladder. In fact, a SVD of reshaped matrix $A_{\sigma,\sigma'}$ allows us to introduce truly site-local $A$-matrices, which have dimension $(1\times 2)$ on odd and $(2\times 1)$ on even sites:
\begin{equation}
A^{\uparrow_{2i-1}} = [ 1\ \ 0] \quad
A^{\downarrow_{2i-1}} = [ 0\ \ -1] \quad
A^{\uparrow_{2i}} = [ 0\ \ 1/\sqrt{2}]^T \quad
A^{\downarrow_{2i}} = [ 1/\sqrt{2} \ \ 0]^T \quad .
\end{equation}
In order to apply the pure state time evolution algorithm, it remains to find the MPO.
 
\subsubsection{MPO for mixed state evolution}

\begin{figure}
\centering\includegraphics[width=250pt]{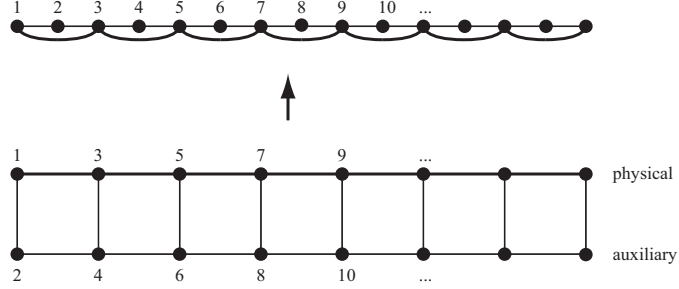}
\caption{The time-evolution of mixed states on a chain can be seen as that of a pure state on a ladder, where the physical sites sit on the first leg and additional auxiliary sites on the second leg. This ladder is mapped to a chain. As the time evolution acts only on the physical states, next-nearest neighbour interactions arise.}
\label{fig:laddersetup}
\end{figure}

The ladder appearing in mixed state simulations can be mapped to a chain (Fig.~\ref{fig:laddersetup}), where the physical Hamiltonian acts only on the odd sites, $1,3,5,\ldots$, and the auxiliary sites are even, $2,4,6,\ldots$. Then the only non-trivial time-evolution connects $(1,3), (3,5), (5,7)$. There are several ways of dealing with such longer-ranged interactions, one explicitly constructing the longer-ranged interaction, the other using so-called swap gates, reducing it to a nearest-neighbour interaction.

The direct MPO description of the ``longer-ranged'' interaction $(1,3)$ involves necessarily a non-trivial tensor on site $2$, whereas the site $4$ is inert. Similarly for $(3,5)$, there is a non-trivial tensor on $4$, but site $6$ is inert. This suggests a Trotter decomposition $(1,2,3,4), (5,6,7,8),\ldots$ in the ``odd'' and $(3,4,5,6),(7,8,9,10),\ldots$ in the ``even'' steps.

The four-site evolution operator on sites 1 through 4 then reads
\begin{equation}
O^{(\sigma_1\sigma_2\sigma_3\sigma_4), (\sigma'_1\sigma'_2\sigma'_3\sigma'_4)} =
O^{(\sigma_1\sigma_2\sigma_3), (\sigma'_1\sigma'_2\sigma'_3)}\cdot \delta_{\sigma_4,\sigma'_4}
\end{equation}
and we can build the three-site unitary with only a slight modification of the two-site unitary which contains the actual physical time evolution:
\begin{equation}
O^{(\sigma_1\sigma_2\sigma_3), (\sigma'_1\sigma'_2\sigma'_3)} = O^{(\sigma_1\sigma_3), (\sigma'_1\sigma'_3)} \cdot \delta_{\sigma_2,\sigma'_2} .
\end{equation}
This three-site unitary is now subjected to two SVDs. For the notation, we first shift down the indices and reorder sitewise. Reshaping with subsequent SVDs then iteratively isolates $\sigma_3,\sigma'_3$, $\sigma_2,\sigma'_2$, and $\sigma_1,\sigma'_1$: 
\begin{eqnarray*}
& & O_{(\sigma_1\sigma_2\sigma_3), (\sigma'_1\sigma'_2\sigma'_3)} \\
&=& P_{(\sigma_1 \sigma'_1\sigma_2\sigma'_2),(\sigma_3\sigma'_3)} \\
&=& \sum_{k_2} U_{(\sigma_1 \sigma'_1\sigma_2\sigma'_2),k_2} S^{[2]}_{k_2,k_2} (V^\dagger_{23})_{k_2,(\sigma_3\sigma'_3)} \\
&=&  \sum_{k_2} U_{(\sigma_1 \sigma'_1),(\sigma_2\sigma'_2 k_2)} S^{[2]}_{k_2,k_2} (V^\dagger_{23})_{k_2,(\sigma_3\sigma'_3)} \\
&=& \sum_{k_1,k_2} U_{(\sigma_1 \sigma'_1),k_1} S^{[1]}_{k_1,k_1} (V^\dagger_{12})_{k_1, (\sigma_2\sigma'_2 k_2)} S^{[2]}_{k_2,k_2} (V^\dagger_{23})_{k_2,(\sigma_3\sigma'_3)} \\
&=& \sum_{k_1,k_2} W^{\sigma_1 \sigma'_1}_{1,k_1} W^{\sigma_2 \sigma'_2}_{k_1,k_2} W^{\sigma_3 \sigma'_3}_{k_2,1} 
\end{eqnarray*}
where, with the introduction of dummy indices and the inert tensor on site 4:
\begin{eqnarray}
W^{\sigma_1 \sigma'_1}_{1,k_1} &=& U_{(\sigma_1 \sigma'_1),k_1} \sqrt{S^{[1]}_{k_1,k_1}} \\
W^{\sigma_2 \sigma'_2}_{k_1,k_2} &=& \sqrt{S^{[1]}_{k_1,k_1}} (V^\dagger_{12})_{k_1, (\sigma_2\sigma'_2 k_2)} \sqrt{S^{[2]}_{k_2,k_2}} \\
W^{\sigma_3 \sigma'_3}_{k_2,1} &=& \sqrt{S^{[2]}_{k_2,k_2}} (V^\dagger_{23})_{k_2,(\sigma_3\sigma'_3)} \\
W^{\sigma_4 \sigma'_4}_{1,1} &=& \delta_{\sigma_4,\sigma'_4}
\end{eqnarray}
From this, MPOs for the entire chain can be formed as for the pure state time evolution. We have done nothing but the iterative decomposition of an MPO on 4 sites. Again, this is still manageable, as only 4 sites are involved. 

Obviously, it is a straightforward step to write an evolution operator acting on all four bonds $(1,2)$, $(1,3)$, $(2,4)$ and $(3,4)$ and subject it to a similar sequence of SVDs, which would allow to consider pure state time-evolution on a real ladder. 

An alternative approach to carry out interactions beyond immediate neighbours is provided by the use of {\em swap gates}\cite{Stoudenmire10}. Let us take the example of a real ladder with interactions on four bonds, two of which [$(1,3)$ and $(2,4)$] are next-nearest-neighbour interactions. But if we swapped states on sites $2 \leftrightarrow 3$, they would be nearest-neighbour interaction. Time-evolution on the ladder would then be done as follows: (i) evolve bonds $(1,2)$ and $(3,4)$; (ii) swap states on sites 2 and 3, (iii) evolve ``new'' bonds $(1,2)$ and $(3,4)$ with the evolution operators for ``old'' bonds $(1,3)$ and $(2,4)$ and (iv) swap states on sites 2 and 3 once again. In other situations, more astute schemes need to be found, preferrably generating a sequence of swaps between nearest neighbours. The swap operator for sites $i$ and $j$ is simply given by
\begin{equation}
\hat{S}_{ij} = \sum_{\sigma_i\sigma'_i \sigma_j\sigma'_j} S^{\sigma_i\sigma_j \sigma'_i \sigma'_j} \ket{\sigma_i\sigma_j}\bra{\sigma'_i\sigma'_j} \quad\quad 
S^{\sigma_i\sigma_j \sigma'_i \sigma'_j} = \delta_{\sigma_i,\sigma'_j} \delta_{\sigma_j,\sigma'_i},
\end{equation}
is unitary and its own inverse. It swaps the physical indices of two sites in an MPS; for swaps between nearest neighbours it is easy to restore the original form of the MPS: assume that the MPS is left-normalized; a unitary applied to sites $i$ and $i+1$ affects this only on these two sites. In particular, the orthonormality of block states $\ket{a_{k<i}}_A$ and $\ket{a_{i+1}}_A$ is not affected.
If we introduce a matrix $M_{(a_{i-1}\sigma_{i+1}),(\sigma_i a_{i+1})} = \sum_{a_i} A^{[i]\sigma_{i+1}}_{a_{i-1},a_i} A^{[i+1]\sigma_i}_{a_i,a_{i+1}}$, we can form $\tilde{M}_{(a_{i-1}\sigma_i),(\sigma_{i+1}a_{i+1})}=M_{(a_{i-1}\sigma_{i+1}),(\sigma_i a_{i+1})}$ and carry out an SVD, where $U_{(a_{i-1}\sigma_i),a_i}$ yields a new left-normalized $\tilde{A}^{\sigma_i}_{a_{i-1},a_i}$ and 
$S_{a_i,a_i} (V^\dagger)_{a_i,(\sigma_{i+1}a_{i+1})}$ a new left-normalized $\tilde{A}^{\sigma_{i+1}}_{a_{i},a_{i+1}}$. That the latter is left-normalized follows from the left-normalization of $\tilde{A}^{\sigma_i}$ and the maintained orthonormality of the $\ket{a_{i+1}}_A$.

Let me conclude this outlook on beyond-nearest-neighbour interactions with the remark that using MPO allows also other Trotter decompositions, e.g. decomposing the Heisenberg Hamiltonian in its $x$, $y$ and $z$-dependent parts, useful for long-range interactions \cite{VerstraetePirvu08}.

\subsection{tDMRG and TEBD compared to MPS time evolution: the little differences}
A bit before time evolution with MPS (tMPS) was developed, two other algorithms were introduced to simulate the real-time dynamics of one-dimensional quantum chains, time-evolving block decimation (TEBD)\cite{Vidal03a,Vidal04b} and real-time or time-dependent DMRG (tDMRG) \cite{Daley04,WhiteFeiguin04}. Both algorithms are also based on MPS, but are different from tMPS, when one looks more closely. Before I get into that, let me stress however that all of them are based on the idea of time-evolving an MPS which was first put forward in \cite{Vidal03a,Vidal04b} and therefore are minor variations on a theme. tDMRG and TEBD are mathematically equivalent, i.e.\ should for exact arithmetic give the same results, whereas numerically they are clearly distinct algorithms, both carrying out operations that have no counterpart in the other method, with their respective advantages and disadvantages. Let us discuss first tDMRG, because its language is closer to that of tMPS, and then TEBD, to see how important (or unimportant) the little differences are.

\subsubsection{Time-dependent DMRG (tDMRG)}
The decomposition of a global time-evolution on an entire lattice into a Trotter sequence of infinitesimal time-evolutions on bonds is the same for all three algorithms discussed here.
Let us therefore focus on one infinitesimal time-evolution $\eul^{-\imag  \hat{h}_{\ell+1} \tau}$ on sites $\ell+1$ and $\ell+2$. 
The evolution operator expressed in the local basis is given by
\begin{equation}
U^{(\sigma_{\ell+1} \sigma_{\ell+2}), (\sigma'_{\ell+1} \sigma'_{\ell+2})} = \bra{\sigma_{\ell+1} \sigma_{\ell+2}} \eul^{-\imag  \hat{h}_{\ell+1} \tau} \ket{\sigma'_{\ell+1} \sigma'_{\ell+2}} .
\end{equation}
The current state $\psi$ is given in the two-site DMRG notation with left- and right normalized matrices as
\begin{equation}
\ket{\psi} = \sum_{\fat{\sigma}} A^{\sigma_1} \ldots A^{\sigma_\ell} \Psi^{\sigma_{\ell+1}\sigma_{\ell+2}} B^{\sigma_{\ell+3}} \ldots B^{\sigma_L} \ket{\fat{\sigma}} .
\end{equation}
The time-evolution turns $\Psi^{\sigma_{\ell+1}\sigma_{\ell+2}}$ into
\begin{equation}
\Phi^{\sigma_{\ell+1}\sigma_{\ell+2}}_{a_\ell, a_{\ell+2}} = \sum_{\sigma'_{\ell+1} \sigma'_{\ell+2}}
U^{(\sigma_{\ell+1} \sigma_{\ell+2}), (\sigma'_{\ell+1} \sigma'_{\ell+2})}  \Psi^{\sigma'_{\ell+1}\sigma'_{\ell+2}}_{a_\ell, a_{\ell+2}} . 
\end{equation} 
This, together with the $A$ and $B$-matrices defines a valid DMRG state we call $\ket{\phi}$. In order to make progress, namely to apply $\eul^{-\imag  \hat{h}_{\ell+3} \tau}$ on the next pair of sites, we have to bring the state into the form 
\begin{equation}
\ket{\phi} = \sum_{\fat{\sigma}} A^{\sigma_1} \ldots A^{\sigma_{\ell+2}} \Phi^{\sigma_{\ell+3}\sigma_{\ell+4}} B^{\sigma_{\ell+5}} \ldots B^{\sigma_L} \ket{\fat{\sigma}} .
\end{equation}
The changes can only concern sites $\ell+1$ through $\ell+4$: on the first two sites because of the action of the evolution operator, on the last two sites because they are brought into DMRG form. Let us first generate the new $A$-matrices on sites $\ell+1$ and $\ell+2$: We reshape $\Phi^{\sigma_{\ell+1}\sigma_{\ell+2}}_{a_\ell, a_{\ell+2}} = \Phi_{(a_\ell \sigma_{\ell+1}), (\sigma_{\ell+2} a_{\ell+2})}$ and subject it to an SVD (DMRG traditionally does this by a density matrix analysis and the DMRG prediction when shifting sites, leading to the same result):
\begin{equation}
\Phi_{(a_\ell \sigma_{\ell+1}),(\sigma_{\ell+2} a_{\ell+2})} = \sum_{a_{\ell+1}} U_{(a_\ell \sigma_{\ell+1}),a_{\ell+1}} S_{a_{\ell+1},a_{\ell+1}} (V^\dagger)_{a_{\ell+1}, (\sigma_{\ell+2} a_{\ell+2})}.
\end{equation}
$U$ can immediately be reshaped into a valid $A$-matrix, but has column dimension up to $dD$, which has to be truncated down to $D$ while maintaining the best approximation to the MPS of dimension $dD$. The answer is provided as always by keeping just the $D$ largest singular values and shrinking the matrices $U$, $S$, $V^\dagger$ accordingly. Here lies the approximation of the method (beyond the obvious Trotter error). This done, we reshape as
\begin{equation}
\sum_{a_{\ell+1}} A^{\sigma_{\ell+1}}_{a_\ell, a_{\ell+1}}  S_{a_{\ell+1},a_{\ell+1}} (V^\dagger)^{\sigma_{\ell+2}}_{a_{\ell+1},  a_{\ell+2}}
\end{equation}
and form $\Phi$ shifted by one site as
\begin{equation}
\Phi^{\sigma_{\ell+2}\sigma_{\ell+3}}_{a_{\ell+1},a_{\ell+3}} = \sum_{a_{\ell+2}} S_{a_{\ell+1},a_{\ell+1}} (V^\dagger)^{\sigma_{\ell+2}}_{a_{\ell+1},  a_{\ell+2}} B^{\sigma_{\ell+3}}_{a_{\ell+2},  a_{\ell+3}} .
\end{equation}
But we have to shift by another site, which we achieve by reshaping $\Phi^{\sigma_{\ell+2}\sigma_{\ell+3}}_{a_{\ell+1},a_{\ell+3}}$ as $\Phi_{(a_{\ell+1}\sigma_{\ell+2}),(\sigma_{\ell+3} a_{\ell+3})}$, carry out an SVD as done before, keep the states corresponding to the $D$ largest out of $dD$ singular values, reshape, note down $A^{\sigma_{\ell+2}}$ and form $\Phi$ shifted by two sites as
\begin{equation}
\Phi^{\sigma_{\ell+3}\sigma_{\ell+4}}_{a_{\ell+2},a_{\ell+4}} = \sum_{a_{\ell+3}} S_{a_{\ell+2},a_{\ell+2}} (V^\dagger)^{\sigma_{\ell+3}}_{a_{\ell+2},  a_{\ell+3}} B^{\sigma_{\ell+4}}_{a_{\ell+3},  a_{\ell+4}} .
\end{equation}
The second SVD and the associated truncation down to $D$ singular values does not lose further information, because there are at most $D$ non-zero singular values, although formally there could be $dD$ of them. The reason is that before the time evolution on sites $\ell+1$ and $\ell+2$, the Schmidt rank across the bond $\ell+2$ was at most $D$ (due to the MPS construction). The Schmidt rank of two states is however identical if they are related by a unitary transformation that acts on either part A or part B. But the infinitesimal time-evolution was a unitary on part A.

We can now continue with the next infinitesimal local time-evolution step, in the spirit of tMPS.   
\subsubsection{Time-evolving block decimation (TEBD)}
Here, we assume that we have $\ket{\psi}$ in the $\Gamma\Lambda$-notation,
\begin{equation}
\ket{\psi} = \sum_{\fat{\sigma}} 
\Gamma^{\sigma_1} \Lambda^{[1]} \Gamma^{\sigma_2} \Lambda^{[2]} \ldots
\Gamma^{\sigma_{\ell}} \Lambda^{[\ell]}  \Gamma^{\sigma_{\ell+1}} \Lambda^{[\ell+1]}
\Gamma^{\sigma_{\ell+2}} \Lambda^{[\ell+2]}  \Gamma^{\sigma_{\ell+3}} \Lambda^{[\ell+3]} \ldots
\Gamma^{\sigma_L}   
\ket{\fat{\sigma}}.
\label{eq:TEBDstart}
\end{equation}
This state can be immediately connected to the two-site DMRG notation. In particular,
\begin{equation}
\Psi^{\sigma_{\ell+1} \sigma_{\ell+2}} = \Lambda^{[\ell]}  \Gamma^{\sigma_{\ell+1}} \Lambda^{[\ell+1]}
\Gamma^{\sigma_{\ell+2}} \Lambda^{[\ell+2]}  .
\end{equation}
This is identical to the DMRG $\Psi$, so is the evolution operator $U^{(\sigma_{\ell+1} \sigma_{\ell+2}), (\sigma'_{\ell+1} \sigma'_{\ell+2})}$, hence also  
\begin{equation}
\Phi^{\sigma_{\ell+1}\sigma_{\ell+2}}_{a_\ell, a_{\ell+2}} = \sum_{\sigma'_{\ell+1} \sigma'_{\ell+2}}
U^{(\sigma_{\ell+1} \sigma_{\ell+2}), (\sigma'_{\ell+1} \sigma'_{\ell+2})}  \Psi^{\sigma_{\ell+1}'\sigma_{\ell+2}'}_{a_\ell, a_{\ell+2}} . 
\end{equation} 
In order to proceed, the $\Gamma\Lambda$-notation has to be restored on the two active sites. In perfect analogy to tDMRG, one obtains by SVD
\begin{equation}
\Phi_{(a_\ell \sigma_{\ell+1}), (\sigma_{\ell+2} a_{\ell+2})} = \sum_{a_{\ell+1}} U_{(a_\ell \sigma_{\ell+1}),a_{\ell+1}} \Lambda^{[\ell+1]}_{a_{\ell+1},a_{\ell+1}} (V^\dagger)_{a_{\ell+1}, (\sigma_{\ell+2} a_{\ell+2})} ,
\end{equation}
What is missing are $\Lambda^{[\ell]}$ and $\Lambda^{[\ell+2]}$. We therefore write (reshaping $U$ and $V^\dagger$ and omitting the $a$-indices)
\begin{equation}
\Phi^{\sigma_{\ell+1}\sigma_{\ell+2}} = \Lambda^{[\ell]} (\Lambda^{[\ell]})^{-1} U^{\sigma_{\ell+1}} 
 \Lambda^{[\ell+1]} V^{\sigma_{\ell+2}\dagger} ( \Lambda^{[\ell+2]})^{-1}  \Lambda^{[\ell+2]} .
\end{equation}
Now, as in tDMRG, there are up to $dD$ singular values in $ \Lambda^{[\ell+1]}$, which we truncate down to the $D$ largest ones, just as in tDMRG, also truncating the neighbouring matrices accordingly. We now introduce
\begin{equation}
\Gamma^{\sigma_{\ell+1}}_{a_\ell,a_{\ell+1}} =  (\Lambda^{[\ell]})^{-1}_{a_\ell,a_\ell} U^{\sigma_{\ell+1}}_{a_\ell,a_{\ell+1}} \quad 
\Gamma^{\sigma_{\ell+2}}_{a_{\ell+1},a_{\ell+2}} =  V^{\sigma_{\ell+2}\dagger}_{a_{\ell+1},a_{\ell+2}} ( \Lambda^{[\ell+2]})^{-1}_{a_{\ell+2},a_{\ell+2}} 
\end{equation}
and obtain
\begin{equation}
\Phi^{\sigma_{\ell+1}\sigma_{\ell+2}} = \Lambda^{[\ell]}  \Gamma^{\sigma_{\ell+1}} 
 \Lambda^{[\ell+1]} \Gamma^{\sigma_{\ell+2}}  \Lambda^{[\ell+2]} ,
\end{equation}
back to the canonical form. In order to consider the time-evolution on the next bond, we have to carry out no SVDs, but just group
\begin{equation}
\Psi^{\sigma_{\ell+3} \sigma_{\ell+4}} = \Lambda^{[\ell+2]}  \Gamma^{\sigma_{\ell+3}} \Lambda^{[\ell+3]}
\Gamma^{\sigma_{\ell+4}} \Lambda^{[\ell+4]}  
\end{equation}
and continue. As in tDMRG, no loss of information is associated with this step, but this is more explicit here.

When $D$ becomes very large in high-precision calculations, singular values will tend to be very small, and dividing by them is, as mentioned previously, a source of numerical instability. In the context of the thermodynamic limit iTEBD method, which we will discuss later, Hastings has proposed an elegant workaround that comes at very low numerical cost\cite{Hastings09}, but it can be easily adapted to finite-system TEBD. Let us assume that we start with a state in representation (\ref{eq:TEBDstart}). We then group all pairs into right-normalized $B$-matrices,
\begin{equation}
B^{\sigma_i} = \Gamma^{\sigma_i} \Lambda^{[i]} ,
\end{equation}
but remember the $\Lambda^{[i]}$ for later use. We then form
\begin{equation}
\overline{\Psi}^{\sigma_{\ell+1} \sigma_{\ell+2}} = B^{\sigma_{\ell+1}} B^{\sigma_{\ell+2}} =  \Gamma^{\sigma_{\ell+1}} \Lambda^{[\ell+1]}
\Gamma^{\sigma_{\ell+2}} \Lambda^{[\ell+2]} ,
\end{equation}
hence $\Psi^{\sigma_{\ell+1} \sigma_{\ell+2}} = \Lambda^{[\ell]} \overline{\Psi}^{\sigma_{\ell+1} \sigma_{\ell+2}}$. We carry out the time-evolution on $\overline{\Psi}^{\sigma_{\ell+1} \sigma_{\ell+2}}$ to obtain
\begin{equation}
\overline{\Phi}^{\sigma_{\ell+1}\sigma_{\ell+2}}_{a_\ell, a_{\ell+2}} = \sum_{\sigma'_{\ell+1} \sigma'_{\ell+2}}
U^{(\sigma_{\ell+1} \sigma_{\ell+2}), (\sigma'_{\ell+1} \sigma'_{\ell+2})}  \overline{\Psi}^{\sigma_{\ell+1}'\sigma_{\ell+2}'}_{a_\ell, a_{\ell+2}} .
\end{equation} 
Then $\Phi^{\sigma_{\ell+1} \sigma_{\ell+2}} = \Lambda^{[\ell]} \overline{\Phi}^{\sigma_{\ell+1} \sigma_{\ell+2}}$. As before, we carry out an SVD on $\Phi^{\sigma_{\ell+1} \sigma_{\ell+2}}$, to obtain 
\begin{equation}
\Phi_{(a_\ell \sigma_{\ell+1}), (\sigma_{\ell+2} a_{\ell+2})} = \sum_{a_{\ell+1}} U_{(a_\ell \sigma_{\ell+1}),a_{\ell+1}} \Lambda^{[\ell+1]}_{a_{\ell+1},a_{\ell+1}} (V^\dagger)_{a_{\ell+1}, (\sigma_{\ell+2} a_{\ell+2})} = A^{\sigma_{\ell+1}} \Lambda^{[\ell+1]} B^{\sigma_{\ell+2}} .
\end{equation}
Truncating down to the $D$ largest singular values, we have found the new $\Lambda^{[\ell+1]}$, to be retained for further usage, and the new $B^{\sigma_{\ell+2}}$. The new $B^{\sigma_{\ell+1}}$
is given by 
\begin{equation}
B^{\sigma_{\ell+1}} = \sum_{\sigma_{\ell+2}}  \overline{\Phi}^{\sigma_{\ell+1}\sigma_{\ell+2}}
B^{\sigma_{\ell+2}\dagger},
\end{equation}
hence costs a simple matrix multiplication; divisions have been avoided. For the last equation, we use right-normalization of $B^{\sigma_{\ell+2}}$, hence $\sum_{\sigma_{\ell+2}} \Phi^{\sigma_{\ell+1}\sigma_{\ell+2}} B^{\sigma_{\ell+2}\dagger} =  A^{\sigma_{\ell+1}} \Lambda^{[\ell+1]}$. At the same time, $B^{\sigma_{\ell+1}} = \Gamma^{\sigma_{\ell+1}} \Lambda^{[\ell+1]} = (\Lambda^{[\ell]})^{-1} A^{\sigma_{\ell+1}}  \Lambda^{[\ell+1]}$. Combining these two identities with $\Phi^{\sigma_{\ell+1} \sigma_{\ell+2}} = \Lambda^{[\ell]} \overline{\Phi}^{\sigma_{\ell+1} \sigma_{\ell+2}}$ gives the result.

\subsubsection{Comparing the algorithms}
Comparing TEBD and tDMRG step by step, one sees immediately the complete mathematical equivalence of the methods. The second SVD in tDMRG does nothing but shifting the boundary between left- and right-normalized matrices, which in TEBD is simply achieved by a rebracketing of $\Gamma$ and $\Lambda$. Nevertheless, there are differences: tDMRG carries out two costly SVD decompositions (or density matrix analyses, which is equivalent) per bond evolution, where TEBD does only one. On the other hand, TEBD encounters divisions by potentially very small singular values, which is a strong source of potential numerical inaccuracies; but these can be eliminated \cite{Hastings09} at low numerical cost. From a numerical point of view, tDMRG is not just a translation of TEBD, which came first, but an algorithm of its own, with strengths and weaknesses.

Both methods share the central feature that time evolution and truncation are intertwined: after each bond evolution, there is a truncation by SVD. By contrast, tMPS evolves all bonds first, and then truncates the entire state by compression of matrix dimensions $d^2 D \rightarrow D$ by SVD or iteratively. 

tMPS is the cleaner approach, but it can also be shown to be more precise. In fact, for real-time evolution it relates to tDMRG or TEBD exactly as iterative variational compression to compression by SVD, which implies that for small state changes (e.g.\ for very small time steps) the difference goes down, as the interdependence of truncations becomes less severe, there being only very benign truncations. That the above relationship exists can be seen from compressing a tMPS state not variationally, but by SVD only: 

Take $\ket{\psi}$ to be right-canonical, and do a tDMRG/TEBD step on the first bond or tMPS steps on all odd bonds. The truncation is now to be carried out by SVD and my claim is that SVD does not see a difference between the two very different time-evolved states. On the first bond itself, all methods produce the same structure, but they differ on all other sites. Whereas
\begin{equation}
\ket{\phi}_{{\rm tDMRG}} = \sum_{\fat{\sigma}} \Psi^{\sigma_1\sigma_2} B^{\sigma_3} B^{\sigma_4} \ldots \ket{\fat{\sigma}}  
\end{equation}
is 
\begin{equation}
\ket{\phi}_{{\rm tMPS}} = \sum_{\fat{\sigma}} \Psi^{\sigma_1\sigma_2} \overline{B}^{\sigma_3} \overline{B}^{\sigma_4} \ldots \ket{\fat{\sigma}}  ,
\end{equation}
where the $\overline{B}$-matrices come from the contraction with the time evolution bond operators. The SVDs on the first bonds are equivalent for both states provided both sets $\{ B \}$ and $\{ \overline{B} \}$ generate sets of orthonormal states. This is indeed the case, because the $B$-matrices do this by definition, and the states generated by the $\overline{B}$-matrices are related to the first set of orthonormal states by a unitary transformation (real-time evolution!). This observation of the equivalence of methods also holds for bonds further down the chain.

Hence, the difference between the three algorithms becomes only visible at the level of variational compression.
 
\subsection{How far can we go?} 
In this section, I have described basic algorithms for the time evolution of pure and mixed states. There were two sources of error. One of them is the Trotter decomposition, which for an $n$th order decomposition generated an error $O(\tau^{n+1})$ for each time step $\tau$. As there are $t/\tau$ time steps, the error will ultimately be $O(\tau^n t)$, i.e.\ linear in time. This means it is only growing moderately in time and can be scaled down by smaller time steps and/or higher-order decompositions. This is common to all current methods \cite{Vidal03a,Vidal04b,Daley04,WhiteFeiguin04,VerstraeteRipoll04}.  In fact, there are other methods of calculating matrix exponentials such as the Krylov method\cite{Schmitteckert04} or lookahead procedures such as in \cite{Feiguin05}, which reduce this error even more. In any case, it is not very worrisome in the long run. 

On the other hand, there is the error due to the truncation of the blown-up bond dimensions of the MPS after each time step. This error is serious; early on it could be shown to lead to errors exponentially blowing up in time \cite{Gobert05}. Yet truncation errors are only the symptom, not the fundamental problem: the real reason is that -- following the Lieb-Robertson theorem -- entanglement $S$ can grow up to linearly in time for an out-of-equilibrium evolution of a quantum state: $S(t) \leq S(0) + c t$, where $c$ is some constant related to the propagation speed of excitations in the lattice\cite{Calabrese06}. This linear bound is actually reached for many quantum quenches, where a Hamiltonian parameter is abruptly changed such that the global energy changes extensively.
Both from $D \sim 2^S$ and from a rigorous analysis \cite{Osborne06} it follows that in such cases the matrix dimensions will have to go up exponentially in time, $D(t) \sim 2^t$, or that for fixed matrix dimensions precision will deteriorate exponentially. 

Nevertheless, in many circumstances matrix size growth is slow enough that numerical resources are sufficient to observe the time-dependent phenomenon of interest: Time-dependent DMRG has been used extensively in the meantime and found to open completely new perspectives on the non-equilibrium behaviour of strongly correlated one-dimensional systems (to name a few:\cite{WhiteFeiguin04,Feiguin05a,Feiguin05,Gobert05,Kollath05,Trebst06,Cramer08,Barthel09a,Kollath06}).
 
\section{Time-dependent Simulation: Extending The Range}
\label{sec:beyondtime}

Time evolution -- whether it is done by TEBD, tDMRG or tMPS, to give the historical order -- is fundamentally limited by the times that can be reached. The underlying reason is the (at worst) linear buildup of entanglement in time in an out-of-equilibrium quantum state, that translates itself into an (at worst) exponential growth of bond dimensions $D(t)$ if a given precision is desired. A ``time wall'' is hit exponentially fast. Can one push it further into the future? A similar issue arises for finite-temperature calculations. While they are not necessarily about dynamics, seen as imaginary time evolutions they raise their own problems, regarding the $T\rightarrow 0$ limit for static or thermodynamic calculations, and regarding dynamics, there in particular at high temperatures.

A second issue that we have not covered so far concerns {\em dissipative} time evolution where we are not looking at a closed quantum system as in pure Hamiltonian dynamics but at an open quantum system. If the dynamics is Markovian (i.e.\ the bath has no memory, a highly non-trivial assumption in many cases), then the most general dynamics is given by the Lindblad equation. While it is easy to show that formally this can be simulated using MPS quite easily, in actual practice this is numerically involved and simpler schemes are highly desirable. 

In this section we will first consider attempts to extend the time range of simulation by different schemes for evaluating the time evolution tensor networks. As we have already seen for simple examples like the evaluation of wave function overlaps, the order of contractions may hugely change the computational effort. In a second step, we will look at a prediction method that picks up on numerical raw data and extrapolates them very successfully over an order of magnitude, provided they meet a certain mathematical form, taking the case of finite temperature as an example. In a third step, we will look at an altogether different way of finite temperature simulations. In a last step, I will take up the issue of dissipative dynamics and show neat progress made in that field.

\subsection{Orders of contraction: light cones, Heisenberg picture and transverse folding}
\begin{figure}
\centering\includegraphics[scale=0.7]{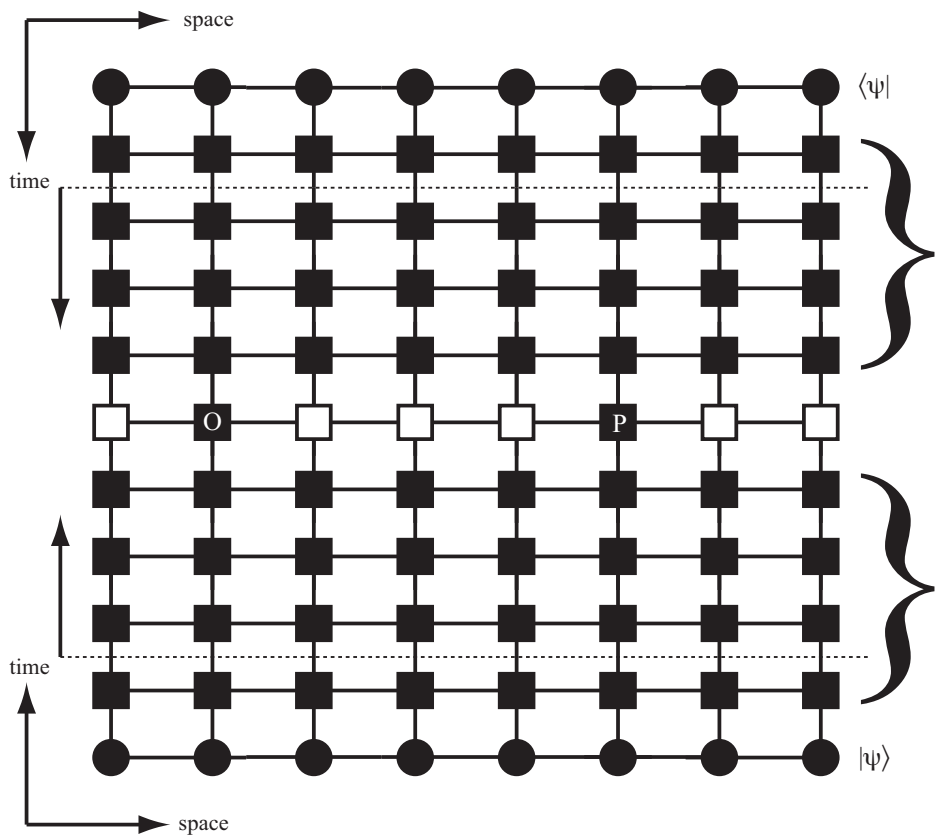}
\caption{Two-dimensional tensor network contracted for time-dependent expectation values: the top and bottom lines represent $\bra{\psi}$ and $\ket{\psi}$. The two arrays of MPOs (indicated by brackets) represent $\eul^{+\imag\hat{H}t}$ and $\eul^{-\imag\hat{H}t}$ in Trotterized form; I do not distinguish between different local MPOs such as identity operators which show up on some sites of the left- and rightmost columns. In the central line, we put identity operators (white squares) and the operators to be evaluated. The dashed line indicates the buildup of contractions in time direction.}
\label{fig:2Dcontractiontime}
\end{figure}

Let us consider the calculation of the time-dependent expectation value
\begin{equation}
\bra{\psi(t)} \hat{O} \hat{P} \ket{\psi(t)} = \bra{\psi} \eul^{+\imag\hat{H}t} \hat{O} \hat{P} \eul^{-\imag\hat{H}t} \ket{\psi} .
\end{equation}
Starting with $\ket{\psi}$, we evolve it up to time $t$, obtaining $\ket{\psi(t)}$. The expectation value then is calculated by sandwiching the two operators between $ \bra{\psi(t)}$ and $\ket{\psi(t)}$, as discussed before. But we can represent this procedure also as the (approximate) contraction over a two-dimensional tensor network as shown in Fig.~\ref{fig:2Dcontractiontime}, which is then contracted line by line along the time direction, moving inwards. 

Assuming $t=N\Delta t$ and $n$ MPOs per Trotter step (e.g.\ 2 in first order), we have a lattice of $L \times (2nN+3)$ sites, i.e. of width $L$ and odd height $2nN+3$. If we call $T^{[i,j]}$ the tensor located on the site in row $i$ and column $j$ (like in a matrix), and if we label indices by up $u$, down $d$, left $l$ and right $r$, and write $T^{[i,j]}$ in analogy to MPOs with indices $T^{[i,j]u,d}_{l,r}$, then we can identify, for example, for $i=1$ (location of the bra state):
\begin{equation}
T^{[1,1] 1,d}_{1,r} = A^{[1]d*}_{1,r} \quad
T^{[1,j] 1,d}_{l,r} = A^{[j]d*}_{l,r} \quad
T^{[1,L] 1,d}_{l,1} = A^{[L]d*}_{l,1}
\end{equation}
where $1<j<L$. Similarly for $i=2nN+3$ (location of the ket state):
\begin{equation}
T^{[2nN+3,1] u,1}_{1,r} = A^{[1]u}_{1,r} \quad
T^{[2nN+3,j] u,1}_{l,r} = A^{[j]u}_{l,r} \quad
T^{[2nN+3,L] u,1}_{l,1} = A^{[L]u}_{l,1}
\end{equation}
and on row $i=nN+2$ (location of the operators):
\begin{equation}
T^{[nN+2,j] u,d}_{1,1} = \left\{ \begin{array}{cl} \hat{O}^{u,d} & {\rm on\ operator\ location\ } j \\
\delta_{u,d} & {\rm else} \end{array} \right\} .
\end{equation}
In this row, horizontally the network is a product of scalars, hence the $(1,1)$. On all other rows, the tensors $T^{[i,j]}$ are given by local MPOs such that $T^{[i,j]u,d}_{l,r}= W^{[\alpha]u,d}_{l,r}$ on all rows $nN+2 < i < 2nN+3$ (with the type $\alpha$ depending on the chosen decomposition) and 
$T^{[i,j]u,d}_{l,r}= W^{[\alpha]d,u*}_{l,r}$ on all rows $1 < i < nN+2$, which correspond to the time evolution of the bra.

\subsubsection{Light cones in time evolution}
\begin{figure}
\centering\includegraphics[scale=0.25]{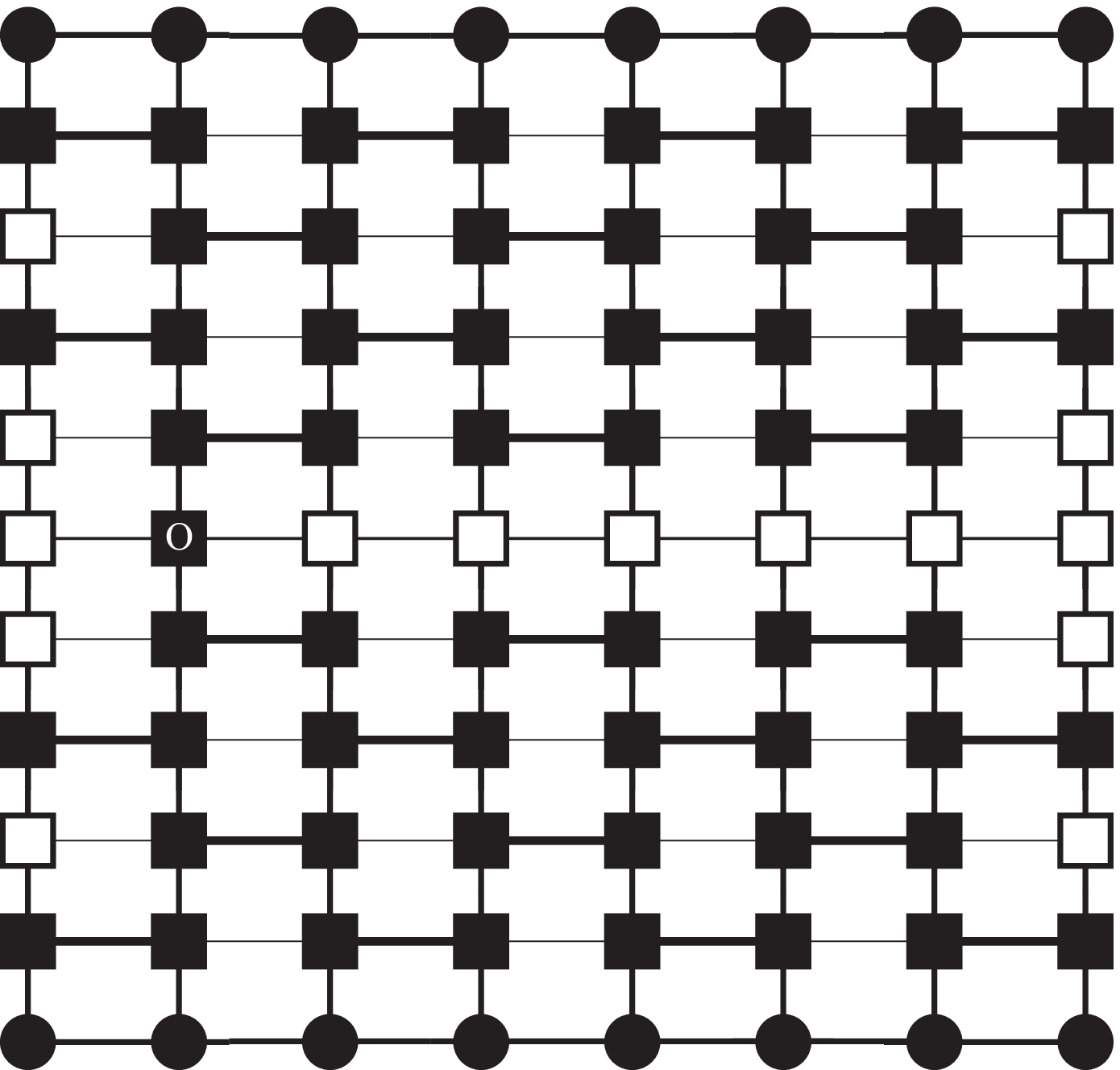}$\quad\quad$\includegraphics[scale=0.25]{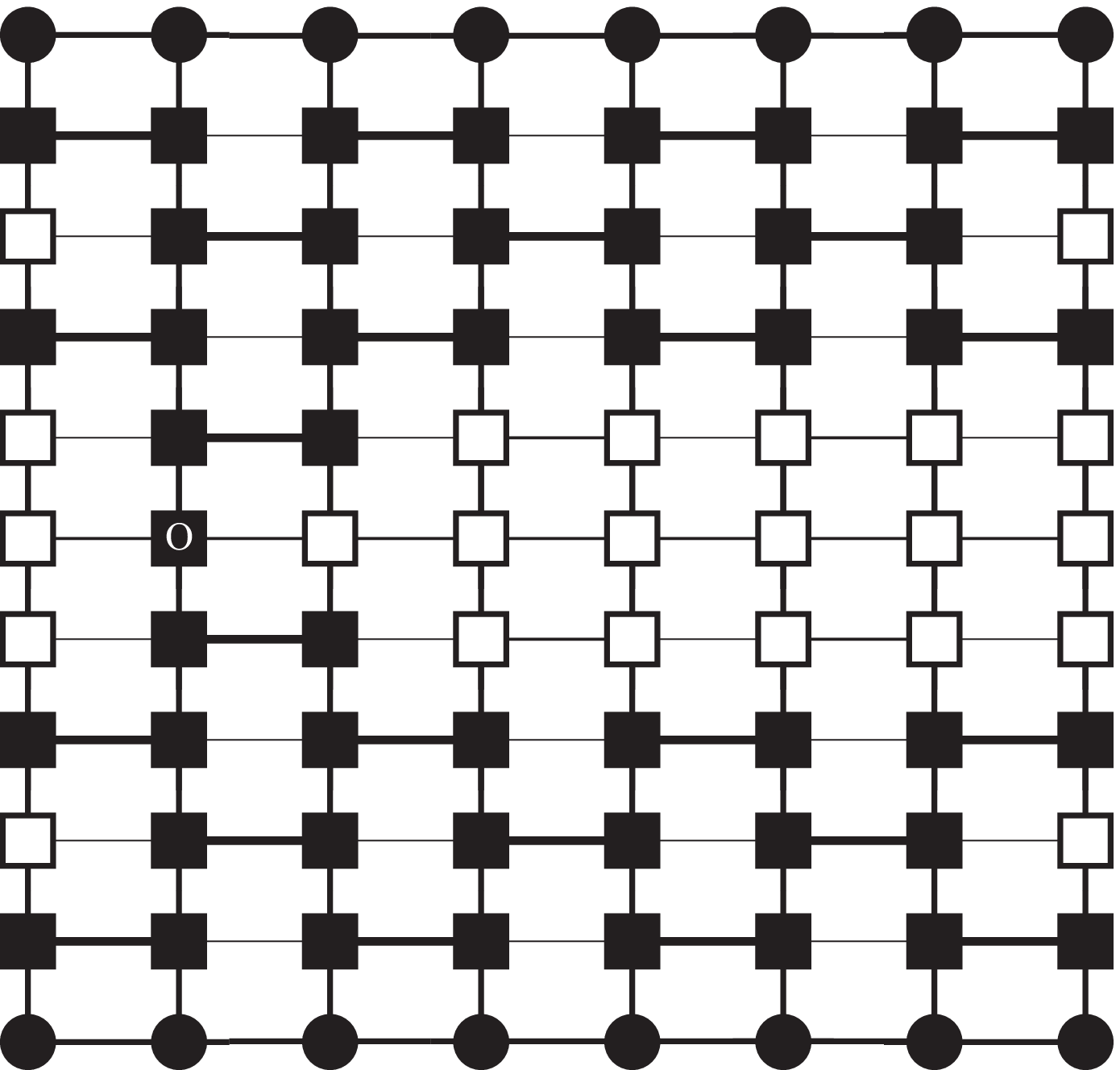}
\vskip 1cm
\centering\includegraphics[scale=0.25]{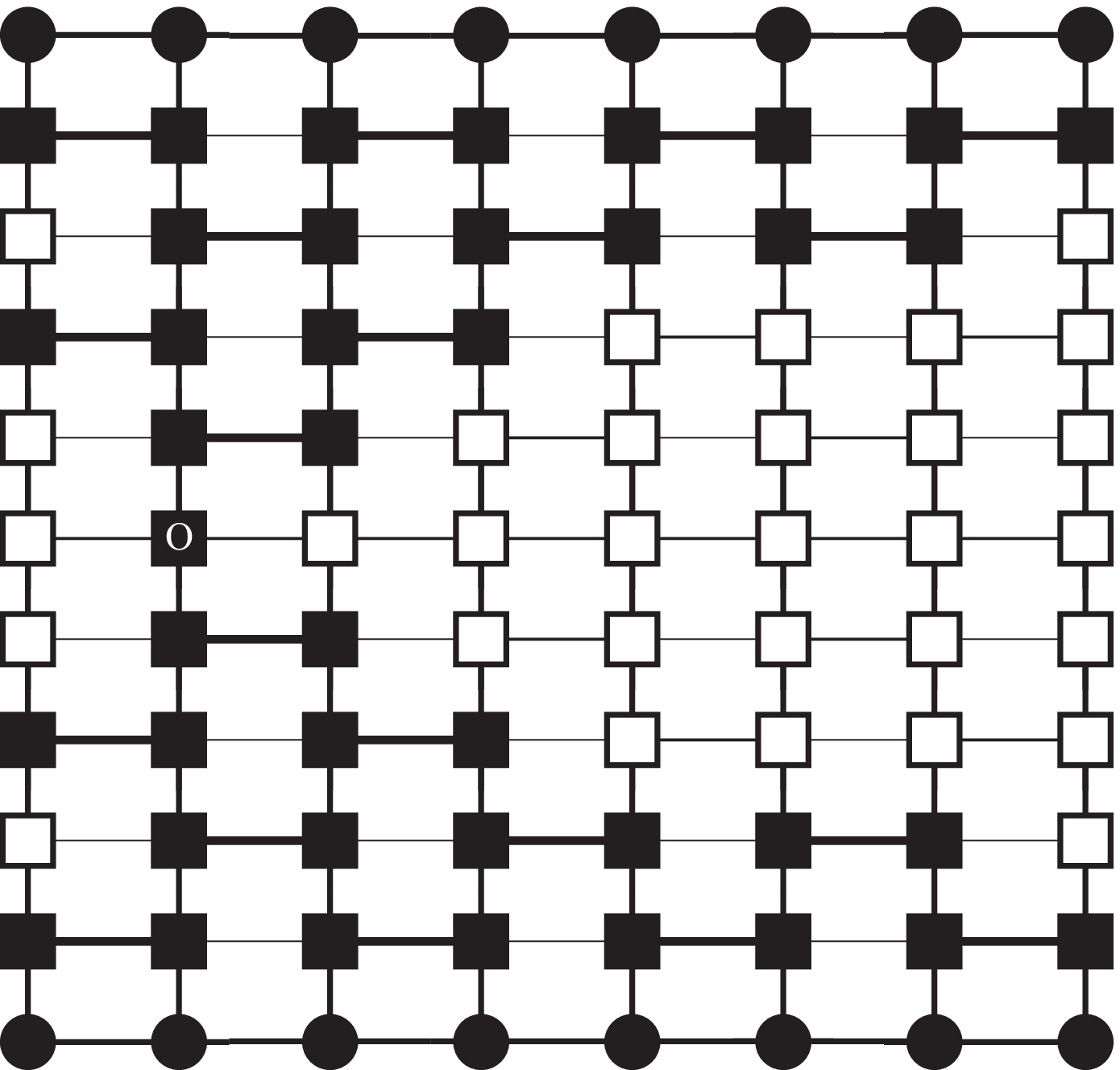}$\quad\quad$\includegraphics[scale=0.25]{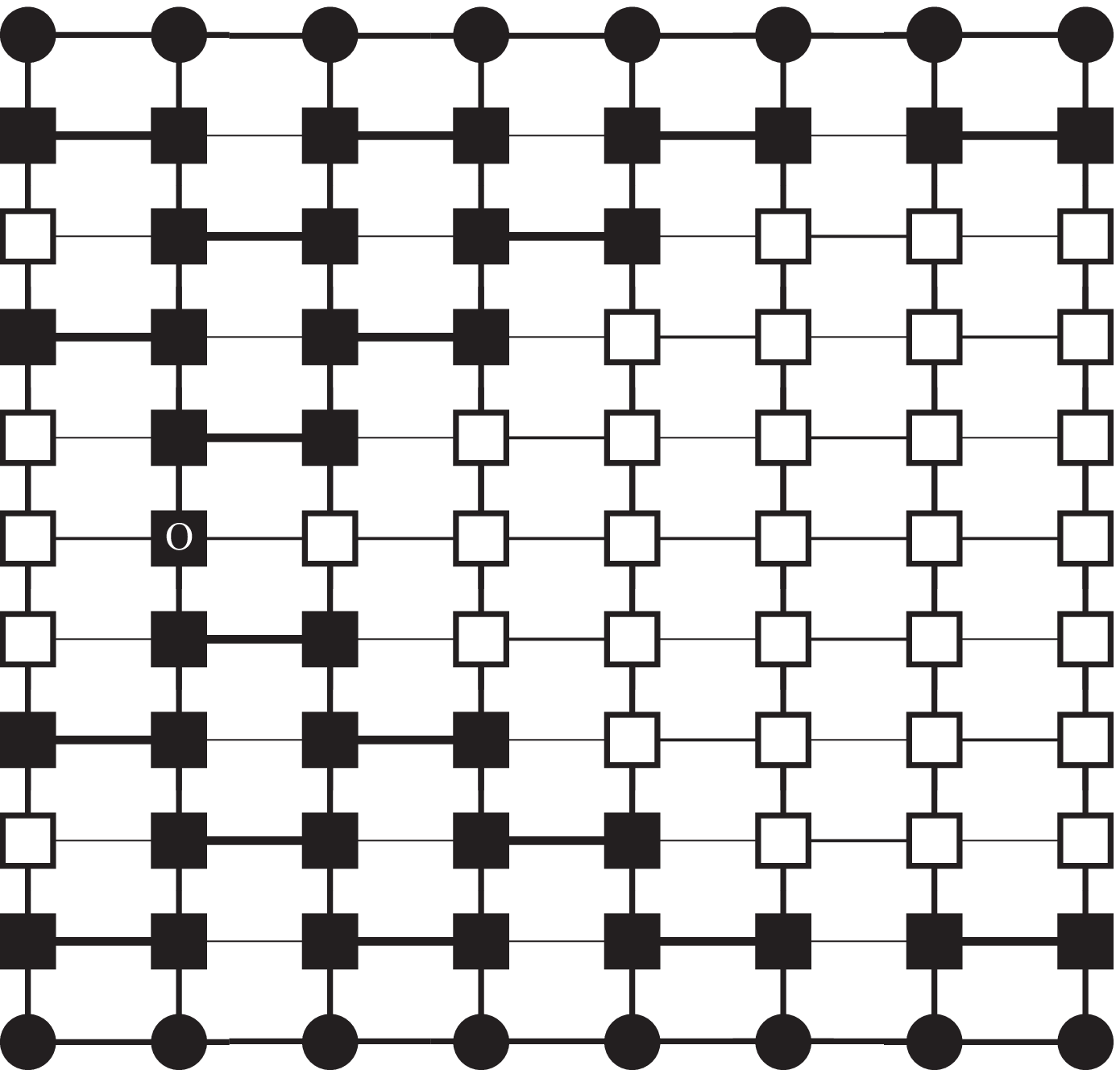}
$\quad\quad$\includegraphics[scale=0.25]{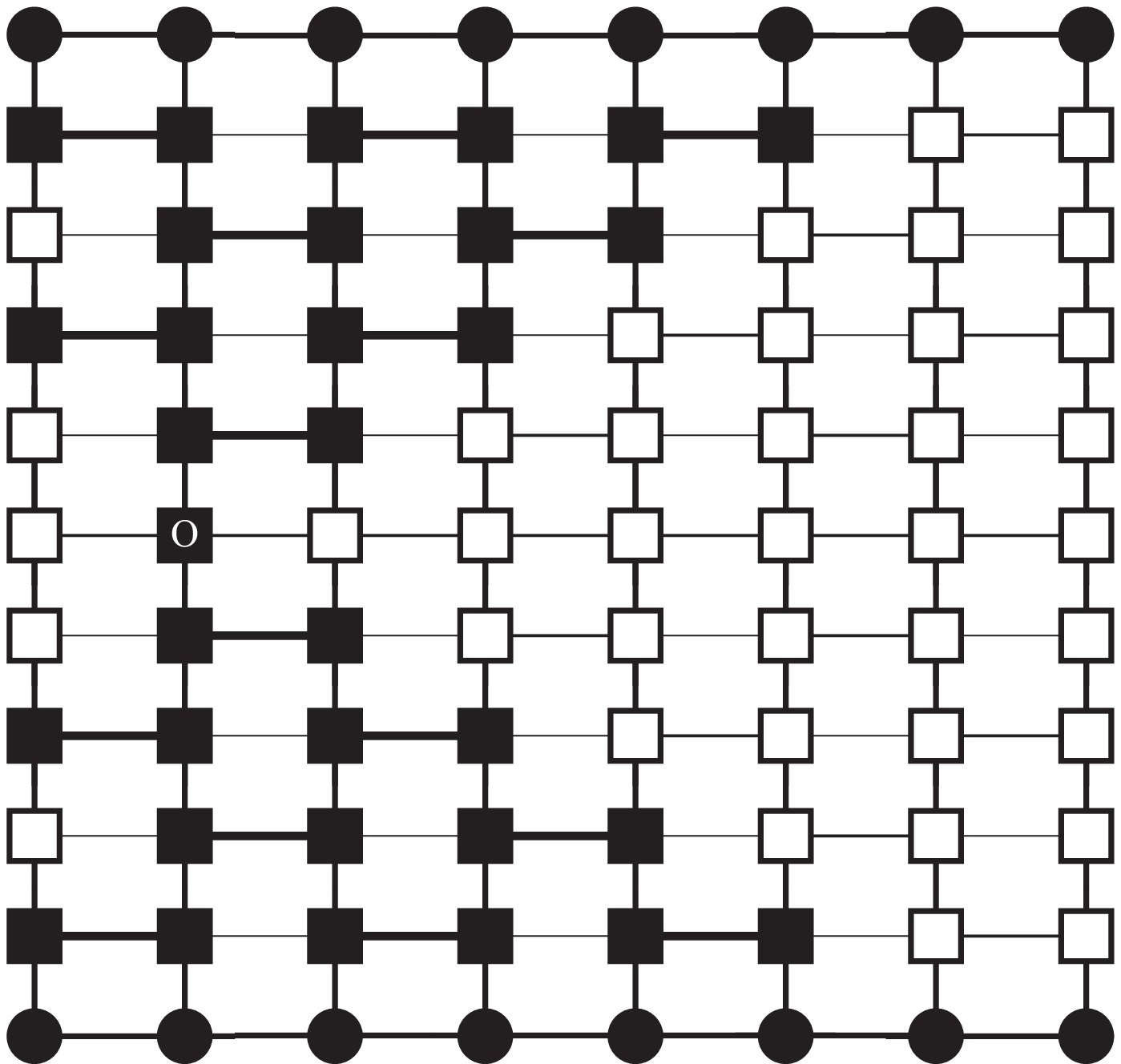}

\caption{Light cone: The diagrams are to be read from left to right, in the top and then the bottom line. On the top left, an operator is sandwiched between two time-evolved states; taking Trotter evolution into account, identity operators are marked by white squares. Working our way outward from the center towards the top and bottom, bond evolution operators cancel each other to identities, provided they do not sandwich the operator or (therefore) surviving bond operators. The result is a light cone of evolution operators, surrounded by numerically trivial identities.}
\label{fig:lightcone1}
\end{figure}

Considering the time evolution of bra and ket together in fact allows important simplifications. In Fig.~\ref{fig:lightcone1} I have restored the alternating pattern of bond evolutions in a first order Trotter decomposition and explicitly marked the position of unit operators by white squares. We would like to calculate $\langle \hat{O}(t) \rangle$, where the operator sits on site 2. Let us look at the last Trotter steps (rows 5 and 7). In row 5 there are several evolution operators $\eul^{+\imag \hat{h} \Delta t}$ with corresponding operators  $\eul^{-\imag \hat{h} \Delta t}$ in row 7. But this means that they cancel each other and can be replaced by unit operators, except in columns 2 and 3 because they have $\hat{O}$ interposed. If, in turn, we look now at rows 4 and 8, there are now evolution operators cancelling each other in columns 5 through 8; the other ones do not cancel, as they sandwich non-identity operators. Like this, we work our way towards the bra and ket states, until no cancellation is possible anymore. The resulting tensor network shows a large degree of replacements of complicated tensors of bond (row) dimensions larger than 1 by identity tensors with bond dimension 1, which means that contractions become trivial and no compression is needed. There remains an algorithmic {\em light cone} of evolution operators that ``see'' the presence of a non-trivial operator $\hat{O}$ to be evaluated. Note that this algorithmic light cone is not to be confused with a physical light cone: if we send $\Delta t\rightarrow 0$, the algorithmic light cone becomes infinitely wide for any fixed time $t$. Physically, the Lieb-Robinson theorem states that beyond a ``light cone'' of width $x=2ct$, where $c$ is some problem-specific ``velocity'', correlations decay exponentially fast, as opposed to the hard cut imposed by a relativistic light cone. The physical light cone and the special decay of correlations is at the basis of very interesting algorithmic extensions of the MPS/tDMRG/TEBD algorithms of the last section by Hastings\cite{Hastings09,Hastings08}, which I will not pursue here.

While this structure becomes more complicated if we look, e.g.\ at $n$-point correlators, we may look at a huge algorithmic saving, even though we have to pay the price that for different locations of operators, new networks have to be considered. 

What is the prize for calculating at different times, e.g.\ $\langle \hat{O} (t_1) \rangle$, $\langle \hat{O} (t_2) \rangle$ and so on? This is a very natural question, as we might be interested in the time evolution of, say, some local density. If we do not use the light cone, then we simply calculate the contraction moving inwards, calculate some average, retrieve the stored result of the contraction up to the line with the operators, add more Trotter steps, contract, calculate some average, and so on. This is exactly what we have been doing all along. Of course, the light cone generated by the operator acting at time $t_2 > t_1$ is different from and larger than that generated by the operator acting at time $t_1$. But if the Hamiltonian is time-independent, the larger light cone contains the smaller one at its tip. It therefore makes numerical sense to reverse the time evolution, and work from the future towards the past. But this corresponds to nothing else but a switch to the Heisenberg picture.

\subsubsection{Heisenberg picture}
\begin{figure}
\centering\includegraphics[scale=0.5]{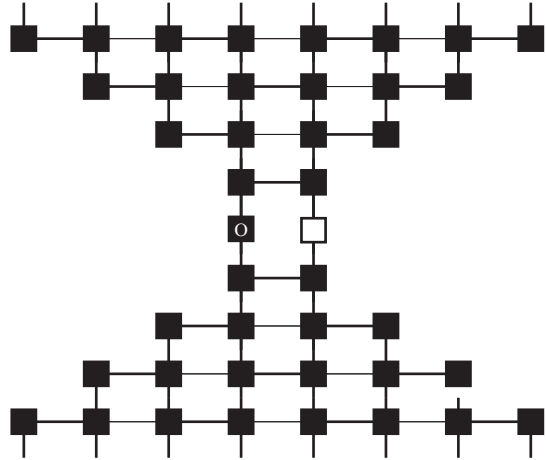}
\caption{Time evolution of an operator $\hat{O}$ in the Heisenberg picture, translated to the MPS/MPO representation. Each symmetric set of layers around the operator corresponds to one time step (more precisely, part of a Trotter time step). The light cone widens as a larger and larger section of the lattice is affected by $\hat{O}$. The identity operator (white square) is inserted for reasons of symmetry, but without explicit function.}
\label{fig:Heisenbergtime}
\end{figure}

Mathematically, the switch to the Heisenberg picture is nothing but a rebracketing:
\begin{equation}
\langle \hat{O}(t) \rangle = \bra{\psi(t)} \hat{O} \ket{\psi(t)} = \bra{\psi} \eul^{+\imag \hat{H} t} \hat{O} \eul^{-\imag \hat{H} t} \ket{\psi} = \bra{\psi} \hat{O}(t) \ket{\psi},
\end{equation}
where we have introduced the time-dependent operator
\begin{equation}
\hat{O}(t) =  \eul^{+\imag \hat{H} t} \hat{O} \eul^{-\imag \hat{H} t} .
\end{equation}
If we Trotterize the time evolutions present, we arrive exactly at the light cone structure of the last section, except that it has not been contracted yet with bra and ket; Fig.~\ref{fig:Heisenbergtime}.

This allows to set up time evolution in the Heisenberg picture \cite{Prosen09,Hartmann09}.Technically, one constructs a spatially growing MPO from MPO-MPO-multiplications as encountered e.g.\ in the calculation of the action of $\hat{H}^2$. If the current MPO of the time-evolved operator consists of local MPOs of the form $O^{\sigma_i \sigma'_i}_{a_{i-1},a_i}$ and bond dimension $D$ (on sites $i$ where it actually has to be considered), and if the time evolution MPO (for $\eul^{-\imag \hat{h} t}$) reads
$W^{\sigma_i \sigma'_i}_{b_{i-1},b_i}$ with bond dimensions $D_W$, then the operator reads after (part of) the time step
\begin{equation}
O^{\sigma_i, \sigma'_i}_{(b_{i-1},a_{i-1},c_{i-1}), (b_i,a_i,c_i)}
= \sum_{\sigma^{''}_i} W^{\sigma^{''}_i,\sigma_i *}_{b_{i-1},b_i} \left( \sum_{\sigma^{'''}_i}  O^{\sigma^{''}_i \sigma^{'''}_i}_{a_{i-1},a_i} W^{\sigma^{'''}_i,\sigma'_i}_{c_{i-1},c_i} \right) . 
\end{equation}
with bond dimensions $DD_W^2$. This operator is then compressed down to bond dimensions $D$ as explained earlier for MPOs, essentially using the method for compressing MPS.

This sets up a ``conventional'' time evolution: instead of a state, an operator in MPO form is subjected to time evolution by MPOs, and compressed to manageable matrix dimensions after each time step. We can basically recycle most of the algorithm. 

What are potential advantages and disadvantages of this formulation? First of all, the savings due to the algorithmic light cone are immediately incorporated. Second, we may hope that truncations are smaller: while the network contracted over is identical in the Heisenberg and Schr\"{o}dinger picture, {\em truncations} in the Schr\"{o}dinger picture do not take into account the operator and hence are less specific - one may enviseage that for ``simple'' operators like local density a lot of the fine structure of the state evolution is not really needed, and evolving the operator itself tells us which information is needed specifically for this operator. 

A corresponding disadvantage is of course that calculations need to be redone for different operators, which in the Schr\"{o}dinger picture may be evaluated whenever one likes, provided the time-evolved wave function is stored. Of course, here the corresponding advantage is that for different states one may evaluate whenever one likes, provided the time-evolved operator is stored. 

At the moment of writing it seems indeed that for simple operators the reachable time spans can be extended substantially, but I would find it hard to commit to some rule of thumb. 

\subsubsection{Transverse contraction and folding}

\begin{figure}
\centering\includegraphics[scale=0.5]{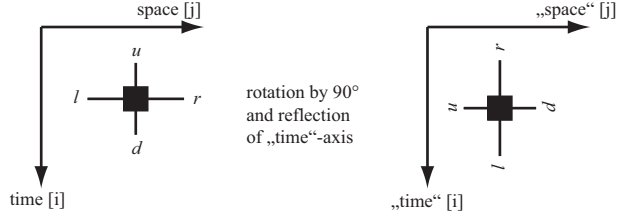}
\caption{Rotating the lattice for code reusage: Assuming a space-time labeling $[i,j]$ with time $i$ and space $j$ (which after rotation refers to ficticious ``time'' and ``space'') tensor indices $u$,$d$ and $l$,$r$ exchange places as shown in the figure.}
\label{fig:latticerotation}
\end{figure}

Of course, the iterative build up of the state as it evolves in time appeals to our intuition about the world, but there is nothing that prevents us to contract the same network in the spatial direction, i.e.\ column by column; as the order of contraction may influence efficiency quite strongly, maybe it helps. In order to recycle existing programs, one may simply rotate the current network by 90 degrees counterclockwise, and obtains a lattice of width $(2nN+3)$ and height $L$. If we continue to label tensors $T^{[i,j]}$ by the vertical before the horizontal and in the row-column logic of a matrix, then tensors in the new lattice read
\begin{equation}
\overline{T}^{[i,j] u,d}_{l,r} = T^{[L+1-j,i] r,l}_{u,d} ,
\end{equation}
as can be seen from Fig.~\ref{fig:latticerotation}. Then we can contract again line by line.

\begin{figure}
\centering\includegraphics[scale=0.5]{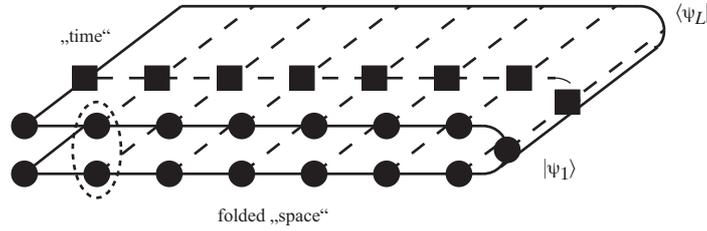}
\caption{The rotated lattice is now folded on the ``space'' (former time) axis which has $2nN+3 \equiv L^{'}$ sites. Sites corresponding to same times come to cover each other (indicated by an ellipse); the line on which operators are evaluated at final time remains single at the bend.}
\label{fig:latticefolding}
\end{figure}

As it turns out, a simple rotation (or transverse contraction) does not extend the reachable timescale. It is by an additional {\em folding} step that a strong extension of the timescale is possible \cite{Banuls09}. The folding happens parallel to the new ``time'' (i.e. real space) axis, and halves the extent of the new ``space'' (i.e. real time) domain (see Fig.~\ref{fig:latticefolding}). Instead of sites 1 through $2nN+3$ we then have double sites 1 through $nN+2$, where double site 1 comprises old sites 1 and $2nM+3$, double site 2 comprises old sites 2 and $2nN+2$; generally $i$ comprises old sites $i$ and $2nN+4-i$, up to the pair $(nN+1, nN+3)$. The site $nN+2$ is special: as we are folding an odd number of sites, one site remains single. This is site $nN+2$, which corresponds to the line that contained the operators to be evaluated at time $t$. On all the other sites, we fold tensors onto each other that correspond to ``identical'' timesteps, one forward and one backward in time.

The expectation is that this folding of forward and backward timesteps leads to cancellations in entanglement buildup, such that larger times can be reached (the growth in $D$ is not as fast).

To simplify notation, we define $L' = 2nN+3 \equiv 2\ell +1$.
If we read the bottom end of the folding as an MPS, the folded state also starts with an MPS whose matrices are formed as
\begin{equation}
M^{f[i]\Sigma_i}_{a^f_{i-1},a^f_i} = M^{f[i]\sigma_i, \sigma_{L'+1-i}}_{(a_{i-1},a_{L'+1-i}),(a_i,a_{L'-i})}
= \overline{T}^{[L,i]\sigma_i}_{a_{i-1},a_i} \overline{T}^{[L,L'+1-i]\sigma_{L'+1-i}}_{a_{L'-i},a_{L'+1-i}} 
\end{equation}  
for all $1 \leq i \leq \ell$ and 
\begin{equation}
M^{f[\ell+1]\Sigma_{\ell+1}}_{a^f_{\ell},1} = M^{f[\ell+1]\sigma_{\ell+1}}_{(a_{\ell},a_{\ell+1}),1} 
=\overline{T}^{[L,\ell+1]\sigma_{\ell+1}}_{a_{\ell},a_{\ell+1}}.
\end{equation}  
We have defined $d^2$ ``fat'' local states $\ket{\Sigma_i} = \ket{\sigma_i}\ket{\sigma_{L'+1-i}}$ on each site, except site $\ell+1$, where it remains of dimension $d$ (for programming, one may of course introduce a dummy site). Similarly, we construct the new tensors for the folded MPOs.

\begin{figure}
\centering\includegraphics[scale=0.5]{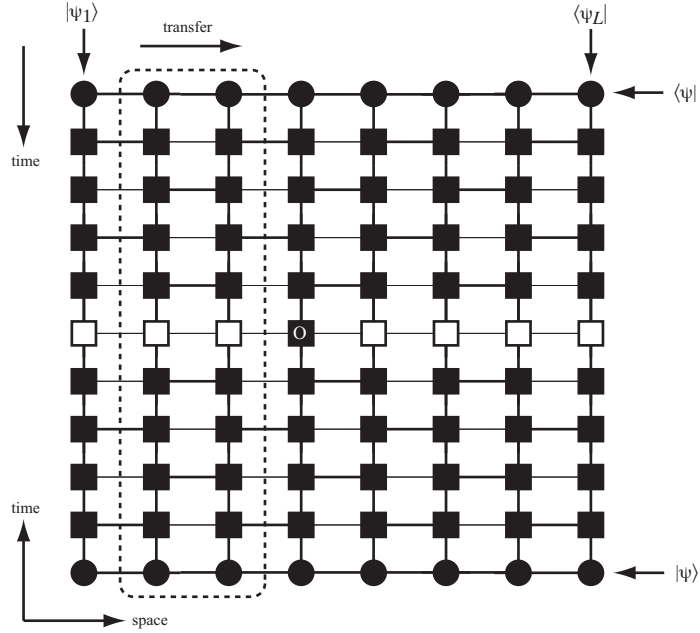}
\caption{Representation of the transfer operator $E$ (dashed rectangle; note direction of action) in the unfolded, unrotated representation of the time evolution network. It repeats throughout the lattice by spatial translation by two sites, except on the sites with evaluation of an operator, where it is modified accordingly.}
\label{fig:bigtransferoperator}
\end{figure}

If we assume that the original lattice and the Hamiltonian acting on it were translationally invariant, at least for translations by an even number of lattice sites, we can write the contractions conveniently using a transfer operator. If we call the state on the first line (first ``time'' slice) of the folded lattice $\ket{\psi_L}$ (corresponding to site $L$ of the original lattice) and the one on the bottom line (last ``time'' slice) $\bra{\psi_1}$, then ($i$ odd for simplicity)
\begin{equation}
\langle \hat{O}_i \rangle = \frac{\bra{\psi_L} E^{(i-3)/2} E_O E^{(L-i-1)/2} \ket{\psi_1}}{ \bra{\psi_L}  E^{(L-2)/2} \ket{\psi_1}} .
\end{equation}
Here, we have introduced the transfer operators $E$ and $E_O$ on stripes of length $\ell$ and width 2, as represented in Fig.~\ref{fig:bigtransferoperator} (in unfolded, unrotated form for simplicity of representation).  $E_O$ is derived from $E$ by inserting $\hat{O}$ instead of the identity at site $\ell+1$.

This can be evaluated by iterative contractions and compressions for spatially finite lattices, but one can also take the thermodynamic limit. Let us assume an eigenvector decomposition of $E$ as
\begin{equation}
E = \sum_{i} \lambda_i \ket{i} \bra{i} .
\end{equation}
Note that $E$ is not hermitian, hence $\ket{i}$ and $\bra{i}$ are not adjoint, but distinct right and left eigenvectors. From the biorthonormality of those, 
\begin{equation}
\lim_{L\rightarrow\infty} E^L = \lim_{L\rightarrow\infty} 
\sum_i \lambda_i^L \ket{i} \bra{i} = \lambda_0^L \ket{R}\bra{L},
\end{equation}
where I have assumed that the largest eigenvalue $\lambda_0$ is non-degenerate (which is usually the case) and changed notation to $\ket{R}$ and $\bra{L}$ for the associated right and left eigenvectors.

We then obtain in the thermodynamic limit as expectation value
\begin{equation}
\langle \hat{O}_i \rangle =  \frac{\bra{L} E_O \ket{R}}{\lambda_0} ,
\end{equation}
where $\lambda_0 = \bra{L} E \ket{R}$. 2-point correlators would then be given by
\begin{equation}
\langle \hat{O}_i \hat{P}_j \rangle =  \frac{\bra{L} E_O E^r E_P \ket{R}}{\lambda_0^{r+2}},
\end{equation}
where $r$ is the number of transfer operators between the sites $i$ and $j$. 

In order to evaluate such expressions, we obviously need $\lambda_0$, $\ket{R}$ and $\bra{L}$. As the components of $E$ are explicitly available, we can construct its transpose equally easily, hence reduce all to the determination of two right eigenvectors. As we are looking for the largest eigenvalue, the power method (iterative application of $E$ to a guess vector) will work. But one can equally read $E$ as the MPO representation of some non-hermitian operator, and reuse iterative ground state search techniques, with two modifications: the search for the lowest eigenvalue is replaced by the highest eigenvalue and the conventional Lanczos algorithm has to be replaced by non-hermitian methods, either the biorthogonal Lanczos algorithm or the Arnoldi method. 

While the coding is more involved than for standard time evolution, the timescales reachable are extended substantially, factors 3 to 5 seem easily possible.
  
\subsection{Linear prediction and spectral functions}
Spectral functions $S(k,\omega)$ are among the most important theoretical and experimental quantities in many-body physics. While there are very accurate ways of calculating them directly at $T=0$ \cite{Hallberg95,Ramasesha97,Kuhner99,Jeckelmann02}, there is also an indirect approach, pioneered in \cite{WhiteFeiguin04}, to calculate real-time real-space correlators like $\langle \hat{S}^+_i(t) \hat{S}^-_j(0) \rangle$, and to carry out a double Fourier transform to momentum and frequency space. This approach has the advantage to extend to finite $T$ seamlessly, but suffers from the limitations of reachable length and time scales.

Of these, the limitations in time are much more serious, because of the rapid growth of entanglement in time. The time scales reachable are mostly so limited that a naive Fourier transform gives strong aliasing or that one has to introduce a windowing of the raw data that smears out spectral information quite strongly.  This limitation can be circumvented however at very low numerical cost by a linear prediction technique both at $T=0$\cite{Pereira08,White08} and $T>0$\cite{Barthel09b} that extends reachable $t$ and thereby greatly refines results in the frequency domain.

For a time series of complex data $x_0, x_1, \ldots, x_n, \ldots, x_N$ at equidistant points in time $t_n=n\*\Delta t$ (and maximal time $t_\obs:= N \Delta t$) obtained by DMRG one makes a prediction of $x_{N+1}, x_{N+2}, \ldots$. 
For the data points beyond $t=t_\obs$, linear prediction makes the ansatz
\begin{equation}\label{eq:predictionAnsatz}
\tilde{x}_n = - \sum_{i=1}^p a_i x_{n-i} .
\end{equation}
The (predicted) value $\tilde{x}_n$ at time step $n$ is assumed to be a linear combination of $p$ previous values $\{x_{n-1},\dots,x_{n-p}\}$. Once the $a_i$ are determined from known data, they are used to calculate (an approximation of) all $x_n$ with $n>N$.

The coefficients $a_i$ are determined by minimizing the least square error in the predictions over a subinterval $t_n \in (t_\obs-t_\fit,t_\obs]$ of the known data (corresponding to a set $\Omega = \{ n\, |\, t_\obs - t_\fit < n\Delta t \leq t_\obs \}$), i.e.\ we minimize in the simplest approach $E\equiv\sum_{n\in\Omega} |\tilde x_n-x_n|^2$. $t_\fit=t_\obs/2$ is often a robust choice to have little short-time influence and enough data points. Minimization of $E$ with respect to $a_i$ yields the linear system
\begin{equation}
\label{eq:linPred:stationaryError}
R \vec{a} = -\vec{r},
\end{equation}
where $R$ and $\vec{r}$ are the autocorrelations 
$R_{ji} = \sum_{n\in\Omega} x^*_{n-j} x_{n-i}$ and $r_{j} = \sum_{n\in\Omega} x^*_{n-j} x_n$.
Eq.~(\ref{eq:linPred:stationaryError}) is solved by $\vec{a}=-R^{-1}\vec{r}$.

One may wonder why extrapolation towards infinite time is possible in this fashion. 
As demonstrated below, linear prediction generates a superposition of oscillating and exponentially decaying (or growing) terms, a type of time-dependence that emerges naturally in many-body physics:
Green's functions of the typical form $G(k,\omega) = (\omega - \epsilon_k - \Sigma(k,\omega))^{-1}$ are in time-momentum representation dominated by the poles; e.g.\ for a single simple pole at $\omega=\omega_1 -\imag \eta_1$ with residue $c_1$, Green's function will read
$G(k,t) = c_1 \eul^{-\imag \omega_1 t - \eta_1 t}$, and similarly it will be a superposition of such terms for more complicated pole structures. Often only few poles matter, and the ansatz of the linear prediction is well suited for the typical properties of the response quantities we are interested in. Where such an ansatz does not hold, the method is probably inadequate.

To see the special form of time-series generated by the prediction, 
we introduce vectors $\vec{x}_n := [x_{n}, \ldots, x_{n-p+1}]^T$ such that (\ref{eq:predictionAnsatz}) takes the form
\begin{equation}
\tilde{\vec{x}}_{n+1} = A \* \vec{x}_n,	
\end{equation}
 with 
\begin{equation}\label{eq:predictionMatrix}
A\equiv
\left[
\begin{array}{ccccc}
 -a_1&-a_2&-a_3&\cdots&-a_p\\
  1  & 0  & 0  &\cdots& 0  \\
  0  & 1  & 0  &\cdots& 0  \\
\vdots&\ddots&\ddots&\ddots&\vdots\\
  0   &\cdots& 0  & 1  & 0
\end{array}  
\right] ,
\end{equation}
with the $a_i$ as the elements of the vector $\vec{a}$ found above.
Prediction therefore corresponds to applying powers of $A$ to the initial vector $\vec{x}_N$. A (non-hermitian) eigenvector decomposition of $A$ with eigenvalues $\alpha_i$ leads to 
\begin{equation}
\tilde x_{N+m} = [A^m \* \vec{x}_N]_1=\sum_{i=1}^p c_i \alpha_i^m,
\end{equation}
where coefficients $c_i$ are determined from $\vec{x}_N$ and the eigenvectors of $A$. The eigenvalues $\alpha_i$ encode the physical resonance frequencies and dampings. The connection is given as
$\alpha_i = \eul^{\imag  \omega_i \Delta t - \eta_i \Delta t}$.
Spurious $|\alpha_i|\geq 1$ may appear, but can be dealt with\cite{Barthel09b}.

At $T=0$, critical one-dimensional systems exhibit power-law decays in their time-dependent correlators. The superposition of exponential decays is then taken to mimic these power-laws \cite{Pereira08}. At finite temperatures, time-dependent correlators $S(k,t)$ decay typically exponentially for large times (due to thermal broadening), making linear prediction especially well-suited for this situation. This is also close to typical experimental situations, like inelastic neutron scattering off one-dimensional magnetic chains.

\begin{figure}
\centering\includegraphics[width=250pt]{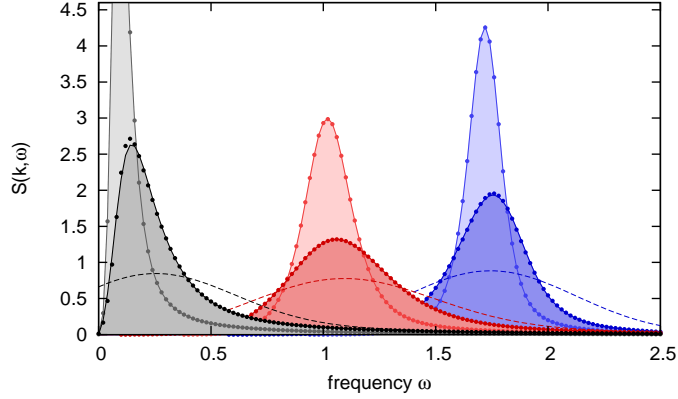}
\caption{Lines and dots represent exact analytical and numerical solutions for the lineshape of the spectral function $S^{+-}(k,\omega)$ of an $XY$-chain at temperatures $\beta=10$ and $\beta=50$ (broad and narrow lineshapes) for (from the left) $k=\pi/8$, $k=\pi/3 $, $k=3\pi/4 $. The dashed lines are the shapes that would optimally be extracted from the $\beta=10$ simulation without prediction using some windowing of the raw data before Fourier transformation. Adapted from \cite{Barthel09b}.}
\label{fig:xychain}
\end{figure}

As example, let us consider a field-free Heisenberg antiferromagnet with $J^z=0$ ($XY$-chain) and $J^z=1$. The former case allows for an exact analytical solution. It turns out that prediction allows to extend time series $S(k,t)$ by over an order of magnitude without appreciable loss of precision. In frequency  space, this corresponds to extremely high-precision spectral lineshapes (Figure \ref{fig:xychain}). 

\begin{figure}
\centering\includegraphics[height=200pt]{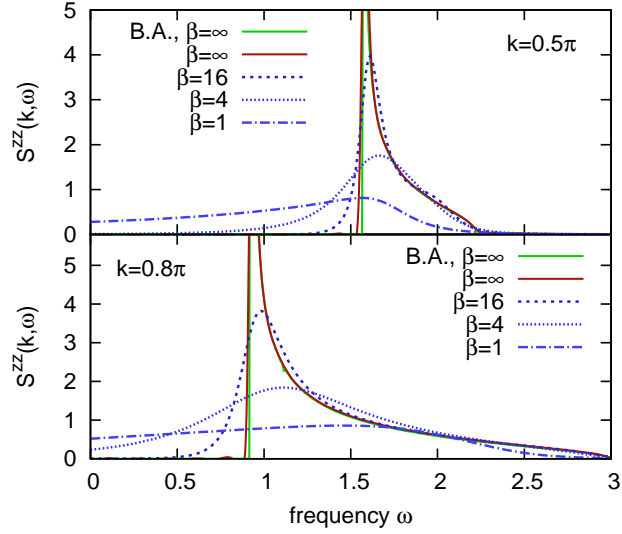}
\caption{Spectral function of the isotropic Heisenberg chain at two momenta, and four different temperatures. At $T=0$, Bethe ansatz (B.A.) and numerics agree extremely well. Adapted from \cite{Barthel09b}.}
\label{fig:USHAFM}
\end{figure}

As the dispersion relation of the $XY$-chain is just a simple magnon line, its self-energy structure is very simple, hence the prediction method easily applicable. As a more demanding example, we consider the spinon continuum of an isotropic $S=1/2$ chain; Fig.~\ref{fig:USHAFM}. In the zero-temperature limit, results agree extremely well with Bethe-ansatz results (where remaining differences are hard to attribute: the Bethe ansatz here can only be evaluated approximately \cite{Caux06}). At finite temperatures, simulations at different precision indicate that results are fully converged and essentially exact. This lets us expect that this method will be a powerful tool in e.g. simulating the results of neutron scattering experiments.

\subsection{Minimally entangled typical thermal states}
Simulating thermal density operators for the calculation of static and dynamic properties works very well. Theoretically, there are no limits to this method. In practice, one encounters several limitations. On the one hand, simulations become difficult at very low temperatures $T\rightarrow 0$. In this limit, the mixed state living on the physical system P will evolve towards the pure state projector on the ground state, $\hat{\rho}^P_\infty = \ket{\psi_\infty}_P \phantom{\,}_P\bra{\psi_\infty}$ (here I use $\ket{\psi_\infty}_P$ as the ground state of the physical Hamiltonian $\hat{H}$, to refer to $\beta=\infty$ as in the purification section). But in this limit P is not entangled with the auxiliary system Q anymore, and we simulate a product of two pure states: assume for simplicity that the ground state energy is set to 0. Consider now the reduced density operator for the auxiliary system. Up to an irrelevant norm, 
\begin{equation}
\hat{\rho}^Q_{\beta\rightarrow\infty} = \lim_{\beta\rightarrow\infty} \tr_P \eul^{-\beta\hat{H}/2} \ket{\psi_0} \bra{\psi_0} \eul^{-\beta\hat{H}/2} = \phantom{\,}_P\braket{\psi_\infty}{\psi_0} \braket{\psi_0}{\psi_\infty}_P ,
\end{equation}
because in the limit $\beta\rightarrow\infty$ the trace reduces to the ground state contribution,
$\tr_P \rightarrow \phantom{\,}_P\bra{\psi_\infty} \cdot \ket{\psi_\infty}_P$. With the $\beta=0$ purification of the density operator, $\ket{\psi_0} = d^{-L/2} \sum_{\fat{\sigma}} \ket{\fat{\sigma}}_P \ket{\fat{\sigma}}_Q$, the last expression reduces, again up to an irrelevant norm, to
\begin{equation}
\hat{\rho}^Q_{\beta\rightarrow\infty} = \sum_{\fat{\sigma}\fat{\sigma}'} \psi^{\fat{\sigma}*}_\infty \ket{\fat{\sigma}}_Q \phantom{\,}_Q \bra{\fat{\sigma}'} \psi^{\fat{\sigma}'}_\infty =
\ket{\psi_\infty}_Q \phantom{\,}_Q \bra{\psi_\infty} ,
\end{equation}
where the $\psi^{\fat{\sigma}}_\infty$ are the expansion coefficients of the ground state. Hence, the zero temperature purification is a product state 
\begin{equation}
\ket{\psi_\infty} = \ket{\psi_\infty}_P \ket{\psi_\infty}_Q,
\end{equation}
where the latter state is just the physical ground state defined on the auxiliary state space.
Assuming that it can be described with sufficient precision using matrix dimension $D$, the product will be described by matrices of dimension $D^2$. Effectively this means that our algorithm scales with the sixth instead of the third power of the characteristic matrix dimension for the problem under study. On the other hand, and this is the issue prediction has tried to address, we encounter long-time simulation problems in particular at high temperatures $T\rightarrow\infty$: many states contribute at similar, but not identical, weight and MPS are not efficient at encoding this wealth of contributing states.

As White \cite{Stoudenmire10,White09} has pointed out, one can avoid the purification approach entirely by sampling over a cleverly chosen set of thermal states, the so-called {\em minimally entangled typical thermal states} (METTS). This approach has already been shown to alleviate strongly the first limitation, while not much is known yet about the second limitation. 

A thermal average is given by
\begin{equation}
\langle \hat{A} \rangle = \frac{1}{Z} \tr \eul^{-\beta \hat{H}} \hat{A} = \frac{1}{Z} \sum_n \eul^{-\beta E_n} \bra{n} \hat{A} \ket{n} ,
\end{equation}
where I have chosen, like all textbooks do, the energy representation of the thermal density operator. As already pointed out by Schr\"{o}dinger many decades ago, this is mathematically correct, but unphysical in the sense that real systems at finite temperature will usually not be in a statistical mixtures of eigenstates, as eigenstates are highly fragile under coupling to an environment. But the choice of the basis for taking the trace is arbitrary, and one may also write
\begin{equation}
\langle A \rangle = \frac{1}{Z} \sum_i \bra{i} \eul^{-\beta\hat{H}/2} \hat{A} \eul^{-\beta\hat{H}/2} \ket{i}
= \frac{1}{Z} \sum_i P(i) \bra{\phi(i)} \hat{A} \ket{\phi(i)}
\end{equation}
where $\{ \ket{i} \}$ is an arbitrary orthonormal basis and $\ket{\phi(i)} = P(i)^{-1/2} \eul^{-\beta\hat{H}/2} \ket{i}$. With $P(i) = \bra{i} \eul^{-\beta \hat{H}} \ket{i}$, we recognize $\ket{\phi_{i}}$ to be normalized. It is easy to see that $\sum_i P(i) = Z$, hence the $P(i)/Z$ are probabilities. One can therefore statistically estimate $\langle \hat{A}\rangle$ by {\em sampling} $\ket{\phi(i)}$ with probabilities $P(i)/Z$ and average over $\bra{\phi(i)} \hat{A} \ket{\phi(i)}$.

Several questions arise before this can be turned into a practical algorithm. How can we sample correctly given that we do not know the complicated probability distribution? Can we choose a set of states such that averages converge most rapidly? Given that an imaginary time evolution will be part of the algorithm, can we find a low-entanglement basis $\{ \ket{i} \}$ such that time evolution can be done with modest $D$, i.e.\ runs fast?

To address these issues, White chooses as orthonormal basis the computational basis formed from product states,
\begin{equation}
\ket{i} = \ket{i_1}\ket{i_2}\ket{i_3} \ldots \ket{i_L} ,
\end{equation}
classical (unentangled) product states (CPS) that can be represented exactly with $D=1$. The corresponding states
\begin{equation}
\ket{\phi(i)} = \frac{1}{\sqrt{P(i)}} \eul^{-\beta \hat{H}/2} \ket{i}
\end{equation}
are so-called {\em minimally entangled typical thermal states} (METTS): The hope is that while the imaginary time evolution introduces entanglement due to the action of the Hamiltonian, it is a reasonable expectation that the final entanglement will be lower than for similar evolutions of already entangled states. While this is not totally true in a strict mathematical sense, in a practical sense it seems to be! Compared to purification, this will be a much faster computation, in particular as the factorization issue of purification will not appear.

In order to sample the $\ket{\phi(i)}$ with the correct probability distribution, which we cannot calculate, one uses the same trick as in Monte Carlo and generates a Markov chain of states,
$\ket{i_1} \rightarrow \ket{i_2} \rightarrow \ket{i_3} \rightarrow \ldots$ such that the correct probability distribution is reproduced. From this distribution, we can generate $\ket{\phi(i_1)}$, $\ket{\phi(i_2)}$, $\ket{\phi(i_3)}$, $\ldots$ for calculating the average.

The algorithm runs as follows: we start with a random CPS $\ket{i}$. From this, we repeat the following three steps until the statistics of the result is good enough:
\begin{itemize}  
\item Calculate $\eul^{-\beta\hat{H}/2} \ket{i}$ by imaginary time evolution and normalize the state (the squared norm is $P(i)$, but we won't need it in the algorithm). 
\item Evaluate desired quantities as $\bra{\phi(i)} \hat{A} \ket{\phi(i)}$ for averaging.
\item Collapse the state $\ket{\phi(i)}$ to a new CPS $\ket{i'}$ by quantum measurements with probability $p(i\rightarrow i') = |\braket{i'}{\phi(i)}|^2$, and restart with this new state.
\end{itemize}

Let us convince ourselves that this gives the correct sampling, following \cite{Stoudenmire10}. As the $\ket{\phi(i)}$ follow the same distribution as the $\ket{i}$, one only has to show that the latter are sampled correctly. Asking with which probability one collapes into some $\ket{j}$ provided the previous CPS $\ket{i}$ was chosen with the right probability $P(i)/Z$, one finds 
\begin{equation}
\sum_i \frac{P(i)}{Z} p(i\rightarrow j) = \sum_i \frac{P(i)}{Z} | \braket{j}{\phi(i)} |^2 = \sum_i
\frac{1}{Z} | \bra{j} \eul^{-\beta \hat{H}/2} \ket{i}|^2 = \frac{1}{Z} \bra{j} \eul^{-\beta\hat{H}} \ket{j} = \frac{P(j)}{Z}.
\end{equation}
This shows that the desired distribution is a fixpoint of the update procedure. It is therefore valid, but it is of course sensible to discard, as in Monte Carlo, a number of early data points, to eliminate the bias due to the initial CPS.  It turns out that - after discarding the first few METTS, to eliminate effects of the initial choice - averaging quantities over only a hundred or so allows to calculate local static quantities (magnetizations, bond energies) with high accuracy.

While we already know how to do an imaginary time evolution, we still have to discuss the collapse procedure. As it turns out, the structure of MPS can be exploited to make this part of the algorithm extremely fast compared to the imaginary time evolution.

For each site $i$, we choose an {\em arbitrary} $d$-dimensional orthonormal basis $\{ \ket{\tilde{\sigma}_i} \}$, to be distinguished from the computational basis $\{ \ket{\sigma_i} \}$. From this we can form projectors $\hat{P}^{\tilde{\sigma}_i} = \ket{\tilde{\sigma}_i} \bra{\tilde{\sigma}_i}$ with the standard quantum mechanical probability of a local collapse into state $\ket{\tilde{\sigma}_i}$ given by
$p_{\tilde{\sigma}_i} =\bra{\psi} \hat{P}^{\tilde{\sigma}_i} \ket{\psi}$. If we collapse $\ket{\psi}$ into the CPS
$\ket{\psi'} = \ket{ \tilde{\sigma}_1} \ket{ \tilde{\sigma}_2} \ldots \ket{ \tilde{\sigma}_L}$, the probability is given by  $\bra{\psi} \hat{P}^{\tilde{\sigma}_1} \ldots \hat{P}^{\tilde{\sigma}_L} \ket{\psi} = | \braket{\psi'}{\psi} |^2$, as demanded by the algorithm. After a single-site collapse, the wave function reads
\begin{equation}
\ket{\psi} \rightarrow  p_{\tilde{\sigma}_i}^{-1/2} \hat{P}^{\tilde{\sigma}_i}\ket{\psi} ,
\end{equation}
where the prefactor ensures proper normalization of the collapsed state as in elementary quantum mechanics. To give an example, for $S=\frac{1}{2}$ spins measured along an arbitrary axis ${\mathbf n}$, the projectors would read
\begin{equation}
\hat{P}^{\uparrow_n,\downarrow_n} = \frac{1}{2} \pm {\mathbf n} \cdot \hat{{\mathbf S}}_i .
\end{equation}

Such a sequence of local measurements and collapses on all sites can be done very efficiently, as pointed out by \cite{Stoudenmire10}, if one exploits two features of MPS and CPS: (i) local expectation values can be evaluated very efficiently if they are on explicit sites (in DMRG language) or on sites between left- and right-normalized sites of a mixed-canonical state (in MPS language) and (ii) after the collapse, the resulting state is a product state of local states on all collapsed sites and the uncollapsed remainder.  

Assume that $\ket{\psi}$ is right-canonical (with the relaxation that the normalization on site 1 is irrelevant). Then the evaluation of $p_{\tilde{\sigma}_1} =\bra{\psi} \hat{P}^{\tilde{\sigma}_1} \ket{\psi}$ trivializes because the contraction of the expectation value network over sites 2 through $L$ just yields $\delta_{a_1,a'_1}$. Hence
\begin{equation}
\bra{\psi} \hat{P}^{\tilde{\sigma}_i} \ket{\psi} = \sum_{a_1,\sigma_1,\sigma'_1} B^{\sigma_1 *}_{a_1} \bra{\sigma_1} \hat{P}^{\tilde{\sigma}_1} \ket{\sigma'_1} B^{\sigma'_1}_{a_1} = 
\sum_{a_1} \left[ \sum_{\sigma_1} B^{\sigma_1*}_{a_1} \braket{\sigma_1}{\tilde{\sigma}_1} \right]
\left[ \sum_{\sigma'_1} B^{\sigma'_1}_{a_1} \braket{\tilde{\sigma}_1}{\sigma'_1} \right] .
\label{eq:probsinmetts}
\end{equation}
This expression looks very specific to the first site (because of the open boundary), but as we will see it is not! 

Once the probabilites for the collapse on site 1 are calculated, one particular collapse is chosen randomly according to the distribution just generated, $\ket{\tilde{\sigma}_1}$. The state after collapse will be of the form $\ket{\tilde{\sigma}_1} \ket{\psi_{{\rm rest}}}$, hence a product state. Therefore, the new matrices (which we call $A^{\sigma_1}$) on site 1 must all be scalars, i.e.\ $D=1$ matrices. From
$\ket{\tilde{\sigma}} = \sum_{\sigma} \ket{\sigma} \braket{\sigma}{\tilde{\sigma}} = \sum_\sigma \ket{\sigma} A^{\sigma}$ they are given by
\begin{equation}
A^{\sigma_1}_{1,1} = \braket{\sigma_1}{\tilde{\sigma}_1} ;
\end{equation}
it is easy to see that left-normalization is trivially ensured, hence the labelling by $A$. But this change in the dimension of $A^{\sigma_1}$ means that $B^{\sigma_2}$ has to be changed too, namely
\begin{equation}
B^{\sigma_2}_{a_1,a_2} \rightarrow M^{\sigma_2}_{\tilde{\sigma}_1,a_2} =
p_{\tilde{\sigma}_1}^{-1/2} \sum_{\sigma_1,a_1} \braket{\tilde{\sigma}_1}{\sigma_1} B^{\sigma_1}_{a_1} B^{\sigma_2}_{a_1 a_2} .
\end{equation}
As the label $\tilde{\sigma}_1$ takes a definite value, it is just a dummy index, and the row dimension of $M^{\sigma_2}$ is just 1, like for the matrices on the first site. Hence, Eq.~(\ref{eq:probsinmetts}) generalizes to all sites, and the most costly step is the update of $B^{\sigma_2}$, which scales as $D^2 d^2$, but not as $D^3$, as time evolution does.

To see the substitution, we express $\hat{P}^{\tilde{\sigma}_1}$ as an MPO, $W^{\sigma_1 \sigma'_1} = \braket{\sigma_1}{\tilde{\sigma}_1} \braket{\tilde{\sigma}_1}{\sigma'_1}$. Hence, the collapsed $\ket{\psi}$ reads
\begin{eqnarray*}
& & p_{\tilde{\sigma}_1}^{-1/2} \sum_{\fat{\sigma}} \sum_{\sigma'_1} \braket{\sigma_1}{\tilde{\sigma}_1} \braket{\tilde{\sigma}_1}{\sigma'_1} B^{\sigma'_1} B^{\sigma_2} \ldots \ket{\fat{\sigma}} = \\
& & \sum_{\fat{\sigma}} \sum_{a_2} A^{\sigma_1}_{1,1} \left( \sum_{\sigma'_1 a_1}  p_{\tilde{\sigma}_1}^{-1/2}  \braket{\tilde{\sigma}_1}{\sigma'_1} B^{\sigma'_1}_{1,a_1} B^{\sigma_2}_{a_1,a_2} \right) (B^{\sigma_3} \ldots B^{\sigma_L})_{a_2,1} \ket{\fat{\sigma}} ,
\end{eqnarray*}
which yields the substitution.

A few more comments are in order. At each site, the measurement basis can be chosen randomly, and in order to obtain short autocorrelation ``times'' of the Markov chain, i.e.\ high quality of the sampling, this is certainly excellent, but also much more costly than collapsing always into the same basis, which however generates ergodicity problems. The proposal is to switch alternatingly between two bases where for each basis projectors are maximally mixed in the other basis (e.g.\ if we measure spins alternatingly along the $x$- and $z$- (or $y$-)axis). Autocorrelation times then may go down to 5 steps or so \cite{Stoudenmire10}. For estimating the statistical error, in the simplest cases it is enough to calculate averages over bins larger than the autocorrelation time, and to look at the statistical distribution of these bin averages to get an error bar. 

Intriguing questions remain, concerning both the potential and the foundation of the algorithm: how well will it perform for longer-ranged correlators, as needed for structure functions? Dynamical quantities can be accessed easily, as the time-evolution of the weakly entangled METTS is not costly - but will the efficiency of averaging over only a few ``typical" states continue to hold? 
\subsection{Dissipative dynamics: quantum jumps}
Dissipative (i.e.\ non-Hamiltonian) dynamics occurs when our physical system A is coupling to some environment B such that A is an {\em open} quantum system. This is a very rich field of physics, so let me review a few core results useful here. The time evolution of the density operator of the system can always be written in the Kraus representation as
\begin{equation}
\hat{\rho}_A (t) = \sum_j \hat{E}_A^j(t) \hat{\rho}_A (0) \hat{E}_A^{j\dagger}(t),
\end{equation}
where the Kraus operators meet the condition 
\begin{equation}
\sum_j \hat{E}_A^{j\dagger}(t) \hat{E}_A^j(t) = \hat{I}_A .
\end{equation}
If the dynamics is without memory (Markovian), it depends only on the density operator at an infinitesimally earlier time, and a master equation, the Lindblad equation, can be derived. In the limit $\diff t\rightarrow 0$, the environment remains unchanged with probability $p_0 \rightarrow 1$ and changes (quantum jumps) with a probability linear in $\diff t$. If we associate Kraus operator $\hat{E}_A^0(t)$ with the absence of change, a meaningful ansatz scaling out time is
\begin{equation}
\hat{E}_A^0(\diff t) = \hat{I}_A + O(\diff t) \quad\quad \hat{E}_A^j(\diff t) = \sqrt{\diff t} \hat{L}_A^j  \quad (j>0)
\end{equation}
or more precisely
\begin{equation}
\hat{E}_A^0(\diff t) = \hat{I}_A + (\hat{K}_A - \imag \hat{H}_A) \diff t ,
\end{equation}
with two Hermitian operators $\hat{K}_A$ and $\hat{H}_A$. The normalization condition of the Kraus operators entails 
\begin{equation}
\hat{K}_A = -\frac{1}{2} \sum_{j>0} \hat{L}_A^{j\dagger} \hat{L}_A^j .
\end{equation}
These ansatzes allow to derive a differential equation from the Kraus evolution formula, which is the Lindblad equation
\begin{equation}
\frac{\diff \hat{\rho}}{\diff t} = - \imag [\hat{H},\hat{\rho}] + \sum_{j>0} \left( \hat{L}^j \hat{\rho} \hat{L}^{j\dagger} - \frac{1}{2} \{ \hat{L}^{j\dagger} \hat{L}^j, \hat{\rho} \} \right) ,
\end{equation} 
where I have dropped the indices A. Indeed, in the absence of  quantum jumps ($j=0$ only), one recovers the von Neumann equation. At the price of non-hermiticity, this equation can be simplified. If we introduce $\hat{H}_{{\rm eff}} := \hat{H} + \imag \hat{K}$, then the last term disappears and we have
\begin{equation}
\frac{\diff \hat{\rho}}{\diff t} = - \imag [\hat{H}_{{\rm eff}}\hat{\rho}-\hat{\rho}\hat{H}_{{\rm eff}}^\dagger] + \sum_{j>0} \hat{L}^j \hat{\rho} \hat{L}^{j\dagger} .
\label{eq:Lindblad2}
\end{equation} 

The simulation of Lindblad equations is possible quite easily in the MPS formalism, in particular using MPOs \cite{VerstraeteRipoll04}, but also in the form of a superoperator formalism \cite{VidalZwolak04}. The problem with this approach is that it is numerically more costly compared to the Hamiltonian evolution of a state. A very attractive alternative, which allows maximal reusage of available pure state codes, has been proposed by \cite{Daley09}, which combines pure state time evolution with the method of quantum trajectories. 

The method of quantum trajectories has been widely applied in quantum optics\cite{Plenio98}. Instead of using the Lindblad equation directly (which takes into account both the probabilistic distribution of initial states through $\hat{\rho}(0)$ and all possible sequences of quantum jumps), the quantum trajectory approach samples over the distribution of initial states, and for each of this sample states carries out a pure state time evolution where random quantum jumps occur at random times. They are chosen such that if one averages physical quantities over this distribution of time-evolving states, the result of the Lindblad equation is recovered. Let us ignore the sampling over initial states, assume that it is always the same, and instead focus on the conceptually more difficult averaging over quantum jumps.

The algorithm then proceeds by generating ${\mathcal N}$ quantum trajectories in a time interval $[0,T]$ (where $T$ is the final time of the simulation) as follows: 
\begin{itemize}
\item Generate a starting state $\ket{\psi(0)}$; it either samples the $t=0$ density operator correctly or is simply always the same, depending on the physical problem.
\item Choose a uniformly distributed random number $p_0$ in $[0,1]$.
\item Carry out, using one of our pure state time evolution methods, the time evolution of $\ket{\psi(0)}$ under $\hat{H}_{{\rm eff}}$. As the effective Hamiltonian is non-hermitian, the norm of the state will decrease over time. Stop the time evolution at time $t_1$, which is defined by 
$\braket{\psi(t_1)}{\psi(t_1)} = p_0$; this is the time of the first quantum jump. Note that if $T<t_1$, our simulation stops at $T$ and we have a trajectory without jump, and we normalize the final state.
\item To carry out the quantum jump at $t_1$, we calculate
\begin{equation}
\tilde{p}_j = \bra{\psi(t_1)} \hat{L}^{j\dagger} \hat{L}^j \ket{\psi(t_1)} \quad\quad p_j = \frac{\tilde{p}_j}{\sum_{j>0} \tilde{p}_j} \quad (j>0)
\end{equation}
and choose a $j$ according to the normalized probability distribution $\{ p_j \}$.
\item We carry out this jump and normalize the state,
\begin{equation}
\ket{\psi(t_1^+)} = \frac{\hat{L}^j \ket{\psi(t_1)}}{\| \hat{L}^j \ket{\psi(t_1)} \|} .
\end{equation}
\item After this, we continue with finding a new $p_0$, from which time evolution of $\ket{\psi(t_1^+)}$ with $\hat{H}_{{\rm eff}}$ generates $t_2$, the location of the second quantum jump, and so on, until $T$ is exceeded.
\end{itemize}
Physical quantities up to time $T$ are now averaged over the ${\mathcal N}$ quantum trajectories that have been generated. The correct probabilities are produced if all states are normalized at all times; as this is not the case in the algorithm, norms at say time $t$ have to be taken into account. Obviously, a careful analysis of convergence in ${\mathcal N} \rightarrow \infty$ has to be carried out, but it seems that for a small number of jump operators, even a few 100 trajectories may give highly reliable results \cite{Daley09}.

The observation that this sampling reproduces the dynamics of the Lindblad equation is part of the standard literature on quantum trajactories. The proof can be done in two steps, which I just sketch here. In a first step, one considers fixed time steps $\diff t$, and calculates probabilities for no jump vs. jump $j$ in this time interval ($p_j = \diff t \bra{\psi(t)} \hat{L}^{j\dagger} \hat{L}^j \ket{\psi(t)}$, $p_0 = 1 - \sum_j p_j$). One then either time-evolves under $\hat{H}_{{\rm eff}}$ over $\diff t$ and normalizes, or does the jump and normalizes, according to the generated distribution. One can show that this reproduces the Lindblad equation. In a second step, one shows that the distributions of quantum jumps generated in this way and the one we use in the algorithm are identical.

\section{DMRG and NRG}
\subsection{Wilson's numerical renormalization group (NRG) and MPS}
Wilson's Numerical Renormalization Group (NRG) \cite{Wilson75,Bulla08,Krishnamurthy80} originates in attempts to explain why metals with a small concentration of magnetic impurities exhibit a non-monotonic behaviour of resistivity.
It was found that an adequate minimal model is provided by 
\begin{equation}
\hat{H}_A = \sum_{k\sigma} \epsilon_k \hat{c}^\dagger_{k\sigma} \hat{c}^{\phantom\dagger}_{k\sigma} + \sum_{k\sigma} V_k (\hat{f}^\dagger_\sigma \hat{c}^{\phantom\dagger}_{k\sigma} + {\rm h.c.}) + U (\hat{n}_{f\uparrow}-1/2) (\hat{n}_{f\downarrow}-1/2).
\end{equation}
This single-impurity Anderson model contains an impurity site that can be occupied by up to two electrons (operators $\hat{f}^\dagger_\sigma$) with on-site repulsion $U$ and which couples to a conduction band (operators $\hat{c}^\dagger_{k\sigma}$) with energy dispersion $\epsilon_k$ through some hybridization function $V_k$.

In order to make it tractable, one changes from momentum to energy representation, assuming that only low-energy isotropic $s$-wave scattering matters, and introduces logarithmic discretization: the band is represented by band segments of an energy width that decreases exponentially close to the Fermi energy $\epsilon_F$. This accounts for the observation that the decisive feature of quantum impurity physics, namely the appearance of a very narrow resonance peak at the Fermi energy in the local impurity spectral function, is linked exponentially strongly to the states close to the Fermi energy. Logarithmic discretization is however also required to make NRG work at all on a technical level!

After further manipulations, for which I refer to \cite{Wilson75,Bulla08}, the Anderson Hamiltonian is finally mapped to a semi-infinite chain of non-interacting sites with the exception of the first one:
\begin{equation} 
\hat{H}= U (\hat{n}_{f\uparrow}-1/2) (\hat{n}_{f\downarrow}-1/2) + t_{-1} \sum_\sigma (\hat{f}^\dagger_\sigma \hat{d}_{0\sigma} + {\rm h.c.}) + \sum_{\sigma,n=0}^\infty t_n (\hat{d}^\dagger_{n\sigma} \hat{d}_{n+1,\sigma} + {\rm h.c.}),
\end{equation}
where the $\hat{d}_{\sigma}$ are fermionic operators.  The crucial point is that the $t_n$ decay exponentially, $t_n \sim \Lambda^{-n}$, where $\Lambda$ is the shrinkage factor of the energy bands in the logarithmic discretization, usually a value of the order 1.5 to 2. This is obviously a model that is amenable to our methods, e.g. a ground state search -- as the hoppings decay exponentially, we will not have to consider a truly infinite chain. 

NRG builds on the observation that the exponential decay leads to a separation of energy scales: assuming we know the spectrum of the partial chain up to some length, all remaining sites will only make exponentially small corrections to it because of the exponentially small energy scales further down the chain. Finding the ground state (and more generally the low lying spectrum) is now achieved by  iterative exact diagonalization: assume that we have an effective $D$-dimensional eigenspace for some left-end part of the chain. Then the next-larger chain has state space dimension $dD=4D$; in order to avoid exponential growth, we have to truncate down to $D$ states. The NRG prescription is to diagonalize that system and to retain the $D$ lowest-lying eigenstates. Starting out from very short chains that can still be done exactly, this procedure resolves the lowest-lying states exponentially well and is justified by the separation of energy scales: the decision which states to retain at some step would not be drastically changed with hindsight, as all further sites in the chain interact at much smaller energies. The obtained eigenspectra at different energy scales (chain lengths) can then be used to extract RG flow information or calculate thermodynamic or dynamic quantities for the impurity problem. 

Given that the building block $A^\sigma$ of an MPS can be interpreted as encoding a decimation step upon growing a block by one site, irrespective of the decimation prescription, it is immediately obvious that NRG, like DMRG, can be seen as operating on MPS\cite{Weichselbaum09}. This closes a historical loop as in fact the analysis of failures of NRG  naively applied to Heisenberg and Hubbard models gave rise to the development of DMRG. A NRG  state would look like
\begin{equation}
\ket{a_\ell} = \sum_{\sigma_1,\ldots,\sigma_\ell} (A^{\sigma_1} \ldots  A^{\sigma_\ell})_{a_\ell} \ket{\sigma_1\ldots\sigma_\ell}.
\end{equation}
At each length $\ell$, we get a spectrum of $D$ states.

Given that DMRG is variational over the MPS ansatz space, it is reasonable to expect that at least some improvement must be possible over the NRG method. In fact this is the case \cite{Weichselbaum09}; in the next section, I am going to discuss some improvements which are already firmly established and others which are more speculative, i.e.\ where benchmarking on relevant complex problems is still lacking. 

\subsection{Going beyond the numerical renormalization group}
In fact, considering an MPS formulation of NRG helps even without resorting to the connection to variational methods like DMRG, as exemplified by the strict enforcement of certain sum rules \cite{Weichselbaum07,Peters06}, but this is outside the topic of this review paper. 

What we can do more, however, is to subject the final MPS construction generated by NRG to DMRG-like sweeping. This will somewhat improve the quality of the ground state, but above all, the truncation procedure for high energies (short chains) will learn about truncation at low energies and vice versa. As opposed to NRG, there is now a feedback between energy scales. In that sense, NRG for an impurity problem is a similar conceptual step as the warm-up procedure infinite-system DMRG provides for variational finite-system DMRG. 

For logarithmic discretization, energy scale separation is big enough that this effect is minor and for a simple single impurity problem with a focus on the Abrikosov-Kondo-Suhl resonance the ultimate improvement is very limited, as NRG is geared to describe this feature optimally. The essential point is that energy scale separation can now be abandoned altogether due to feedback, hence also logarithmic discretization, and we may choose a more fine-grained resolution of the energy band wherever it is physically suitable. This could find a variety of applications.

\begin{figure}
\centering\includegraphics[height=200pt]{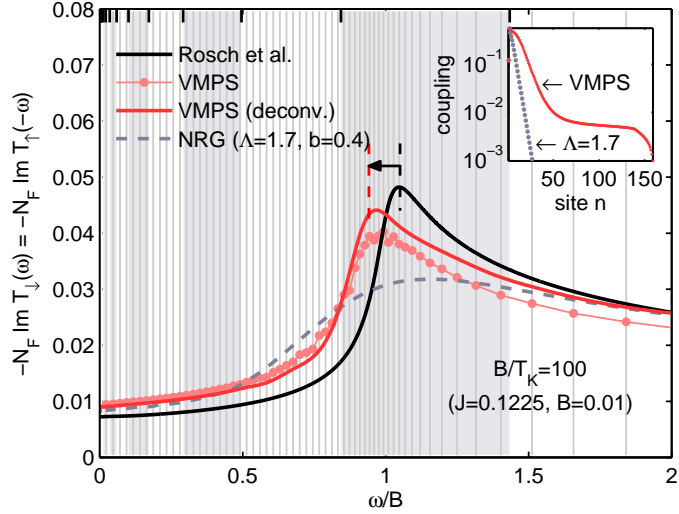}
\caption{Impurity spectral function for a Kondo model in an external field. Thin vertical lines indicate the energy intervals: below a certain energy, logarithmic discretization is turned linear. Variational MPS calculations (with and without deconvolution) reveal a peak missed by NRG, where the peak position is in very good agreement with a calculation by Rosch {\em et al.} \cite{Rosch03} combined with a perturbatively found peak shift. Taken from \cite{Weichselbaum09}.}
\label{fig:NRGDMRG}
\end{figure}

In one application, variational calculus over MPS was applied to an impurity problem in an external field. The external field leads to a splitting of the peak into two spin-dependent ones, shifted above and below the Fermi energy. In Figure \ref{fig:NRGDMRG} we consider one of these peaks, using three techniques, NRG, an analytical approach\cite{Rosch03,Garst05}, and variational MPS (DMRG) calculus. NRG due to logarithmic discretization focuses on $\epsilon_F$ and does not see the field-dependent peak at all. Relaxing logarithmic discretization and providing sufficiently fine energy intervals around the expected peak positions away from $\epsilon_F$ the shifted resonance can be resolved clearly and even in very good agreement with analytics.

A second interesting application of this could be to replace NRG as an impurity solver in the context of the dynamical mean-field theory (DMFT) \cite{Georges96,Nishimoto04,Garcia04,Raas04,Raas05}. In that case, information beyond the metallic resonance at the Fermi energy is required such that improving spectral resolution on other energy scales would be highly desirable. 

As the semi-infinite chain is non-interacting but on the first site, one can think about unfolding it into an infinite chain of spin-$1/2$, with the impurity at the center and the presence or absence of spin-up or spin-down fermions corresponding to the 2 spin states, the left half of the chain corresponding to the spin-up fermions and the right half to the spin-down fermions \cite{Raas04}. Similar energies are now no longer grouped together, but in a DMRG-like approach this does not matter anymore! The intuition that spins that interact only through the central impurity might be essentially unentangled is corroborated by actual calculations. This is important as this means we will not pay a strong price by increased matrix dimensions. On the contrary: if in the NRG approach we are essentially looking at two uncoupled spin chains parallel to each other, this means that the corresponding MPS has dimension $O(D^2)$ if the spin chain has dimension $D$. We can therefore expect that a NRG calculation with state number $D$ can be replaced by a faster DMRG calculation with a state number $O(\sqrt{D})$.  

Beyond this speedup, unfolding can of course also be applied if the impurity couples to multiple bands, where NRG becomes exponentially complex\cite{Weichselbaum09}. The central site, of course, remains the same, and its numerical treatment can become extremely costly, such that new strategies have to be designed for that site. Much work remains to be done here, but first interesting follow-ups on these ideas have been made\cite{Saberi08,Holzner10}.

\section{Infinite-size algorithms}
\label{sec:infinite}

\subsection{Reformulation of infinite-system DMRG in MPS language}

After the extensive discussion of finite-system algorithms, let us now reformulate infinite-system DMRG entirely in MPS language. It is convenient to label local states a bit differently to account for the iterative insertion of sites; we call the states $\ket{\sigma_1^A} \ket{\sigma_2^A} \ldots \ket{\sigma_\ell^A}\ket{\sigma_\ell^B} \ldots \ket{\sigma_2^B}\ket{\sigma_1^B}$. Moreover, it will be very useful to give two labels to the matrices $A$ and $B$, 
because the link between the matrices and the site on which they were generated will disappear.

\begin{figure}
\centering\includegraphics[scale=0.7]{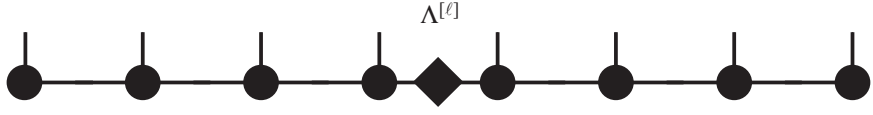}
\caption{MPS structure generated by infinite-system DMRG at step $\ell$: a string of left-normalized $A$, a string of right-normalized $B$, joined by a diagonal singular value matrix $\Lambda^{[\ell]}$. Note that structurally the central unit does not repeat.}
\label{fig:infiniteDMRG_AB}
\end{figure}

Starting from blocks of size 0 (i.e.\ a chain of 2 sites), the ground state wavefunction is $\ket{\psi_1}=\sum_{\sigma_1^A \sigma_1^B} \Psi^{\sigma_1^A \sigma_1^B} \ket{\sigma_1^A} \ket{\sigma_1^B}$. Reading $\Psi^{\sigma_1^A\sigma_1^B}$ as matrix $(\Psi^1)_{\sigma_1^A,\sigma_1^B}$, it is singular-value decomposed as  $\Psi^1=U_1 \Lambda^{[1]} V^\dagger_1$. From this we read off
\begin{equation}
A^{[1]\sigma_1^A}_{1,a_1} = (U_1)_{\sigma_1^A,a_1} \quad\quad B^{[1]\sigma_1^B}_{a_1,1} = (V^\dagger_1)_{a_1,\sigma_1^B} .
\end{equation} 
$A$ and $B$ inherit left and right-normalization properties from $U$ and $V^\dagger$, and the state takes the form
\begin{equation}
\ket{\psi_1} = \sum_{\sigma_1^A \sigma_1^B} A^{[1]\sigma_1^A} \Lambda^{[1]} B^{[1]\sigma_1^B} \ket{\sigma_1^A \sigma_1^B} .
\end{equation}
If we now insert two sites, and minimize the energy with respect to $\hat{H}$, we obtain
\begin{equation}
\ket{\psi_2} = \sum_{\sigma_1^A \sigma_2^A \sigma_2^B \sigma_1^B} A^{[1]\sigma_1^A} 
\Psi^{\sigma_2^A \sigma_2^B} 
B^{[1]\sigma_1^B} \ket{\sigma_1^A \sigma_2^A \sigma_2^B \sigma_1^B},
\end{equation}
where each $\Psi^{\sigma_2^A \sigma_2^B}$ is a matrix with dimensions to match those of $A$ and $B$, implicit matrix multiplications $A\Psi B$ assumed. Reshaping this set of $\Psi$-matrices into one,
\begin{equation}
(\Psi_2)_{(a_1^A \sigma_2^A),(a_1^B \sigma_2^B)} =  (\Psi^{\sigma_2^A \sigma_2^B})_{a_1^A,a_1^B},
\end{equation}
SVD gives $\Psi_2 = U_2 \Lambda^{[2]} V^\dagger_2$, from which we can form
\begin{equation}
A^{[2]\sigma_2^A}_{a_1^A,a_2^A} = U_{(a_1^A \sigma_2^A), a_2^A} \quad\quad  
B^{[2]\sigma_2^B}_{a_2^B,a_1^B} = V^\dagger_{(a_1^B \sigma_2^B), a_2^B}
\end{equation}
such that 
\begin{equation}
\ket{\psi_2} = \sum_{\sigma_1^A \sigma_2^A \sigma_2^B \sigma_1^B} 
A^{[1]\sigma_1^A}A^{[2]\sigma_2^A} \Lambda^{[2]} B^{[2]\sigma_2^B}B^{[1]\sigma_1^B} 
\ket{\sigma_1^A \sigma_2^A \sigma_2^B \sigma_1^B}.
\end{equation}
At the $\ell$th step, the wavefunction will read
\begin{equation}
\ket{\psi_\ell} = \sum_{\sigma_1^A \ldots \sigma_\ell^A \sigma_\ell^B \ldots \sigma_1^B} 
A^{[1]\sigma_1^A} \ldots A^{[\ell]\sigma_\ell^A} \Lambda^{[\ell]} B^{[\ell]\sigma_\ell^B} \ldots B^{[1]\sigma_1^B}  
\ket{\sigma_1^A \ldots \sigma_\ell^A \sigma_\ell^B \ldots \sigma_1^B}
\end{equation}
and look like in Fig.~\ref{fig:infiniteDMRG_AB}.

Of course, at each step we discard the smallest singular values and their associated singular vectors once matrix dimensions exceed $D$, which is nothing but the density-matrix based truncation in the original formulation.
At each step (new chain length) we can write down $\hat{H}$ for that length as an MPO and do the energy minimization. Other operators find similar representations as in the finite-size case. 

\begin{figure}
\centering\includegraphics[scale=0.7]{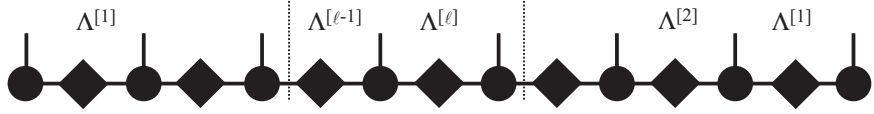}
\caption{MPS structure generated by infinite-system DMRG at step $\ell$ in the $\Gamma\Lambda$-notation: $\Gamma$ and $\Lambda$ matrices alternate, and a possible identification of a (repetitive) two-site building block is given.}
\label{fig:infiniteDMRG_GL}
\end{figure}

Let me briefly go through the reformulation of this algorithm in the $\Gamma\Lambda$-notation. In the first step we simply rename $A$ and $B$ into $\Gamma$, in line with the translation of boundary sites in the finite-system case:
\begin{equation}
\ket{\psi_1} =  \sum_{\sigma_1^A \sigma_1^B} \Gamma^{[1]\sigma_1^A} \Lambda^{[1]} \Gamma^{[1]\sigma_1^B} \ket{\sigma_1^A \sigma_1^B} .
\end{equation}
We then minimize $\Psi^{\sigma_2^A \sigma_2^B}$ in 
\begin{equation}
\ket{\psi_2} = \sum_{\sigma_1^A \sigma_2^A \sigma_2^B \sigma_1^B} \Gamma^{[1]\sigma_1^A} 
\Psi^{\sigma_2^A \sigma_2^B} 
\Gamma^{[1]\sigma_1^B} \ket{\sigma_1^A \sigma_2^A \sigma_2^B \sigma_1^B},
\end{equation}
and decompose it -- as before -- into $A^{[2]\sigma_2^A} \Lambda^{[2]} B^{[2]\sigma_2^B}$. Now we define (and due to the labelling, there is a slight change for the $B$-matrices compared to the finite-system setup)
\begin{equation}
\Lambda^{[1]}_a \Gamma^{\sigma_2^A}_{ab} = A^{[2]\sigma_2^A}_{ab} \quad\quad
\Gamma^{\sigma_2^B}_{ab} \Lambda^{[1]}_b = B^{[2]\sigma_2^B}_{ab}
\end{equation}
to arrive at
\begin{equation}
\ket{\psi_2} =  \sum_{\sigma_1^A \sigma_2^A \sigma_1^B \sigma_2^B} \Gamma^{[1]\sigma_1^A} \Lambda^{[1]} \Gamma^{[2]\sigma_2^A} \Lambda^{[2]} \Gamma^{[2]\sigma_2^B} \Lambda^{[1]} \Gamma^{[1]\sigma_1^B}   \ket{\sigma_1^A \sigma_2^A \sigma_2^B \sigma_1^B} ,
\end{equation}
as represented in Fig.~\ref{fig:infiniteDMRG_GL}. 

We can now ask two questions: (i) in DMRG, finding the ground state by an iterative solver like Lanczos is the most time-consuming part. Can we find a speed-up by providing a good initial guess? In finite-system DMRG the MPS formulation automatically yielded White's prediction method, whereas attempts at speeding up infinite-system DMRG have been made in the past, meeting with mixed success\cite{Schollwoeck98,Sun02}. (ii) Can we use the information at the chain center to build a translationally invariant state (up to period 2) in the thermodynamic limit, find its ground state or evolve it in time?

The answer is yes to both questions, and builds on the identification of a two-site repetititve structure in the states.  As opposed to the $A,B$-notation, where the central unit does not repeat itself even in the thermodynamic limit, it is very easy to read off a two-site repeat unit in the $\Gamma\Lambda$-notation, given by
\begin{equation}
\Lambda^{[\ell-1]} \Gamma^{[\ell]\sigma_\ell^A} \Lambda^{[\ell]} \Gamma^{[\ell]\sigma_\ell^B} .
\end{equation}
Using the translation rules it can be translated into the $A$, $B$-language:
\begin{equation}
A^{[\ell]\sigma_\ell^A} \Lambda^{[\ell]} B^{[\ell]\sigma_\ell^B} (\Lambda^{[\ell-1]})^{-1} .
\end{equation}
This result can also be obtained directly from the $A$, $B$ notation, but the argument is more complicated than in the $\Gamma\Lambda$ notation. It is of course to be understood that repeating these state fragments does not generate the state they were taken from; the claim is just that in the thermodynamic limit $\ell\rightarrow\infty$, when all sites are created equal, this repetition can come close. In any case, they are an educated guess about what the state will look like! 

We will now put this state fragment to multiple use, first on finite systems generated by infinite-system DMRG and then on thermodynamic limit states, both in the context of ground state searches and time evolutions. In the former case, it will provide a highly efficient guess for the next quantum state; the evaluation of observables on this state proceed exactly as in the other finite systems.
In the second case, both ground state and time evolution algorithms can be formulated (iDMRG and iTEBD), which however necessitate both an (identical) analysis of the issue of orthonormality of states in the thermodynamic limit.

\subsection{State prediction in infinite-system DMRG} 
The identification of the ``unit cell'' of the state allows us to define a good initial guess for infinite-system DMRG\cite{McCulloch08}, which avoids all the problems encountered by previous authors and leads to a dramatic speed-up even for small chains, where the underlying assumption that the chain center is representative of the physics of the thermodynamic limit state is certainly wrong: in order to grow the chain, we simply insert one unit cell, even though for small chains the idea that the state is just a repetition of these unit cells is not well verified -- but even then so much better than a random guess. Starting from
\begin{equation}
\ket{\psi_{\ell}} =  \sum_{\fat{\sigma}} A^{[1]\sigma_1^A} \ldots A^{[\ell-1]\sigma_{\ell-1}^A} (A^{[\ell]\sigma_\ell^A} \Lambda^{[\ell]} B^{[\ell]\sigma_\ell^B} [\Lambda^{[\ell-1]}]^{-1}) \Lambda^{[\ell-1]} B^{[\ell-1]\sigma_{\ell-1}^B} \ldots B^{[1]\sigma_1^B} \ket{\fat{\sigma}},
\end{equation} 
where the repeat unit has been bracketed out, the guess will then read
\begin{eqnarray}
\ket{\psi_{\ell+1}^{{\rm guess}}} &=& \sum_{\fat{\sigma}} A^{[1]\sigma_1^A} \ldots  A^{[\ell-1]\sigma_{\ell-1}^A} \times \nonumber \\
& & ( A^{[\ell]\sigma_\ell^A} \Lambda^{[\ell]} B^{[\ell]\sigma_{\ell+1}^A} [\Lambda^{[\ell-1]}]^{-1} ) (A^{[\ell]\sigma_{\ell+1}^B} \Lambda^{[\ell]} B^{[\ell]\sigma_\ell^B}  
[\Lambda^{[\ell-1]}]^{-1} ) \times \nonumber \\
& &  \Lambda^{[\ell-1]} B^{[\ell-1]\sigma_{\ell-1}^B} \ldots B^{[1]\sigma_1^B} \ket{\fat{\sigma}} 
\end{eqnarray}
or, multiplying out, 
\begin{equation}
\ket{\psi_{\ell+1}^{{\rm guess}}} = \sum_{\fat{\sigma}} A^{[1]\sigma_1^A} \ldots A^{[\ell]\sigma_\ell^A} \Lambda^{[\ell]} B^{[\ell]\sigma_{\ell+1}^A} [\Lambda^{[\ell-1]}]^{-1} A^{[\ell]\sigma_{\ell+1}^B} \Lambda^{[\ell]} B^{[\ell]\sigma_\ell^B} B^{[\ell-1]\sigma_{\ell-1}^B} \ldots B^{[1]\sigma_1^B} \ket{\fat{\sigma}} .
\end{equation}
In this ansatz, we can now identify a guess for $\Psi^{\sigma_{\ell+1}^A\sigma_{\ell+1}^B}$ as
\begin{equation}
\Psi^{\sigma_{\ell+1}^A\sigma_{\ell+1}^B}_{{\rm guess}} =
\Lambda^{[\ell]} B^{[\ell]\sigma_{\ell+1}^A} [\Lambda^{[\ell-1]}]^{-1} A^{[\ell]\sigma_{\ell+1}^B} \Lambda^{[\ell]} .
\label{eq:guess2sites}
\end{equation}
From this ansatz, we can then iteratively find the $\Psi^{\sigma_{\ell+1}^A\sigma_{\ell+1}^B}$ that minimizes the energy in the infinite-system DMRG framework, generating from it $A^{[\ell+1]\sigma_{\ell+1}^A}$, $\Lambda^{[\ell+1]}$, and $B^{[\ell+1]\sigma_{\ell+1}^B}$.

\begin{figure}
\centering\includegraphics[scale=0.7]{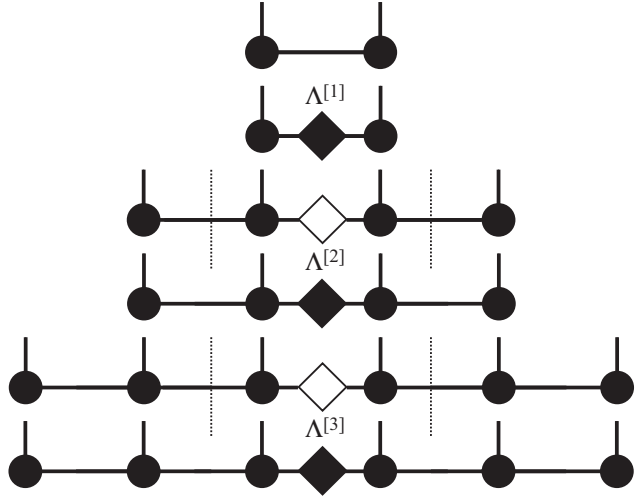}
\caption{Summarizing representation of the infinite system algorithm including prediction. At each step, two more sites are introduced, with an ansatz (white) calculated from the last iteration, leading to a new singular value decomposition (black).}
\label{fig:infinitebuildup}
\end{figure}

Alternatively, the ansatz can be brought into in a more elegant form. At the moment, $B$-matrices show up on the A-side of the lattice and vice versa. But we can exploit our ability to canonize MPS, and canonize $\Lambda^{[\ell]} B^{[\ell]\sigma_{\ell+1}^A}$ by SVD to
$A^{[\ell+1]\sigma_{\ell+1}^A} \Lambda_R^{[\ell]}$, where $A$ is derived from $U$ and $\Lambda$ from $DV^\dagger$ in the way described before ($\Lambda_a B^\sigma_{ab} = M_{a\sigma,b} = \sum_k U_{a\sigma,k} D_k (V^\dagger)_{k,b} = A^\sigma_{ak} \Lambda_{kb}$). Similarly, we do a canonization from the right on  $A^{[\ell]\sigma_{\ell+1}^B} \Lambda^{[\ell]}$ to obtain $\Lambda_L^{[\ell]} B^{[\ell+1]\sigma_{\ell+1}^B}$, where $B$ is from $V^\dagger$. Then we have an ansatz
\begin{equation}
\ket{\psi_{\ell+1}^{{\rm guess}}} = \sum_{\fat{\sigma}} A^{[1]\sigma_1} \ldots  A^{[\ell+1]\sigma_{\ell+1}} \Lambda^{[\ell+1]}_{{\rm guess}} B^{[\ell+1]\sigma_{\ell+1}}  \ldots B^{[1]\sigma_1} \ket{\fat{\sigma}} ,
\end{equation}
where 
\begin{equation}
\Lambda^{[\ell+1]}_{{\rm guess}} =  \Lambda_R^{[\ell]} [\Lambda^{[\ell-1]}]^{-1} \Lambda^{[\ell]}_L .
\end{equation}
From this ansatz, we can then iteratively find the $\Lambda^{[\ell+1]}$ that minimizes the energy, slightly modifying the minimization part of variational MPS for a single site. In general, the result will not have the diagonal form resulting from an SVD, because $\Lambda_R$ and $\Lambda_L$ are not diagonal to begin with. But an SVD on it yields two unitaries that can be absorbed into the neighbouring $A$ and $B$ without affecting their normalization properties, such that the final $\Lambda^{[\ell+1]}$ is diagonal. In this form, the algorithm can be represented as in Fig.~\ref{fig:infinitebuildup}.

As shown by McCulloch\cite{McCulloch08}, this prediction leads to a dramatic speedup of infinite-system DMRG which complements nicely prediction algorithms of finite-system DMRG: the overlap between the predicted and calculated state often approaches unity up to $10^{-10}$ or so!

\subsection{iDMRG: variational ground state search in the thermodynamic limit}
Using the ideas of the preceding sections, it is very simple now to turn infinite-system DMRG into a performing and precise algorithm, called {\em iDMRG}, referring to the thermodynamic limit version of DMRG:
\begin{itemize}
\item Set up an infinite-system DMRG algorithm.
\item Add the prediction procedure of the last section to the minimization algorithm.
\item Run the modified algorithm until convergence is achieved. Convergence of the wave function
to the thermodynamic limit can be judged by considering the relationship Eq.~(\ref{eq:rhoarecursion})
\begin{equation}
\sum_{\sigma_\ell} A^{[\ell]\sigma_\ell} \rho_A^{[\ell]} A^{[\ell]\sigma_\ell\dagger} = \rho_A^{[\ell-1]} ,
\end{equation}
where $\rho_A^{[\ell]} = \Lambda^{[\ell]}\Lambda^{[\ell]\dagger}$ and $\rho_A^{[\ell-1]} = \Lambda^{[\ell-1]}\Lambda^{[\ell-1]\dagger}$ are the reduced density operators of the left half of the system. Note that while this relationship holds always between reduced density operators in the same finite system, here they originate from systems of two different lengths $2(\ell-1)$ and $2\ell$, such that this relationship is expected to hold only as a fixed point relationship for $\ell\rightarrow\infty$. Following the same argument as for generating the ansatz for the larger system, we may transform $A^{[\ell]\sigma_\ell}\Lambda^{[\ell]}$ to $\Lambda^{[\ell]}_L B^{[\ell+1]\sigma_\ell}$. Then the left-hand side of the fixed point relationship simplifies, using right normalization, to $\Lambda^{[\ell]}_L\Lambda^{[\ell]\dagger}_L \equiv \hat{\rho}^{[\ell]}_L$, and it becomes $\hat{\rho}^{[\ell]}_L = \rho_A^{[\ell-1]}$. If this relationship holds to high accuracy, the thermodynamic fixed point has been reached.  
One way of measuring the closeness of the two density operators is given by the fidelity \cite{McCulloch08}
\begin{equation}
F(\hat{\rho}^{[\ell]}_L, \hat{\rho}_A^{[\ell-1]}) = \tr \sqrt{\sqrt{\hat{\rho}^{[\ell]}_L}\hat{\rho}_A^{[\ell-1]} \sqrt{\hat{\rho}^{[\ell]}_L}} .
\end{equation}
Inserting the definitions and using cyclicity properties of the trace, one can show that $F(\hat{\rho}^{[\ell]}_L, \hat{\rho}^{[\ell-1]}_A) = \sum_i s_i$, where $s_i$ are the singular values of  $\Lambda^{[\ell]}_L \Lambda^{[\ell-1]\dagger}$. 
\end{itemize}
Of course the algorithm can be stopped at any time, but then we have a finite system result which definitely can be improved by using finite-system DMRG. The convergence criterion given really gives us access to the thermodynamic limit state, which we might write down formally as
\begin{equation}
\ket{\psi} = \sum_{\fat{\sigma}} \ldots A^{[\ell]\sigma_i}\Lambda^{[\ell]} B^{[\ell]\sigma_{i+1}} [\Lambda^{[\ell-1]}]^{-1}A^{[\ell]\sigma_{i+2}}\Lambda^{[\ell]} B^{[\ell]\sigma_{i+3}} [\Lambda^{[\ell-1]}]^{-1}A^{[\ell]\sigma_{i+4}}\Lambda^{[\ell]} B^{[\ell]\sigma_{i+5}} [\Lambda^{[\ell-1]}]^{-1} \ldots 
 \ket{\fat{\sigma}} ,
\end{equation}
where we take $\ell$ to be the iteration step when the convergence criterion is met. The question is now how to evaluate expectation values. Obviously, we cannot write down a finite network contraction as before; it will be of infinite size and therefore cannot be contracted naively. A contraction can only work if we can reduce the number to a finite number of contractions. For finite-system networks we saw that left- and right-orthonormality allow to eliminate most contractions: For observables on sites $i$ and $j$, one only had to consider the contractions on and between these two sites. There is however no reason why the thermodynamic limit state should meet normalization criteria; in fact, usually it does not. We therefore need an orthonormalization procedure. After that, expectation values can be evaluated as for a finite lattice with left- and right-orthonormalization. Because this procedure is also important for the next algorithm and conceptually a bit more advanced, I postpone it to an extra section. 

At this point it should be mentioned that iDMRG can be related to earlier algorithmic approaches under the name of PWFRG (product wave function renormalization group) \cite{NishinoOkunishi95,Hieida97,Ueda06} which already contain part of the above ideas; iDMRG takes them to their natural completion \cite{Ueda08,Ueda10}.

\subsection{iTEBD: time evolution in the thermodynamic limit}
In this section, I switch to the $\Gamma\Lambda$ notation, although the formulae can be easily translated into the $A,B$-formulation. Using our state fragment $\Lambda^{[\ell-1]} \Gamma^{[\ell]\sigma_\ell^A} \Lambda^{[\ell]} \Gamma^{[\ell]\sigma_\ell^B}$, we can set up an infinite chain
\begin{equation}
\ket{\psi} = \sum_{\fat{\sigma}} \ldots (\Lambda^A \Gamma^{A \sigma_i} \Lambda^B \Gamma^{B \sigma_{i+1}})(\Lambda^A \Gamma^{A \sigma_{i+2}} \Lambda^B \Gamma^{B \sigma_{i+3}}) \ldots \ket{\fat{\sigma}} ,
\end{equation}
just like in the previous section, where $\Gamma^{A\sigma}= \Gamma^{[\ell]\sigma_\ell^A}$,
$\Gamma^{B\sigma}= \Gamma^{[\ell]\sigma_\ell^B}$, $\Lambda^A=\Lambda^{[\ell-1]}$ and 
$\Lambda^B=\Lambda^{[\ell]}$. The fragment may be the result of a converged ground state calculation or from some simple starting state that one can construct exactly. 

We can now write down a time evolution in the Trotter form by applying an infinitesimal time step to all odd bonds (which I will refer to as AB) and then on all even bonds (which I will refer to as BA). The bond evolution operators will be exactly as in the tMPS/tDMRG/TEBD cases, I will refer to them as $U_{AB} = \sum_{\sigma_A\sigma_B\sigma'_A\sigma'_B} U_{AB}^ {\sigma_A\sigma_B, \sigma'_A\sigma'_B} \ket{\sigma_A\sigma_B} \bra{\sigma'_A\sigma'_B}$ and similarly $U_{BA}$.

As we have already seen [cf.\ Eq.~(\ref{eq:guess2sites})], a full two-site fragment consists of a product of five matrices, 
$\Lambda^A \Gamma^{A \sigma_A} \Lambda^B \Gamma^{B \sigma_{B}} \Lambda^A$. Then time evolution on bond AB yields a set of matrices
\begin{equation}
M^{\sigma_A\sigma_{B}} = \sum_{\sigma'_A\sigma'_{B}} U_{AB}^ {\sigma_A\sigma_{B}, \sigma'_A\sigma'_{B}}  \Lambda^A \Gamma^{A \sigma'_A} \Lambda^B \Gamma^{B \sigma'_{B}} \Lambda^A .
\end{equation}
Upon the by now standard reshaping we obtain by SVD 
\begin{equation}
M^{\sigma_A\sigma_{B}}=A^{\sigma_A} \Lambda^B B^{\sigma_{B}} ,
\end{equation}
where the new $\Lambda^B$ (and correspondingly $A^{\sigma_A}$ and $B^{\sigma_{B}}$) are truncated as in tDMRG or TEBD, to replace the old one. Using $\Lambda^A$ (still from the last iteration), we can define new $\Gamma^{A \sigma_A}$ and $\Gamma^{B \sigma_{B}} $ (via $A^{\sigma_A} = \Lambda^A \Gamma^{A \sigma_A}$ and
$B^{\sigma_{B}} = \Gamma^{B \sigma_{B}} \Lambda^A$). This defines a new ``unit cell''. 

If we write it down and attach another one, we can read off the bond BA in the center of the two AB unit cells as
$ \Lambda^B \Gamma^{B \sigma_{B}} \Lambda^A \Gamma^{A \sigma_{A}} \Lambda^B$, for which time evolution gives
\begin{equation}
N^{\sigma_{B}\sigma_{A}} = \sum_{\sigma'_{B}\sigma'_{A}} U_{BA}^ {\sigma_{B}\sigma_{A}, \sigma'_{B}\sigma'_{A}}  \Lambda^B \Gamma^{B \sigma'_{B}} \Lambda^A \Gamma^{A \sigma'_{A}} \Lambda^B .
\end{equation}
Reshaping and SVD gives us
\begin{equation}
N^{\sigma_{B}\sigma_{A}}=A^{\sigma_{B}} \Lambda^A B^{\sigma_{A}} ,
\end{equation}
where again $\Lambda^A$ (and correspondingly the other matrices) are truncated and replace the old ones. Using $\Lambda^B$ (still from the last iteration), we can define new $\Gamma^{B \sigma_{B}}$ and $\Gamma^{A \sigma_{A}} $ (via $B^{\sigma_{A}} =   \Gamma^{A \sigma_{A}} \Lambda^B$ and $A^{\sigma_{B}} =\Lambda^B \Gamma^{B \sigma_{B}}$). The problematic division by small singular values can be avoided by the simple modification already discussed for TEBD \cite{Hastings09}.

By applying sequences of infinitesimal bond evolution operators we can therefore set up a real or imaginary time evolution for the thermodynamic limit. This algorithm is referred to as {\em iTEBD}, because it provides the infinite-size generalization of TEBD \cite{Vidal07}.

Again, the question of orthonormality arises \cite{Orus08}. Let us assume that the initial state was meeting orthonormality criteria. A pure real-time evolution generates a sequence of unitaries acting on the state, which preserves orthonormality properties. But the inevitable truncation after each time step spoils this property, even though truncation may only be small. To turn this method into a viable algorithm, we have to address the issue of orthogonalization in the thermodynamic limit, as for iDMRG, after each step.

Let me mention here that McCulloch\cite{McCulloch08} has shown that iDMRG can be turned into iTEBD  by replacing the minimization on the central bond by a time-evolution on the central bond, with some conceptual advantages over the original iTEBD algorithm.

Let me conclude this section by a few words on extrapolation. In finite-system DMRG (or MPS), the recipe was to extrapolate for each finite length $L$ results in $D$ to maximize accuracy, and then to extrapolate these results in $L$ to the thermodynamic limit. Here, we are working directly in the thermodynamic limit (assuming that iDMRG has been taken to convergence), and the extrapolation in $D$ remains. Interestingly, at criticality, this extrapolation in $D$ has profound and useful connections to entanglement entropy scaling and critical exponents \cite{Tagliacozzo08}. Effectively, the finite matrix dimension introduces a finite correlation length into the critical system, not unlike the finite-system DMRG case, where the length scale on which the MPS-typical superposition of exponentials mimicks a power-law properly also scales with some power of $D$ \cite{Andersson99}.

\subsection{Orthogonalization of thermodynamic limit states}

Within iDMRG on a finite system, $A$- and $B$-matrices retain left- and right-normalization; this implies that the left and right block states are orthogonal among each other, as shown previously. We will call a state with this property {\em orthogonal} in a slight abuse of conventional usage. As we have seen in the previous section, we may use a fragment 
\begin{equation}
A^{[\ell]\sigma_\ell^A}\Lambda^{[\ell]} B^{[\ell]\sigma_\ell^B} (\Lambda^{[\ell-1]})^{-1}
\end{equation}
that we can repeat to build up an infinitely long chain,
\begin{equation}
\ket{\psi} = \sum_{\fat{\sigma}} \ldots A^{\sigma_i}\Lambda^{[\ell]} B^{\sigma_{i+1}} (\Lambda^{[\ell-1]})^{-1}A^{\sigma_{i+2}}\Lambda^{[\ell]} B^{\sigma_{i+3}} (\Lambda^{[\ell-1]})^{-1}A^{\sigma_{i+4}}\Lambda^{[\ell]} B^{\sigma_{i+5}} (\Lambda^{[\ell-1]})^{-1} \ldots \ket{\fat{\sigma}} ,
\end{equation}
where I have simplified the notation of $A$, $B$. The problem with these states is that, for an arbitrary bipartition into two blocks, the states on the left and right blocks will in general {\em not be orthogonal}: if we transform $\Lambda^{[\ell]} B$ into $\tilde{A} \Lambda_R^{[\ell]}$ as described above, the chain will read
\begin{equation}
\ket{\psi} = \sum_{\fat{\sigma}} \ldots A^{\sigma_1}\tilde{A}^{\sigma_{2}} P
A^{\sigma_{3}} \tilde{A}^{\sigma_{4}} P A^{\sigma_{5}}\tilde{A}^{\sigma_{6}} P \ldots \ket{\fat{\sigma}} ,
\end{equation}
where $P= \Lambda_R^{[\ell]} (\Lambda^{[\ell-1]})^{-1}$. If we absorb $P$ into the $\tilde{A}$ to its left, $\tilde{A}P \rightarrow \overline{A}$, the normalization condition becomes
\begin{equation}
\sum_\sigma \overline{A}^{\sigma\dagger} \overline{A}^{\sigma} = P^\dagger \sum_\sigma \tilde{A}^{\sigma\dagger} \tilde{A}^\sigma P = P^\dagger P.
\end{equation}
In general, however, $P^\dagger P \neq I$. This is not only the case if $\ell$ is small and we are far from the infinite-system fixed point. It is also the case at the fixed point as long as the discarded state weight is finite, which is usually the case in DMRG calculations, even if it is very small.

\begin{figure}
\centering\includegraphics[scale=0.7]{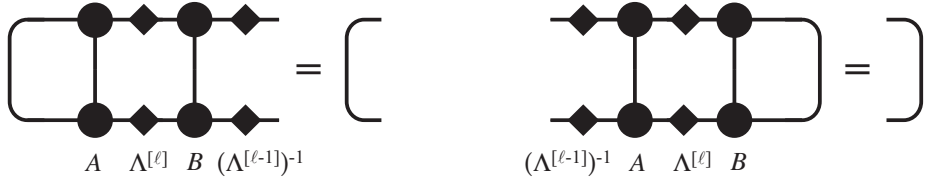}
\caption{If matrices are properly normalized, the thermodynamic limit ansatz generates both a left- and right-normalization condition. Note that the two transfer operators are defined on two differing two-site unit cells of the thermodynamic limit state.}
\label{fig:infiniteorthogonalization}
\end{figure}

As pointed out by Orus and Vidal\cite{Orus08} -- in the presentation I follow \cite{McCulloch08} --, a condition to detect orthonormality is to check whether the expectation value of the unit operator between two block states $\ket{a}$, $\ket{a'}$ is $\delta_{a,a'}$ (see Fig.~\ref{fig:infiniteorthogonalization}). Let us consider an expectation value contraction as for a finite system and assume we have carried it out up to site $0$, coming from the left, $-\infty$. The result will be a matrix-like object $C_{a'a}$, corresponding to the open legs. Let us now consider the operation $E_L(C)$, which carries the contraction two steps further, i.e. over sites 1 and 2. This {\em transfer operator} reads [cf.\ Sec.~\ref{subsubsec:transferoperator}]
\begin{equation}
E_L(C) = \sum_{\sigma_1\sigma_2} P^\dagger \tilde{A}^{\sigma_2\dagger} A^{\sigma_1\dagger} C A^{\sigma_1} \tilde{A}^{\sigma_2} P.
\end{equation}
For an orthonormal state, we want that $E_L(I) = I$, which is just the expectation value matrix the unit operator produces for orthonormal block states. What we get, however, is, using the left-normalization condition, $E_L(I)=P^\dagger P$. As the system extends to infinity, $E_L(I) = I$ must be associated with the largest eigenvalue; normalizability of the entire state implies that the largest eigenvalue must be 1. The ``quadratic'' form of $E_L$ implies that the associated eigenmatrix $V_L$ is hermitian and non-negative. An eigenvalue or singular value decomposition allows to decompose $V_L = X^\dagger X$, where $X$ is invertible. We can insert $X^{-1}X$ after each $P$, such that the unit cell becomes $X A^{\sigma_1} \tilde{A}^{\sigma_2} P X^{-1}$ and the new transfer operator reads
\begin{equation}
E'_L(C) = \sum_{\sigma_1\sigma_2} X^{-1\dagger}P^\dagger \tilde{A}^{\sigma_2\dagger} A^{\sigma_1\dagger} X^\dagger C X A^{\sigma_1} \tilde{A}^{\sigma_2} P X^{-1}.
\end{equation}
Then 
\begin{equation}
E'_L(I) = X^{-1\dagger} V_L X^{-1} = I
\end{equation}
from the eigenmatrix properties of $V_L$ with respect to $E_L$.
(If the largest eigenvalue of $E_L$ happens not to be 1, $X$ must be suitably scaled.)

Inserting the definition of $P$ in $X A^{\sigma_1}\tilde{A}^{\sigma_2}P X^{-1}$, undoing the transformation to $\tilde{A}$, and transforming $A^{\sigma_1} \Lambda^{[\ell]} \rightarrow \Lambda^{[\ell]}_L \tilde{B}^{\sigma_1}$, the unit cell becomes $X \Lambda^{[\ell]}_L \tilde{B}^{\sigma_1} B^{\sigma_2} (\Lambda^{[\ell-1]})^{-1} X^{-1}$. Shifting the starting point of the unit cell it becomes $(\Lambda^{[\ell-1]})^{-1} X^{-1} X \Lambda^{[\ell]}_L \tilde{B}^{\sigma_1} B^{\sigma_2} = Q \tilde{B}^{\sigma_1} B^{\sigma_2}$, where $Q = (\Lambda^{[\ell-1]})^{-1} X^{-1} X \Lambda_L^{[\ell]} = (\Lambda^{[\ell-1]})^{-1} \Lambda_L^{[\ell]}$, independent of $X$. Calculating a contraction from the right leads to a transfer operator
\begin{equation}
E_R(C) = \sum_{\sigma_1\sigma_2} Q \tilde{B}^{\sigma_1} B^{\sigma_2} C B^{\sigma_2\dagger} \tilde{B}^{\sigma_1\dagger} Q^\dagger .
\end{equation}
The same eigenvalue/eigenmatrix argument as before leads to the dominant eigenmatrix $V_R = YY^\dagger$, $Y$ invertible, and a unit cell $Y^{-1}Q \tilde{B}^{\sigma_1} B^{\sigma_2} Y$.
This in turn leads to $E'_R(C)$ with $E'_R(I)=I$. 

If we insert the definition of $Q$ into the unit cell, return from $\tilde{B}^{\sigma_1}$ to $A^{\sigma_1}$ and make $Q$ explicit, the unit cell reads 
$Y^{-1}  (\Lambda^{[\ell-1]})^{-1} X^{-1} X A^{\sigma_1} \Lambda^{[\ell]} B^{\sigma_2} Y$, shifting its origin we obtain
\begin{equation}
X A^{\sigma_1} \Lambda^{[\ell]} B^{\sigma_2} Y Y^{-1}  (\Lambda^{[\ell-1]})^{-1} X^{-1} ,
\end{equation}
which can be brought back to the original form of the unit cell by setting 
$A^{\sigma_1} \leftarrow XA^{\sigma_1}$, $B^{\sigma_2} \leftarrow B^{\sigma_2}Y$ and $\Lambda^{[\ell-1]} \leftarrow X\Lambda^{[\ell-1]}Y$, but now with {\em proper left- and right-normalization ensured.} 

More precisely, the new unit cell leads to $E_L(I)=I$ and $E_R(I)=I$. But note that $E_L$ and $E_R$ are constructed from slightly shifted unit cells, namely $A^{\sigma_1} \Lambda^{[\ell]} B^{\sigma_2} (\Lambda^{[\ell-1]})^{-1}$ for $E_L$ and $(\Lambda^{[\ell-1]})^{-1} A^{\sigma_1} \Lambda^{[\ell]} B^{\sigma_2}$ for $E_R$, as shown in the pictorial representation. We can lump together the first and second unit cells into left- and right-normalized two-site matrices $A^{\sigma_1\sigma_2}$ and $B^{\sigma_1\sigma_2}$. These can now be decomposed into left- and right-normalized matrices in the standard way, giving $A^{[1]\sigma_1}A^{[2]\sigma_2}$ and $B^{[1]\sigma_1}B^{[2]\sigma_2}$. Note that these matrices are of course different from those we had originally.

The thermodynamic limit state can now be written as 
\begin{equation}
\ket{\psi} = \sum_{\fat{\sigma}} \ldots A^{[1]\sigma_1} A^{[2]\sigma_2} A^{[1]\sigma_3} A^{[2]\sigma_4} \ldots \ket{\fat{\sigma}}
\end{equation}
or analogously using $B^{[1,2]}$, for left- and right-canonical form. Of particular interest for expectation values is the mixed-canonical form, with $A$-matrices on the left and $B$-matrices on the right. If we consider the two underlying unit cells, we see that at the boundary, both want to incorporate the same $(\Lambda^{[\ell-1]})^{-1}$ to generate $A$ and $B$-matrices. This problem can be solved by inserting the identity $I = \Lambda^{[\ell-1]} (\Lambda^{[\ell-1]})^{-1}$ after the problematic $(\Lambda^{[\ell-1]})^{-1}$. Then we can immediately write down a mixed-canonical form as 
\begin{equation}
\ket{\psi} = \sum_{\fat{\sigma}} \ldots A^{[1]\sigma_1} A^{[2]\sigma_2} A^{[1]\sigma_3} A^{[2]\sigma_4} \Lambda^{[\ell-1]} 
B^{[1]\sigma_5} B^{[2]\sigma_6} B^{[1]\sigma_7} B^{[2]\sigma_8}
\ldots \ket{\fat{\sigma}} .
\label{eq:infinitemixedcanonical}
\end{equation}
\subsubsection{Calculation of expectation values in the thermodynamic limit}
In finite systems, we have seen how an expectation value can be calculated by transferring an object $C^{[i]}$, starting as a dummy scalar $C^{[0]}=1$ from $C^{[i]}$ to $C^{[i+1]}$ by means of a transfer operator $E^{[i]}_O$, where $hat{O}$ is the locally acting operator in the expectation value structure (mostly the identity, except, say, two sites if we are looking at a two-point correlator). In the case of the identity operator, for left-normalized matrices, the transfer operator mapping from left to right maps the identity to the identity; similarly, the transfer operator mapping from right to left maps the identity to the identity if formed from right-normalized matrices. 

\begin{figure}
\centering\includegraphics[scale=0.7]{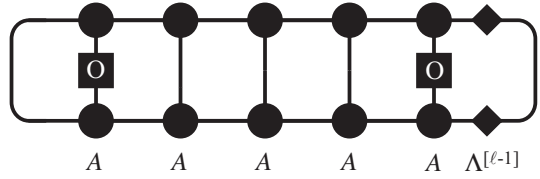}
\caption{Exploiting left- and right-normalization, the evaluation of an infinite contraction can be reduced to a finite contraction that can be dealt with as for any finite MPS.}
\label{fig:finitecontractionremainder}
\end{figure}

The same structure has been found in the last section for the thermodynamic limit state and its two two-site transfer operators $E_L$ and $E_R$. This allows a direct transfer of the old results. Assume we want to calculate $\bra{\psi} \hat{O}_1 \hat{O}_i \ket{\psi}$; then we bring the state into the mixed-canonical form of Eq.~(\ref{eq:infinitemixedcanonical}), with $A$-matrices up to site $i$ or $i+1$ (depending on the odd-even structure), then $\Lambda^{[\ell-1]} $, followed by $B$-matrices up to infinity. Contracting from the left over all $A$-matrices up to 0 and from the right all $B$-matrices, we obtain a remaining finite network as in Fig.~\ref{fig:finitecontractionremainder}.
This expectation value is then evaluated as in a finite network; if we use the $C^{[i]}$ and transfer operator notation, it starts from $C^{[0]}=I$. The difference is that at the end, $\Lambda^{[\ell-1]}$ shows up in the contraction and the final reduction to a scalar is done by the closing $\delta_{aa'}$ line (which can be read as a trace): assuming the last site is $i$, then the final expectation value is given in the last step as
\begin{equation}
\bra{\psi} O^1 \otimes \ldots O^i \ket{\psi} = \tr  \Lambda^{[\ell-1]\dagger} C^{[i]} \Lambda^{[\ell-1]} = 
\tr   \Lambda^{[\ell-1]} \Lambda^{[\ell-1]\dagger} C^{[i]} = \tr \rho_A C^{[i]} ,
\end{equation}
where we have used the relationship between $\Lambda$-matrices and reduced density operators in canonical representations.

This calculation can be reduced easily to the case of the overlap of two states, which is just the matrix element of the unit operator between them. Assume that both states are in translationally invariant form, e.g. by using left-normalized matrices $A^{[1]}$ and $A^{[2]}$ (and $\tilde{A}^{[1]}$, $\tilde{A}^{[2]}$ respectively). We now carry forward an infinite overlap calculation by two sites (say 1 and 2) towards the right using $E_L$: If the current overlap matrix is $C$, it is carried forward as
\begin{equation}
E_L (C) = \sum_{\sigma_1\sigma_2} \tilde{A}^{[2]\sigma_2\dagger} \tilde{A}^{[1]\sigma_1\dagger} C A^{[1[\sigma_1} A^{[2]\sigma_2} .
\end{equation}  
If we decompose $C$ in the eigenmatrices of $E_L$, in the thermodynamic limit only the largest eigenvalue contribution will survive. For an orthonormal state, for the overlap with itself, $C=I$ and $\lambda=1$ are the dominant eigenpair. A smaller $\lambda$ in the overlap of two states can be interpreted as an overlap per site, while of course the two states are orthogonal with respect to each other in the thermodynamic limit (overlap $\lim_{L\rightarrow\infty} \lambda^L = 0$). Such thermodynamic overlaps (or fidelities) per site can be used very nicely to detect quantum phase transitions by overlapping ground states for two Hamiltonians with slightly different parameters\cite{McCulloch08}.
 
\section{Conclusion: other topics to watch}
After moving through this long list of topics, focussing on the fundamental algorithmic building blocks, ground state searches, thermodynamic limit algorithms and a wealth of real and imaginary time methods at zero and finite temperature, the possibilities of DMRG and MPS-based algorithms are far from being exhausted. A few topics that I have not touched upon, but which I would like to mention briefly (again in a non-exhaustive list), are: transfer matrix DMRG methods (cf.\ the introductory section for references), DMRG and MPS with periodic boundary conditions, and as the most recent addition, MPS for continuous space\cite{Verstraete10}, which emerge as a beautiful generalization of coherent states and should allow for interesting applications in field theories. For {\em periodic boundary conditions} quite a lot of results already exist, so let me give just a brief overview. PBC have already been treated in the DMRG framework by introducing one long-ranged interaction between sites 1 and $L$ on an open-boundary chain (see e.g.\ \cite{Schmitteckert96,Rapsch99,Meden03a,Meden03b}; however, the scaling of accuracy was consistently found to be much worse than for open boundary conditions. The underlying reason is (roughly speaking) that on a ring the surface between A and B doubles, hence the entanglement; given the exponential relationship to the MPS dimension, this means that resources have to go up from $D$ to up to $D^2$, meaning that for similar accuracy, the algorithm needs the square of time (sometimes referred to as $D^6$-scaling, referring to the open boundary condition $D$). The physically adequate ansatz for MPS for periodic boundary conditions is given by Eq. (\ref{eq:MPSforPBC}); one needs roughly the same $D$ as for OBC, but rerunning the variational ground state search algorithm on it scales as $D^5$ (because the simplification of vectors instead of matrices on sites 1 and $L$ does not occur) \cite{VerstraetePorras04}. At the same time, the simplification of the generalized to a standard eigenvalue problem does not occur, which may lead to bad conditioning. A nice feature of the MPS representation for PBC is that one can generate eigenstates of momentum: For $k = n (2\pi/L)$ and a (non-translationally invariant) MPS $\ket{\psi} = \sum_{{\fat{\sigma}}} \tr (A^{[1]\sigma_1} \ldots A^{[L]\sigma_L}) \ket{\fat{\sigma}}$, the following state is a translationally invariant eigenstate of momentum $k$: \cite{Porras06}
\begin{equation}
\ket{\psi_k} = \sum_{n=0}^{L-1} \eul^{\imag k n}  \tr (A^{[1]\sigma_{1+n}} \ldots A^{[L]\sigma_{L+n}}) \ket{\fat{\sigma}} .
\end{equation}
Recently, interesting proposals to improve the $D^5$ scaling have been made \cite{Pippan10,Pirvu10}, and this is a field of ongoing interest. Reference \cite{VerstraeteMurg08} discusses this topic quite extensively.

I think one may conclude by saying that while the fundamental framework of MPS is by now very well established, and while DMRG has come of age as one of the most powerful numerical methods available for strongly correlated quantum systems, even in the well-established field of one-dimensional systems many of the algorithms presented will still allow further improvement, bringing new applications into our reach. It is in fact quite surprising that for quite a few of the methods presented (and also the others) very little is known about their detailed behaviour in real-world problems, analyzing which might give interesting further ideas. Also, the ratio between applications done and applications doable seems very favourable for future exciting research.

\section{Acknowledgments}
I would like to thank the Aspen Center for Physics, the International Center for Theoretical Physics in Trieste, Italy, and the Wissenschaftskolleg (Institute for Advanced Study), Berlin, where major parts of this work were carried out, for their hospitality. Their combination of stimulating discussions and room for withdrawal to intensive work are a physicist's dream. Thanks go of course to the many people I have discussed these subjects with and from whom I have learned a lot: Mari-Carmen Banuls, Thomas Barthel, Ignacio Cirac, Andrew Daley, Jan von Delft, Jens Eisert, Adrian Feiguin, Juanjo Garcia-Ripoll, Fabian Heidrich-Meisner, Corinna Kollath, Ian McCulloch, Valentin Murg, Volker Nebendahl, Tomotoshi Nishino, Reinhard Noack, Tobias Osborne, Lode Pollet, Matthias Troyer, Frank Verstraete, Guifre Vidal, Andreas Weichselbaum, Steve White and Tao Xiang, to name but a few. Special thanks go to Matt Hastings for pointing out an important paper of his and to Stefan Depenbrock for a very careful reading of an almost final version of this review (all remaining errors were added afterwards).

\end{document}